%% file: main.tex
\documentclass[journal]{IEEEtran}

\usepackage{docmute}
\usepackage{ifpdf}
\usepackage{cite}
\usepackage{amsthm}
\usepackage{amssymb}
\usepackage{booktabs}
\usepackage{url}
\usepackage{color}
\usepackage{multirow}
\usepackage{graphics}
\usepackage{siunitx}
\usepackage{amsmath,graphicx}
\usepackage{amsfonts}
\usepackage{algorithm, algpseudocode}
\usepackage[geometry]{ifsym}
\usepackage{array}
\usepackage{tabularx}

\def\x{{\mathbf x}}
\def\y{{\mathbf y}}
\def\z{{\mathbf z}}

\def\argmin{\mathop{\rm argmin}\limits}

\def\tilHr{\widetilde{\mathbf{h}}_{r}}
\def\tilHt{\widetilde{\mathbf{h}}_{t}}

\def\hatHr{\widehat{\mathbf{h}}_{r}}
\def\hatHt{\widehat{\mathbf{h}}_{t}}
\def\Hr{\mathbf{h}_{r}}
\def\Ht{\mathbf{h}_{t}}
\def\Lr{\mathbf{l}_{r}}
\def\Lt{\mathbf{l}_{t}}

\def\nh {\mathbf{n}_{h}}
\def\nl {\mathbf{n}_{l}}
\def\sh {\mathbf{s}_{h}}
\def\sl {\mathbf{s}_{l}}

\def\tilshr{\widetilde{\mathbf{s}}_{hr}}
\def\tilslr{\widetilde{\mathbf{s}}_{lr}}
\def\tilslt{\widetilde{\mathbf{s}}_{lt}}

\def\onenorm#1{\| #1 \|_1}
\def\twonorm#1{\| #1 \|_2}
\def\pnorm#1{\| #1 \|_p}
\def\onetwonorm#1{\| #1 \|_{1,2}}
\def\O#1{\mathcal{O}\left( #1 \right)}

\def \R{\mathbb{R}}
\def \D{\mathbf{D}}
\def \S{\mathbf{S}}
\def \B{\mathbf{B}}
\def \I{\mathbf{I}}

\def \z{\mathbf{z}}
\def \u{\mathbf{u}}
\def \v{\mathbf{v}}
\def \w{\mathbf{w}}

\def \prox{\mathrm{prox}}

\begin{document}
\bstctlcite{IEEEexample:BSTcontrol}
\title{Robust Spatiotemporal Fusion of Satellite Images: A Constrained Convex Optimization Approach}

\author{Ryosuke~Isono~\IEEEmembership{Student~Member,~IEEE,},~Kazuki~Naganuma,~\IEEEmembership{Student~Member,~IEEE,}
        Shunsuke~Ono,~\IEEEmembership{Member,~IEEE,}
\thanks{R. Isono is with the Department of Computer Science, Tokyo Institute of Technology, Yokohama, 226-8503, Japan (e-mail: isono.r.ac@m.titech.ac.jp).}
\thanks{K. Naganuma is with the Department of Computer Science, Tokyo Institute of Technology, Yokohama, 226-8503, Japan (e-mail: naganuma.k.aa@m.titech.ac.jp).}
\thanks{S. Ono is with the Department of Computer Science, Tokyo Institute of Technology, Yokohama, 226-8503, Japan (e-mail: ono@c.titech.ac.jp).}
\thanks{This work was supported in part by JST ACT-X Grant Number JPMJAX23CJ, JST PRESTO under Grant JPMJPR21C4, and JST AdCORP under Grant JPMJKB2307, in part by JSPS KAKENHI under Grant 22H03610, 22H00512, and 23H01415, and in part by Grant-in-Aid for JSPS Fellows under Grant 23KJ0912.}}

\markboth{IEEE TRANSACTIONS ON GEOSCIENCE AND REMOTE SENSING}%
{Shell \MakeLowercase{\textit{et al.}}: Bare Demo of IEEEtran.cls for Journals}

\maketitle

\begin{abstract}
\input{abstract.tex}
\end{abstract}



\IEEEpeerreviewmaketitle

\section{Introduction}
\label{sec:intro}
\input{introduction}

\section{Preliminaries}
\label{sec:preli}
\vspace{-1mm}
\input{preliminaries}

\section{Proposed Method}
\label{sec:proposed}
\input{proposedmethod}

\section{Experiments}
\label{sec:exp}
\input{Experiments/experiments}

\section{Conclusion}
\label{sec:conclusion}
\input{conclusion}

\ifCLASSOPTIONcaptionsoff
  \newpage
\fi

\input{main.bbl}

\begin{IEEEbiography}[{\includegraphics[width=1in,height=1.25in,clip,keepaspectratio]{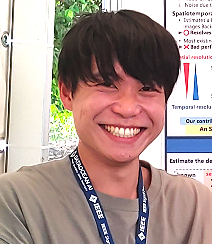}}]{Ryosuke Isono} (S’23) received a B.E. degree in Information and Computer Sciences in 2022 from the Osaka University. 

He is currently pursuing an M.E. degree at the Department of Computer Science in the Tokyo Institute of Technology.
\end{IEEEbiography}

\begin{IEEEbiography}[{\includegraphics[width=1in,height=1.25in,clip,keepaspectratio]{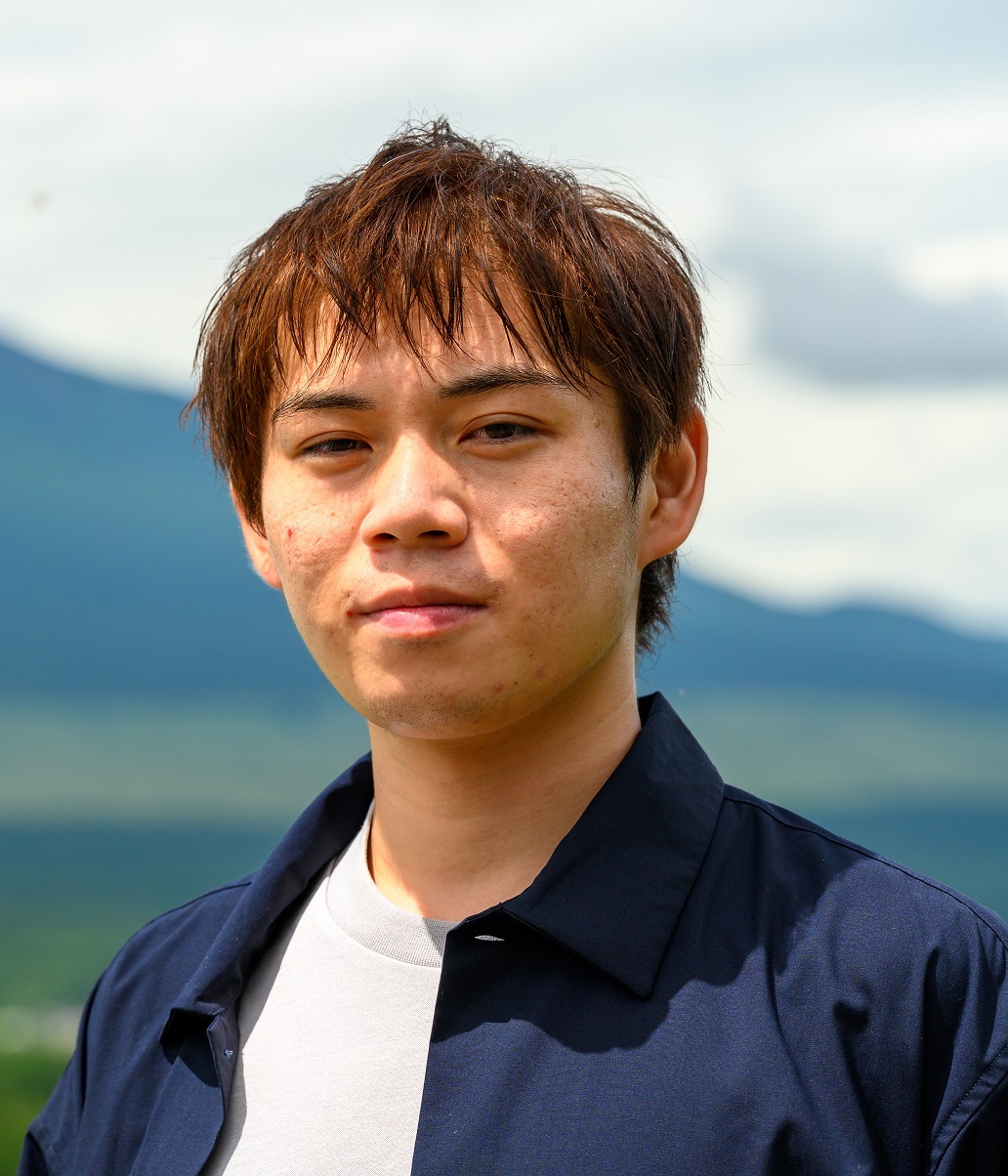}}]{Kazuki Naganuma}
(S’21) received a B.E. degree and M.E. degree in Information and Computer Sciences in 2020 from the Kanagawa Institute of Technology and from the Tokyo Institute of Technology, respectively.
	
	He is currently pursuing an Ph.D. degree at the Department of Computer Science in the Tokyo Institute of Technology.
	From April 2023 and October 2023 to present, he is a Research Fellow (DC2) of the Japan Society for the Promotion of Science (JSPS) and a Researcher of ACT-X of the Japan Science and Technology Corporation (JST), Tokyo, Japan.
	His current research interests are in signal and image processing and optimization theory.
	
	Mr. Naganuma received the Student Conference Paper Award from IEEE SPS Japan Chapter in 2023.
\end{IEEEbiography}

\begin{IEEEbiography}[{\includegraphics[width=1in,height=1.25in,clip,keepaspectratio]{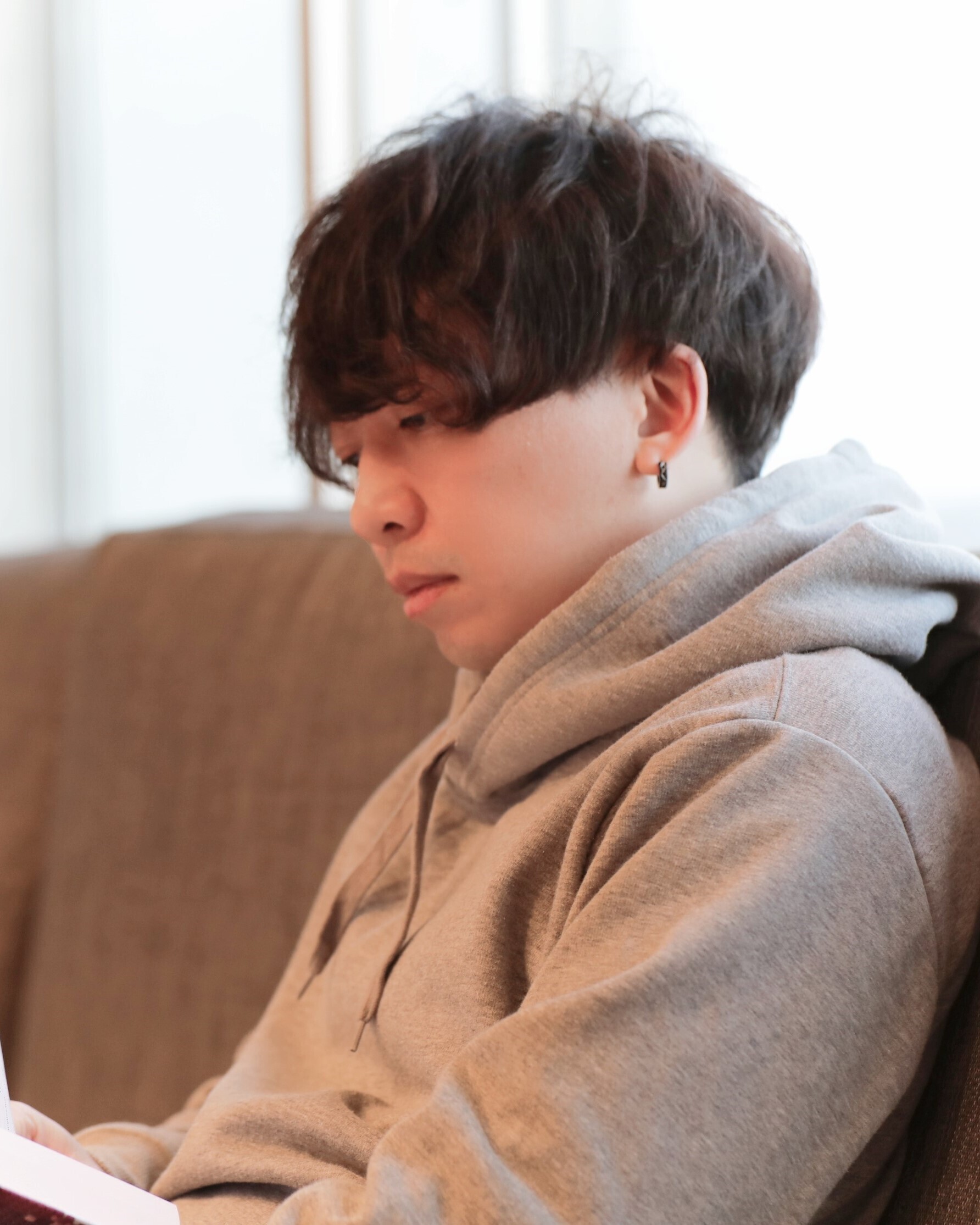}}]{Shunsuke Ono}
(S’11–M’15) received a B.E. degree in Computer Science in 2010 and M.E. and Ph.D. degrees in Communications and Computer Engineering in 2012 and 2014 from the Tokyo Institute of Technology, respectively.

From April 2012 to September 2014, he was a Research Fellow (DC1) of the Japan Society for the Promotion of Science (JSPS). He is currently an Associate Professor in the Department of Computer Science, School of Computing, Tokyo Institute of Technology. From October 2016 to March 2020 and from October 2021 to present, he was/is a Researcher of Precursory Research for Embryonic Science and Technology (PRESTO), Japan Science and Technology Corporation (JST), Tokyo, Japan. His research interests include signal processing, image analysis, remote sensing, mathematical optimization, and data science.

Dr. Ono received the Young Researchers’ Award and the Excellent Paper Award from the IEICE in 2013 and 2014, respectively, the Outstanding Student Journal Paper Award and the Young Author Best Paper Award from the IEEE SPS Japan Chapter in 2014 and 2020, respectively, the Funai Research Award from the Funai Foundation in 2017, the Ando Incentive Prize from the Foundation of Ando Laboratory in 2021, and the Young Scientists’ Award from MEXT in 2022. He has been an Associate Editor of IEEE TRANSACTIONS ON SIGNAL AND INFORMATION PROCESSING OVER NETWORKS since 2019.
\end{IEEEbiography}
\end{document}

%% file: abstract.tex
This paper proposes a novel spatiotemporal (ST) fusion framework for satellite images, named Robust Optimization-based Spatiotemporal Fusion (ROSTF). ST fusion is a promising approach to resolve a trade-off between the temporal and spatial resolution of satellite images. Although many ST fusion methods have been proposed, most of them are not designed to explicitly account for noise in observed images, despite the inevitable influence of noise caused by the measurement equipment and environment. Our ROSTF addresses this challenge by formulating noise removal and ST fusion as a unified optimization problem. Specifically, first, we define observation models for satellite images possibly contaminated with random noise, outliers, and/or missing values, and then introduce certain assumptions that would naturally hold between the observed images and the target high-resolution image. Then, based on these models and assumptions, we formulate the fusion problem as a constrained optimization problem and develop an efficient algorithm based on a preconditioned primal-dual splitting method for solving the problem. The performance of ROSTF was verified using simulated and real data. The results show that ROSTF performs comparably to several state-of-the-art ST fusion methods in noiseless cases and outperforms them in noisy cases.

%% file: introduction.tex
\IEEEPARstart{T}{he} analysis of temporal image series is necessary and important in many remote sensing applications, such as vegetation/crop monitoring and estimation \cite{vegetation}, evapotranspiration estimation \cite{evapotranspiration}, atmosphere monitoring \cite{atmosphere}, land-cover/land-use change detection \cite{landuse}, surface dynamic mapping \cite{mapping}, ecosystem monitoring \cite{ecosystem}, soil water content analysis \cite{soil}, and detailed analysis of human-nature interactions \cite{human}. These applications require time series of high spatial resolution images to properly model the ground surface. In addition, time series of high temporal resolution images are also needed to capture the changes in the ground surface that occur over short periods of time. 

However, there is a trade-off between the temporal and spatial resolution of satellite sensors, and no single sensor can satisfy both requirements. For example, the Landsat sensors can acquire images with a high spatial resolution of 30-m,  but they have a revisit period of up to 16 dates. On the other hand, the Moderate resolution Imaging Spectroradiometer (MODIS) sensors can acquire images for the same scene at least once per date, but the images are at a low spatial resolution of 500-m~\cite{STARFM}. Therefore, the simultaneous acquisition of image series of high spatial and high temporal resolution is a major challenge in the remote sensing community~\cite{STfusion_survey}. The direct solution to this challenge is simply to estimate an unobserved high spatial resolution (HR) image from the single corresponding low spatial resolution (LR) image by super-resolution~\cite{LIIF, MetaSR}. However, this is too difficult because the spatial resolution gap between the two satellite images is often quite large. 

\textit{Spatiotemporal fusion (ST fusion)} addresses this challenge by referring to pairs of HR and LR images on reference dates that are temporally close to the target date. Specifically, the unobserved HR image on the target date is estimated by combining detailed spatial structure extracted from the HR images on the reference dates and spectral changes captured from the differences between the LR images on the reference and target dates.
In ideal situations where a large number of reliable reference images are available, accurate ST fusion would be easy to achieve because the correct spatial structure and spectral changes are readily available. In real-world applications, however, such situations are very rare. Therefore, to achieve the desired ST fusion in real-world applications, the following two requirements are important.
\begin{itemize}
    \item \textit{Minimal reference dates:} ST fusion methods that can be used with minimal reference dates are preferable. In many remote sensing applications, only one pair on a reference date may be available due to cloud contamination, time inconsistency of image acquisition, or other reasons. In addition, in some cases, it is very time-consuming to prepare another pair of images. Therefore, ST fusion methods using a single pair of HR and LR images on a reference date are applicable to much more cases than those using multiple pairs, although such a situation is obviously challenging~\cite{one-pair}.
    \item \textit{Robustness to noise:} Due to the measurement equipment and/or environment, satellite images are often contaminated with various types of noise, such as random noise, outliers, and missing values~\cite{noise, sparsenoise}. An estimated HR image that remains noisy would significantly affect subsequent processing. Therefore, it is crucial to develop a method that is robust to noise.\end{itemize}

\subsection{Prior Research}
\label{ssec: prior research}
Many ST fusion methods have been proposed over the past decades. They are roughly categorized into five groups~\cite{review}: unmixing-based, weight-function-based, Bayesian-based, learning-based, and hybrid methods. Unmixing-based methods estimate the pixel values of an HR image through unmixing the pixels of the input LR images based on the linear spectral mixing theory~\cite{MMT},\cite{UBDF}. Weight-function-based methods generate an HR image by combining information of all the input images based on weight functions~\cite{STARFM},\cite{ESTARFM}. Bayesian-based methods use Bayesian estimation theory to fuse the input images in a probabilistic manner~\cite{Bayesian1, Bayesian2}. In the Bayesian framework, ST fusion can be considered as a maximum a posteriori (MAP) problem, i.e., the goal is to obtain an HR image on the target date by maximizing its conditional probability relative to the input HR and LR images. Learning-based methods model the relationship between observed HR and LR image pairs and then predict an unobserved HR image using machine learning algorithms such as sparse representation learning~\cite{dict, RobOt}, random forest~\cite{random_forest}, and neural networks~\cite{CNN, aritifical_NN, extreme, RSFN, GAN-STFM, DISTF, SWIN}. Hybrid methods integrate two or more techniques from the above four categories~\cite{FSDAF, VIPSTF}.

Some of the unmixing-based, weight-function-based, and Bayesian-based methods allow an arbitrary number of reference dates, i.e., they can handle the cases with a single reference date. However, they are sensitive to noise because they estimate an HR image at the pixel level based entirely on reference images, which may be noisy. Thus, if the input images are noisy, the output image will be severely degraded.

On the other hand, in the context of learning-based methods, a robust spatiotemporal fusion network (RSFN)~\cite{RSFN} is established that accounts for Gaussian noise. RSFN automatically filters noise and prevents predictions from being corrupted by using convolutional neural networks (CNNs), generative adversarial networks (GANs), and an attention mechanism. Specifically, RSFN uses the attention mechanism to ignore noisy pixels in two reference HR images and focus on clean pixels. This method only works in situations where noisy pixels in the two reference HR images do not appear at the same location. In real-world measurements, however, such situations are rare because noise in general contaminates the entire image, not just parts of it. Furthermore, as mentioned above, RSFN requires two reference dates.

\vspace{-2mm}
\subsection{Contributions and Paper Organization}
\label{ssec: contributions}
Now, a natural question arises: \textit{How to achieve robust ST fusion with only a single reference date?} In this paper, we propose a novel ST fusion framework for satellite images, named \textit{robust optimization-based ST fusion (ROSTF)}, to estimate an HR image on the target date while simultaneously denoising an HR image on a single reference date. 

Before formulating our optimization problem, we newly define two observation models (they will be detailed in Sec.~\ref{ssec: model}). 
\begin{itemize}
    \item The first model describes the relationship between an observed noisy image and the oracle noiseless image. The model is designed under the assumption that the observed images are not only contaminated with random noise but also with outliers and missing values. Specifically, random noise is modeled by Gaussian noise while outliers and missing values are modeled by sparse noise. 
    
    \item The second model represents the relationship between an HR image and the corresponding LR image, based on a super-resolution model \cite{relationshipmodel}.
\end{itemize}
We also introduce the following two assumptions about satellite images (they will be detailed in Sec.~\ref{ssec: formulation}). 
\begin{itemize}
    \item[i)] The reflectance may change significantly between the reference and target dates, but the land structure (the locations of the edges) does not. This is a very natural assumption in the context of ST fusion.
    \item[ii)] An HR image and the corresponding LR image have similar brightness. This assumption is necessarily true if the HR and LR sensors have similar spectral resolution, such as Landsat and MODIS~\cite{STARFM}.
\end{itemize}

Based on the observation models and assumptions, we formulate the fusion problem as a constrained convex optimization problem and develop an efficient algorithm based on the preconditioned primal-dual splitting method (P-PDS)~\cite{P-PDS} with an operator-norm-based design method of variable-wise diagonal preconditioning, named OVDP~\cite{P-PDS_OVDP}, which can automatically determine the appropriate stepsizes for solving the optimization problem.

The main contributions of the paper are given as follows.
\begin{itemize}
    \item[1)] \textit{Robustness to Random Noise, Outliers, and Missing Values:} As described above, no existing ST fusion methods can handle random noise, outliers, and missing values, although this type of noise contaminates satellite images due to the measurement equipment and environment. Thanks to the formulation that incorporates the first observation model developed in Sec.~\ref{ssec: model}, ROSTF is robust to such mixed noise.
    
    \item[2)] \textit{A Single Reference Date:} Assumption i) is very simple but effective for ST fusion. By incorporating this assumption as a constraint in the optimization problem, we realize the mechanism to promote the estimated HR image on the target date and the denoised HR image on the reference date to be consistent with the assumption, i.e., to have edges at similar locations. Thanks to such a mechanism, ROSTF performs well even based on a single reference date.
    
    \item[3)] \textit{Facilitation of Parameter Adjustment:} The objective function of the optimization problem of ROSTF consists only of image regularization terms to promote spatial piecewise smoothness. The other components, corresponding to data fidelity based on the two observation models and our two assumptions, are imposed as hard constraints. Such a formulation using constraints instead of adding terms to the objective function has the advantage of simplifying parameter setting~\cite{constrained1,constrained2,constrained3,constrained4,constrained5}: the appropriate parameters in the constraints do not depend on each other and can be determined independently for each constraint. 
    
    \item[4)] \textit{Automatic StepsizeAdjustment:} In order to solve our optimization problem for ST fusion, we develop an efficient algorithm based on P-PDS with OVDP. The appropriate stepsizes of the standard PDS~\cite{PDS} (and most other optimization methods) would be different depending on the problem structure, which means that we have to select them manually. On the other hand, P-PDS with OVDP can automatically determine the appropriate stepsizes based on the problem structure and thus our algorithm is free from such a troublesome task.
\end{itemize}

In the following sections, we first cover the mathematical preliminaries for our method in Sec.~\ref{sec:preli} and then move on to the establishment of our method in Sec.~\ref{sec:proposed}. In Sec.~\ref{sec:exp}, we demonstrate the performance of ROSTF and the effectiveness of each key component of ROSTF through comparative experiments and ablation studies, respectively. Experimental results show that ROSTF performs comparably to several state-of-the-art ST fusion methods for noiseless images and better than them for noisy images, thanks to the effective work of each key component.

The preliminary version of this work, without considering sparse noise, mathematical details, comprehensive experimental comparison, deeper discussion, or implementation using P-PDS with OVDP, has appeared in conference proceedings~\cite{ICASSP}.

%% file: preliminaries.tex
\subsection{Notations}
\label{ssec:Notation}
\vspace{-0.5mm}
Let $\mathbb{R}$ be the set of all real numbers. Vectors and matrices are denoted by bold lower and upper case letters, respectively. We treat a multispectral image with spatial resolution $N_1 \times N_2$ and spectral resolution (the number of bands) $B$ as a vector $\mathbf{x} = ([\mathbf{x}]_1^{\top},\cdots,[\mathbf{x}]_B^{\top})^{\top}\in \mathbb{R}^{NB} (N=N_1 N_2)$, where $[\mathbf{x}]_b \in \mathbb{R}^{N}$ is the vector of the $b$-th band pixel values whose $n$-th value is $x_{b,n} \in \mathbb{R}$. Let $\Gamma_0(\mathbb{R}^{NB})$ be the set of all proper lower-semicontinuous convex functions defined on $\mathbb{R}^{NB}$.
The $\ell_1$-norm, the $\ell_2$-norm, and the $\ell_{1,2}$-norm of $\mathbf{x}$ are defined as $\onenorm{\x}:=\sum_{b}\sum_{n}|x_{b,n}|$, $\twonorm{\x}:=\sqrt{\sum_{b}\sum_{n}|x_{b,n}|^{2}}$, and $\onetwonorm{\x}:=\sum_{n}\sqrt{\sum_{b}|x_{b,n}|^{2}}$, respectively.
For an image $\x \in \R^{NB}$, let $\D_v$ and $\D_h\in \mathbb{R}^{NB \times NB}$denote the matrices for computing the difference between vertical/horizontal adjacent pixels, respectively, and let $\D := (\D_v^{\top} ~ \D_h^{\top})^{\top} \in \mathbb{R}^{2NB \times NB}$.
The hyperslab with the center $\omega$ and radius $\alpha$ is denoted as
\begin{equation}
	\label{eq: hyperslab}
	S_{\alpha}^{\omega} := \{ \mathbf{z} | \,| \omega - \mathbf{1}^{\top}\mathbf{z} | \leq \alpha  \}.
\end{equation}
Also, the $\ell_p$-norm ball ($p=1, 2$) with the center $\mathbf{c}$ and radius $\varepsilon$  is denoted as 
\begin{equation}
	\label{eq: lpnorm_ball}
	B_{p, \varepsilon}^{\mathbf{c}} := \{ \mathbf{z} | \| \mathbf{z} - \mathbf{c} \|_p \leq \varepsilon \}.
\end{equation} 
The indicator function $\iota_C : \mathbb{R}^N \rightarrow (-\infty,\infty]$ of a nonempty closed convex set $C$ is defined as
\begin{eqnarray}
\label{eq: indicator func}
    \iota_C := 
    \begin{cases}
        0, & \mathrm{if} \, \mathbf{x} \in C, \\
        \infty, & \mathrm{otherwise}.
    \end{cases}
\end{eqnarray}

\subsection{Proximal Tools}
\label{ssec: prox}
\vspace{-0.5mm}
The optimization problem of ROSTF that will be formulated in Sec.~\ref{ssec: formulation} consists of nonsmooth convex functions. To solve such a problem, we introduce the notion of the {\it proximity operator} of index $\gamma > 0$ of $f \in \Gamma_0(\mathbb{R}^{NB})$ as follows: 
\begin{equation}
  \label{eq: prox}
  \mathrm{prox}_{\gamma f}: \mathbb{R}^{NB} \rightarrow \mathbb{R}^{NB} : 
  \mathbf{x} \mapsto \argmin_{\mathbf{y} \in \mathbb{R}^{NB}} f(\mathbf{y}) + \frac{1}{2\gamma}\|\mathbf{x} - \mathbf{y}\|_2^2.
\end{equation}

The Fenchell-Rockerfellar conjugate function $f^*$ of the function $f \in \Gamma_0(\mathbb{R}^{NB})$ is denoted by
\begin{equation}
\label{eq: conjugate}
f^*(\mathbf{y}):= \sup_{\mathbf{x}\in \mathbb{R}^{NB}} \left\{ \langle\mathbf{x},\mathbf{y}\rangle - f(\mathbf{x}) \right\}.
\end{equation}
Thanks to Moreus's identity~\cite{Moreau}, the proximity operator of $f^*$ is efficiently calculated as 
\begin{equation}
\label{eq: prox_conjugate}
\prox_{\gamma f^*}(\x) = \x - \gamma \prox_{\frac{1}{\gamma}f}(\tfrac{1}{\gamma}\x).
\end{equation}

Below, we show the specific proximity operators of the functions that we use in this paper.
The proximity operator of the $\ell_{1,2}$-norm is given by
\begin{equation}
    \label{eq: prox_l1,2norm}
    [\mathrm{prox}_{\gamma \|\cdot\|_{1,2}}(\mathbf{x})]_{b,n}  = \max \left\{ 1 - \frac{\gamma}{\sqrt{\sum_{b'=1}^{B} |x_{b',n}|^2}},0 \right\} x_{b,n}. 
\end{equation}
The proximity operator of the indicator function of the hyperslab in \eqref{eq: hyperslab} is expressed as follows: 
\begin{eqnarray}
    \label{eq: prox_hyperslab}
    &\mathrm{prox}_{\gamma \iota_{S_{\alpha}^{\omega}}}(\mathbf{x})& =
    \begin{cases}
    \mathbf{x} + \frac{\eta_1 - \mathbf{1}^{\top}\mathbf{x}}{NB}\mathbf{1}, \; \; & \mathrm{if} \, \mathbf{1}^{\top}\mathbf{x} < \eta_1, \\
    \mathbf{x} + \frac{\eta_2 - \mathbf{1}^{\top}\mathbf{x}}{NB}\mathbf{1}, \; \; & \mathrm{if} \, \mathbf{1}^{\top}\mathbf{x} > \eta_2, \\
    \mathbf{x}, & \mathrm{otherwise},
    \end{cases} \nonumber \\
\end{eqnarray}
where $\eta_1 = \omega - \alpha$ and $\eta_2 = \omega + \alpha$.
The proximity operators of the indicator functions of the $\ell_2$-norm ball and the $\ell_1$-norm ball in \eqref{eq: lpnorm_ball} are calculated by 
\begin{eqnarray}
    \label{eq: prox_l2ball}
    &\mathrm{prox}_{\gamma \iota_{B_{2, \varepsilon}^{\mathbf{c}}}}(\mathbf{x})& =
    \begin{cases}
    \mathbf{x}, & \mathrm{if} \, \mathbf{x} \in \iota_{B_{2, \varepsilon}^{\mathbf{c}}}, \\
    \mathbf{c} + \frac{\varepsilon (\mathbf{x} - \mathbf{c})}{\| \mathbf{x} - \mathbf{c} \|_2}, & \mathrm{otherwise},
    \end{cases} 
\end{eqnarray}
and a fast $\ell_1$-ball projection algorithm~\cite{ell1ball_projection}, respectively.

\begin{figure*}[t]
	\begin{center}
		\begin{minipage}{0.97\hsize} 
			\centerline{\includegraphics[width=\hsize]{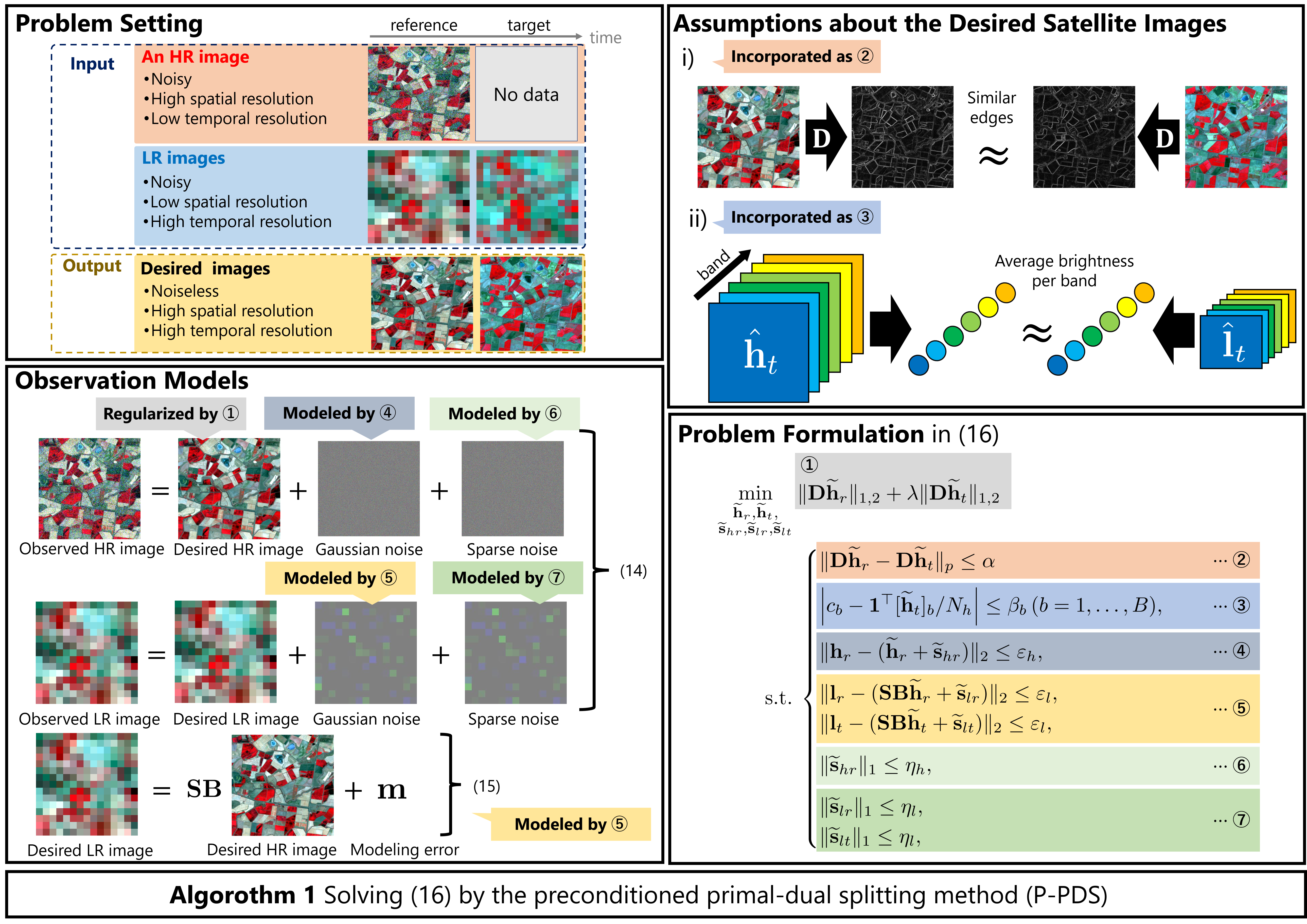}} 
		\end{minipage}
	\end{center}
        \vspace{-2mm}
	\caption{Illustration of our method, i.e., ROSTF.}
        \label{fig: illustration of ROSTF}
\end{figure*}

\subsection{P-PDS with OVDP}
\label{ssec : p-pds}
The standard PDS~\cite{PDS} is a versatile and efficient proximal algorithm that can solve a wide class of nonsmooth convex optimization problems without using matrix inversion. However, it is troublesome to manually set the appropriate stepsizes of the standard PDS. Therefore, we adopt P-PDS~\cite{P-PDS} with OVDP~\cite{P-PDS_OVDP}, a method that automatically determines the appropriate stepsizes according to the problem structure.

Let $\y_i \in \mathbb{R}^{K_i}~(i = 1,\dots, N)$ and $\z_j \in \mathbb{R}^{L_j}~(j = 1,\dots, M)$. Consider convex optimization problems of the following form:
\begin{align}
  \label{eq: general_PDS}
  \min_{\substack{\y_1,\dots,\y_N \\ \z_1,\dots,\z_M}} 
  & \sum_{i=1}^{N}g_i(\y_i)+\sum_{j=1}^{M}h_j(\z_j), \nonumber\\
  {\rm s.t.} & 
  \begin{cases}
      \z_1 = \sum\nolimits_{i=1}^{N} \mathbf{G}_{1,i}\y_i, \\
      \vdots \\
      \z_M = \sum\nolimits_{i=1}^{N} \mathbf{G}_{M,i}\y_i,
  \end{cases}
\end{align}
where $g_i\in \Gamma_0(\mathbb{R}^{K_i})~(i = 1,\dots, N), h_j\in \Gamma_0(\mathbb{R}^{L_j})~(j = 1,\dots, M)$, and $\mathbf{G}_{j,i}: \mathbb{R}^{K_i} \rightarrow\mathbb{R}^{L_j}~(i=1,\dots,N,~j=1,\dots,M)$ are linear operators. P-PDS with OVDP solves (\ref{eq: general_PDS}) by the following iterative procedures:
\begin{align}
\left\lfloor
\begin{aligned}
&\bar{\y}_{1}^{(n)} &\leftarrow &\y_{1}^{(n)}-\gamma_{1,1}\sum\nolimits_{j=1}^{M}\mathbf{G}_{j,1}^{\top}\z_{j}^{(n)}, \nonumber\\
&\y_{1}^{(n+1)} &\leftarrow &\prox_{\gamma_{1,1}g_1}(\bar{\y}_{1}^{(n)}), \nonumber\\
&&\vdots& \nonumber\\
&\bar{\y}_{N}^{(n)} &\leftarrow &\y_{N}^{(n)}-\gamma_{1,N}\sum\nolimits_{j=1}^{M}\mathbf{G}_{j,N}^{\top}\z_{j}^{(n)},\nonumber\\
&\y_{N}^{(n+1)} &\leftarrow &\prox_{\gamma_{1,N}g_N}(\bar{\y}_{N}^{(n)}), \nonumber\\
&\bar{\z}_{1}^{(n)} &\leftarrow &\z_{1}^{(n)} + \gamma_{2,1}\sum\nolimits_{i=1}^{N}\mathbf{G}_{1,i}( 2\y_{i}^{(n+1)}-\y_{i}^{(n)}),\nonumber\\
&\z_{1}^{(n+1)} &\leftarrow &\prox_{\gamma_{2,1}h_{1}^{*}}(\bar{\z}_{1}^{(n)}), \nonumber\\
&&\vdots& \nonumber\\
&\bar{\z}_{M}^{(n)} &\leftarrow &\z_{M}^{(n)} + \gamma_{2,M}\sum\nolimits_{i=1}^{N}\mathbf{G}_{M,i}( 2\y_{i}^{(n+1)}-\y_{i}^{(n)}),\nonumber\\
&\z_{M}^{(n+1)} &\leftarrow &\prox_{\gamma_{2,M}h_{M}^{*}}(\bar{\z}_{M}^{(n)}),
\end{aligned}
\right.
\end{align}
where $\gamma_{1,i}~(i=1,\dots,N)$ and $\gamma_{2,j}~(j=1,\dots,M)$ are stepsize parameters. The stepsize parameters can be determined automatically as follows~\cite{P-PDS_OVDP}:
\begin{equation}
\label{eq: P-PDS stepsizes calculation}
\gamma_{1,i} = \frac{1}{\sum\nolimits_{j=1}^{M} \| \mathbf{G}_{j,i}\|_{\mathrm{op}}^2}, ~
\gamma_{2,j} = \frac{1}{N},
\end{equation}
where $\|\cdot\|_{\mathrm{op}}$ is the operator norm defined by
\begin{equation}
\|\mathbf{G}\|_{\mathrm{op}} := \sup_{\x\neq\mathbf{0}}\frac{\|\mathbf{G}\x\|_2}{\|\x\|_2}.
\end{equation}

%% file: proposedmethod.tex
In what follows, we focus on the cases with a single reference date. Specifically, we consider a situation where both HR and LR sensors observe the same scene on the single reference date, but on the target date, only the LR sensor observes that scene and not the HR sensor. Let the HR image on the reference date, the LR image on the reference date, and the LR image on the target date be $\Hr\in \mathbb{R}^{N_h B}$ , $\Lr\in \mathbb{R}^{N_l B}$ , and $\Lt\in \mathbb{R}^{N_l B}$ , respectively. Our method, ROSTF, estimates the desired noiseless HR image on the target date, denoted by $\hatHt\in \mathbb{R}^{N_h B}$, based on these three observed images, while simultaneously denoising $\Hr$. A general diagram of ROSTF is shown in Fig.~\ref{fig: illustration of ROSTF}.


\subsection{Observation Models}
\label{ssec: model}
Let $\widehat{\mathbf{h}} \in \mathbb{R}^{N_h B}$ and $\widehat{\mathbf{l}} \in \mathbb{R}^{N_l B}$ be a noiseless HR image and a noiseless LR image, respectively, taken on the same date. We introduce observation models for a noisy HR image $\mathbf{h} \in \mathbb{R}^{N_h B}$ and a noisy LR image $\mathbf{l} \in \mathbb{R}^{N_l B}$. To be more specific, we consider that the observed satellite images $\mathbf{h}$ and $\mathbf{l}$ are possibly contaminated with random noise, outliers, and missing values. Random noise added to $\widehat{\mathbf{h}}$ and $\widehat{\mathbf{l}}$ is modeled by Gaussian noise $\nh$ and $\nl$ with standard deviation $\sigma_h$ and $\sigma_l$, respectively, while outliers and missing values affecting $\widehat{\mathbf{h}}$ and $\widehat{\mathbf{l}}$ are modeled by sparsely distributed noise $\sh$ and $\sl$ with the superimposition ratio $r_h$ and $r_l$, respectively. By modeling the noise in this manner, the observation models for $\mathbf{h}$ and $\mathbf{l}$ are described as
\begin{align}
\label{eq: noise model}
  \mathbf{h} &= \widehat{\mathbf{h}} + \nh + \sh, \nonumber\\
  \mathbf{l} &=  \widehat{\mathbf{l}} + \nl + \sl.
\end{align}
Here, $\sigma_h > \sigma_l$ and $r_h > r_l$ generally hold since HR images often contain more severe noise than LR images. This is because the amount of light received per pixel decreases as the number of pixels increases~\cite{noise_pixel_relationship}.

On the other hand, $\widehat{\mathbf{l}}$ can be approximated by the image obtained by blurring and down-sampling $\widehat{\mathbf{h}}$, known as a typical super-resolution model \cite{relationshipmodel}, as follows:
\begin{eqnarray}
\label{eq: relationship model}
  \widehat{\mathbf{l}} = \mathbf{S}\mathbf{B}\widehat{\mathbf{h}} + \mathbf{m},
\end{eqnarray}
where $\mathbf{B} \in \mathbb{R}^{N_h B \times N_h B}$ is the spatial spread transform matrix as introduced in \cite{blurring}, $\mathbf{S}\in \mathbb{R}^{N_l B \times N_h B}$ is the down-sampling matrix, and $\mathbf{m} \in \mathbb{R}^{N_l B}$ is the modeling error. Such a model has been used in the ST fusion literature \cite{integrated}.

\subsection{Problem Formulation}
\label{ssec: formulation}
We introduce the following two assumptions about the noiseless HR and LR images on the reference and target dates, i.e., $\widehat{\mathbf{h}}_r$, $\widehat{\mathbf{l}}_r$, $\widehat{\mathbf{h}}_t$, and $\widehat{\mathbf{l}}_t$.

\begin{itemize}
	\item[i)] The reflectance may change significantly between the reference and target dates, but the land structure does not. This is a very natural assumption in ST fusion and is implicitly accepted in previous studies. If the land structure has not changed significantly, the edges of $\widehat{\mathbf{h}}_r$ and $\widehat{\mathbf{h}}_t$ appear at almost the same locations, implying that the difference between $\mathbf{D}\widehat{\mathbf{h}}_r$ and $ \mathbf{D}\widehat{\mathbf{h}}_t$ tends to be small. We measure the similarity of these edges using the $\ell_p$ ($p=1$ or $2$) norm as $\| \mathbf{D}\widehat{\mathbf{h}}_r - \mathbf{D}\widehat{\mathbf{h}}_t \|_{p}$.
 
	\item[ii)] The HR and LR images on the same date have similar average brightness per band. For example, the difference in average brightness of the $b$-th band of $\widehat{\mathbf{h}}_t$ and $\widehat{\mathbf{l}}_t$, i.e.,
    \begin{equation}
        \left|\frac{1}{N_{l}}\mathbf{1}^{\top}[\widehat{\mathbf{l}}_{t}]_{b}-\frac{1}{N_{h}}\mathbf{1}^{\top}[\widehat{\mathbf{h}}_{t}]_{b}\right|,
    \end{equation}
    is expected to be small. This is necessarily true if the HR and LR sensors have similar spectral resolution, as is the case with Landsat and MODIS~\cite{STARFM}.
\end{itemize}

Based on these assumptions and the observation models in (\ref{eq: noise model}) and (\ref{eq: relationship model}), we formulate the fusion problem as the following constrained convex optimization problem: 
\begin{align}
	\min_{\substack{\tilHr,\tilHt, \\ \tilshr,\tilslr,\tilslt}} \:
	& \|\mathbf{D}\tilHr\|_{1,2} + 
	\lambda\|\mathbf{D}\tilHt\|_{1,2} \nonumber\\ 
	\mathrm{s.t.} \:
	& \begin{cases}
        \pnorm{\D\tilHr - \D\tilHt} \leq \alpha, \\
        \left| c_{b}- \mathbf{1}^{\top}[\widetilde{\mathbf{h}}_{t}]_{b}/{N_{h}}\right| \leq \beta_{b} \, (b = 1, \ldots, B), \\
        \twonorm{\Hr - (\tilHr + \tilshr)}\leq\varepsilon_{h}, \\
        \twonorm{\Lr - (\S\B\tilHr + \tilslr)}\leq\varepsilon_{l},\\
	\twonorm{\Lt - (\S\B\tilHt + \tilslt)}\leq\varepsilon_{l},\\
        \onenorm{\tilshr} \leq \eta_h, \\
        \onenorm{\tilslr} \leq \eta_l, \\
        \onenorm{\tilslt} \leq \eta_l,
	\end{cases}
\label{eq: our ST fusion}
\end{align}
where $\lambda$ is a balancing parameter. The variables $\tilHr$ and $\tilHt$ correspond to the estimates of $\hatHr$ and $\hatHt$, respectively, and $\tilshr$, $\tilslr$ and $\tilslt$ correspond to the estimates of sparse noise superimposed on $\Hr$, $\Lr$ and $\Lt$, respectively. Each term in the objective function and each constraint have the following roles.
\begin{itemize}
    
    \item The two terms in the objective function promote spatial piecewise smoothness of $\widetilde{\mathbf{h}}_r$ and $\widetilde{\mathbf{h}}_t$, with the hyperspectral total variation (HTV) \cite{HTV} as regularization.
    
    \item The first constraint encourages $\mathbf{D}\widetilde{\mathbf{h}}_r$ and $ \mathbf{D}\widetilde{\mathbf{h}}_t$ to be similar based on Assumption~i). The parameter $\alpha$ controls the degree of similarity. Hereafter, the constraint is referred to the \textit{edge constraint}.
    
    \item The second constraint is designed based on Assumption ii). Since $\Lt$ is contaminated by noise, we do not use the average brightness of $[\Lt]_{b}$ itself, i.e., $\mathbf{1}^{\top}[\Lt]_{b}/N_{l}$, but the parameter $c_b$, which is determined based on $\mathbf{1}^{\top}[\Lt]_{b}/N_{l}$ and the noise intensity. The parameter $\beta_b$ controls the strength of this constraint. Hereafter, the constraint is referred to the \textit{brightness constraint}.
    
    \item The third constraint serves as data-fidelity based on the observation model in (\ref{eq: noise model}). The parameter $\varepsilon_h$ depends on Gaussian noise intensity on the HR image, i.e., $\sigma_h$.
    
    \item The fourth and fifth constraints are to ensure that the solutions follow the observation model in (\ref{eq: relationship model}). The parameter $\varepsilon_l$ depends on Gaussian noise intensity on the LR images, i.e., $\sigma_l$.
    
    \item The last three constraints characterize the sparse noise using the $\ell_1$-norms. The parameters $\eta_h$ and $\eta_l$ depend on sparse noise intensity on the HR image and the LR images, respectively, i.e., $r_h$ and $r_l$.
\end{itemize}
Using constraints instead of adding terms to the objective function in this way simplifies the parameter setting\cite{constrained1,constrained2,constrained3,constrained4,constrained5}: we can determine the appropriate parameters for each constraint independently because they are decoupled. The detailed setting of these parameters is discussed in Sec.~\ref{ssec: experimental setup}.

\subsection{Optimization}
\input{proposedmethod/algorithm}
For solving (\ref{eq: our ST fusion}) by an algorithm based on P-PDS with OVDP, we need to transform (\ref{eq: our ST fusion}) into (\ref{eq: general_PDS}). First, using the indicator function (see (\ref{eq: indicator func}) for the definition), we reformulate our problem in (\ref{eq: our ST fusion}) as follows:
\begin{align}
  \label{eq: OptForm'}
  \min_{\substack{\tilHr,\tilHt, \\ \tilshr,\tilslr,\tilslt}}
  & \|\mathbf{D}\tilHr\|_{1,2} + 
  \lambda \|\mathbf{D}\tilHt\|_{1,2} + 
  \iota_{B_{p, \alpha}^{\mathbf{0}}}(\D\tilHr - \D\tilHt)\nonumber \\  
  & + \sum_{b=1}^{B}\iota_{S_{\beta_b'}^{c_b'}}([\tilHt]_{b})
  + \iota_{B_{2,\varepsilon_h}^{\Hr}}(\tilHr + \tilshr)\nonumber \\
  & +\iota_{B_{2,\varepsilon_l}^{\Lr}}(\S\B\tilHr + \tilslr) 
  + \iota_{B_{2,\varepsilon_l}^{\Lt}}(\S\B\tilHt + \tilslt)\nonumber \\
  & + \iota_{B_{1, \eta_h}^{\mathbf{0}}}(\tilshr)  
  + \iota_{B_{1, \eta_l}^{\mathbf{0}}}(\tilslr)  
  + \iota_{B_{1, \eta_l}^{\mathbf{0}}}(\tilslt),
\end{align}
where $\beta_b'=\beta_b N_h$ and $c_b'=c_b N_h$ for $b = 1,\ldots,B$. Introducing auxiliary variables $\z_1$, $\z_2$, $\z_3$, $\z_4$, $\z_5$, and $\z_6$, we can transform (\ref{eq: OptForm'}) into the following equivalent problem:   
\begin{align}
  \label{eq: OptForm}
  \min_{\substack{\tilHr,\tilHt, \\ \tilshr,\tilslr,\tilslt}}
  & \|\z_1\|_{1,2} + 
  \lambda \|\z_2\|_{1,2} + 
  \iota_{B_{p, \alpha}^{\mathbf{0}}}(\z_3)  
  + \sum_{b=1}^{B}\iota_{S_{\beta_b'}^{c_b'}}([\tilHt]_{b}) \nonumber \\
  & + \iota_{B_{2,\varepsilon_h}^{\Hr}}(\z_4)
  +\iota_{B_{2,\varepsilon_l}^{\Lr}}(\z_5) 
  + \iota_{B_{2,\varepsilon_l}^{\Lt}}(\z_6)\nonumber \\
  & + \iota_{B_{1, \eta_h}^{\mathbf{0}}}(\tilshr)  
  + \iota_{B_{1, \eta_l}^{\mathbf{0}}}(\tilslr)  
  + \iota_{B_{1, \eta_l}^{\mathbf{0}}}(\tilslt),\\
  \mathrm{s.t.} 
  &\begin{cases}
      \z_1 = \D\tilHr, \nonumber\\
      \z_2 =  \D\tilHt, \nonumber\\
      \z_3 = \D\tilHr - \D\tilHt, \nonumber\\
      \z_4 = \tilHr + \tilshr, \nonumber\\
      \z_5 = \S\B\tilHr + \tilslr, \nonumber\\
      \z_6 = \S\B\tilHt + \tilslt. \nonumber\\
  \end{cases}
\end{align}
Then, by defining 
\begin{align}
    &g_1(\widetilde{\mathbf{h}}_{r})=0,  g_2(\widetilde{\mathbf{h}}_{t})=\sum_{b=1}^{B}\iota_{S_{\beta_b'}^{c_b'}}([\widetilde{\mathbf{h}}_{t}]_{b}), \nonumber\\
    &g_3(\tilshr)=\iota_{B_{1, \eta_h}^{\mathbf{0}}}(\tilshr), 
    g_4(\tilslr)=\iota_{B_{1, \eta_l}^{\mathbf{0}}}(\tilslr), \nonumber\\
    &g_5(\tilslt)=\iota_{B_{1, \eta_l}^{\mathbf{0}}}(\tilslt), 
    h_1(\z_1) = \|\z_1\|_{1,2},  \nonumber\\
    &h_2(\z_2) = \lambda \|\z_2\|_{1,2},  
    h_3(\z_3) = \iota_{B_{p, \alpha}^{\mathbf{0}}}(\z_3), \nonumber\\ 
    &h_4(\z_4) = \iota_{B_{2,\varepsilon_h}^{\Hr}}(\z_4),  
    h_5(\z_5) = \iota_{B_{2,\varepsilon_l}^{\Lr}}(\z_5), \nonumber\\
    &h_6(\z_6) = \iota_{B_{2,\varepsilon_l}^{\Lt}}(\z_6), 
\end{align}
we reduce \eqref{eq: general_PDS} to \eqref{eq: OptForm} , i.e.,  \eqref{eq: our ST fusion}.

The algorithm for solving (\ref{eq: our ST fusion}) is summarized in Algorithm~\ref{algo: PDS_for_OptForm}. The stepsizes are determined based on OVDP~\cite{P-PDS_OVDP} as follows:
\begin{align}
\label{eq: stepsizes setting}
\gamma_{1,1} &= \frac{1}{2\|\D\|_{\mathrm{op}}^2 + \|\I\|_{\mathrm{op}}^2 + \|\S \B\|_{\mathrm{op}}^2} = \frac{1}{18}\nonumber, \\
\gamma_{1,2} &= \frac{1}{2\|\D\|_{\mathrm{op}}^2 + \|\S \B\|_{\mathrm{op}}^2} = \frac{1}{17}\nonumber, \\
\gamma_{1,3} &=\gamma_{1,4} = \gamma_{1,5} = \frac{1}{\|\I\|_{\mathrm{op}}^2} = 1\nonumber, \\
\gamma_{2,i} &= \frac{1}{5}, ~\text{for}~ i = 1,\dots, 6\nonumber. \\
\end{align}

\subsection{Detailed Computations and Their Complexity}
\label{ssec: computational complexity}

\input{proposedmethod/Onotation}
Table~\ref{table: computational complexity} shows the computational complexity (in $\mathcal{O}$-notation) of each operation used in Algorithm~\ref{algo: PDS_for_OptForm}, where the operated vector is $\mathbf{x} = ([\mathbf{x}]_1^{\top},\cdots,[\mathbf{x}]_B^{\top})^{\top}\in \mathbb{R}^{NB}$. According to Table~\ref{table: computational complexity}, the computational complexity of each step in one iteration of Algorithm~\ref{algo: PDS_for_OptForm} is as follows.
\begin{itemize}
    \item Steps 4, 5, 23, and 24: $\mathcal{O}(kN_hB)$.
    \item Steps 6, 7, 14, 15, 16, 19, 20, 21, 22, 25, 26, 27, and 28: $\mathcal{O}(N_h B)$ when $p=2$.
    \item Step 9: $\mathcal{O}(N_h)$.
    \item Steps 17, 18, 29, and 30: $\mathcal{O}(N_l B)$.
    \item Steps 11 and 27: $\mathcal{O}(N_h B \log(N_h B))$ when $p=1$.
    \item Steps 12 and 13: $\mathcal{O}(N_l B \log(N_l B))$.
\end{itemize}
After all, one iteration of the algorithm has an overall computational complexity of $\mathcal{O}(N_h B \log(N_h B))$.

%% file: proposedmethod/algorithm.tex
\begin{algorithm}[t]
  \caption{P-PDS-based solver for (\ref{eq: our ST fusion})}
  \label{algo: PDS_for_OptForm}
  \begin{algorithmic}[1]
      \Require{$\lambda, p, \alpha, \beta_b', c_b', \varepsilon_h, \varepsilon_l, \eta_h, \eta_l$}
      \Ensure{$\tilHr^{(n)}, \tilHt^{(n)}, \tilshr^{(n)}, \tilslr^{(n)}, \tilslt^{(n)}$}
      \State Initialize $\tilHr^{(0)}, \tilHt^{(0)}, \tilshr^{(0)}, \tilslr^{(0)}, \tilslt^{(0)}, \mathbf{z}_{j}^{(0)}(j=1,\dots,6)$;
      \State Set $\gamma_{1,i} (i=1,\cdots, 3), \gamma_{2,j} (j=1,\cdots, 6),$ as in \eqref{eq: stepsizes setting};
      \While {until a stopping criterion is not satisfied}
      \State $\u_{r} \leftarrow \D^{\top}\z_{1}^{(n)} + \D^{\top}\z_{3}^{(n)} + \z_{4}^{(n)} +  \B^{\top}\S^{\top}\z_{5}^{(n)}$
      \State $\u_{t} \leftarrow \D^{\top}\z_{2}^{(n)} - \D^{\top}\z_{3}^{(n)} +  \B^{\top}\S^{\top}\z_{6}^{(n)}$
      \State $\tilHr^{(n+1)} \leftarrow \tilHr^{(n)} - \gamma_{1,1}\u_{r}$
      \State $\tilHt^{(n+1)}\leftarrow \tilHt^{(n)}-\gamma_{1,2}\u_{t}$
      \For {$b=1,\dots, B$}
        \State $[\tilHt^{(n+1)}]_{b}\leftarrow \prox_{\iota_{S_{\beta_b'}^{c_b'}}}([\tilHt^{(n+1)}]_{b})$
      \EndFor
      \State $\tilshr^{(n+1)} \leftarrow \prox_{\gamma_{1,3} \iota_{B_{1,\eta_h}^{0}}}(\tilshr^{(n)} - \gamma_{1,3} \z_{4})$
      \State $\tilslr^{(n+1)} \leftarrow \prox_{\gamma_{1,4} \iota_{B_{1,\eta_l}^{0}}}(\tilslr^{(n)} - \gamma_{1,4} \z_{5})$
      \State $\tilslt^{(n+1)} \leftarrow \prox_{\gamma_{1,5} \iota_{B_{1,\eta_l}^{0}}}(\tilslt^{(n)} - \gamma_{1,5}\z_{6})$
      \State $\v_{r} \leftarrow 2\tilHr^{(n+1)} - \tilHr^{(n)}$
      \State $\v_{t} \leftarrow 2\tilHt^{(n+1)} - \tilHt^{(n)}$
      \State $\w_{hr} \leftarrow 2\tilshr^{(n+1)} - \tilshr^{(n)}$
      \State $\w_{lr} \leftarrow 2\tilslr^{(n+1)} - \tilslr^{(n)}$
      \State $\w_{lt} \leftarrow 2\tilslt^{(n+1)} - \tilslt^{(n)}$
      \State $\z_{1}^{(n)} \leftarrow \z_{1}^{(n)} + \gamma_{2,1}\D\v_{r}$
      \State $\z_{2}^{(n)} \leftarrow \z_{2}^{(n)} + \gamma_{2,2}\D\v_{t}$
      \State $\z_{3}^{(n)} \leftarrow \z_{3}^{(n)} + \gamma_{2,3}(\D\v_{r} - \D\v_{t})$
      \State $\z_{4}^{(n)} \leftarrow \z_{4}^{(n)} + \gamma_{2,4}(\v_{r} + \w_{hr})$
      \State $\z_{5}^{(n)} \leftarrow \z_{5}^{(n)} + \gamma_{2,5}(\S\B\v_{r} + \w_{lr})$
      \State $\z_{6}^{(n)} \leftarrow \z_{6}^{(n)} + \gamma_{2,6}(\S\B\v_{t} + \w_{lt})$
      \State $\z_{1}^{(n+1)} \leftarrow \z_{1}^{(n)} - \gamma_{2,1}\prox_{\frac{1}{\gamma_{2,1}}\|\cdot\|_{1,2}}(\frac{1}{\gamma_{2,1}}\z_{1}^{(n)})$
      \State $\z_{2}^{(n+1)} \leftarrow \z_{2}^{(n)} - \gamma_{2,2}\prox_{\frac{\lambda}{\gamma_{2,2}}\|\cdot\|_{1,2}}(\frac{1}{\gamma_{2,2}}\z_{2}^{(n)})$
      \State $\z_{3}^{(n+1)} \leftarrow \z_{3}^{(n)} - \gamma_{2,3}\prox_{\iota_{B_{p,\alpha}^{\mathbf{0}}}}(\frac{1}{\gamma_{2,3}}\mathbf{z}_{3}^{(n)})$
      \State $\z_{4}^{(n+1)} \leftarrow \z_{4}^{(n)} - \gamma_{2,4}\prox_{\iota_{B_{2,\varepsilon_h}^{\Hr}}}(\frac{1}{\gamma_{2,4}}\mathbf{z}_{4}^{(n)})$
      \State $\z_{5}^{(n+1)} \leftarrow \z_{5}^{(n)} - \gamma_{2,5}\prox_{\iota_{B_{2,\varepsilon_l}^{\Lr}}}(\frac{1}{\gamma_{2,5}}\mathbf{z}_{5}^{(n)})$
      \State $\z_{6}^{(n+1)} \leftarrow \z_{6}^{(n)} - \gamma_{2,6}\prox_{\iota_{B_{2,\varepsilon_l}^{\Lt}}}(\frac{1}{\gamma_{2,6}}\mathbf{z}_{6}^{(n)})$
      \State $n\leftarrow n+1$
      \EndWhile
  \end{algorithmic}
\end{algorithm}

%% file: proposedmethod/Onotation.tex
\begin{table}[t]
	\begin{center}
		\caption{The Computational Complexity of Each Operation}
		\label{table: computational complexity}
		\centering
            \vspace{-2mm}
		\begin{tabular}{cc}
			\toprule
			Operation 
                & $\mathcal{O}$-notation \\
			
			\midrule
			$\D\x$ 
                & $\O{NB}$ \\
                
                $\D^{\top}\x$ 
                & $\O{NB}$ \\
                
                $\S\x$ 
                & $\O{NB/k}$, where $k$ is the window size of $\S$ \\
                
                $\S^{\top}\x$ 
                & $\O{NB/k}$, where $k$ is the window size of $\S$ \\
                
                $\B\x$ 
                & $\O{kNB}$, where $k$ is the filter size of $\B$ \\
                
                $\B^{\top}\x$ 
                & $\O{kNB}$, where $k$ is the filter size of $\B$ \\
                
                $\prox_{\iota_{S_{\alpha}^{\omega}}}\left(\left[\x\right]_b\right)$ 
                in \eqref{eq: prox_hyperslab}
                & $\O{N}$ \\
                
                $\prox_{\iota_{B_{1,\eta}^{c}}}\left(\x\right)$ 
                in \cite{ell1ball_projection}
                & $\O{NB \log\left(NB\right)}$ \\
                
                $\prox_{\iota_{B_{2,\varepsilon}^{c}}}\left(\x\right)$ 
                in \eqref{eq: prox_l2ball}
                & $\O{NB}$ \\
                
                $\prox_{\gamma\|\cdot\|_{1,2}}\left(\x\right)$ 
                in \eqref{eq: prox_l1,2norm}
                & $\O{NB}$ \\
			\bottomrule
		\end{tabular}
	\end{center}
\end{table}

%% file: Experiments/experiments.tex
We demonstrate the effectiveness of our ST fusion method, ROSTF, through comprehensive experiments using simulated and real data for two sites.
Our experiments aim to verify the following three items.
\begin{itemize}
    \item ROSTF is as effective as state-of-the-art ST fusion methods in noiseless cases and performs better than them in noisy cases. We conducted comparative experiments on four cases of noise contamination. The experimental results for simulated data and real data are presented in Sec.~\ref{ssec: SemiSim results} and Sec.~\ref{ssec: Real results}, respectively.
    \item Key components of ROSTF, such as the assumption-based constraints and the denoising mechanism, work effectively as expected. To measure their influence, ROSTF without each key component is compared with the original ROSTF in terms of fusion accuracy and convergence speed in Sec.~\ref{ssec: ablation study}.
    \item ROSTF is practical in terms of computational time. For a fair comparison, only non-deep learning based methods are compared to ROSTF in Sec.~\ref{ssec: computational time cost}.
\end{itemize}

Note that there are two options for ROSTF: $p=1$ or $2$, where $p$ corresponds to the choice of the norm in the first constraint in \eqref{eq: our ST fusion}, i.e., the edge constraint. Hereafter, ROSTF with $p=1$ and ROSTF with $p=2$ are referred to as ROSTF-1 and ROSTF-2, respectively.

\subsection{Data Description}
\label{ssec: Data Description}
We tested our methods both on real data and simulated data. In the case of satellite observations, radiometric and geometric inconsistencies exist between two different image sensors. This means that the fusion capability of each method cannot be accurately evaluated in experiments using real data because these inconsistencies affect performance, as also adressed in \cite{FSDAF}. Therefore, we generated simulated data based on the observation models and verified the performance of each method using the simulated data in addition to the real data. Specifically, in experiments using simulated data, the simulated LR images were generated from the corresponding real HR images according to (\ref{eq: relationship model}) with $\mathbf{m}=\mathbf{0}$ while the real HR images were used as HR images.

We used MODIS and Landsat time-series images for the following two sites in our experiments.
\begin{description}
    \item[Site1:] The first site is located in the Daxing district in the south of Beijing city ($\ang{39.0009}$N, $\ang{115.0986}$E) \cite{review}. In Site1, we used MODIS and Landsat time-series images acquired on May 29, 2019 (a reference date) and June 14, 2019 (a target date).
    \item[Site2:] The second site is located in southern New South Wales, Australia ($\ang{34.0034}$S, $\ang{145.0675}$E) \cite{Australiadataset}. In Site2, MODIS and Landsat time-series images acquired on January 4, 2002 (a reference date) and February 12, 2002 (a target date) were used.
\end{description}

\input{Experiments/existingmethods_table}
\input{Experiments/ParameterTable}
\input{Experiments/SemiSim_experimental_result_metrics}

\subsection{Compared Methods}
Our method was compared with STARFM~\cite{STARFM}, VIPSTF~\cite{VIPSTF}, RSFN~\cite{RSFN}, RobOt~\cite{RobOt}, and SwinSTFM~\cite{SWIN}. 
Table~\ref{table: existingmethods} shows the characteristics of these methods and ROSTF. Note that unlike the other methods, RSFN requires the preparation of input images obtained on two reference dates. Since our experiments assume a situation where there is only a single reference date, the same HR-LR image pair was input as two different reference image pairs for RSFN. 

As the parameters of these existing methods, we used the values recommended in each reference. It should be noted that RSFN and SwinSTFM require much more data than our and other existing methods due to training and validation processes. In the experiments, we trained and validated RSFN and SwinSTFM using a different set of images from those used for the tests described in Sec.~\ref{ssec: Data Description}. Specifically, 24 groups from Site1 and two groups from Site2 were used for training, and one group from Site1 and two groups from Site2 were used for validation.

\subsection{Experimental Setup}
\label{ssec: experimental setup}

Our method, ROSTF, is implemented using MATLAB. The source code is available on the GitHub\footnote{https://github.com/MDI-TokyoTech/ROSTF} platform.
In these experiments, the spatial spread transform matrix $\mathbf{B}$ in (\ref{eq: relationship model}) was implemented as a blurring operator with the averaging filter of size $k$. The downsampling matrix $\mathbf{S}$ in (\ref{eq: relationship model}) was designed to take the top-left pixel value in the $k \times k$ window. The parameter $k$ was set to 20 to account for the difference in spatial resolution between the Landsat and MODIS images.
The balancing parameter $\lambda$ in (\ref{eq: our ST fusion}) was set to 1 since $\tilHr$ and $\tilHt$ have the same number of elements and are expected to be piecewise smooth to the same degree.
The parameters appearing in the constraints in (\ref{eq: our ST fusion}) were set as shown in Table~\ref{table: parameters}.
We set the maximum number of iterations for Algorithm~\ref{algo: PDS_for_OptForm} to 50000 and the stopping criterion of Algorithm~\ref{algo: PDS_for_OptForm} was set to $\|\tilHr^{(n)} - \tilHr^{(n-1)}\|_2/\| \tilHr^{(n-1)} \|_2<10^{-5}$ ,$\|\tilHt^{(n)} - \tilHt^{(n-1)}\|_2/\| \tilHt^{(n-1)} \|_2<10^{-5}$, $\| \Lr - \mathbf{S}\mathbf{B}\tilHr^{(n)}\|_2 \leq \varepsilon_l$, and $\| \Lt - \mathbf{S}\mathbf{B}\tilHt^{(n)}\|_2 \leq \varepsilon_l$.

To verify both the pure fusion capability and the robustness against noise of the existing and our methods, we conducted the following four combinations of Gaussian noise with different standard deviations and sparse noise (salt-and-pepper noise) with different rates.
\begin{description}
    \item[Case1:] The observed HR and LR images are noiseless, i.e., $\sigma_h=\sigma_l=r_h=r_l=0$ in (\ref{eq: noise model}). 
    \item[Case2:] The observed HR images are contaminated with Gaussian noise with standard deviation $\sigma_h=0.05$ while the observed LR images are noiseless, i.e., $\sigma_h=0.05, \sigma_l=r_h=r_l=0$ in (\ref{eq: noise model}).
    \item[Case3:] The observed HR images are contaminated with sparse noise with the superimposition rate $r_h=0.05$ while the observed LR images are noiseless, i.e., $r_h=0.05, \sigma_h=\sigma_l=r_l=0$ in (\ref{eq: noise model}).
    \item[Case4:] The observed HR images are contaminated with Gaussian noise with standard deviation $\sigma_h=0.05$ and sparse noise with the superimposition rate $r_h=0.05$, while the observed LR images are noiseless, i.e., $\sigma_h=r_h=0.05, \sigma_l=r_l=0$ in (\ref{eq: noise model}).
\end{description}

For the quantitative evaluation, we used the following four metrics: the root mean square error (RMSE):
\begin{equation}
    \mathrm{RMSE} = \sqrt{\frac{1}{N_hB} \|\tilHt - \hatHt\|_2^2},
\end{equation}
where $\tilHt$ and $\hatHt$ denote an estimated HR image and a ground-truth HR image, respectively, on the target date; the spectral angle mapper (SAM)~\cite{SAM}:
\begin{equation}
    \mathrm{SAM} = \frac{1}{N_h}\sum_{n=1}^{N_h}\mathrm{arccos}\left(\frac{\left<\widetilde{\mathbf{e}}_{n},\widehat{\mathbf{e}}_{n}\right>}{\|\widetilde{\mathbf{e}}_{n}\|_2 \cdot\|\widehat{\mathbf{e}}_{n}\|_2}\right),
\end{equation}
where $\widetilde{\mathbf{e}}_{n}$ and $\widehat{\mathbf{e}}_{n}\in \R^{B}$ denote the spectral vectors of the $n$-th pixel of $\tilHt$ and $\hatHt$, respectively; the mean structural  similarity overall bands (MSSIM)~\cite{SSIM}:
\begin{equation}
    \mathrm{MSSIM} = \frac{1}{B}\sum_{b = 1}^{B}\mathrm{SSIM}\left([\tilHt]_{b}, [\hatHt]_{b} \right);
\end{equation}
and the correlation coefficient (CC):
\begin{equation}
    \mathrm{CC} = \frac{s_{\tilHt\hatHt}}{s_{\tilHt}s_{\hatHt}},
\end{equation}
where $s_{\tilHt\hatHt}$ denotes the covariance of $\tilHt$ and $\hatHt$, and $s_{\tilHt}$ and $s_{\hatHt}$ denote the standard deviations of $\tilHt$ and $\hatHt$, respectively. We calculated RMSE to gauge the difference between the estimated image and the ground-truth at the pixel level. SAM was used to measure spectral fidelity. Lower RMSE and SAM indicate better estimation performance. We used MSSIM to evaluate the similarity of the overall structure. CC shows the strength of the linear relationship between the estimated image and the ground-truth. Higher MSSIM and CC indicate better estimation performance.

\input{Experiments/results/Site1_SemiSim_Case1}
\input{Experiments/results/Site2_SemiSim_Case1}
\input{Experiments/results/Site2_SemiSim_Case234}
\input{Experiments/results/Site2_SemiSim_scatter}
\input{Experiments/results/Site2_SemiSim_SpectralPlot}

\subsection{Experimental Results with Simulated Data}
\label{ssec: SemiSim results}
Table~\ref{table: simulated metrics} shows the RMSE, SAM, MSSIM, and CC results in experiments with the simulated data. In Case1, STARFM, VIPSTF, RobOt, SwinSTFM, ROSTF-1, and ROSTF-2 perform equally well. In contrast, the results of RSFN are not good for both the Site1 and Site2 data. This may be because the training data described above was not large enough to train RSFN sufficiently. Thus, if more training data were used, RSFN might have produced better results. However, in actual applications, it is difficult to collect noise-free training data, and this is the situation considered in these experiments. Next, we focus on the results in Case2, Case3, and Case4, where the observed reference images are contaminated with noise. While STARFM, VIPSTF, RobOt, SwinSTFM, and RSFN perform significantly worse due to the influence of noise, ROSTF-1 and ROSTF-2 show no significant performance degradation in Case2, Case3, and Case4. This is because ROSTF estimates the target HR image while simultaneously denoising the reference HR image.

Fig.~\ref{fig: Site1 SemiSim Case1 results} and Fig.~\ref{fig: Site2 SemiSim Case1 results} show the estimated results in Case1 for the Site1 and Site2 simulated data, respectively. In the zoomed-in areas, there are large temporal changes in brightness between the reference HR image~$\Hr$ and the target HR image, ground-truth. For the Site1 data, it can be visually seen that ROSTF-1 and ROSTF-2 capture these changes most accurately and estimate the brightness closest to that of the ground-truth compared with the other methods. This is thanks to the brightness constraint and the fifth constraint in (\ref{eq: our ST fusion}), which promote the brightness of the estimated image to be close to that of the observed LR image $\Lt$ on the target date, based on Assumption~ii) and the observation model in \eqref{eq: relationship model}. Following ROSTF, RobOt also captures the temporal changes well. STARFM and VIPSTF are greatly affected by the reference HR image $\Hr$ and estimate closer brightness to that of $\Hr$ than to that of the ground-truth. The result of RSFN is not good due to lack of training data. The SwinSTFM result appears to show spectral distortion. For the Site2 data, 
Fig.~\ref{fig: Site2 SemiSim Case1 results} shows that ROSTF-1 captures the changes in brightness of the zoomed-in area most accurately. ROSTF-2 and RobOt have good estimates, followed by ROSTF-1. STARFM and SwinSTFM do not capture the large temporal changes. The result of VIPSTF has different tints throughout. RSFN estimates brightness closer to that of $\Hr$ than to that of the ground-truth.

\input{Experiments/Real_experimental_result_metrics}
\input{Experiments/results/Site1_Real_Case1}
\input{Experiments/results/Site2_Real_Case1}

Next, we focus on the results in noisy cases. Fig.~\ref{fig: Site2 SemiSim Case234 results} shows the estimated results for the Site2 data in noisy cases, i.e., Case2, Case3, and Case4. The results of STARFM, VIPSTF, and RobOt are contaminated with noise similar to that of $\Hr$ because they estimate the target image at the pixel level based entirely on the noisy reference image $\Hr$. The results of RSFN are corrupted for the unexpected inputs because RSFN was not trained on noisy data. Furthermore, using noisy data to train RSFN may not have produced good results. This is because RSFN is robust to noise only in the situations where noisy pixels in the two reference HR images do not appear at the same location. SwinSTFM generates unstable noisy results due to noisy unexpected inputs. On the other hand, ROSTF-1 and ROSTF-2 provide good estimates even when the observed reference images are noisy, without much loss of accuracy compared to the noiseless case. This is thanks to the framework that simultaneously performs noise removal and ST fusion.

\input{Experiments/results/Site1_Real_Case234}
\input{Experiments/results/Site1_Real_Case234_dif_crop}

The effect of noise on each method can also be seen visually in Fig.~\ref{fig: Site2 SemiSim scatter} and Fig.~\ref{fig: Spectral Plot Site1}. The scatter plots in Fig.~\ref{fig: Site2 SemiSim scatter} reveal the difference between the ground-truth and the estimated values of each method for the simulated Site2 data. STARFM, VIPSTF, RSFN, RobOt, and SwinSTFM present greater variance in Case2 than in Case1 due to the influence of Gaussian noise. In Case3, STARFM, VIPSTF, and RobOt estimate the wrong values close to 0 or 1 affected by sparse noise while the RSFN and SwinSTFM results have no such outliers but show greater variance. Furthermore, in Case 4, the distributions show that STARFM, VIPSTF, RSFN, RobOt, and SwinSTFM are affected by both Gaussian and sparse noise. In contrast, the results of ROSTF-1 and ROSTF-2 have small variance and no outliers, which means that they are robust to Gaussian, sparse, and mixed noise. Spectral profiles of a specific pixel in the results of each method for the Site2 data in Case1 and Case4 are shown in Fig.~\ref{fig: Spectral Plot Site1}. STARFM, VIPSTF, RobOt, SwinSTFM, and ROSTF estimate similarly accurate values in Case1, i.e., they are comparable in the noiseless case. In Case4, STARFM, VIPSTF, RobOt, and SwinSTFM estimate completely wrong values for the third band affected by sparse noise and perform worse for the other bands due to the influence of Gaussian noise. In contrast, ROSTF-1 and ROSTF-2 have accurate estimates for all bands, even in the noisy case.

\subsection{Experimental Results with Real Data}
\label{ssec: Real results}
Table~\ref{table: real metrics} shows the RMSE, SAM, MSSIM, and CC results in experiments with the real data. Compared to the results for the simulated data in Table~\ref{table: simulated metrics}, the performance of ROSTF-1 and ROSTF-2 degrades due to radiometric and geometric inconsistencies between the Landsat and MODIS sensors. Despite such the performance degradation,  ROSTF-1 and ROSTF-2 perform as well as the existing methods in the noiseless case, i.e., Case1, and outperform them in the noisy cases, i.e., Case2, Case3, and Case4, as in the experiments with the real data. Therefore, we can say that ROSTF is robust to noise even for the real data.

Fig.~\ref{fig: Site1 Real Case1 results} and Fig.~\ref{fig: Site2 Real Case1 results} show the estimated results in Case1 for the Site1 and Site2 real data, respectively. Compared to the results for the simulated data in Fig.~\ref{fig: Site1 SemiSim Case1 results} and Fig.~\ref{fig: Site2 SemiSim Case1 results}, ROSTF-1 and ROSTF-2 perform worse because there are modeling errors between the real HR images $\Hr, \Ht$ and the real LR images $\Lr, \Lt$ due to radiometric and geometric inconsistencies.

Fig.~\ref{fig: Site1 Real Case234 results} shows the estimated results for the Site2 data in the noisy cases, i.e., Case2, Case3, and Case4. The results of the existing methods are not good, especially STARFM, VIPSTF, and RobOt generate noisy outputs.
The estimated images of ROSTF seem to be blurred in the zoomed area in Fig.~\ref{fig: Site1 Real Case234 results}. This is due to oversmoothing by the HTV regularization terms in \eqref{eq: our ST fusion}, which might have undesirable effects on some applications. However, according to the difference map in Fig.~\ref{fig: difference map}, the results of ROSTF have the least error, and the accuracy evaluation in Table~\ref{table: real metrics} also shows that ROSTF achieves the best performance in all metrics.

\subsection{Ablation Study}
\label{ssec: ablation study}

We conducted ablation experiments with a focus on the following three items:
\begin{itemize}
    
    \item The edge constraint, $\|\D\tilHr-\D\tilHt\|_{p} \leq \alpha$, to encourage the land structure, i.e. the edges, of the reference and target HR images to be similar based on Assumption~i).

    \item The brightness constraint, $\left| c_{b}- \mathbf{1}^{\top}[\widetilde{\mathbf{h}}_{t}]_{b}/{N_{h}}\right| \leq \beta_{b} \, (b = 1, \ldots, B)$, which is designed based on Assumption~ii), so that the estimated target HR image has a similar average brightness to the target LR image.
    
    \item The denoising mechanism for the reference HR image~$\Hr$, i.e., the first regularization term $\|\D\tilHr\|_{1,2}$ and the third, fourth, sixth, and seventh constraints, $\twonorm{\Hr - (\tilHr + \tilshr)}\leq\varepsilon_{h}$, $\twonorm{\Lr - (\S\B\tilHr + \tilslr)}\leq\varepsilon_{l}$, $\onenorm{\tilshr} \leq \eta_h$, $\onenorm{\tilslr} \leq \eta_l$.
\end{itemize}

We tested ROSTF with each of the above three components removed. In the following, we first present the ablation studies on the two constraints, the edge constraint and the brightness constraint, followed by the ablation study on the denoising mechanism. The hyperparameters in each optimization problem and the stopping criterion of each P-PDS-based algorithm to solve it were set as in the original ROSTF. On the other hand, the stepsizes in each algorithm were set as the values computed according to the operator-norm-based design method of variable-wise diagonal preconditioning in (\ref{eq: P-PDS stepsizes calculation}).

\input{Experiments/AblationStudy/average_convergence_iterations_and_performance_metrics}

\input{Experiments/AblationStudy/compare_convergence_figure}
\input{Experiments/AblationStudy/without_12constraint_figure}

\vspace{2mm}
\subsubsection{Edge and Brightness Constraints}
\label{sssec: ablation study constraints}

First, we measure the effectiveness of the two constraints based on Assumption~i) and Assumption~ii), i.e., the edge constraint and the brightness constraint, in terms of convergence speed and fusion performance.

Table~\ref{table: convergence iterations} shows the average number of iterations spent before each algorithm stopped and the average performance results for all sites (Site1 or Site2), data types (real or simulated data), and noise cases (from Case1 to Case4). Note that each algorithm always stopped at the maximum number of iterations, 50000, even if the stopping criterion was not met. The original ROSTF performs best with the fewest number of iterations on average, indicating that both of the two constraints contribute to achieving higher fusion performance with fewer iterations.


Fig.~\ref{fig: convergence results} shows the transition of the objective function values and the RMSE values for the original ROSTF-2, ROSTF-2 without the edge constraint, and ROSTF-2 without the brightness constraint in the experiment on the Site1 simulated data in Case1. ROSTF-2 without the edge constraint does not meet the stopping criterion until 50000 iterations. This would be because the solution space of the optimization problem without the edge constraint is too large to efficiently reach an optimal solution. This suggests that the edge constraint has the effect of making the solution space moderately small and speeding up convergence. On the other hand, ROSTF-2 without the brightness constraint converges faster than ROSTF-2 without the edge constraint, but shows unstable behavior, especially in the early iterations. This may be because the variable $\tilHt$ is no longer directly updated as a primal variable when the brightness constraint is removed. The update equation corresponding to the brightness constraint is implemented as the only update equation for the primal variable $\tilHt$, as in the lines 8 to 10 of Algorithm~1, while the update equations corresponding to the other components for $\tilHt$ update the dual variables $\mathbf{z}_2$, $\mathbf{z}_3$ and $\mathbf{z}_6$. Directly updating the primal variable leads to faster convergence of the algorithm and more stable behavior than updating the primal variable via the update of the dual variables. By introducing the brightness constraint, the variable $\tilHt$ is updated directly as the primal variable, which improves convergence speed and stability.

Fig.~\ref{fig: without 12const results} shows a visual comparison of the original ROSTF-2 and ROSTF-2 without these constraints. The image estimated by ROSTF-2 without the edge constraint~(a) loses spatial structure, and ROSTF-2 without the brightness constraint~(b) produces an image with incorrect brightness. On the other hand, the original ROSTF-2~(c) estimates brightness close to that of the ground-truth while still preserving spatial structure, indicating that both constraints work effectively.

\vspace{2mm}
\subsubsection{Denoising Mechanism}
\label{sssec: ablation study denoising}

Next, we move on to the ablation study of the denoising mechanism of ROSTF. The optimization problem of ROSTF without the denoising mechanism is formulated as
\begin{align}
	\min_{\substack{\tilHt,\tilslt}} \:
	& \|\mathbf{D}\tilHt\|_{1,2} \nonumber\\ 
	\mathrm{s.t.} \:
	& \begin{cases}
        \pnorm{\D\Hr - \D\tilHt} \leq \alpha, \\
        \left| c_{b}- \mathbf{1}^{\top}[\widetilde{\mathbf{h}}_{t}]_{b}/{N_{h}}\right| \leq \beta_{b} \, (b = 1, \ldots, B), \\
	\twonorm{\Lt - (\S\B\tilHt + \tilslt)}\leq\varepsilon_{l},\\
        \onenorm{\tilslt} \leq \eta_l.
	\end{cases}
\label{eq: ROSTF without denoising}
\end{align}
Since the objective function contains only one regularization term for $\tilHt$, no balancing parameter is needed. Also, note that the edge constraint of \eqref{eq: ROSTF without denoising} is different from that of the original ROSTF in \eqref{eq: our ST fusion} because this optimization problem has no variable $\tilHr$.

\input{Experiments/AblationStudy/without_denoising_metrics}

Table~\ref{table: without denoising} shows the average RMSE, SAM, MSSIM, and CC results for each noise case. In the noisy cases, i.e. Case2, Case3, and Case4, as expected, ROSTF without the denoising mechanism performs significantly worse than the original ROSTF due to the direct effect of noise. This shows that the denoising mechanism works effectively to make ROSTF robust to noise. On the other hand, we newly found that in the noiseless case, i.e., Case1, ROSTF without the denoising mechanism achieves slightly better fusion results than the original one. This result indicates that in noiseless cases, the observed HR image $\Hr$ can be used as the reference as it is, and there is no need to estimate $\tilHr$. Furthermore, by removing the variable $\tilHr$, ROSTF does not need the reference LR image $\Lr$ as input in \eqref{eq: ROSTF without denoising}. Thus, fortunately, if noiseless images can be observed, ROSTF can be applied in situations where only two input images are available, a reference HR image~$\Hr$ and a target LR image~$\Lt$. 

\input{Experiments/AblationStudy/without_denoising_figure}

Fig.~\ref{fig: without denoising} shows the fusion results of ROSTF-1 without the denoising mechanism for the Site2 simulated data. It can also be visually seen that in Case1, ROSTF-1 without the denoising mechanism obtains a good result based on only two input images. On the other hand, in Case2, Case3, and Case4, the results of ROSTF without the denoising mechanism are contaminated with noise. This is because the edge constraint copies not only the true edge or spatial structure but also the noise in the reference HR image to the estimated target HR image. The results of this ablation study confirm that the denoising mechanism plays an effective role in avoiding such noise effects.

\subsection{Computational Cost}
\label{ssec: computational time cost}
We measured the actual runnning times using MATLAB (R2022b) on a Windows 11 computer with an Intel Core i9-13900 1.0GHz processor, 32GB of RAM, and NVIDIA GeForce RTX 4090. We used the Site1 and Site2 data with $400\times400$ pixels and six bands. The measurement results show that the average number of iterations for Algorithm~1 was 59.20 per second. Table~\ref{table: running time} shows the average running times of the non-deep learning-based comparison methods (STARFM, VIPSTF, and RobOt) and our methods (ROSTF-1 and ROSTF-2). For ROSTF, the average number of iterations for Algorithm~1 is also given in parentheses. Note that only the non-deep learning based methods were compared with ROSTF for a fair comparison, because deep learning based methods require training processes in addition to ST fusion processes.

ROSTF-1 and ROSTF-2 took about 4 to 10 minutes. STARFM and VIPSTF took longer than ROSTF because they estimate the target HR image pixel by pixel. On the other hand, RobOt ran much faster than ROSTF, possibly because the Least Absolute Shrinkage and Selection Operator (LASSO) problem in RobOt has a closed-form solution in our experiment with only one reference date. This result indicates that ROSTF is slower than RobOt, but we believe that ROSTF is still practical in terms of computational cost.

\input{Experiments/results/running_time}

\subsection{Summary}
\label{ssec: Summary}
We summarize the insights from the experiments as follows.
\begin{itemize}
    \item From the results of the experiments in Case1, we see that ROSTF is comparable to state-of-the-art ST fusion methods in noiseless cases. Therefore, the observation model in \eqref{eq: relationship model} and the assumptions introduced in Sec.~\ref{ssec: formulation} are valid for ST fusion.
    \item The results of the experiments in Case2, Case3, and Case4 confirm that ROSTF has good performance even when observed HR images are degraded by random noise, missing values, and outliers.
    \item The ablation studies demonstrate that the key components, such as the edge constraint, the brightness constraint, and the denoising mechanism, work effectively as expected.
\end{itemize}

%% file: Experiments/existingmethods_table.tex
\begin{table}[t]
	\begin{center}
		\caption{Existing Methods}
		\label{table: existingmethods}
            \vspace{-2mm}
            \scalebox{0.9}{
		\begin{tabular}{cccc}
			\toprule
                & \hspace{-4mm}  \begin{tabular}{c} Minimum number\\of reference dates \end{tabular}
                & \hspace{-4mm}  \begin{tabular}{c} Robustness to \\Gaussian noise \end{tabular} 
                & \hspace{-4mm}  \begin{tabular}{c} Robustness to \\sparse noise \end{tabular} \hspace{-3mm}  \\
			\cmidrule(lr){1-4}
                \vspace{0.5mm}
			STARFM~\cite{STARFM} 
                & 1
                & --
                & -- \\
                \vspace{0.5mm}
			VIPSTF~\cite{VIPSTF}
                & 1
                & --
                & -- \\
                \vspace{0.5mm}
                RSFN~\cite{RSFN}
                & 2
                & $\checkmark$
                & -- \\
                \vspace{0.5mm}
                RobOt~\cite{RobOt}
                & 1
                & --
                & -- \\
                \vspace{0.5mm}
                SwinSTFM~\cite{SWIN}
                & 1
                & --
                & -- \\
                \vspace{0.5mm}
                \textbf{ROSTF}
                & 1
                & $\checkmark$
                & $\checkmark$ \\
                \bottomrule
		\end{tabular}
            }
	\end{center}
\end{table}

%% file: Experiments/ParameterTable.tex
\begin{table}[t]
	\begin{center}
		\caption{The Parameter Setting}
		\label{table: parameters}
            \vspace{-2mm}
		\begin{tabular}{cc}
			\toprule
			\vspace{1mm}
			$\alpha$ 
                & $\begin{cases}
                2\times10^{-3}N_hB ~(p=1)\nonumber, \\ 1\times10^{-6}N_hB ~(p=2) \nonumber,\\
                \end{cases}$ 
                \vspace{3mm} \\
                $\beta_b$
                &$\left | \frac{1}{N_{l}}\mathbf{1}^{\top}[\mathbf{l}_{r}]_{b}- \frac{1}{N_{h}}\mathbf{1}^{\top}[\mathbf{h}_{r}]_{b} \right |$
                \vspace{3mm} \\
                $c_b$
                & $\frac{1}{N_l}\mathbf{1}^{\top}[\Lt]_{b} - r_l(0.5 - \frac{1}{N_l}\mathbf{1}^{\top}[\Lt]_{b})$
                \vspace{3mm} \\
                $\varepsilon_h$
                & $0.98 \sigma_h \sqrt{N_h B (1 - r_h)}$
                \vspace{3mm} \\
                $\varepsilon_l$
                & $\| \Lr - \mathbf{S}\mathbf{B}\Hr\|_2$ \vspace{1mm} \\
                \bottomrule
		\end{tabular}
	\end{center}
\end{table}

%% file: Experiments/SemiSim_experimental_result_metrics.tex
\begin{table*}[t]
	\begin{center}
		\caption{The RMSE, SAM, MSSIM, and CC Results in the Experiments With Simulated Data}
        \label{table: simulated metrics}
            \vspace{-2mm}
		\centering
            \scalebox{1}{
		\begin{tabular}{p{0.5cm} p{0.7cm} p{0.7cm} >{\centering\arraybackslash}p{1.3cm} >{\centering\arraybackslash}p{1.3cm} >{\centering\arraybackslash}p{1.3cm} >{\centering\arraybackslash}p{1.3cm} >{\centering\arraybackslash}p{1.3cm} >{\centering\arraybackslash}p{1.3cm} >{\centering\arraybackslash}p{1.3cm}}
			\toprule
                
                \multirow{2}{*}{Site} & \multirow{2}{*}{Noise} & \multirow{2}{*}{Metrics} & STARFM & VIPSTF & RSFN & RobOt & SwinSTFM & \textbf{ROSTF-1} & \textbf{ROSTF-2} \\

                & & & \cite{STARFM} & \cite{VIPSTF} & \cite{RSFN} & \cite{RobOt} & \cite{SWIN} & \textbf{(Ours)} & \textbf{(Ours)} \\
                
                
                \midrule 
                \multirow{16}{*}{Site1} 
                & \multirow{4}{*}{Case1} 
                & RMSE 
                & 0.0235 & \textbf{0.0234} & 0.0541 & 0.0246 & 0.0289 & 0.0245 & 0.0238 \\ 
                
                & & SAM 
                & \textbf{0.0711} & 0.0742 & 0.1763 & 0.0748 & 0.0970 & 0.0762 & 0.0740 \\ 
                
                & & SSIM 
                & \textbf{0.9755} & 0.9752 & 0.8610 & 0.9734 & 0.9636 & 0.9734 & 0.9746 \\ 
                
                & & CC 
                & 0.9794 & \textbf{0.9795} & 0.8997 & 0.9776 & 0.9698 & 0.9778 & 0.9789  \\ 
                \cmidrule(lr){2-10} 
                & \multirow{4}{*}{Case2} 
                & RMSE 
                & 0.0554 & 0.0510 & 0.0672 & 0.0546 & 0.0391 & \textbf{0.0298} & 0.0299 \\ 
                
                & & SAM 
                & 0.2431 & 0.2254 & 0.2390 & 0.2380 & 0.1453 & \textbf{0.0966} & 0.0968 \\ 
                
                & & SSIM 
                & 0.8484 & 0.8652 & 0.7882 & 0.8515 & 0.9284 & \textbf{0.9543} & 0.9541 \\ 
                
                & & CC 
                & 0.8984 & 0.9098 & 0.8198 & 0.8982 & 0.9490 & \textbf{0.9667} & 0.9666  \\ 
                \cmidrule(lr){2-10} 
                & \multirow{4}{*}{Case3} 
                & RMSE 
                & 0.1402 & 0.1288 & 0.0542 & 0.1390 & 0.0504 & 0.0314 & \textbf{0.0278} \\ 
                
                & & SAM 
                & 0.2133 & 0.2017 & 0.1769 & 0.2167 & 0.1676 & 0.1173 & \textbf{0.0896} \\ 
                
                & & SSIM 
                & 0.5193 & 0.5578 & 0.8556 & 0.5225 & 0.8767 & 0.9510 & \textbf{0.9666} \\ 
                
                & & CC 
                & 0.6134 & 0.6374 & 0.8937 & 0.6167 & 0.9080 & 0.9633 & \textbf{0.9711}  \\ 
                \cmidrule(lr){2-10} 
                & \multirow{4}{*}{Case4} 
                & RMSE 
                & 0.1484 & 0.1360 & 0.0682 & 0.1469 & 0.0544 & 0.0385 & \textbf{0.0370} \\ 
                
                & & SAM 
                & 0.3502 & 0.3230 & 0.2453 & 0.3467 & 0.1918 & 0.1297 & \textbf{0.1210} \\ 
                
                & & SSIM 
                & 0.4826 & 0.5227 & 0.7799 & 0.4836 & 0.8587 & 0.9282 & \textbf{0.9346} \\ 
                
                & & CC 
                & 0.5911 & 0.6156 & 0.8130 & 0.5890 & 0.8964 & 0.9435 & \textbf{0.9483}  \\ 
                \midrule 
                \multirow{16}{*}{Site2} 
                & \multirow{4}{*}{Case1} 
                & RMSE 
                & 0.0367 & \textbf{0.0312} & 0.0499 & 0.0341 & 0.0375 & 0.0398 & 0.0428 \\ 
                
                & & SAM 
                & 0.1091 & \textbf{0.1040} & 0.1583 & 0.0989 & 0.1290 & 0.1231 & 0.1323 \\ 
                
                & & SSIM 
                & 0.9116 & \textbf{0.9343} & 0.8712 & 0.9223 & 0.9057 & 0.8946 & 0.8816 \\ 
                
                & & CC 
                & 0.9372 & \textbf{0.9523} & 0.9065 & 0.9441 & 0.9308 & 0.9289 & 0.9182  \\ 
                \cmidrule(lr){2-10} 
                & \multirow{4}{*}{Case2} 
                & RMSE 
                & 0.0628 & 0.0462 & 0.0677 & 0.0527 & 0.0449 & \textbf{0.0328} & 0.0364 \\ 
                
                & & SAM 
                & 0.2943 & 0.2091 & 0.2816 & 0.2365 & 0.1672 & \textbf{0.1023} & 0.1132 \\ 
                
                & & SSIM 
                & 0.7281 & 0.8332 & 0.7529 & 0.7908 & 0.8566 & \textbf{0.9240} & 0.9100 \\ 
                
                & & CC 
                & 0.8408 & 0.8978 & 0.7868 & 0.8741 & 0.9006 & \textbf{0.9482} & 0.9371  \\ 
                \cmidrule(lr){2-10} 
                & \multirow{4}{*}{Case3} 
                & RMSE 
                & 0.1389 & 0.0909 & 0.0520 & 0.1103 & 0.0543 & \textbf{0.0391} & 0.0420 \\ 
                
                & & SAM 
                & 0.2668 & 0.2100 & 0.1767 & 0.2280 & 0.2040 & \textbf{0.1244} & 0.1258 \\ 
                
                & & SSIM 
                & 0.4037 & 0.5863 & 0.8575 & 0.5019 & 0.7853 & \textbf{0.8964} & 0.8878 \\ 
                
                & & CC 
                & 0.5585 & 0.7048 & 0.8981 & 0.6395 & 0.8517 & \textbf{0.9281} & 0.9188  \\ 
                \cmidrule(lr){2-10} 
                & \multirow{4}{*}{Case4} 
                & RMSE 
                & 0.1476 & 0.0969 & 0.0683 & 0.1189 & 0.0569 & \textbf{0.0367} & 0.0376 \\ 
                
                & & SAM 
                & 0.4138 & 0.2970 & 0.2828 & 0.3326 & 0.2219 & 0.1300 & \textbf{0.1160} \\ 
                
                & & SSIM 
                & 0.3635 & 0.5463 & 0.7494 & 0.4542 & 0.7650 & 0.9065 & \textbf{0.9068} \\ 
                
                & & CC 
                & 0.5319 & 0.6767 & 0.7832 & 0.6040 & 0.8374 & \textbf{0.9335} & 0.9302  \\ 
                
                \bottomrule

		\end{tabular}
            }
	\end{center}
\end{table*}

%% file: Experiments/results/Site1_SemiSim_Case1.tex
\begin{figure}[t]
	\begin{center}
        \begin{minipage}{0.24\hsize} 
            \centerline{\includegraphics[width=\hsize]{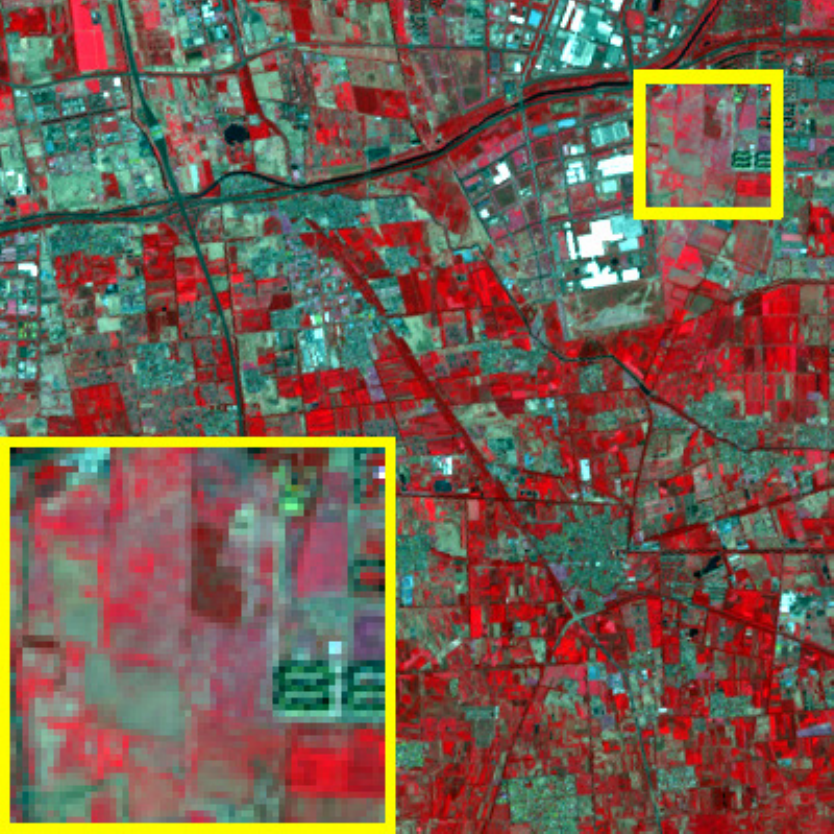}} 
        \end{minipage}
        \begin{minipage}{0.24\hsize}
            \centerline{\includegraphics[width=\hsize]{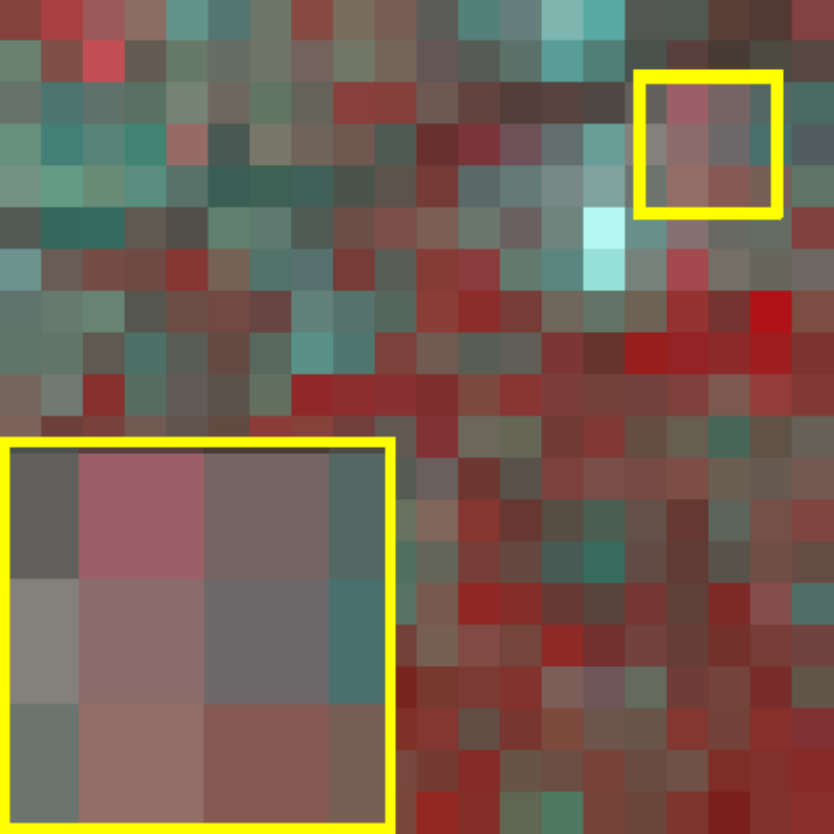}} 
        \end{minipage}
        \begin{minipage}{0.24\hsize}
            \centerline{\includegraphics[width=\hsize]{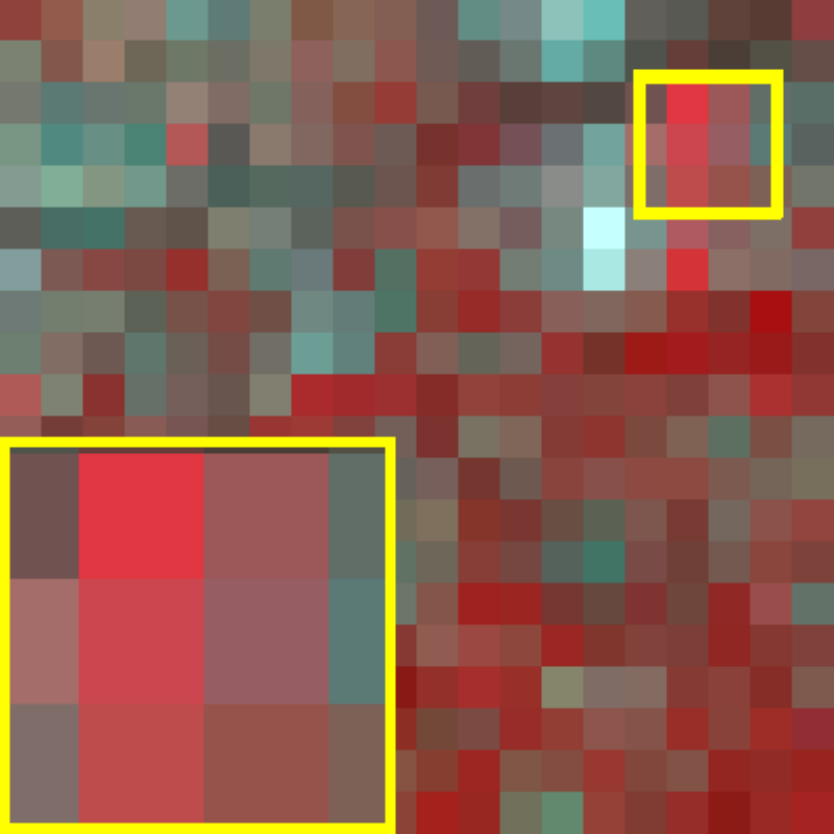}} 
        \end{minipage} \\
        \vspace{1mm}
        \begin{minipage}{0.24\hsize} 
            \centerline{$\Hr$}
        \end{minipage}
        \begin{minipage}{0.24\hsize} 
            \centerline{$\Lr$}
        \end{minipage}
        \begin{minipage}{0.24\hsize} 
            \centerline{$\Lt$}
        \end{minipage} \\
        \vspace{2mm}
        \begin{minipage}{0.24\hsize}
            \centerline{\includegraphics[width=\hsize]{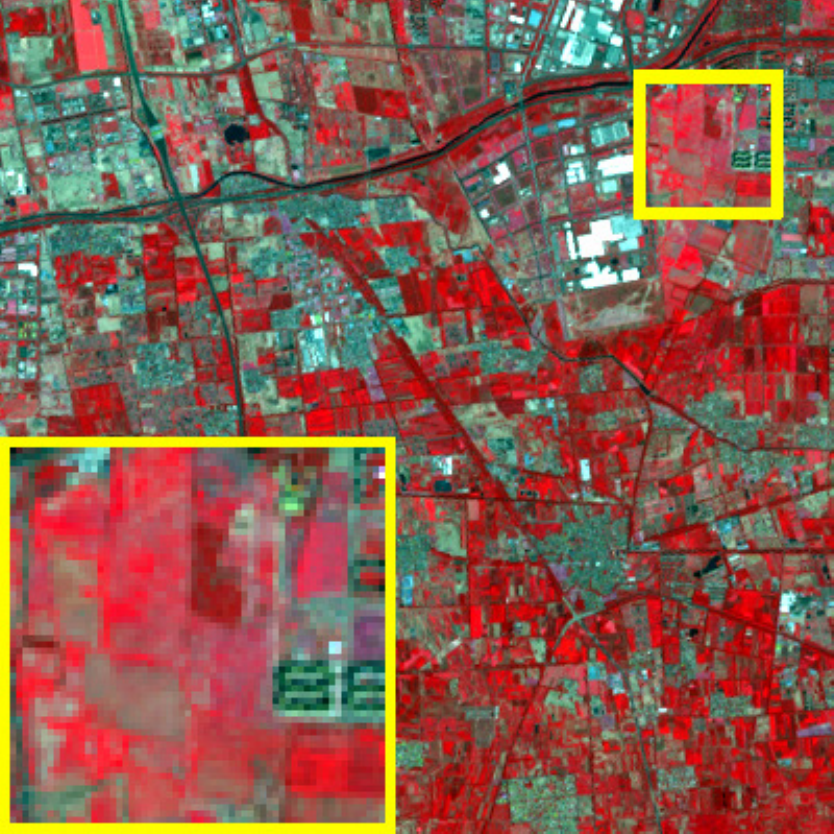}} 
        \end{minipage} 
        \begin{minipage}{0.24\hsize}
            \centerline{\includegraphics[width=\hsize]{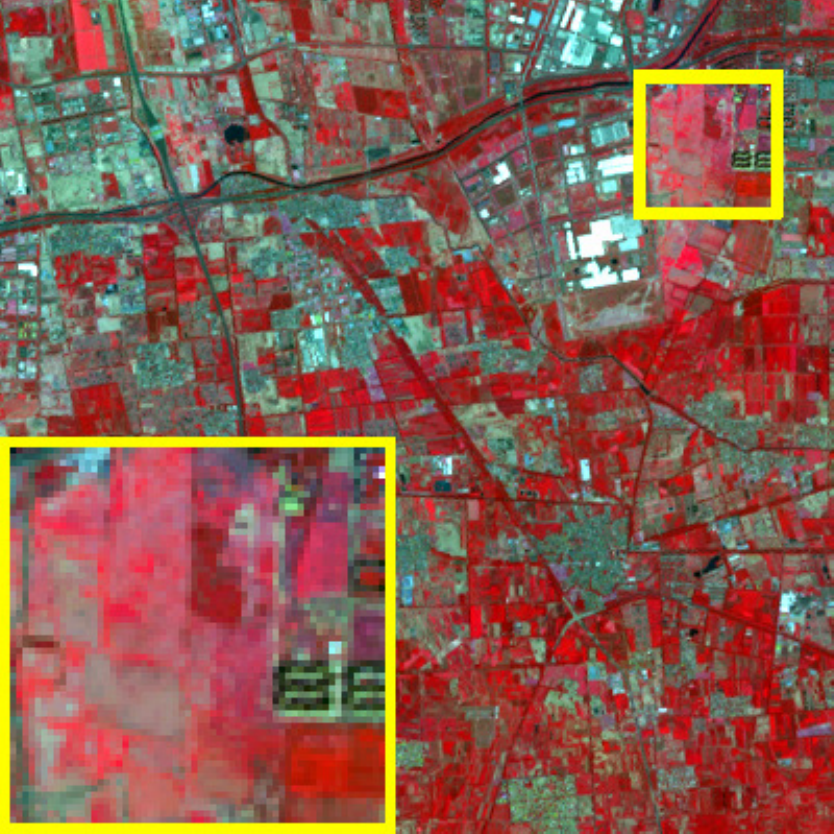}} 
        \end{minipage}
        \begin{minipage}{0.24\hsize}
            \centerline{\includegraphics[width=\hsize]{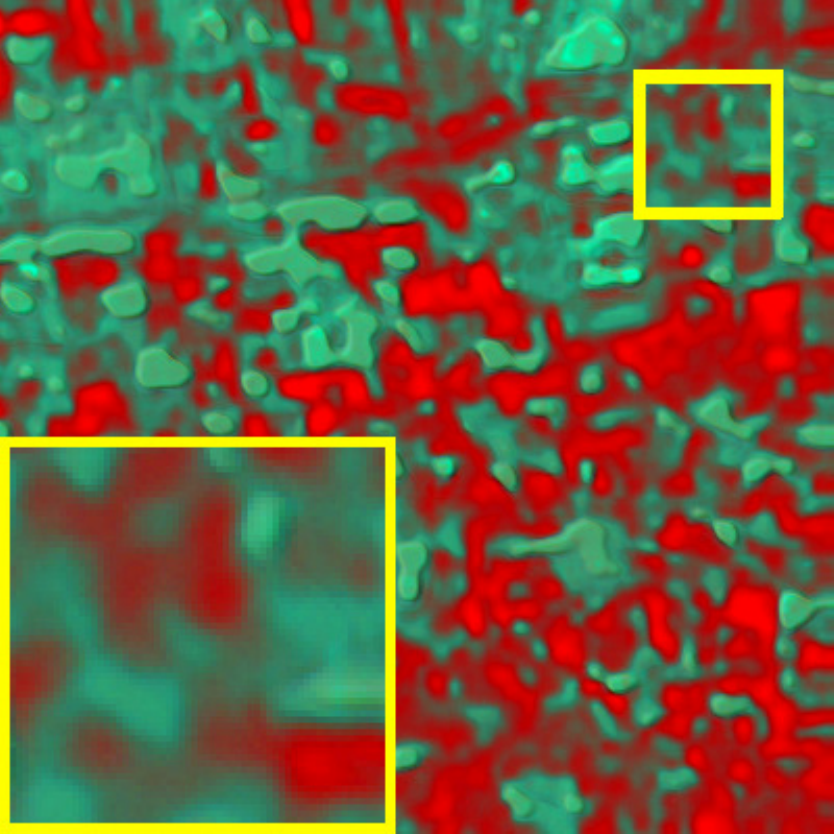}} 
        \end{minipage}
        \begin{minipage}{0.24\hsize}
            \centerline{\includegraphics[width=\hsize]{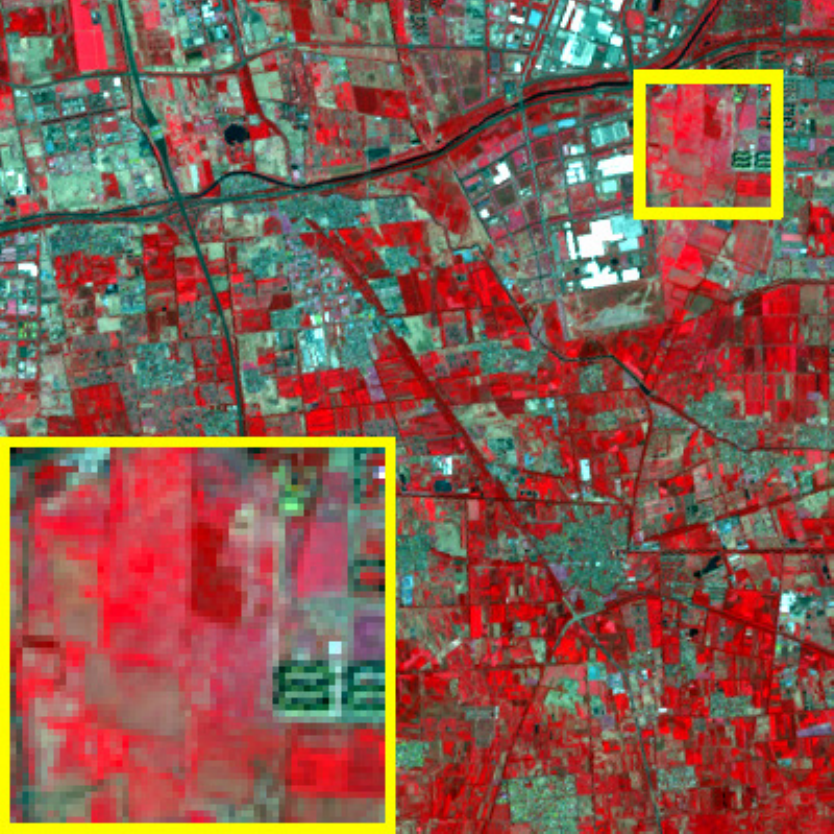}} 
        \end{minipage} \\
        \vspace{1mm}
        \begin{minipage}{0.24\hsize} 
            \centerline{STARFM}
        \end{minipage} 
        \begin{minipage}{0.24\hsize} 
            \centerline{VIPSTF}
        \end{minipage}
        \begin{minipage}{0.24\hsize} 
            \centerline{RSFN}
        \end{minipage}
        \begin{minipage}{0.24\hsize} 
            \centerline{RobOt}
        \end{minipage} \\
        \vspace{2mm}
        \begin{minipage}{0.24\hsize}
            \centerline{\includegraphics[width=\hsize]{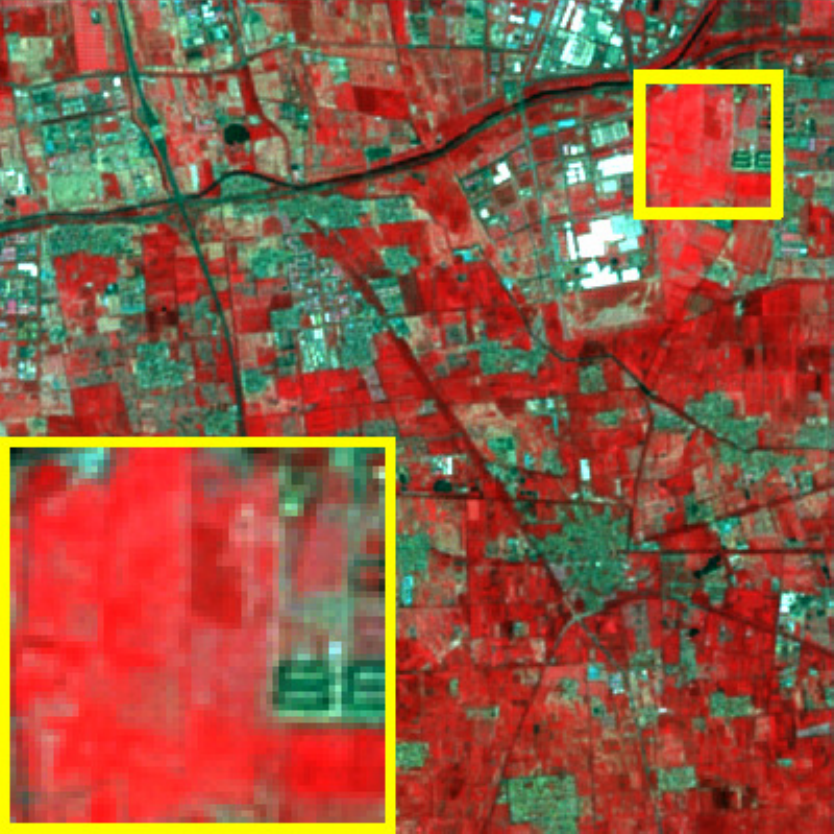}} 
        \end{minipage}
        \begin{minipage}{0.24\hsize}
            \centerline{\includegraphics[width=\hsize]{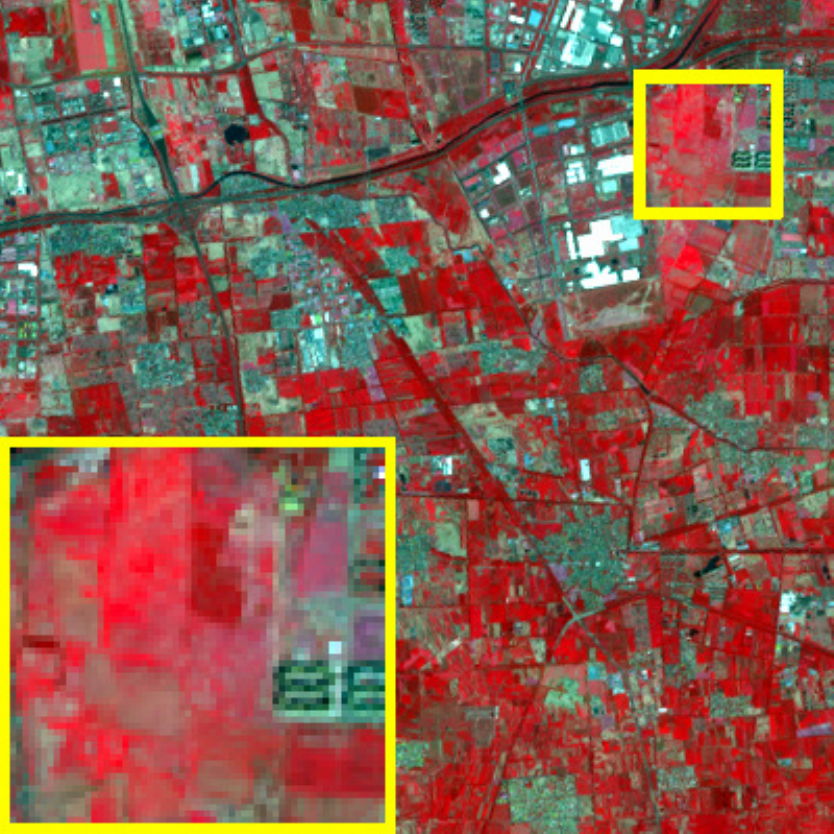}} 
        \end{minipage}
        \begin{minipage}{0.24\hsize}
            \centerline{\includegraphics[width=\hsize]{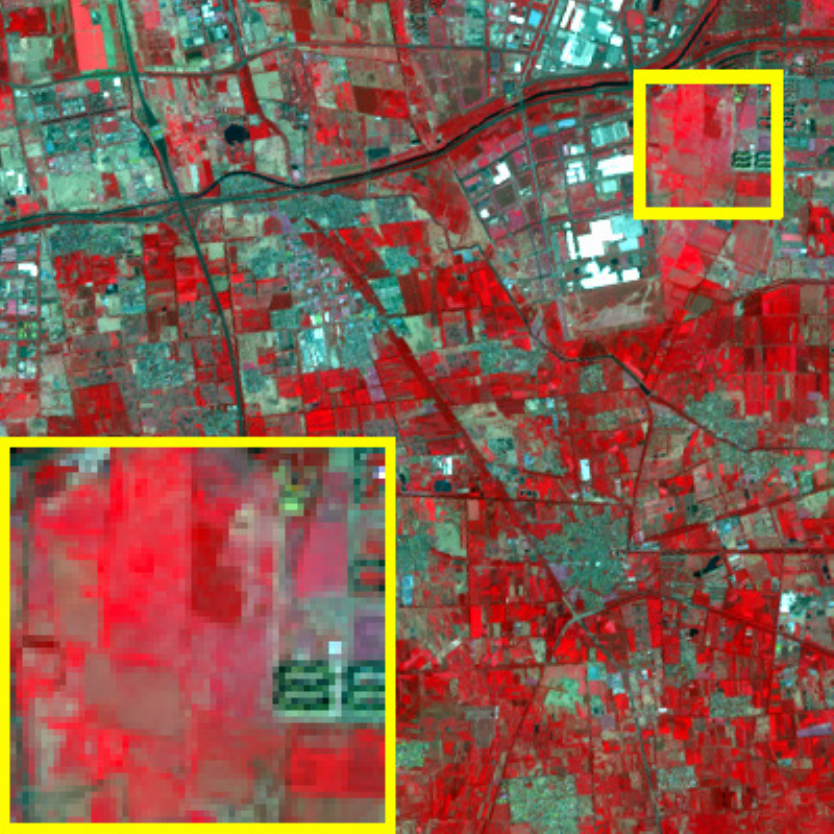}} 
        \end{minipage} 
        \begin{minipage}{0.24\hsize}
            \centerline{\includegraphics[width=\hsize]{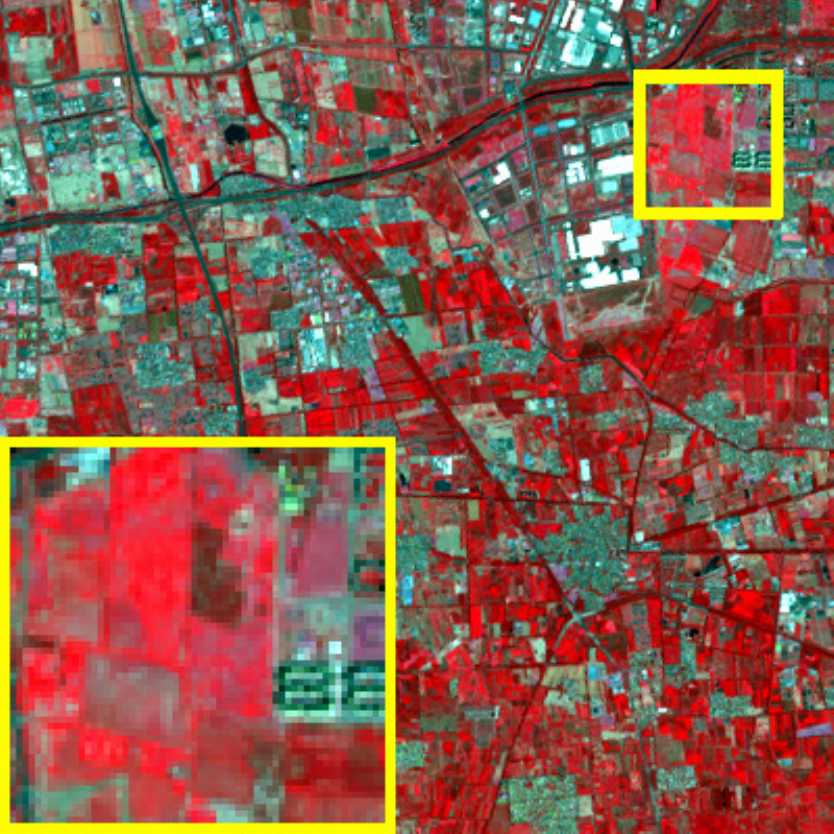}} 
        \end{minipage} \\
        \vspace{1mm}
        \begin{minipage}{0.24\hsize} 
            \centerline{SwinSTFM}
        \end{minipage}
        \begin{minipage}{0.24\hsize} 
            \centerline{\textbf{ROSTF-1}}
        \end{minipage}
        \begin{minipage}{0.24\hsize} 
            \centerline{\textbf{ROSTF-2}}
        \end{minipage}
        \begin{minipage}{0.24\hsize} 
            \centerline{Ground-truth}
        \end{minipage} \\
	\end{center}
        \vspace{-3mm}
	\caption{ST fusion results for the Site1 simulated data in Case1.}
 \label{fig: Site1 SemiSim Case1 results}
\end{figure}

%% file: Experiments/results/Site2_SemiSim_Case1.tex
\begin{figure}[t]
	\begin{center}
        \begin{minipage}{0.24\hsize} 
            \centerline{\includegraphics[width=\hsize]{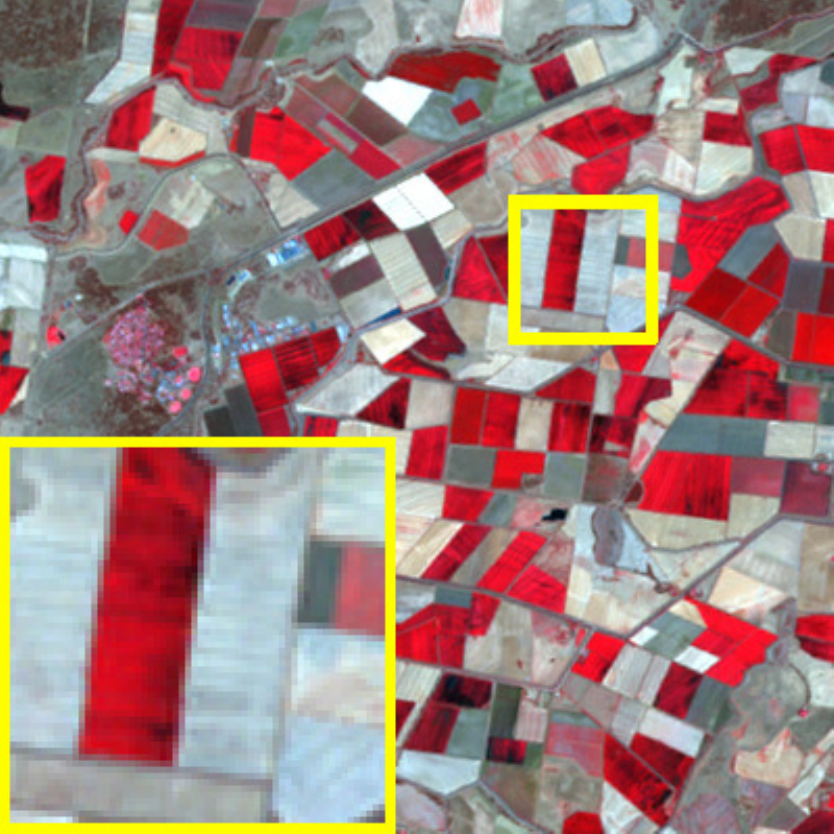}} 
        \end{minipage}
        \begin{minipage}{0.24\hsize}
            \centerline{\includegraphics[width=\hsize]{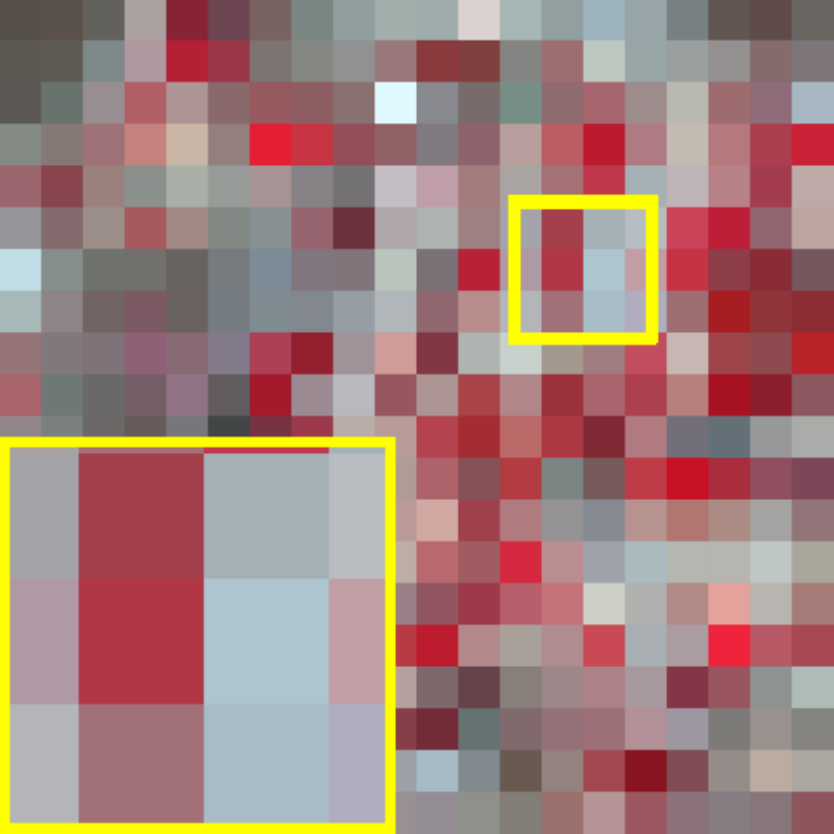}} 
        \end{minipage}
        \begin{minipage}{0.24\hsize}
            \centerline{\includegraphics[width=\hsize]{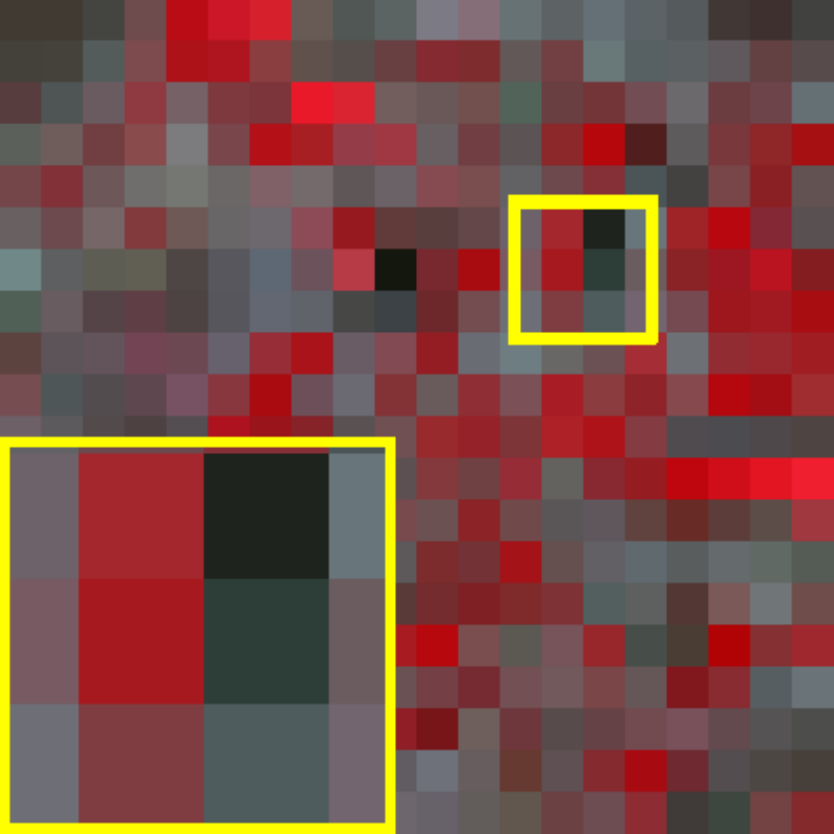}} 
        \end{minipage} \\
        \vspace{1mm}
        \begin{minipage}{0.24\hsize} 
            \centerline{$\Hr$}
        \end{minipage}
        \begin{minipage}{0.24\hsize} 
            \centerline{$\Lr$}
        \end{minipage}
        \begin{minipage}{0.24\hsize} 
            \centerline{$\Lt$}
        \end{minipage} \\
        \vspace{2mm}
        \begin{minipage}{0.24\hsize}
            \centerline{\includegraphics[width=\hsize]{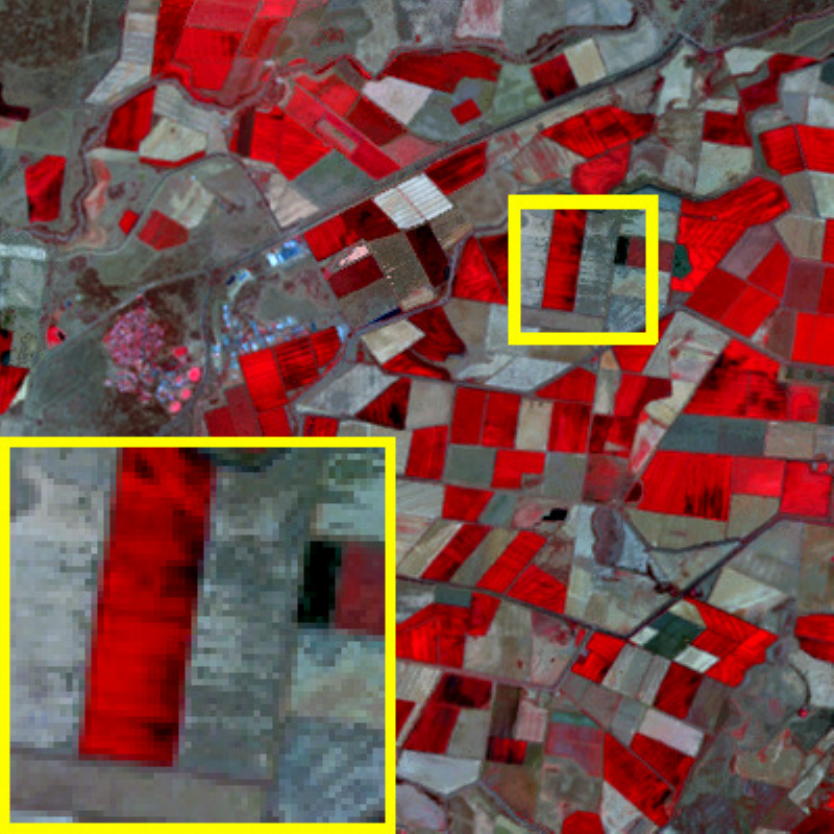}} 
        \end{minipage} 
        \begin{minipage}{0.24\hsize}
            \centerline{\includegraphics[width=\hsize]{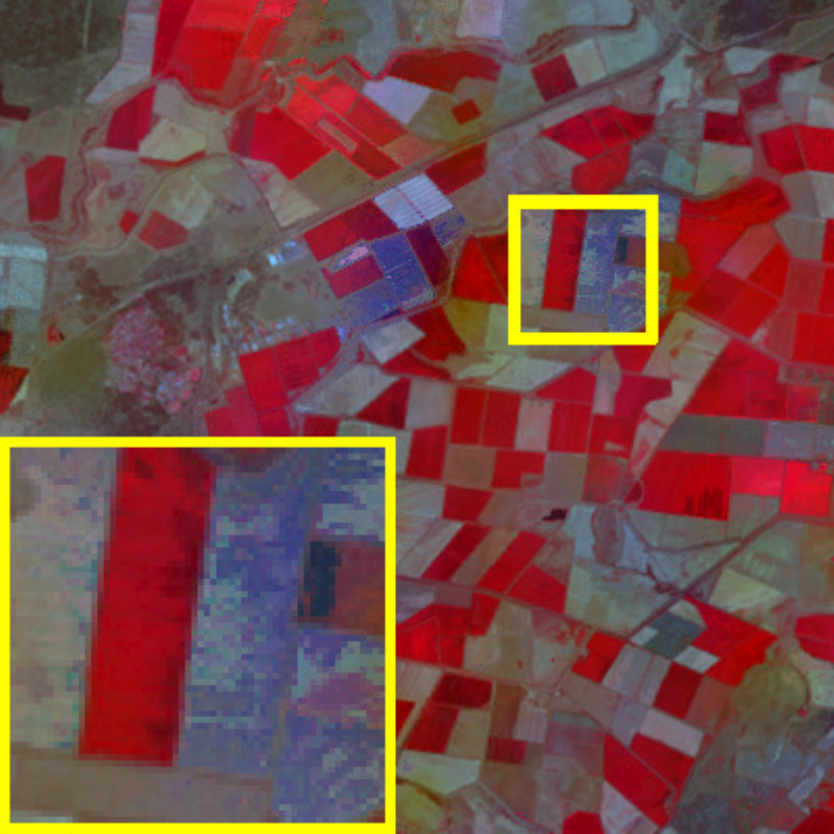}} 
        \end{minipage}
        \begin{minipage}{0.24\hsize}
            \centerline{\includegraphics[width=\hsize]{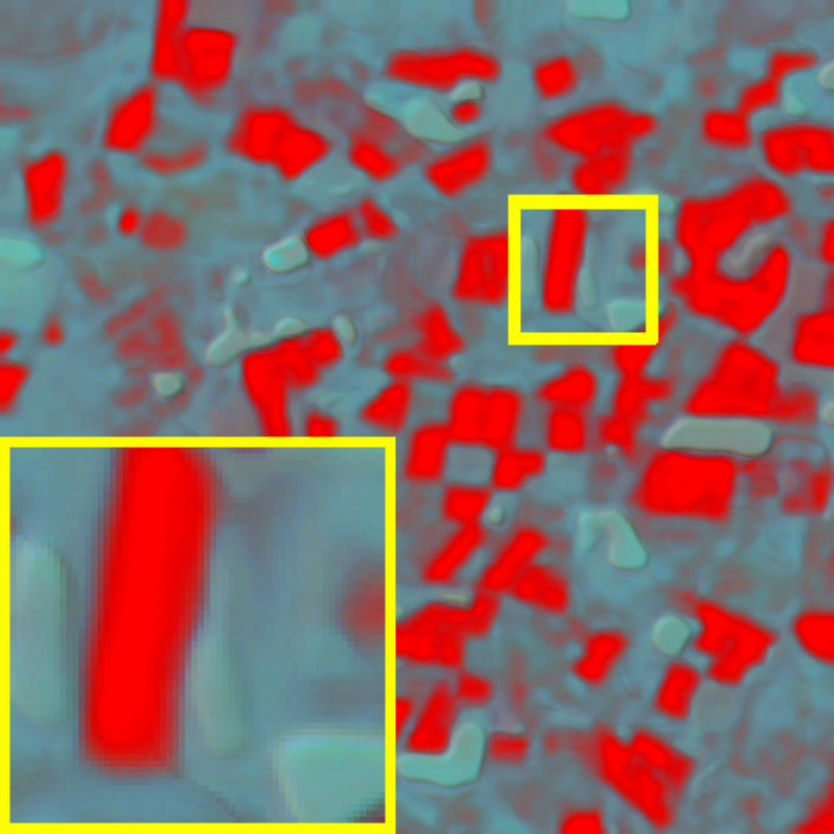}} 
        \end{minipage}
        \begin{minipage}{0.24\hsize}
            \centerline{\includegraphics[width=\hsize]{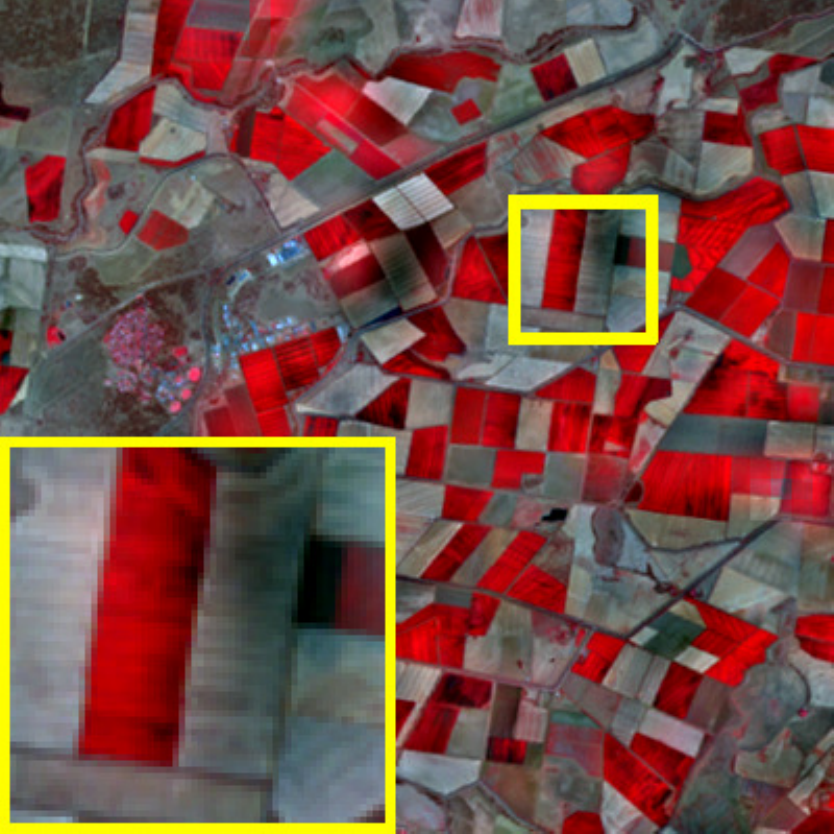}} 
        \end{minipage} \\
        \vspace{1mm}
        \begin{minipage}{0.24\hsize} 
            \centerline{STARFM}
        \end{minipage} 
        \begin{minipage}{0.24\hsize} 
            \centerline{VIPSTF}
        \end{minipage}
        \begin{minipage}{0.24\hsize} 
            \centerline{RSFN}
        \end{minipage}
        \begin{minipage}{0.24\hsize} 
            \centerline{RobOt}
        \end{minipage} \\
        \vspace{2mm}
        \begin{minipage}{0.24\hsize}
            \centerline{\includegraphics[width=\hsize]{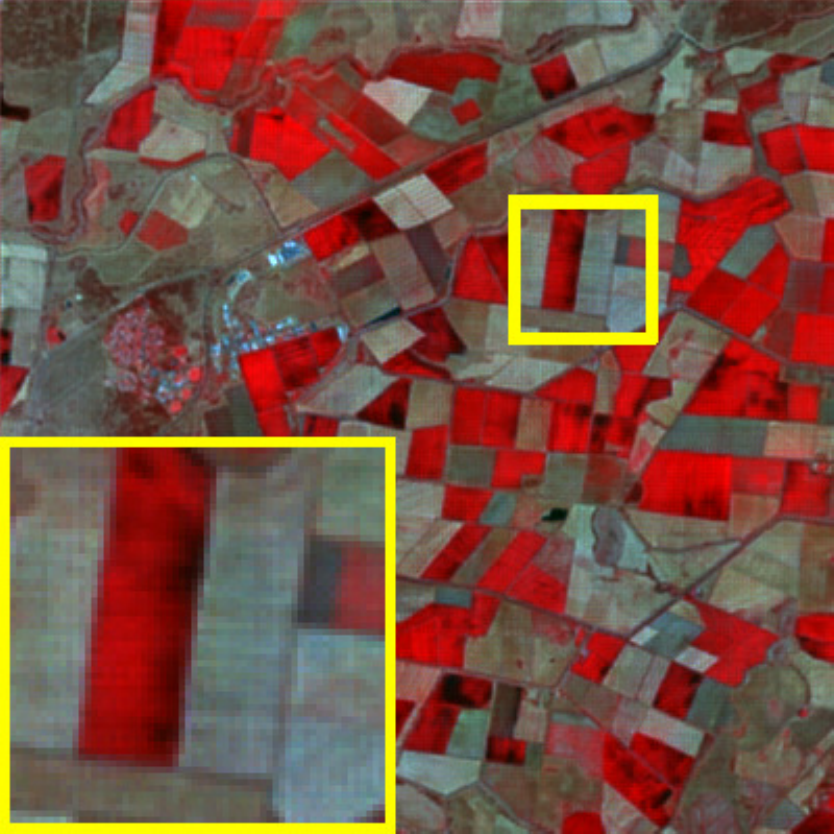}} 
        \end{minipage}
        \begin{minipage}{0.24\hsize}
            \centerline{\includegraphics[width=\hsize]{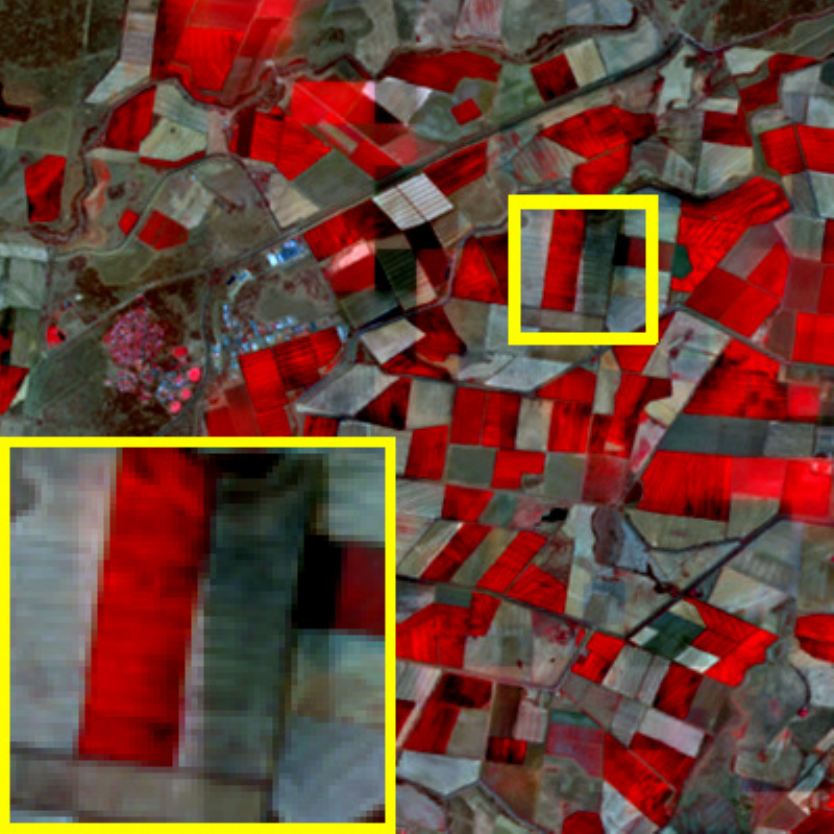}} 
        \end{minipage}
        \begin{minipage}{0.24\hsize}
            \centerline{\includegraphics[width=\hsize]{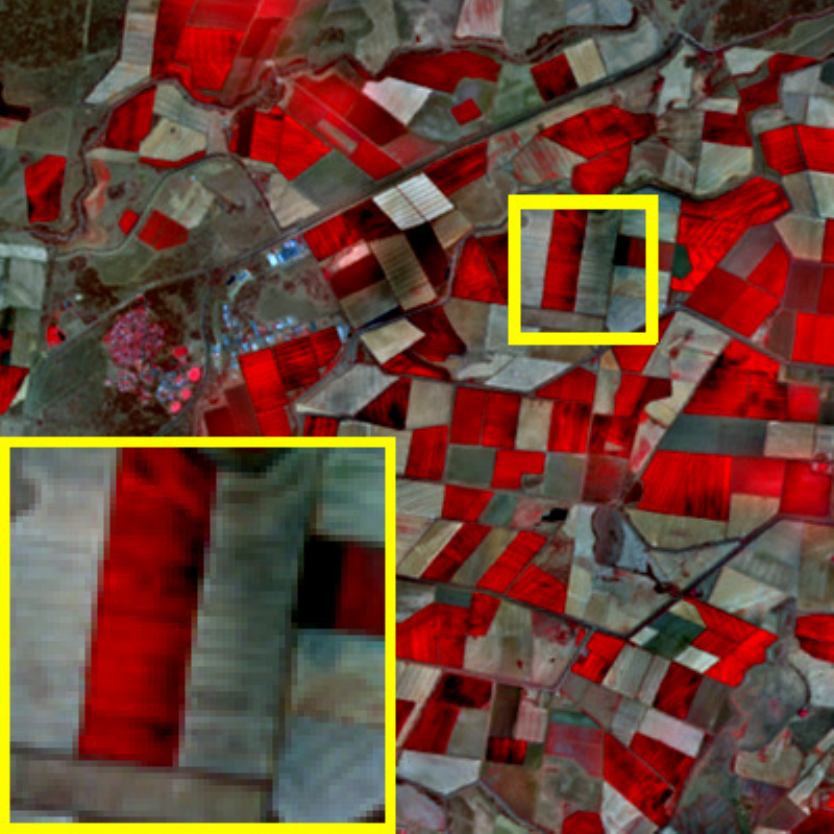}} 
        \end{minipage} 
        \begin{minipage}{0.24\hsize}
            \centerline{\includegraphics[width=\hsize]{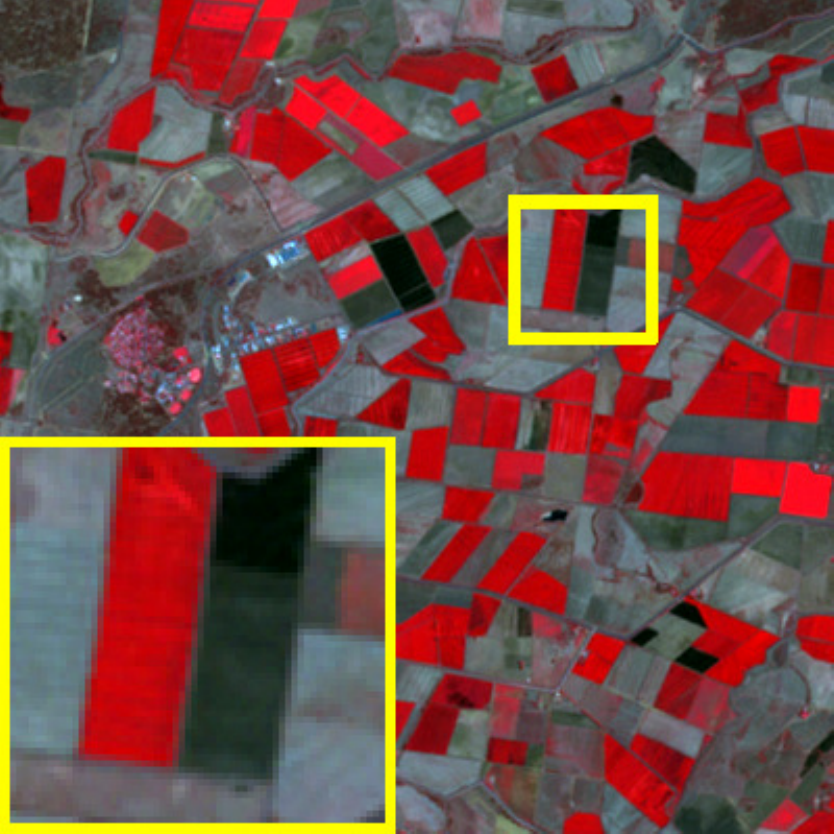}} 
        \end{minipage} \\
        \vspace{1mm}
        \begin{minipage}{0.24\hsize} 
            \centerline{SwinSTFM}
        \end{minipage}
        \begin{minipage}{0.24\hsize} 
            \centerline{\textbf{ROSTF-1}}
        \end{minipage}
        \begin{minipage}{0.24\hsize} 
            \centerline{\textbf{ROSTF-2}}
        \end{minipage}
        \begin{minipage}{0.24\hsize} 
            \centerline{Ground-truth}
        \end{minipage} \\
	\end{center}
        \vspace{-3mm}
	\caption{ST fusion results for the Site2 simulated data in Case1.}
 \label{fig: Site2 SemiSim Case1 results}
\end{figure}

%% file: Experiments/results/Site2_SemiSim_Case234.tex
\begin{figure*}[ht]
	\begin{center}
        \begin{minipage}{0.01\hsize}
            \centerline{\rotatebox{90}{Case2}}
        \end{minipage}
        \begin{minipage}{0.1\hsize} 
            \centerline{\includegraphics[width=\hsize]{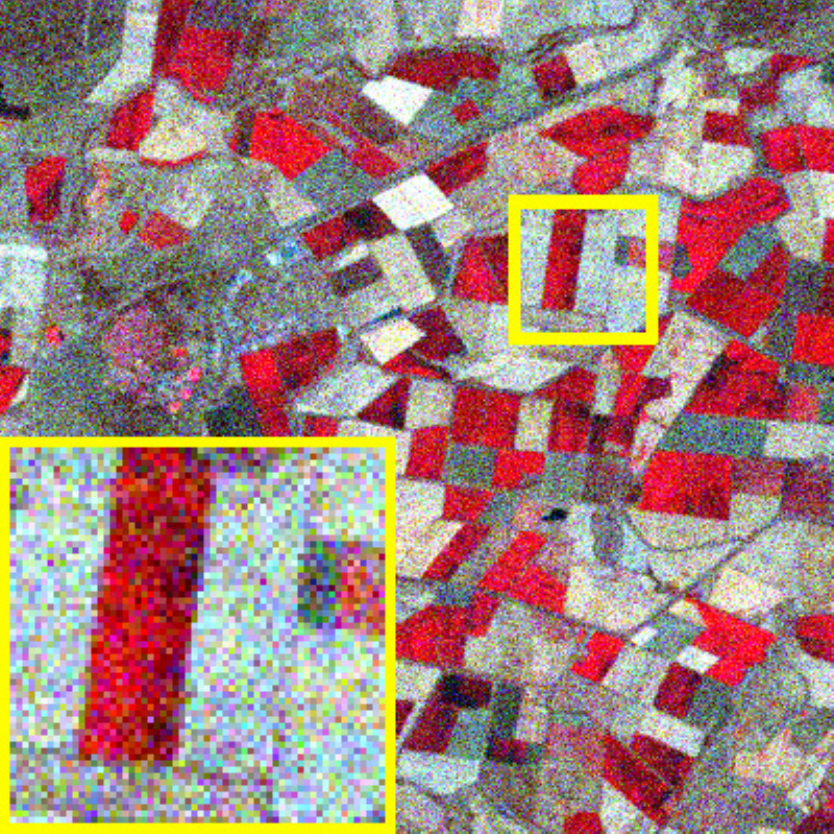}} 
        \end{minipage}
        \begin{minipage}{0.1\hsize}
            \centerline{\includegraphics[width=\hsize]{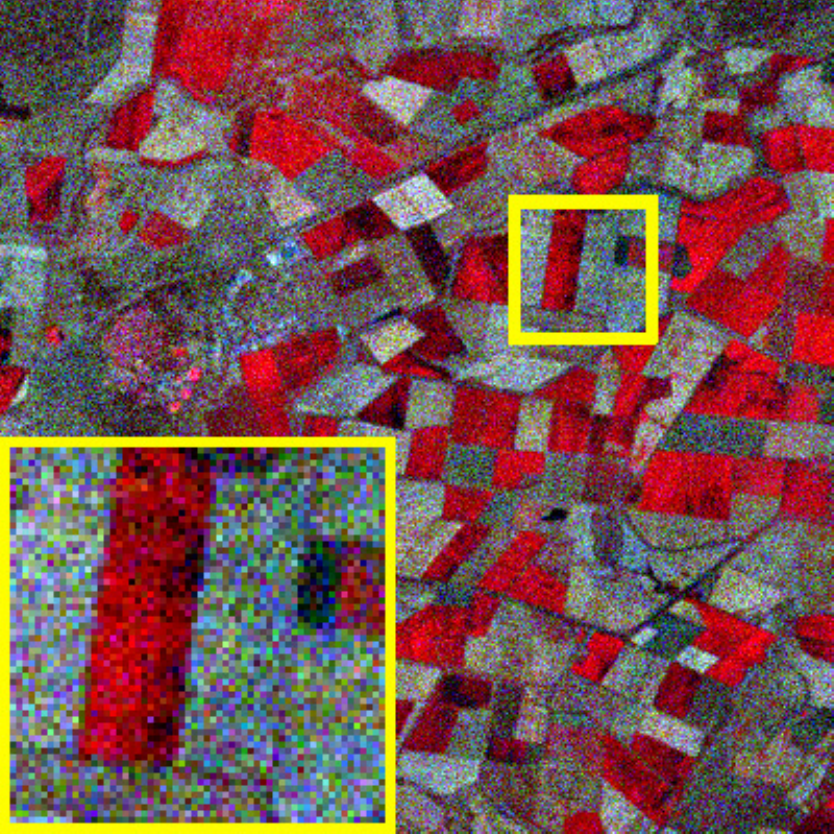}} 
        \end{minipage}
        \begin{minipage}{0.1\hsize}
            \centerline{\includegraphics[width=\hsize]{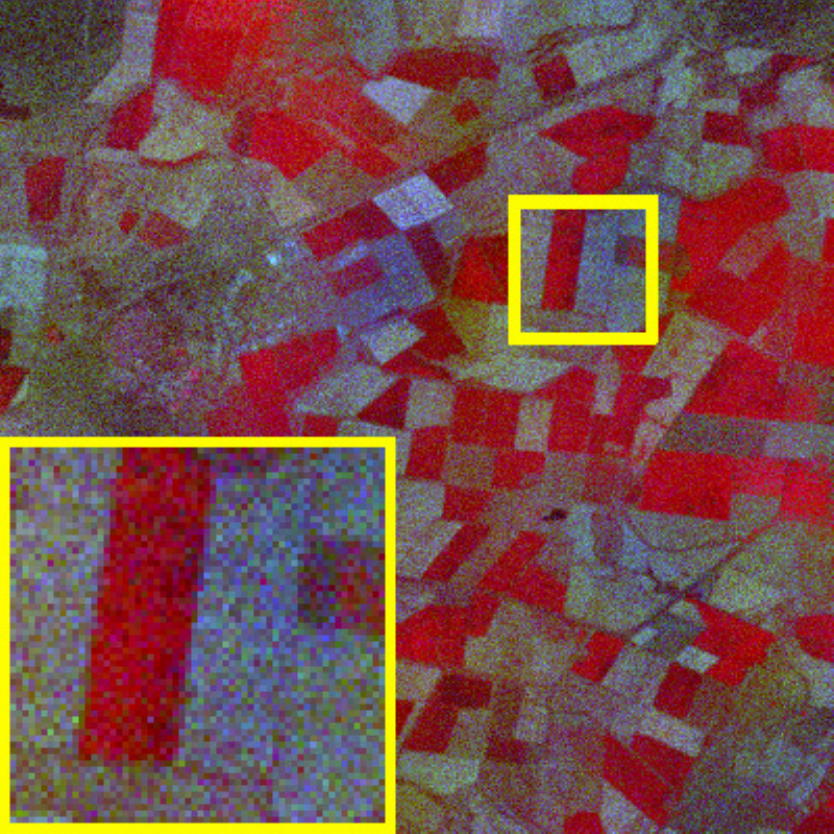}} 
        \end{minipage}
        \begin{minipage}{0.1\hsize}
            \centerline{\includegraphics[width=\hsize]{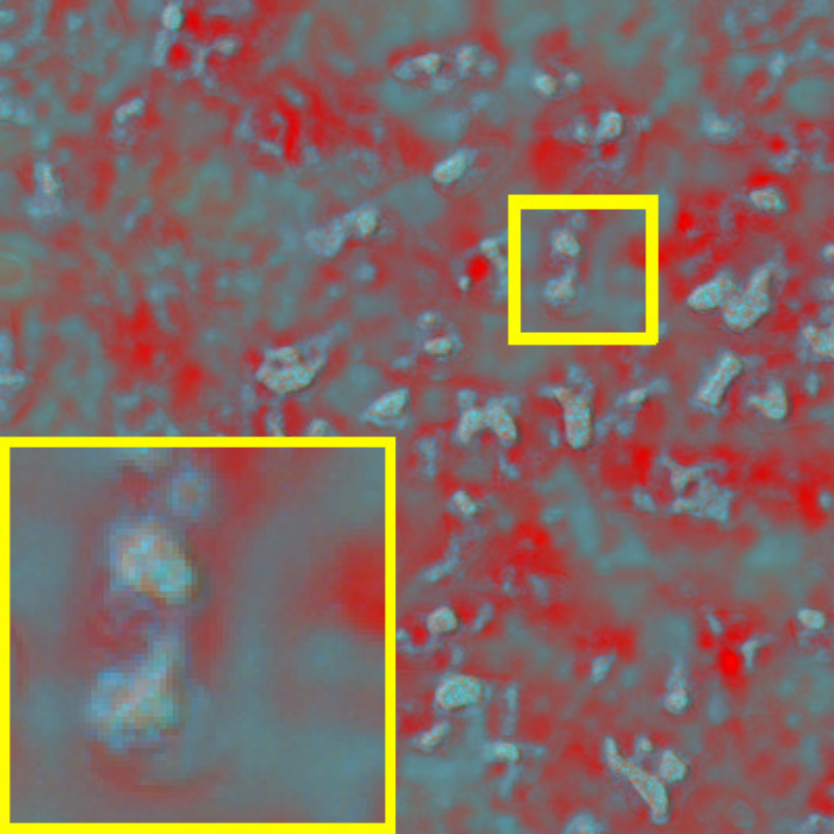}} 
        \end{minipage}
        \begin{minipage}{0.1\hsize}
            \centerline{\includegraphics[width=\hsize]{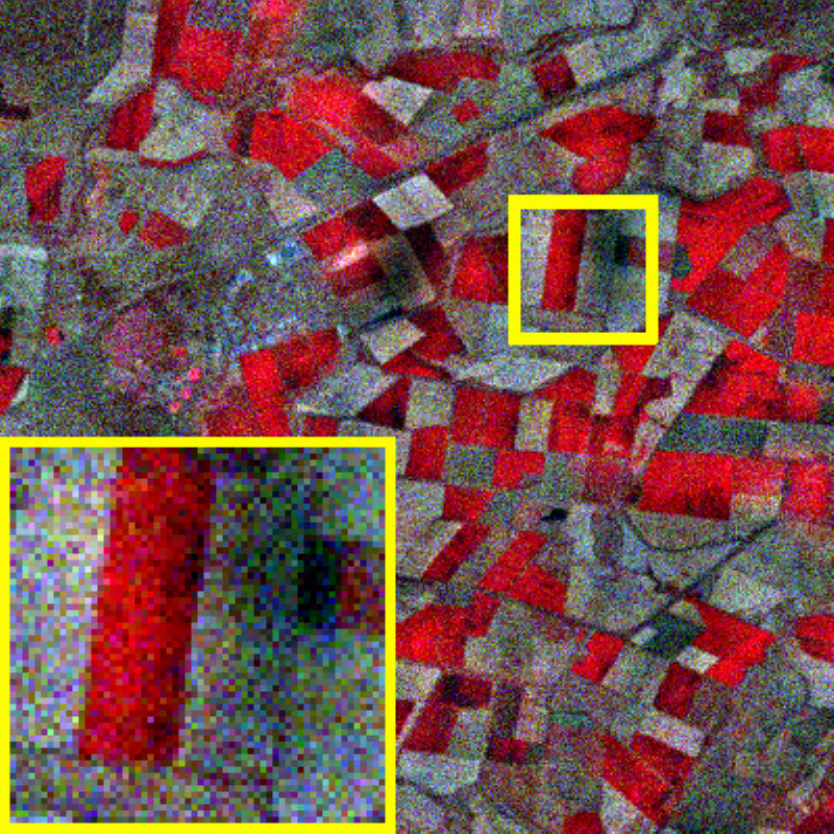}} 
        \end{minipage}
        \begin{minipage}{0.1\hsize}
            \centerline{\includegraphics[width=\hsize]{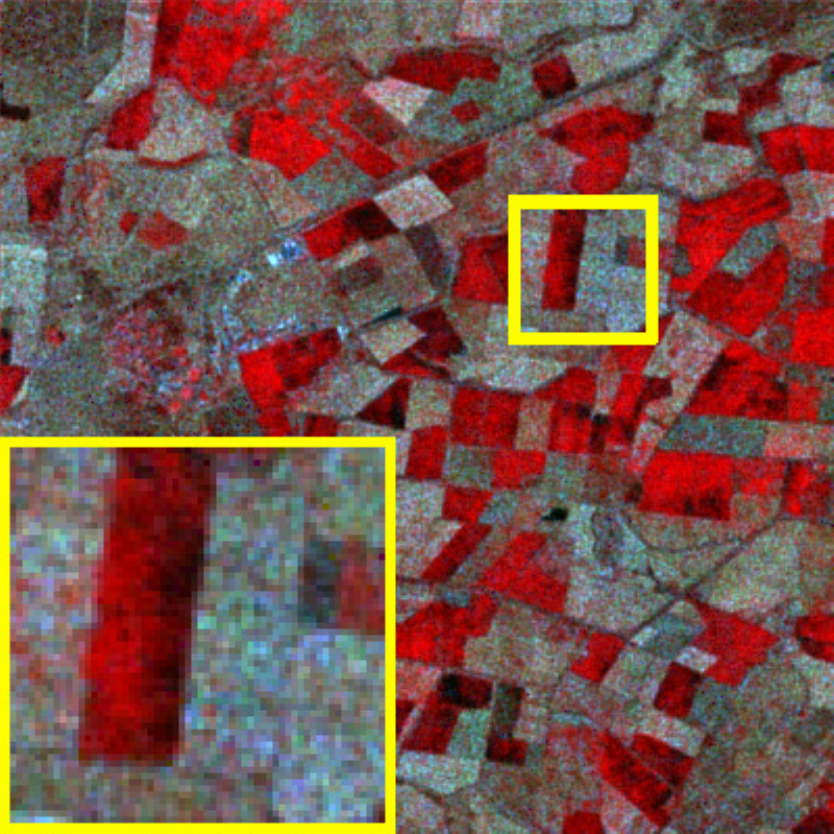}} 
        \end{minipage}
        \begin{minipage}{0.1\hsize}
            \centerline{\includegraphics[width=\hsize]{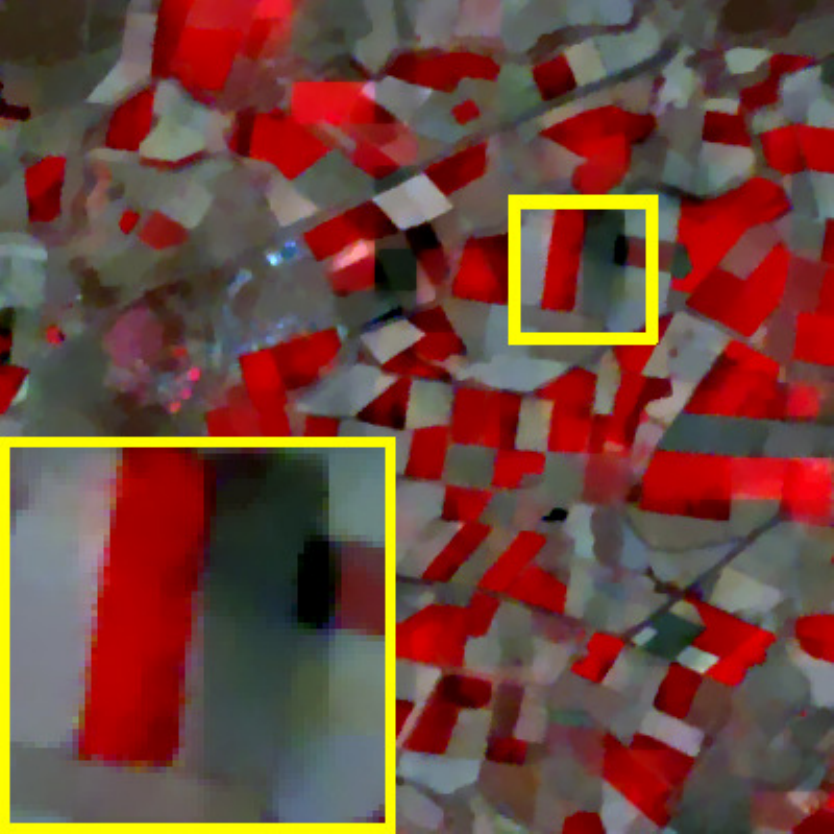}} 
        \end{minipage}
        \begin{minipage}{0.1\hsize}
            \centerline{\includegraphics[width=\hsize]{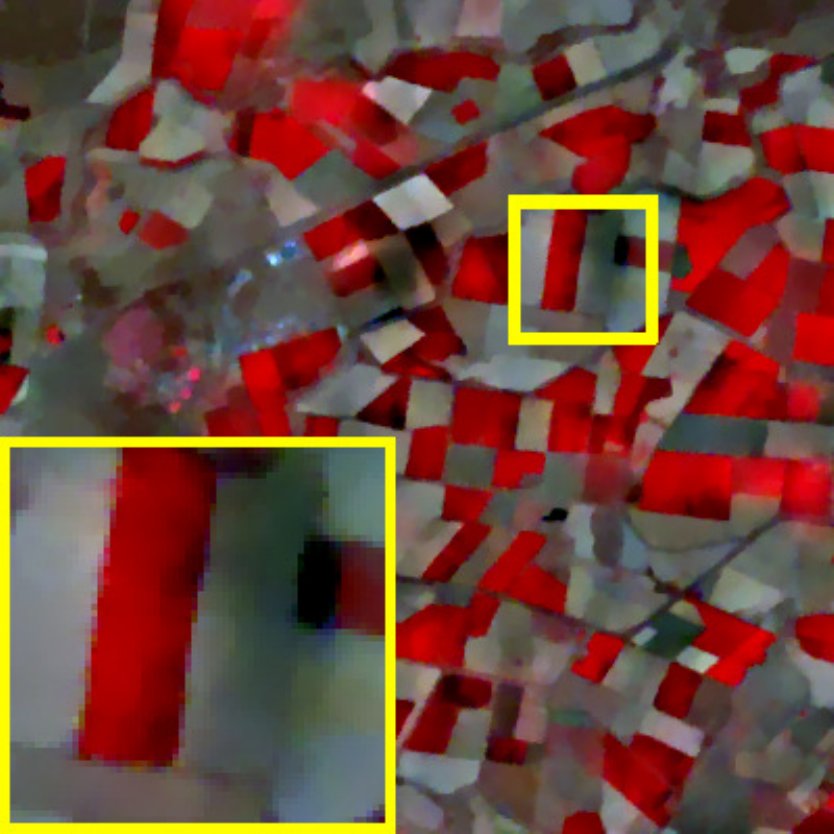}} 
        \end{minipage}
        \begin{minipage}{0.1\hsize}
            \centerline{\includegraphics[width=\hsize]{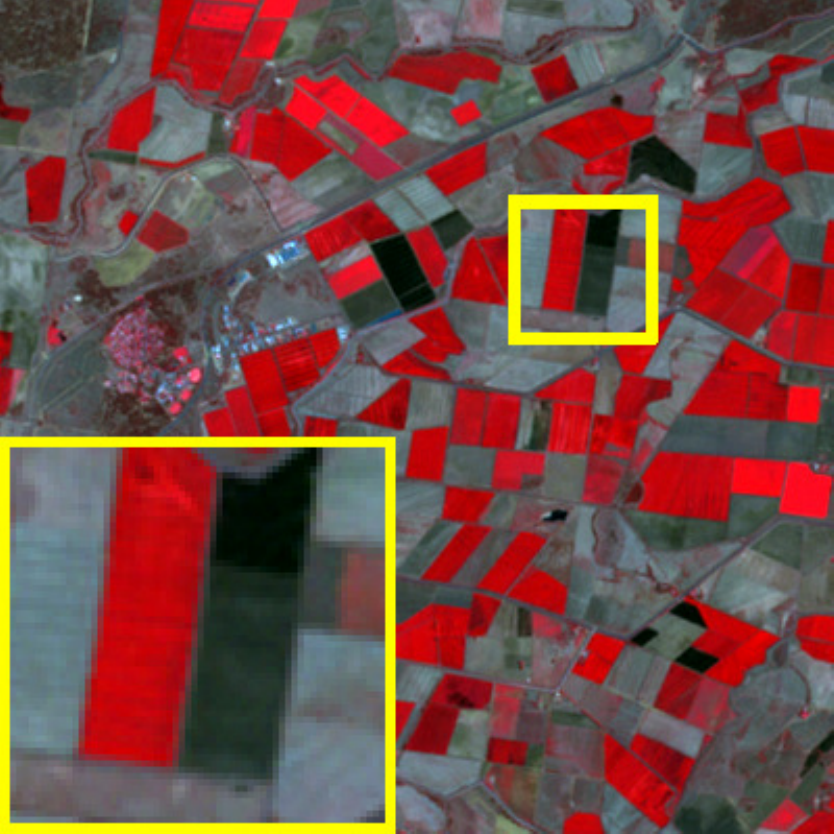}} 
        \end{minipage}  \\

        \vspace{2mm}
        
        \begin{minipage}{0.01\hsize}
			\centerline{\rotatebox{90}{Case3}}
	\end{minipage}
        \begin{minipage}{0.1\hsize} 
            \centerline{\includegraphics[width=\hsize]{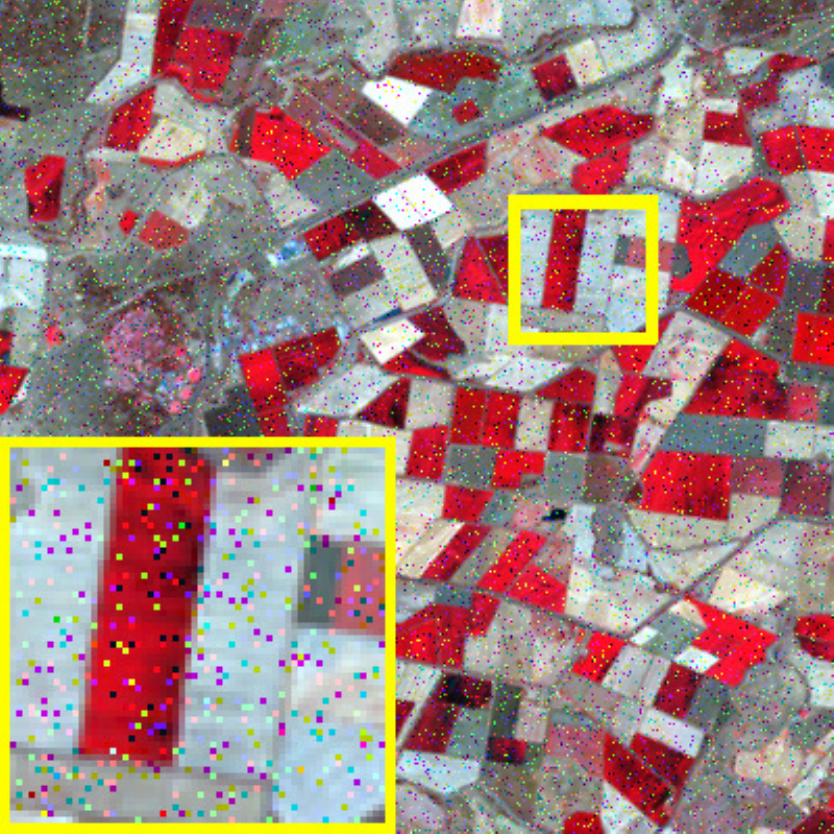}} 
        \end{minipage}
        \begin{minipage}{0.1\hsize}
            \centerline{\includegraphics[width=\hsize]{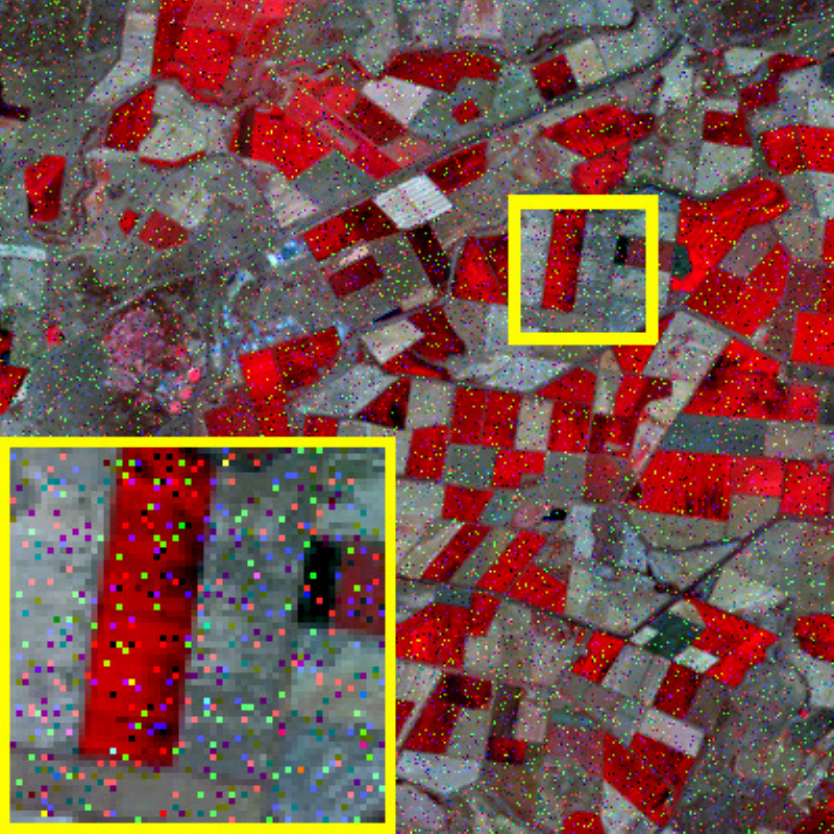}} 
        \end{minipage}
        \begin{minipage}{0.1\hsize}
            \centerline{\includegraphics[width=\hsize]{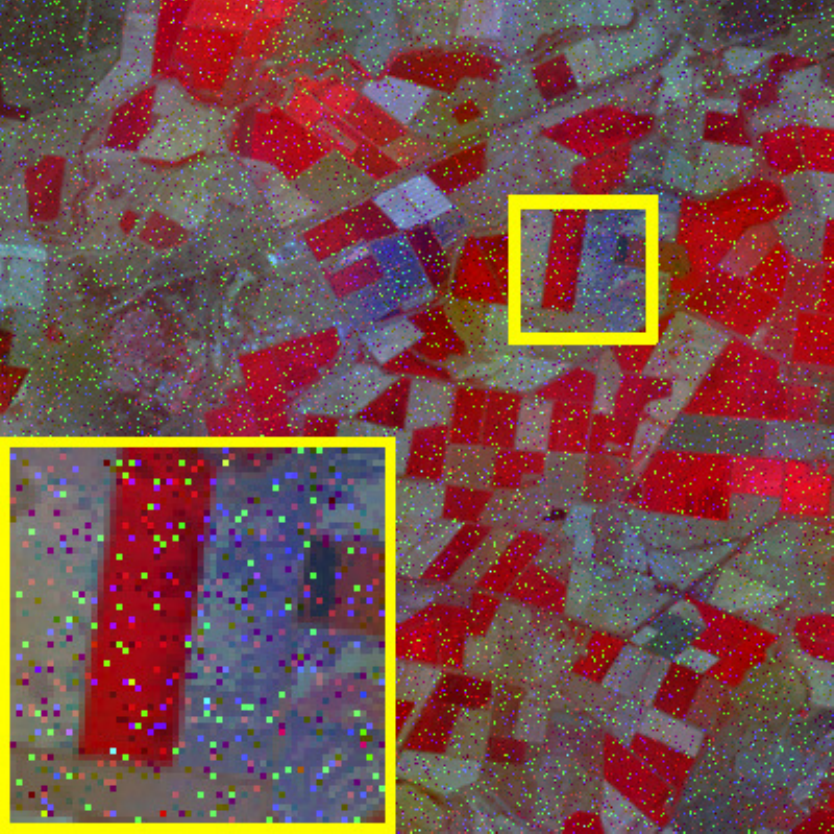}} 
        \end{minipage}
        \begin{minipage}{0.1\hsize}
            \centerline{\includegraphics[width=\hsize]{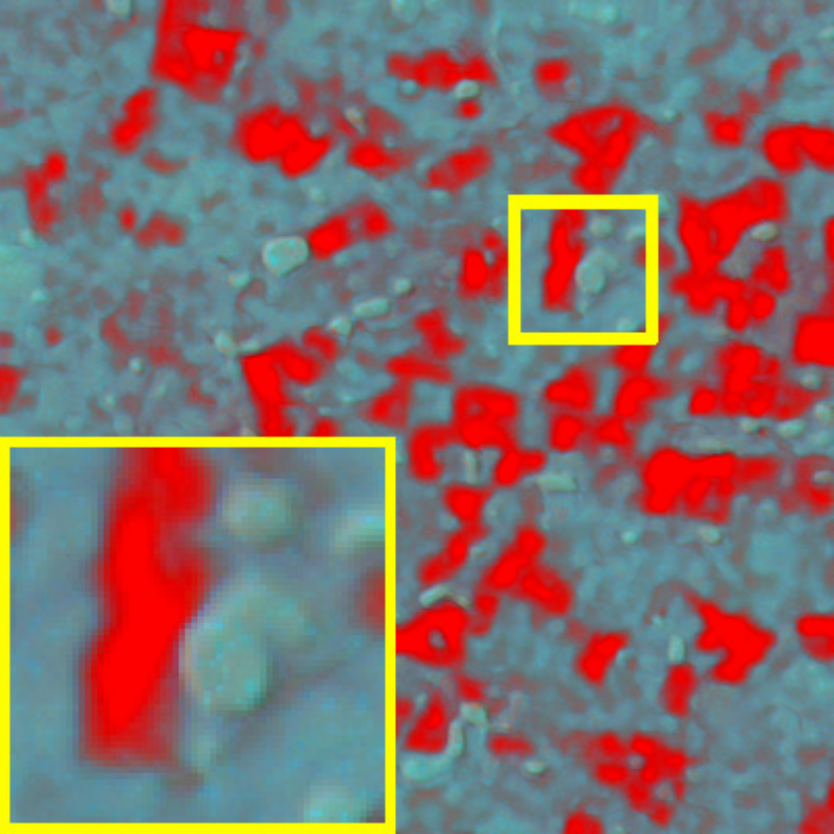}} 
        \end{minipage}
        \begin{minipage}{0.1\hsize}
            \centerline{\includegraphics[width=\hsize]{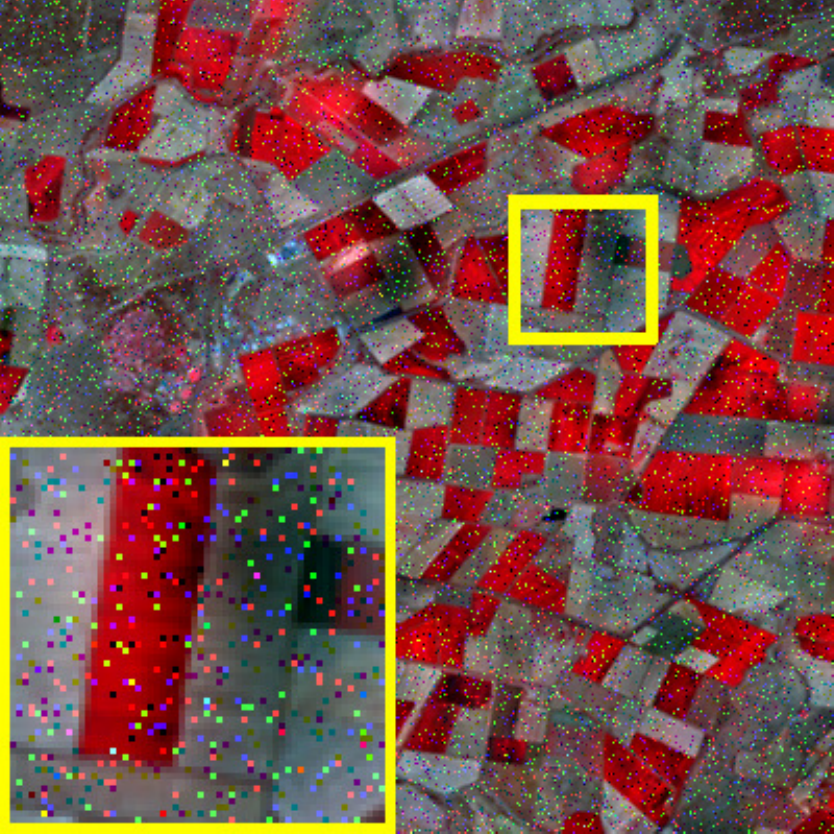}} 
        \end{minipage}
        \begin{minipage}{0.1\hsize}
            \centerline{\includegraphics[width=\hsize]{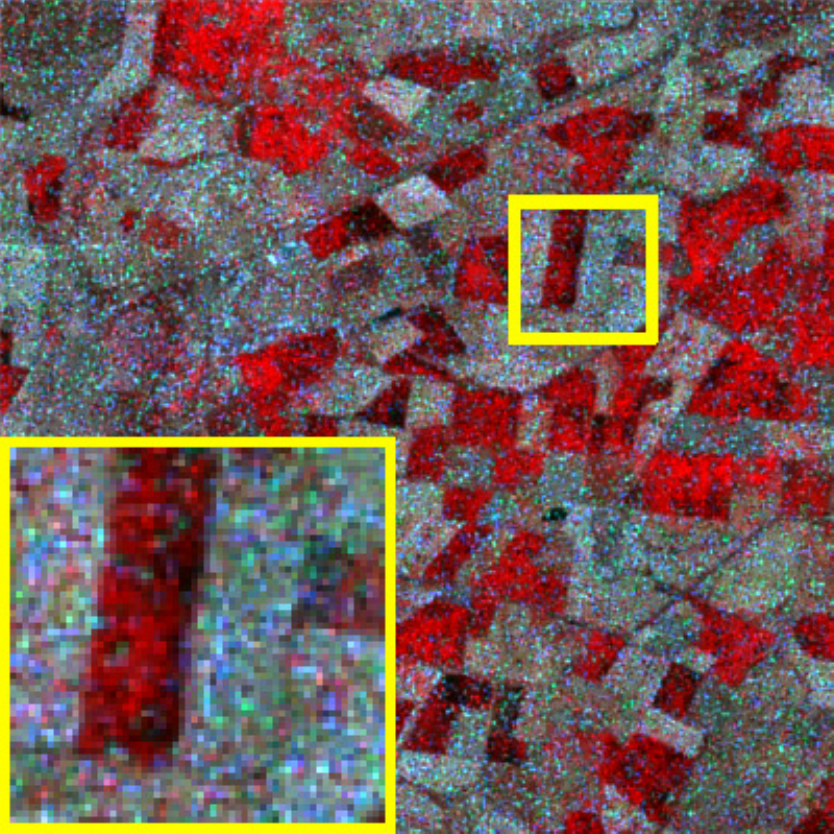}} 
        \end{minipage}
        \begin{minipage}{0.1\hsize}
            \centerline{\includegraphics[width=\hsize]{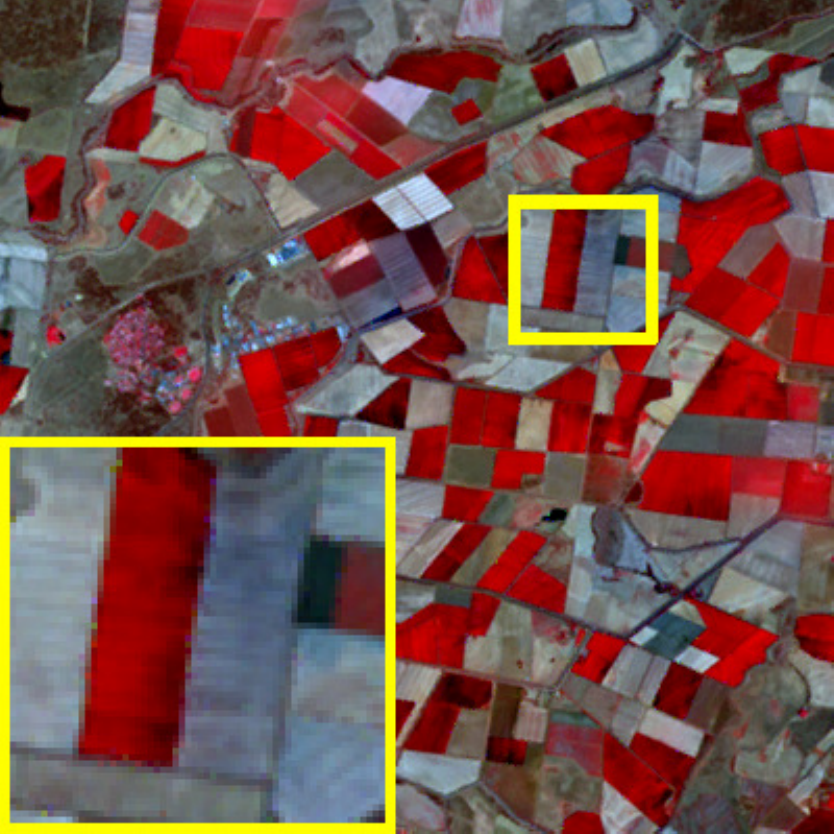}} 
        \end{minipage}
        \begin{minipage}{0.1\hsize}
            \centerline{\includegraphics[width=\hsize]{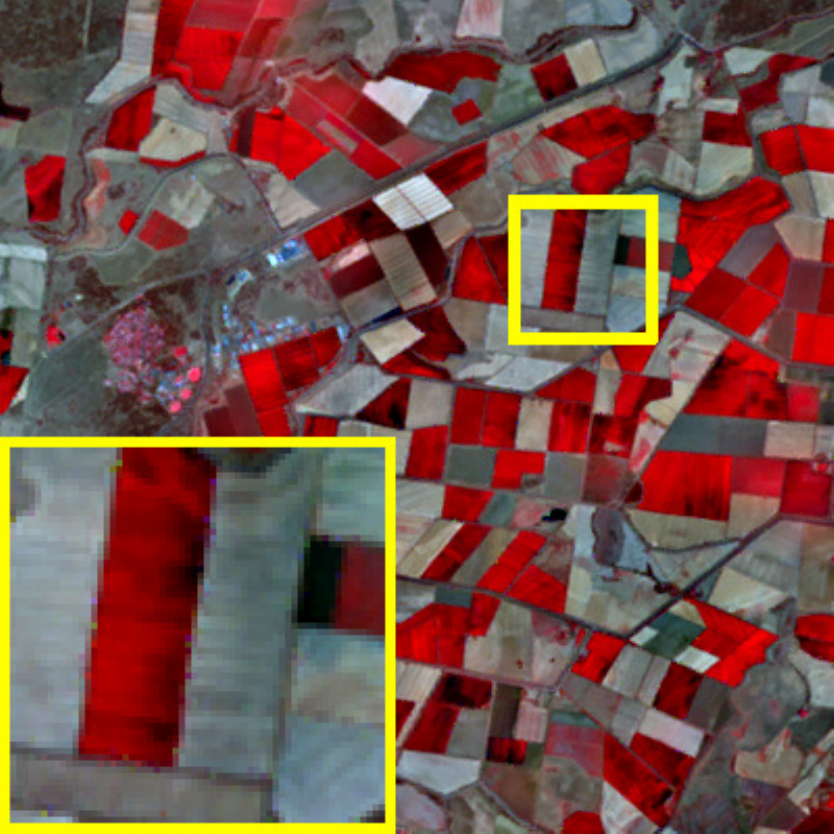}} 
        \end{minipage}
        \begin{minipage}{0.1\hsize}
            \centerline{\includegraphics[width=\hsize]{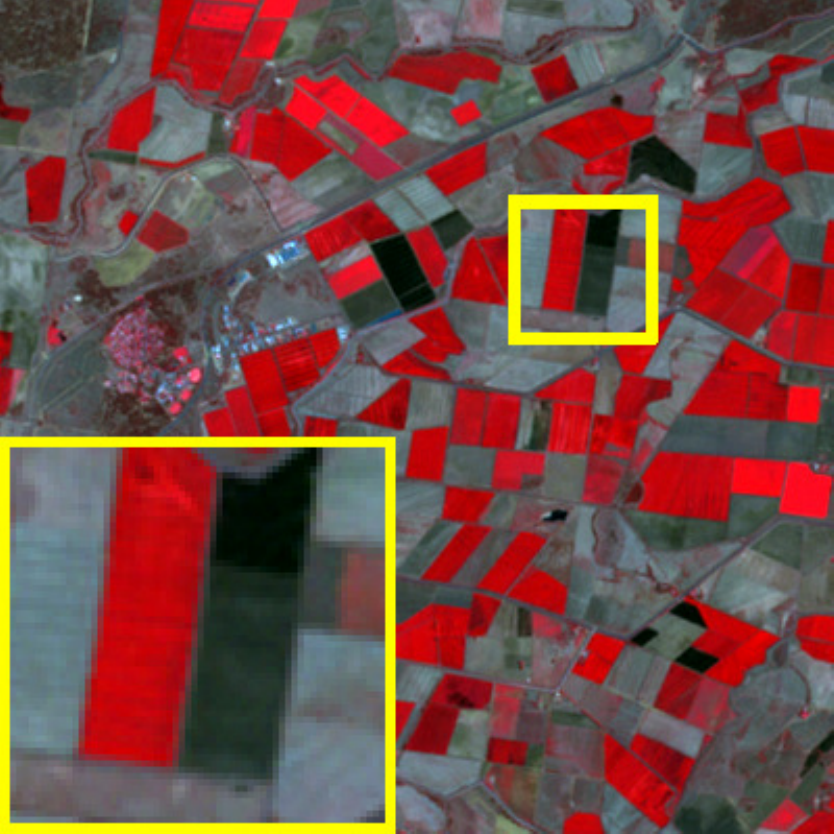}} 
        \end{minipage}  \\

        \vspace{2mm}
  
        \begin{minipage}{0.01\hsize}
			\centerline{\rotatebox{90}{Case4}}
	\end{minipage}
        \begin{minipage}{0.1\hsize} 
            \centerline{\includegraphics[width=\hsize]{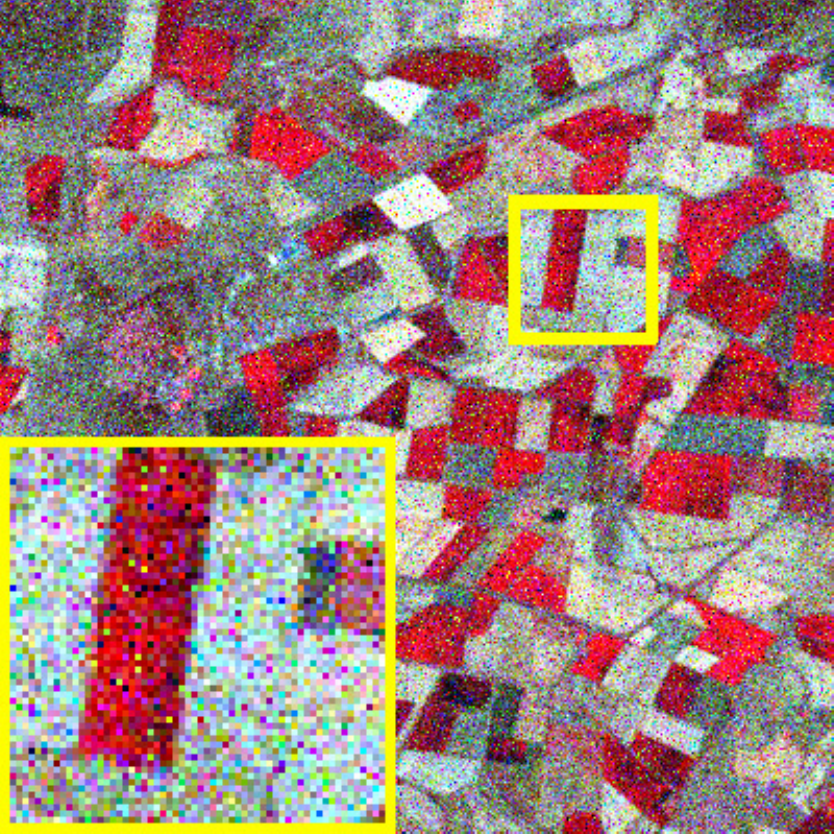}} 
        \end{minipage}
        \begin{minipage}{0.1\hsize}
            \centerline{\includegraphics[width=\hsize]{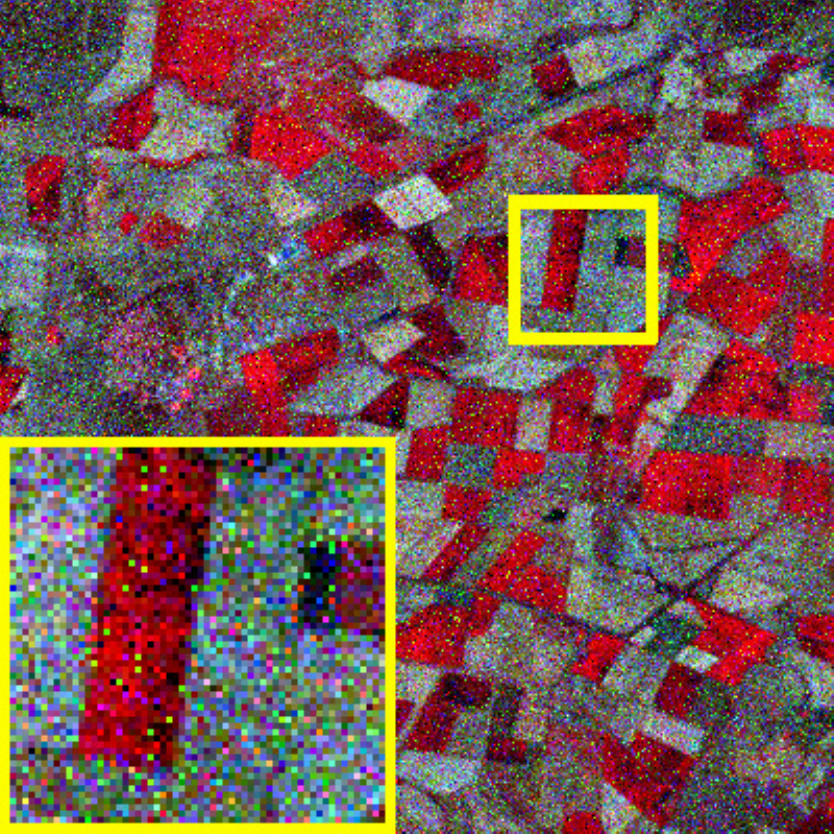}} 
        \end{minipage}
        \begin{minipage}{0.1\hsize}
            \centerline{\includegraphics[width=\hsize]{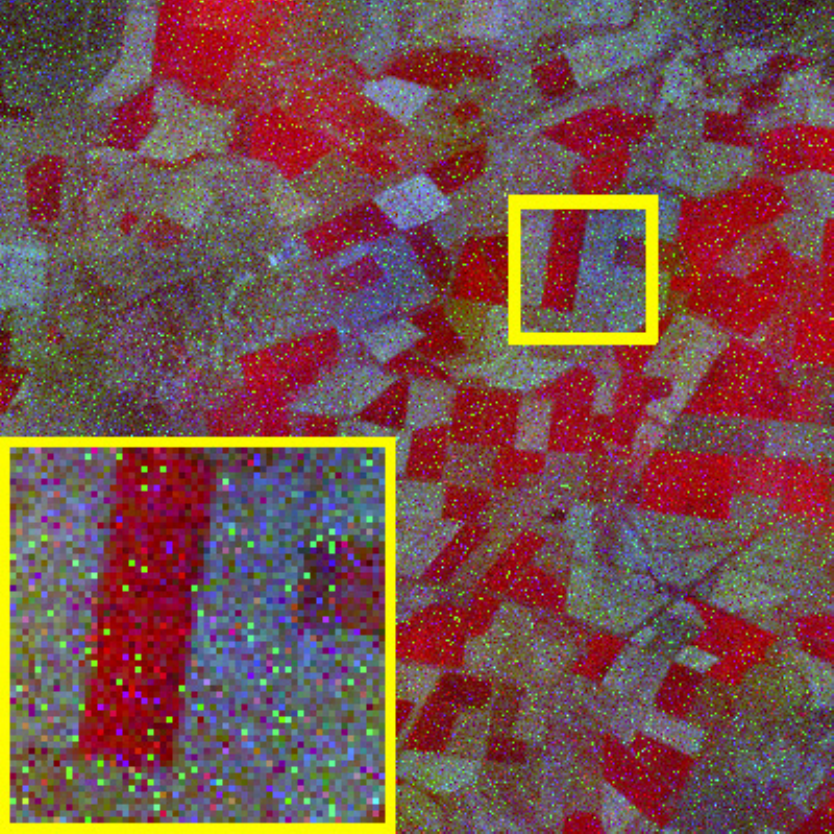}} 
        \end{minipage}
        \begin{minipage}{0.1\hsize}
            \centerline{\includegraphics[width=\hsize]{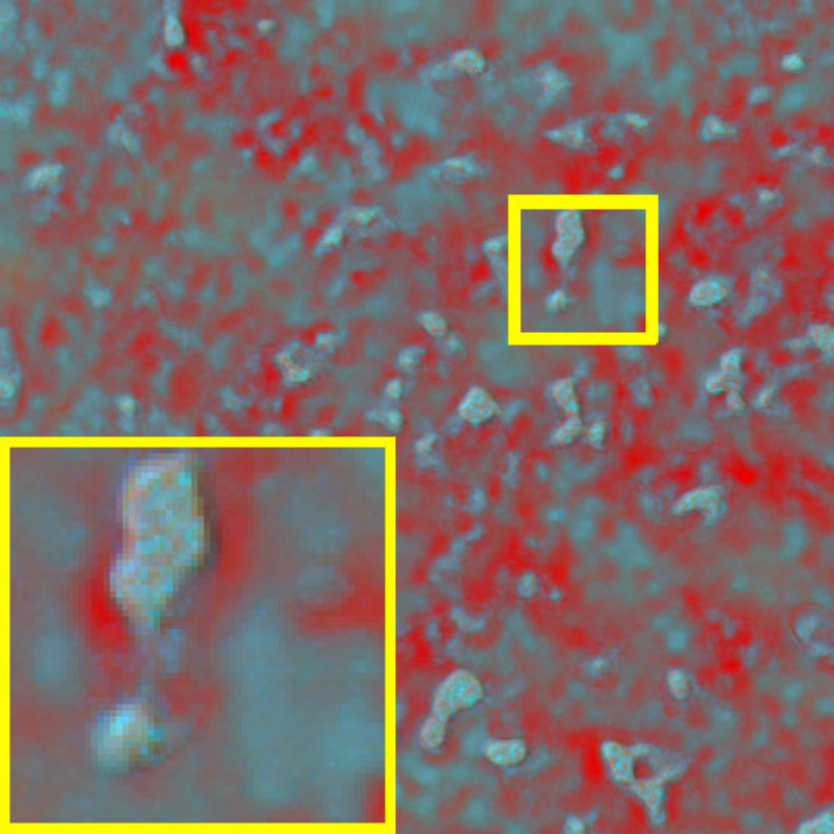}} 
        \end{minipage}
        \begin{minipage}{0.1\hsize}
            \centerline{\includegraphics[width=\hsize]{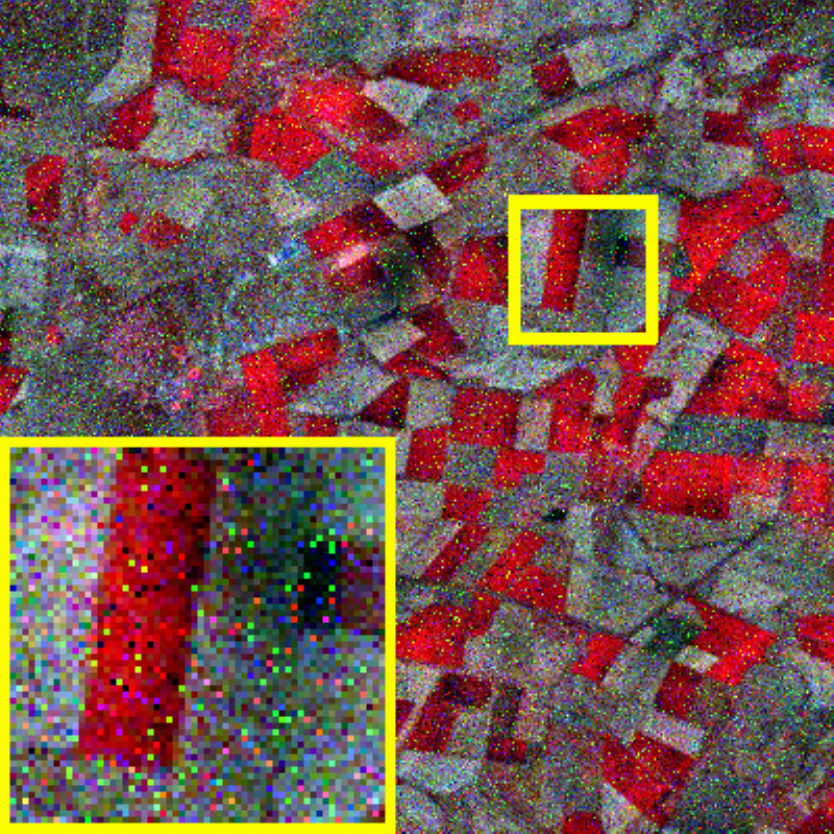}} 
        \end{minipage}
        \begin{minipage}{0.1\hsize}
            \centerline{\includegraphics[width=\hsize]{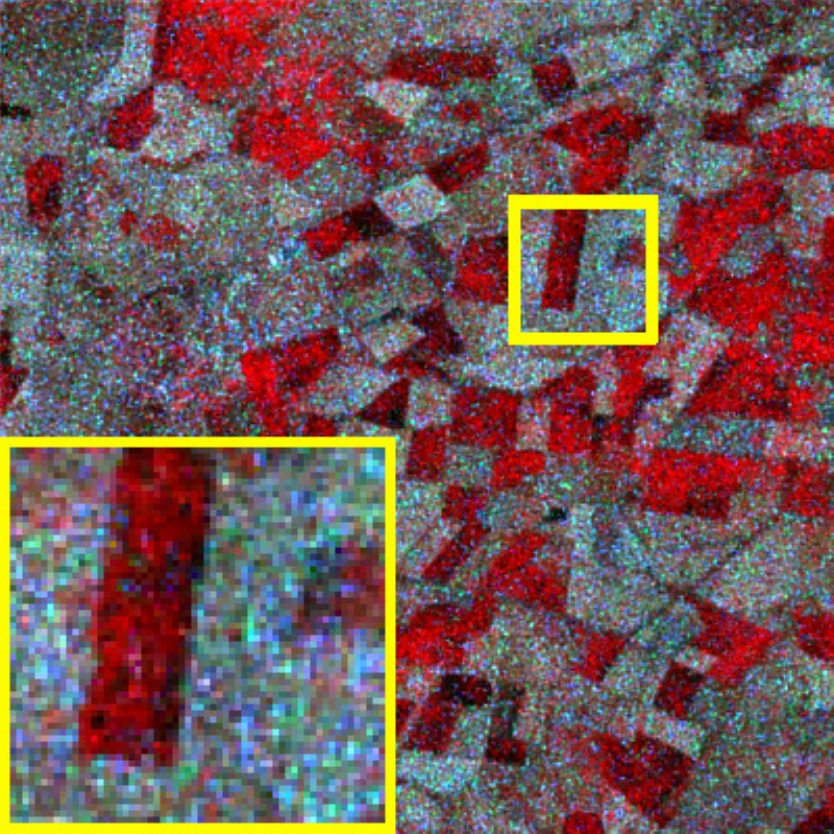}} 
        \end{minipage}
        \begin{minipage}{0.1\hsize}
            \centerline{\includegraphics[width=\hsize]{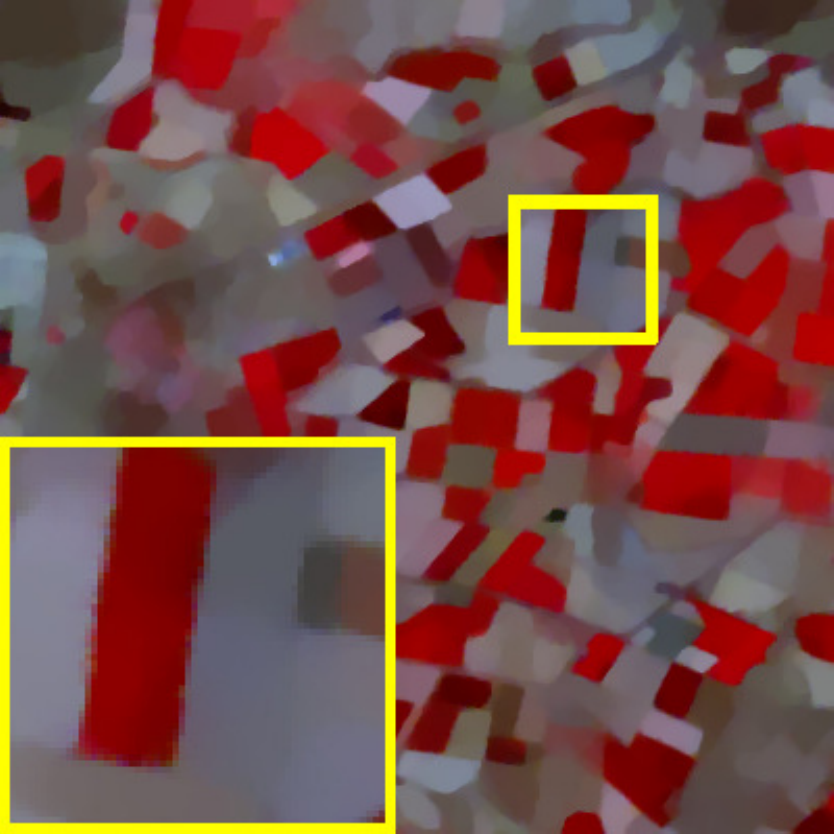}} 
        \end{minipage}
        \begin{minipage}{0.1\hsize}
            \centerline{\includegraphics[width=\hsize]{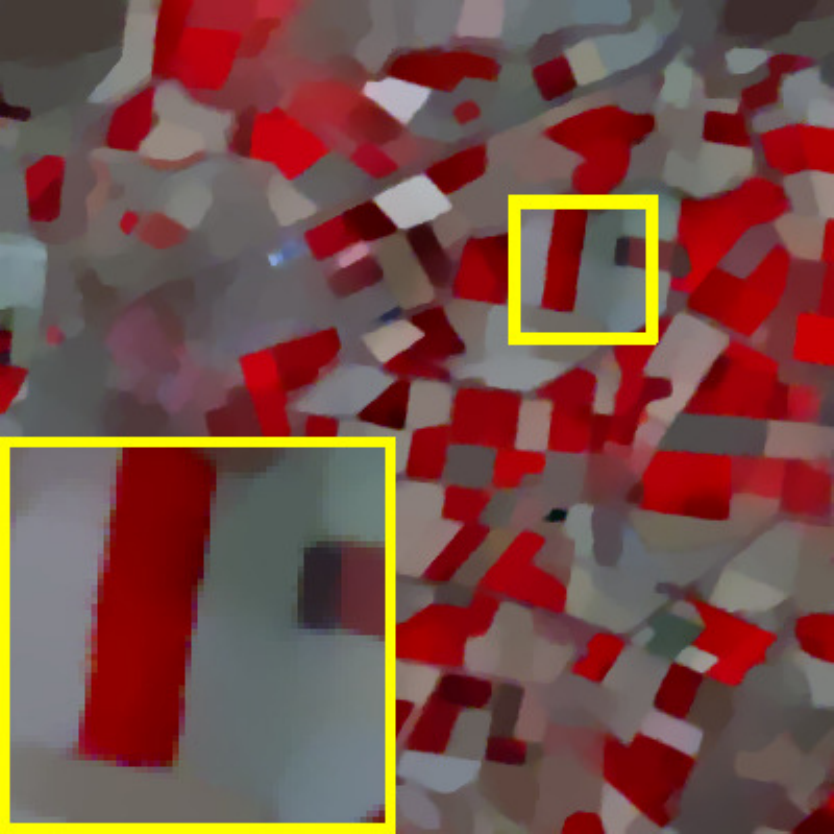}} 
        \end{minipage}
        \begin{minipage}{0.1\hsize}
            \centerline{\includegraphics[width=\hsize]{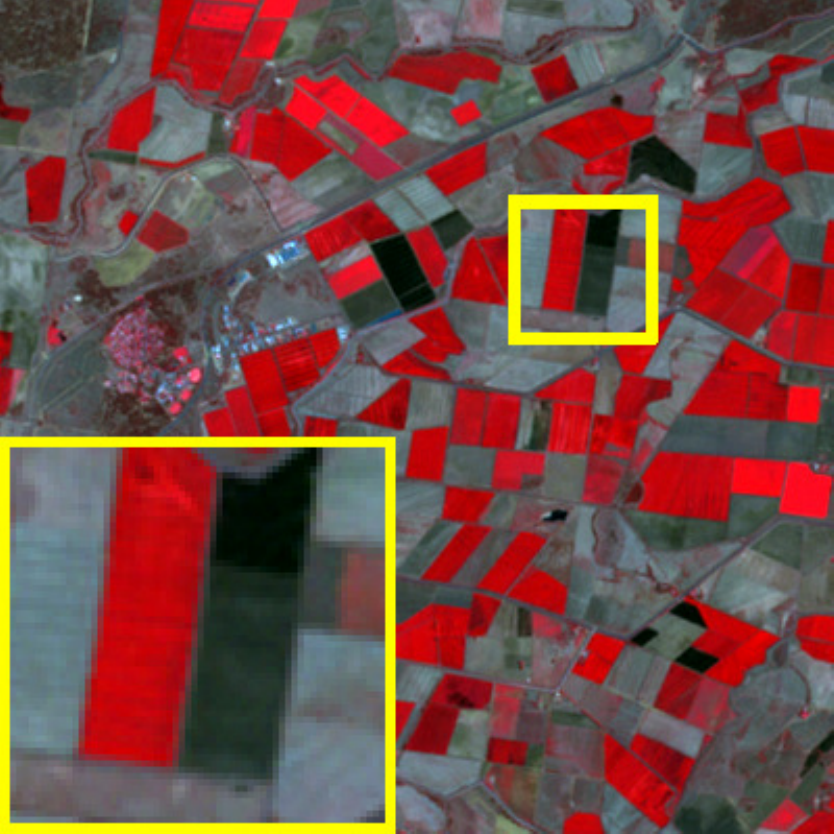}} 
        \end{minipage}  \\
        \vspace{1mm}
        \begin{minipage}{0.01\hsize}
            \centerline{}
        \end{minipage}
        \begin{minipage}{0.1\hsize} 
            \centerline{$\Hr$}
        \end{minipage}
        \begin{minipage}{0.1\hsize} 
            \centerline{STARFM}
        \end{minipage}
        \begin{minipage}{0.1\hsize} 
            \centerline{VIPSTF}
        \end{minipage}
        \begin{minipage}{0.1\hsize} 
            \centerline{RSFN}
        \end{minipage}
        \begin{minipage}{0.1\hsize} 
            \centerline{RobOt}
        \end{minipage}
        \begin{minipage}{0.1\hsize} 
            \centerline{SwinSTFM}
        \end{minipage}
        \begin{minipage}{0.1\hsize} 
            \centerline{\textbf{ROSTF-1}}
        \end{minipage}
        \begin{minipage}{0.1\hsize} 
            \centerline{\textbf{ROSTF-2}}
        \end{minipage}
        \begin{minipage}{0.1\hsize} 
            \centerline{Ground-truth}
        \end{minipage}\\
	\end{center}
        \vspace{-3mm}
	\caption{ST fusion results for the noisy Site2 simulated data. The top, middle, and bottom rows represent the results in Case2, Case3, and Case4, respectively.}
        \label{fig: Site2 SemiSim Case234 results}
\end{figure*}

%% file: Experiments/results/Site2_SemiSim_scatter.tex
\begin{figure*}[ht]
	\begin{center}
        
        \begin{minipage}{0.01\hsize}
            \centerline{\rotatebox{90}{Case1}} 
        \end{minipage}
        \begin{minipage}{0.12\hsize}
            \centerline{\includegraphics[width=\hsize]{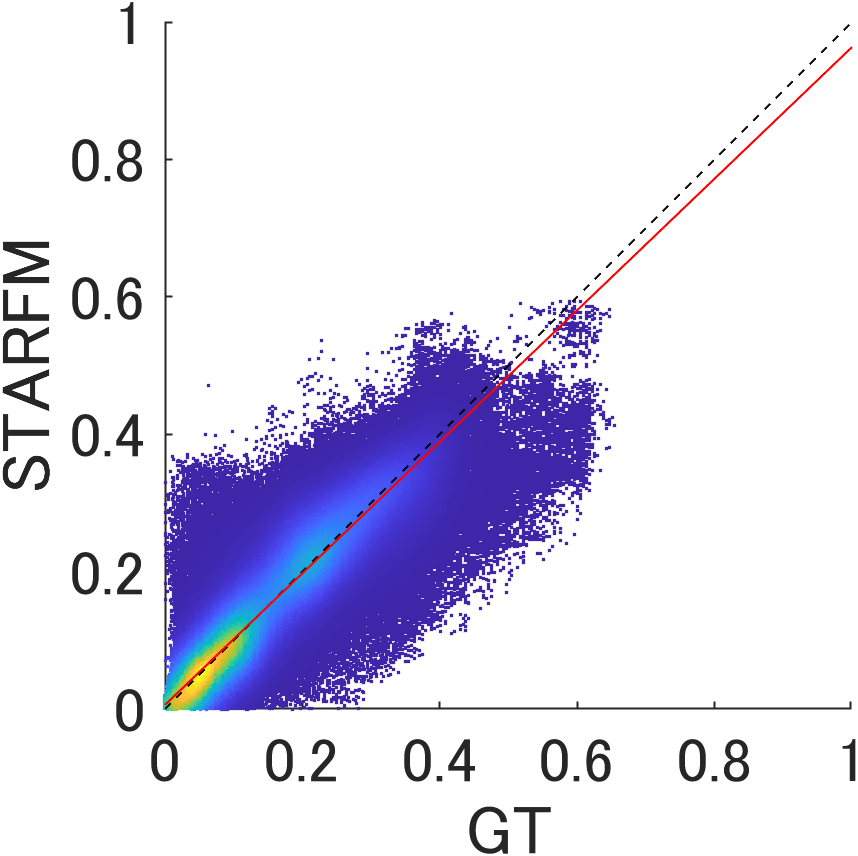}} 
        \end{minipage}
        \begin{minipage}{0.12\hsize}
            \centerline{\includegraphics[width=\hsize]{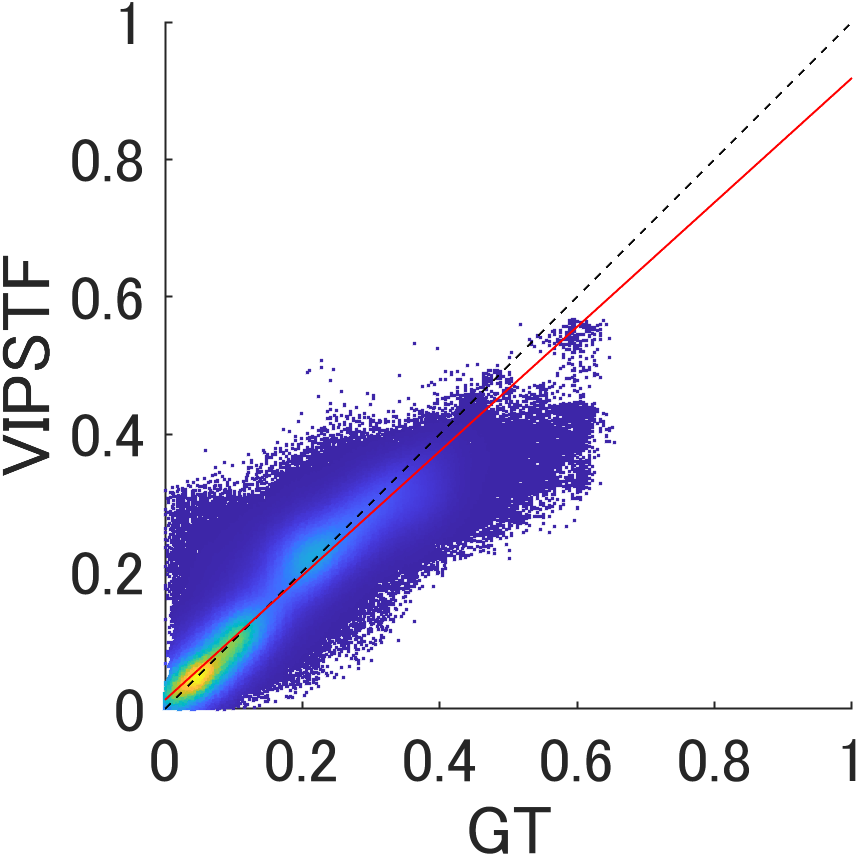}} 
        \end{minipage}
        \begin{minipage}{0.12\hsize}
            \centerline{\includegraphics[width=\hsize]{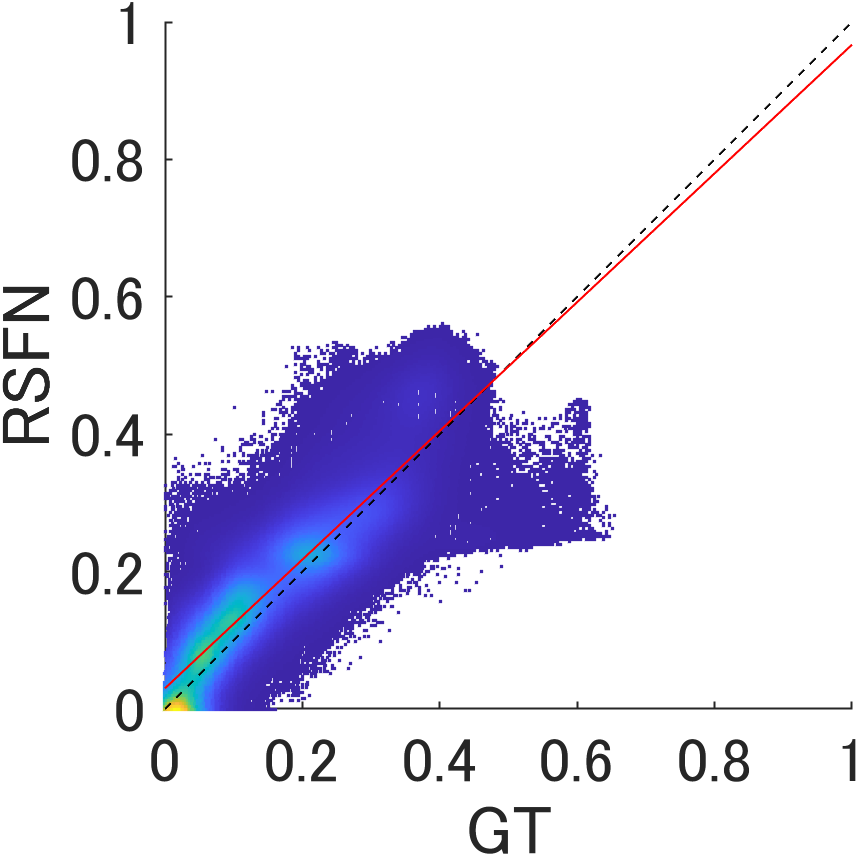}} 
        \end{minipage}
        \begin{minipage}{0.12\hsize}
            \centerline{\includegraphics[width=\hsize]{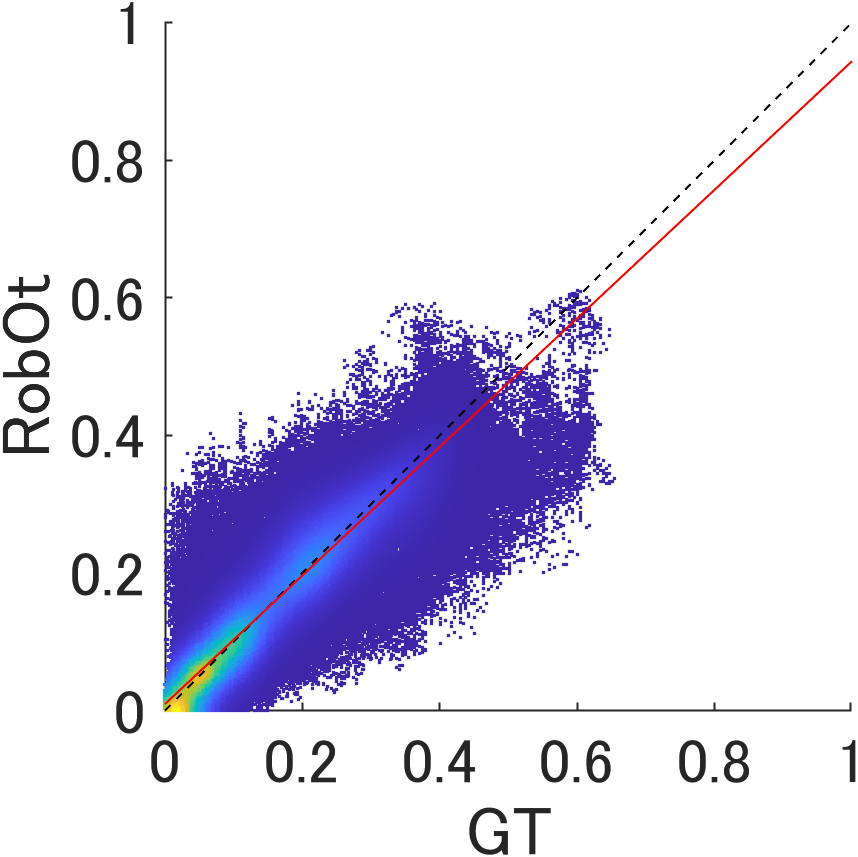}} 
        \end{minipage}
        \begin{minipage}{0.12\hsize}
            \centerline{\includegraphics[width=\hsize]{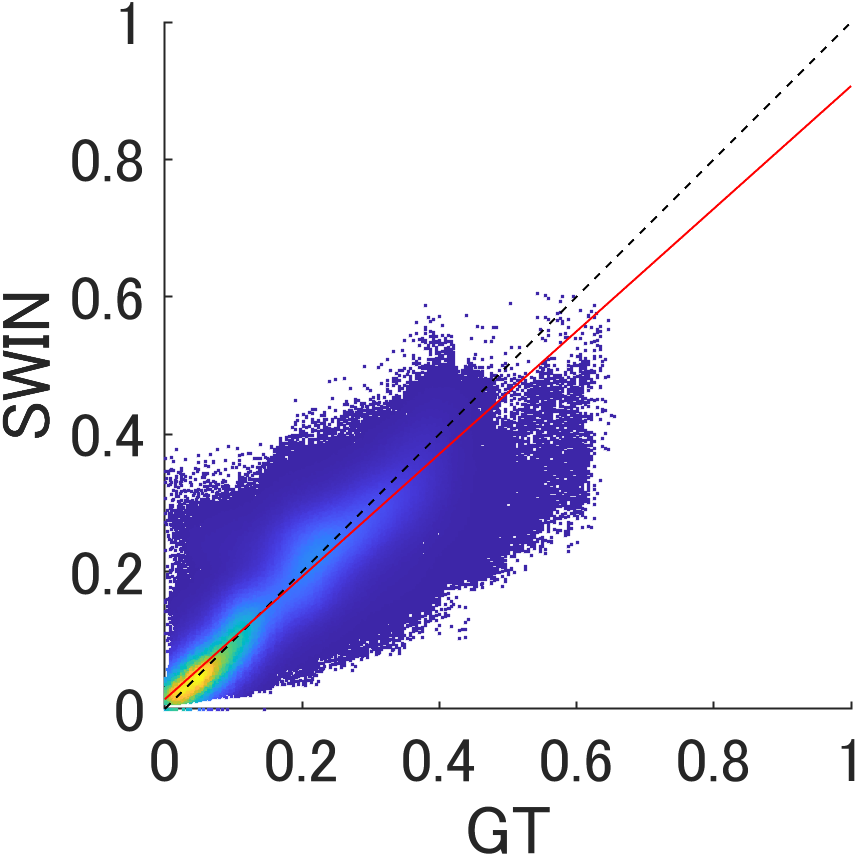}} 
        \end{minipage}
        \begin{minipage}{0.12\hsize}
            \centerline{\includegraphics[width=\hsize]{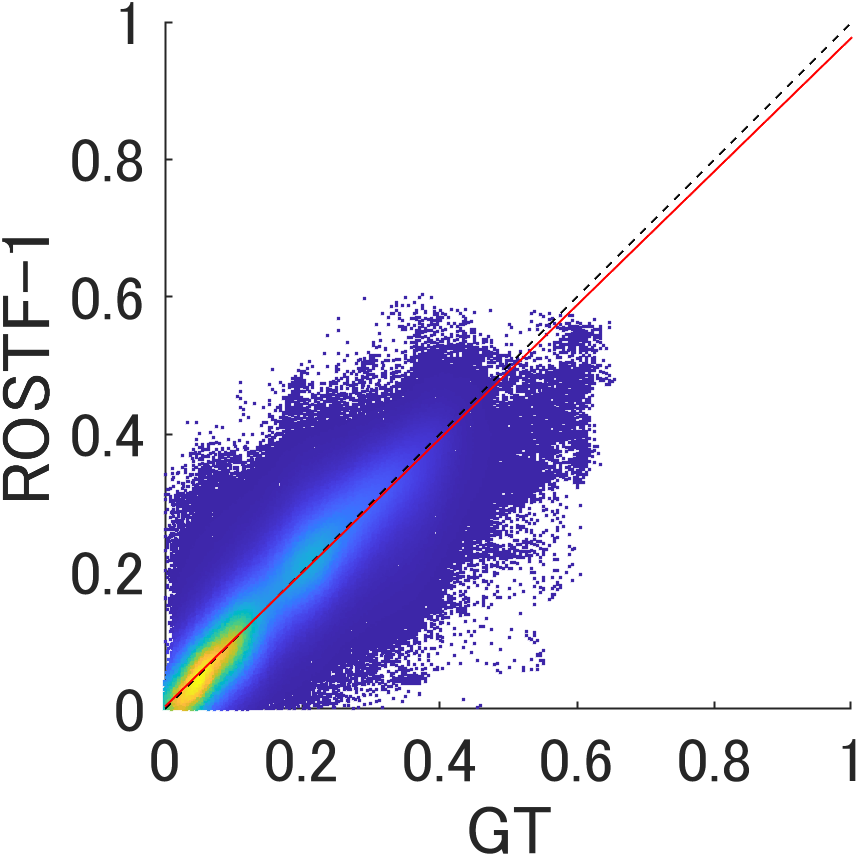}} 
        \end{minipage}
        \begin{minipage}{0.12\hsize}
            \centerline{\includegraphics[width=\hsize]{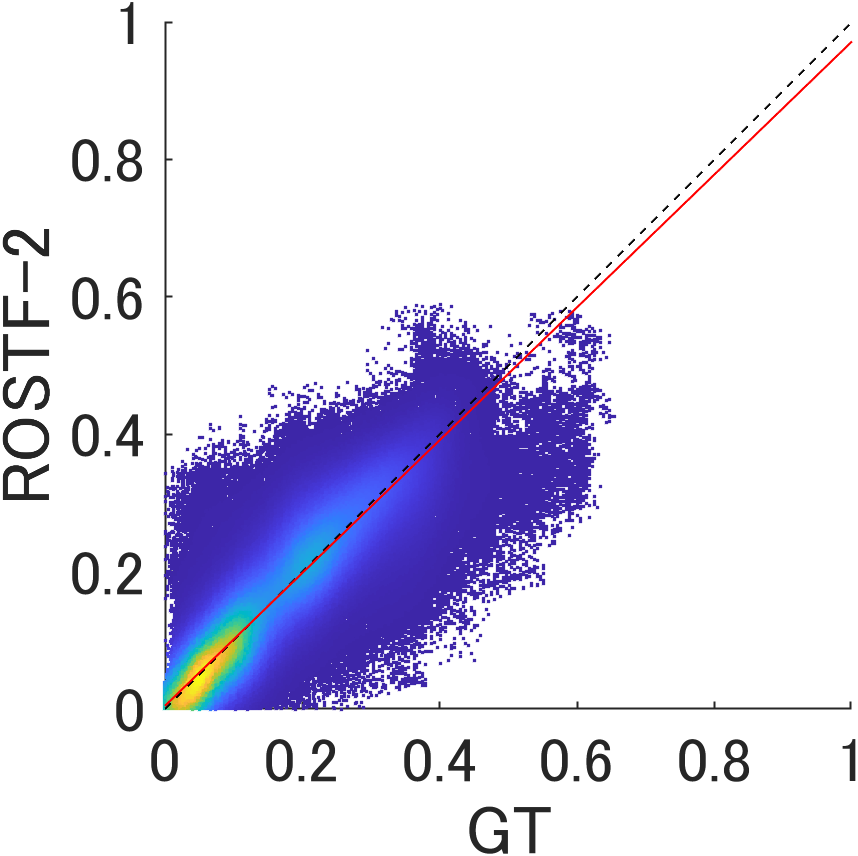}} 
        \end{minipage}  
        \begin{minipage}{0.05\hsize}
            \centerline{\includegraphics[height=70pt]{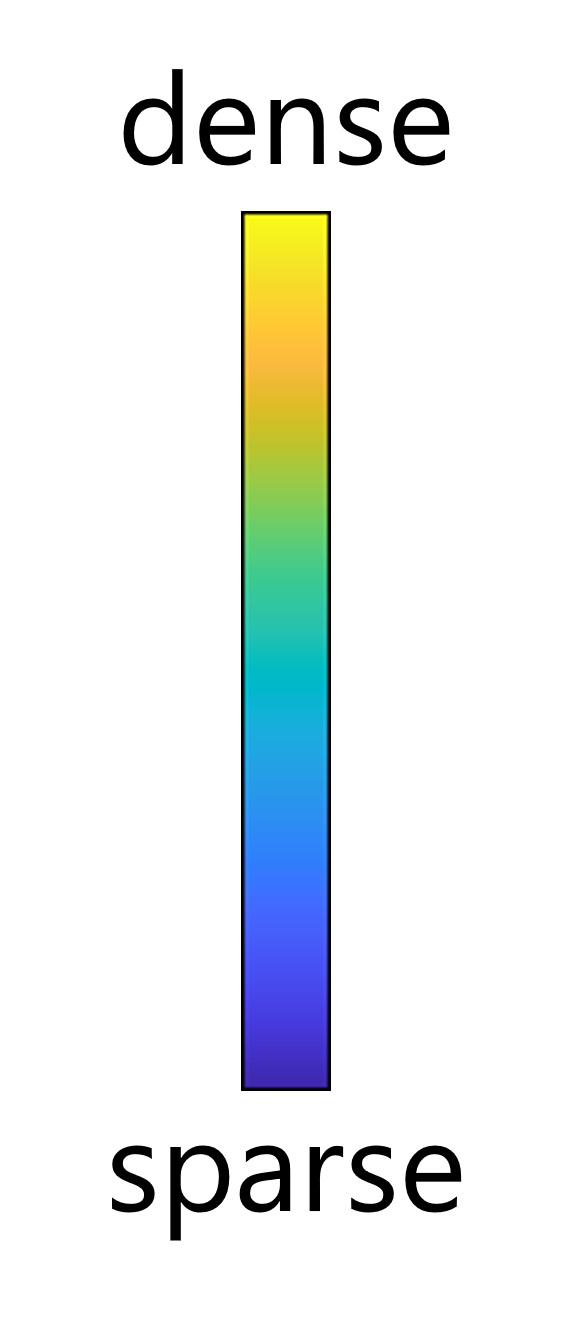}} 
        \end{minipage} \\

        \begin{minipage}{0.01\hsize}
			\centerline{\rotatebox{90}{Case2}} 
        \end{minipage}
	\begin{minipage}{0.12\hsize}
            \centerline{\includegraphics[width=\hsize]{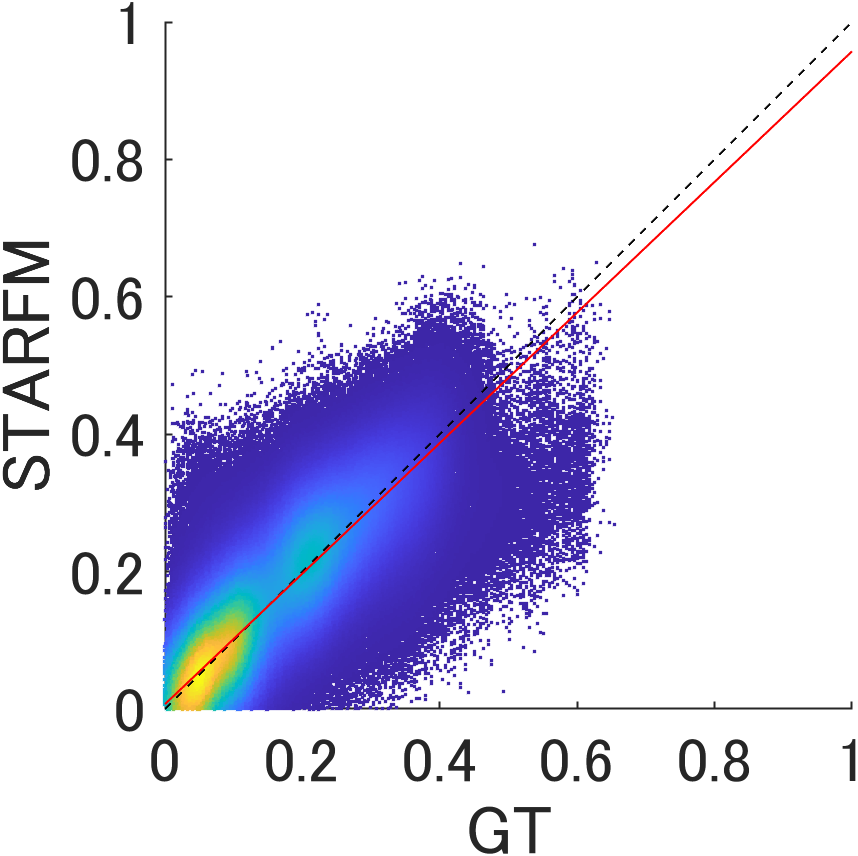}} 
        \end{minipage}
        \begin{minipage}{0.12\hsize}
            \centerline{\includegraphics[width=\hsize]{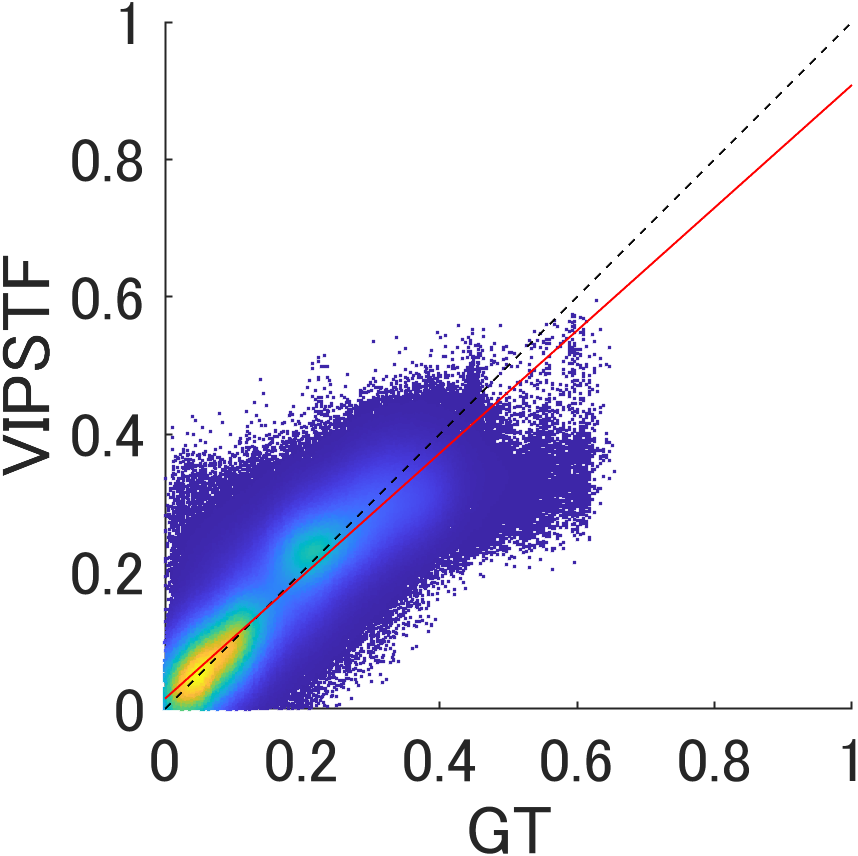}} 
        \end{minipage}
        \begin{minipage}{0.12\hsize}
            \centerline{\includegraphics[width=\hsize]{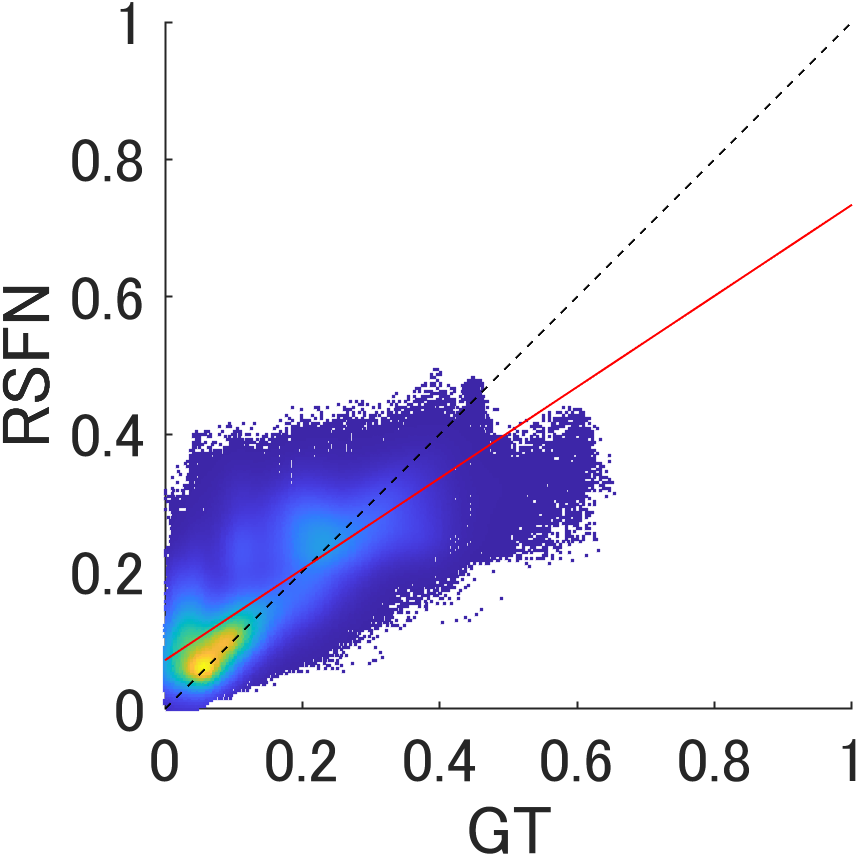}} 
        \end{minipage}
        \begin{minipage}{0.12\hsize}
            \centerline{\includegraphics[width=\hsize]{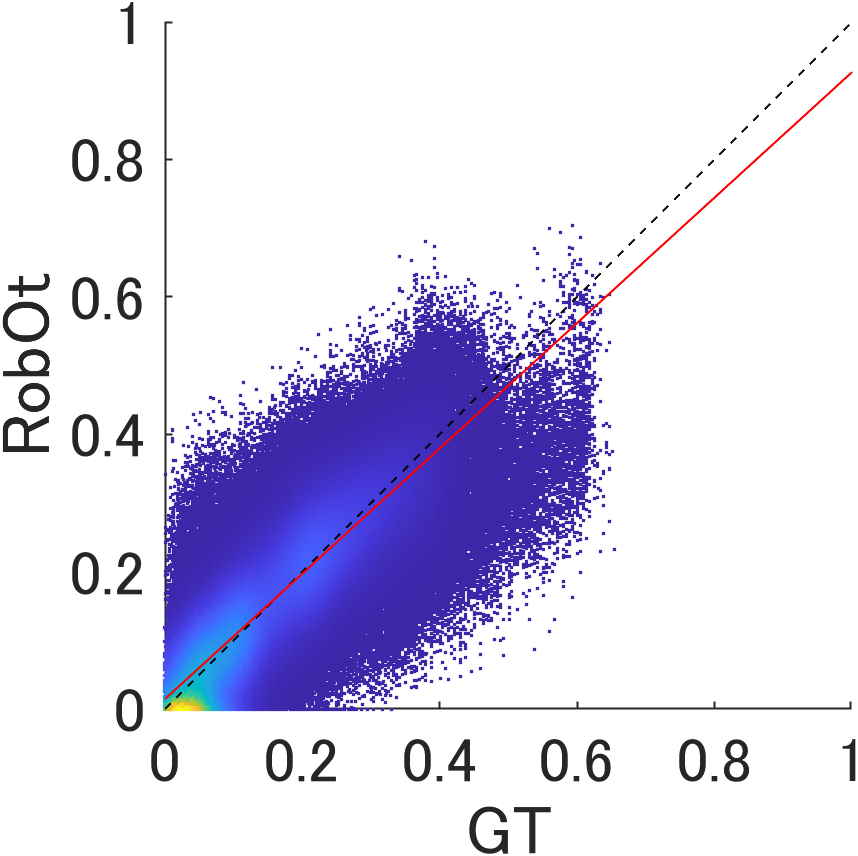}} 
        \end{minipage}
        \begin{minipage}{0.12\hsize}
            \centerline{\includegraphics[width=\hsize]{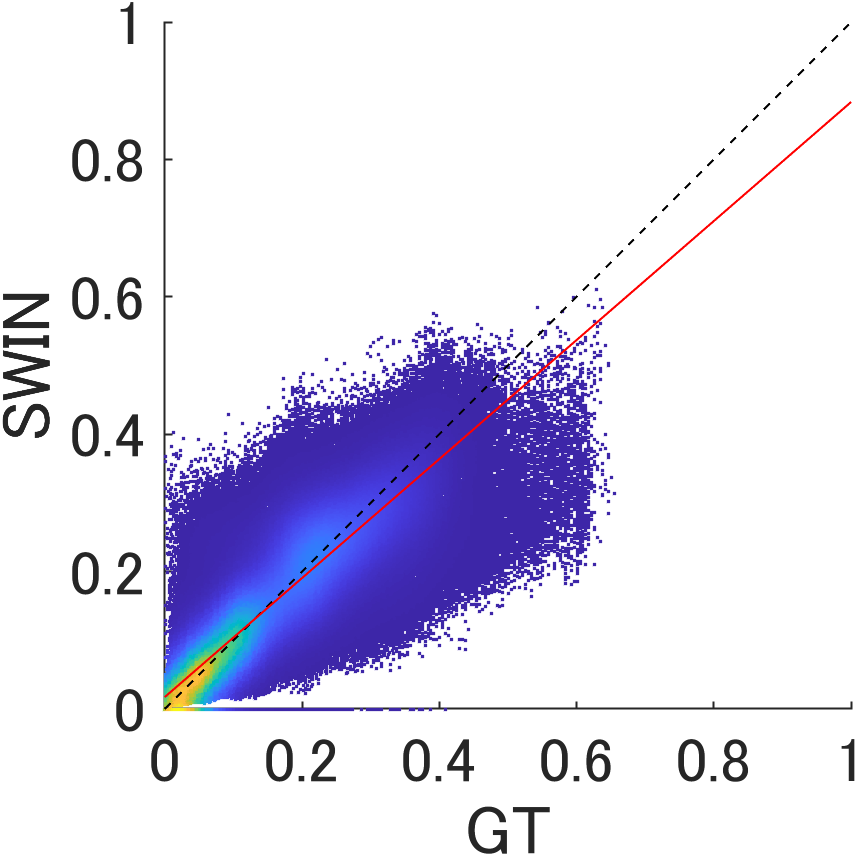}} 
        \end{minipage}
        \begin{minipage}{0.12\hsize}
            \centerline{\includegraphics[width=\hsize]{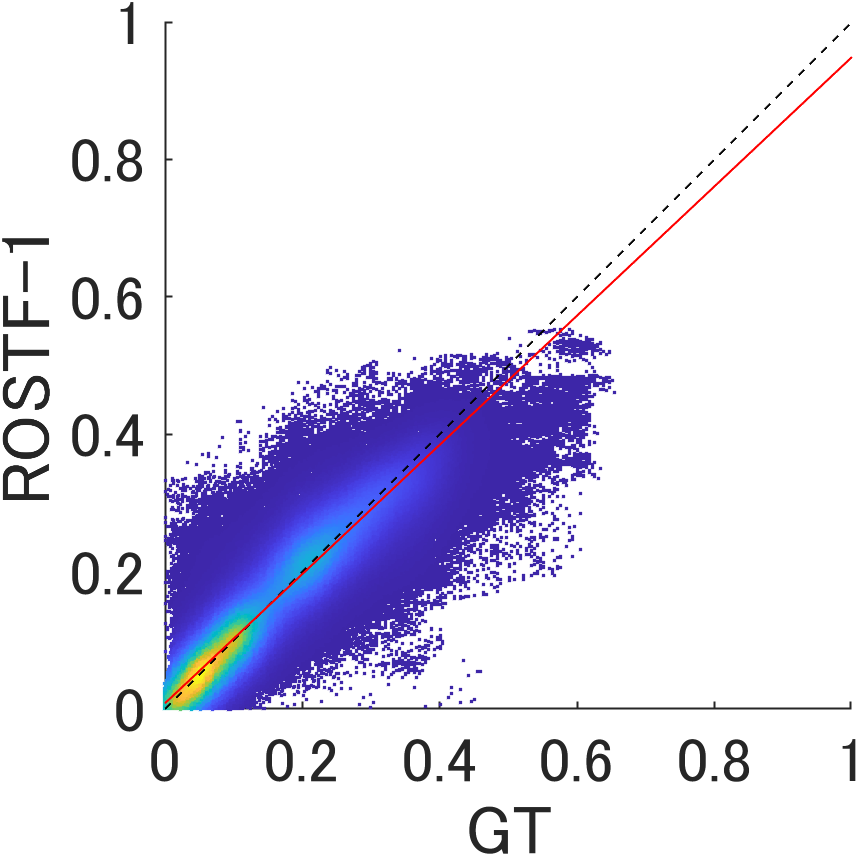}} 
        \end{minipage}
        \begin{minipage}{0.12\hsize}
            \centerline{\includegraphics[width=\hsize]{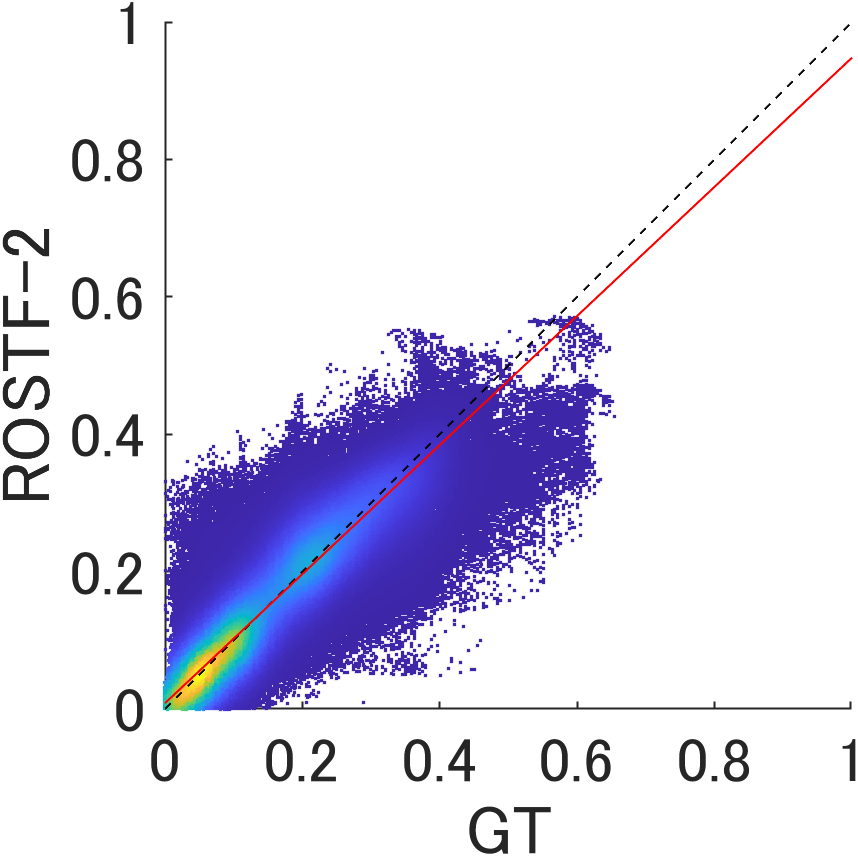}} 
        \end{minipage}  
        \begin{minipage}{0.05\hsize}
            \centerline{\includegraphics[height=70pt]{img/colorbar.png}} 
        \end{minipage} \\

        \begin{minipage}{0.01\hsize}
			\centerline{\rotatebox{90}{Case3}} 
        \end{minipage}
	\begin{minipage}{0.12\hsize}
            \centerline{\includegraphics[width=\hsize]{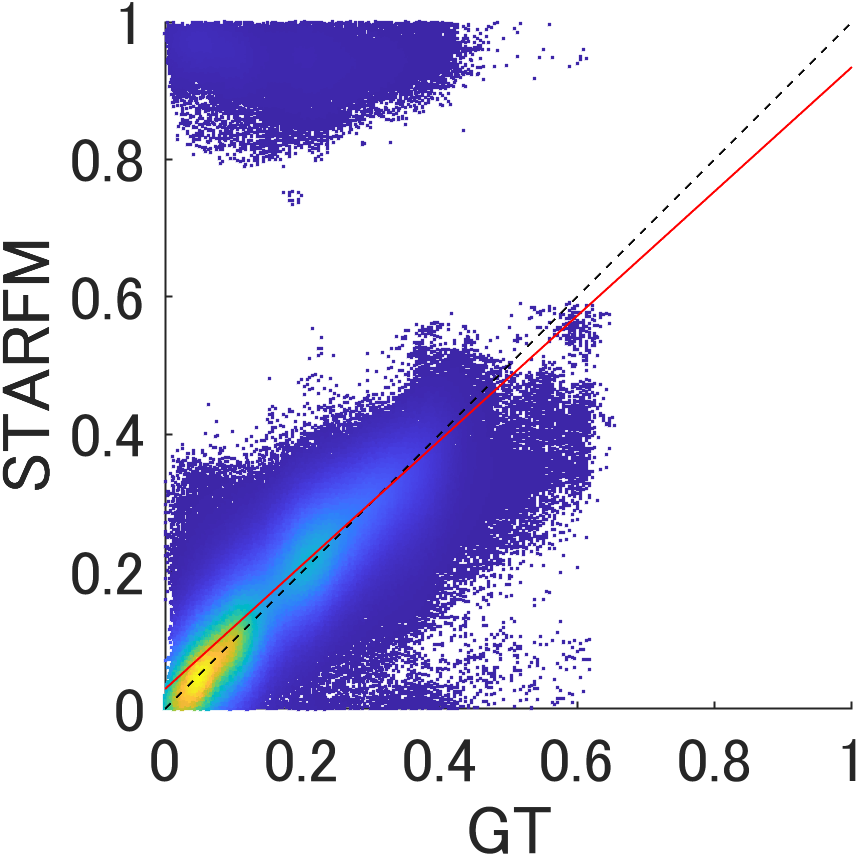}} 
        \end{minipage}
        \begin{minipage}{0.12\hsize}
            \centerline{\includegraphics[width=\hsize]{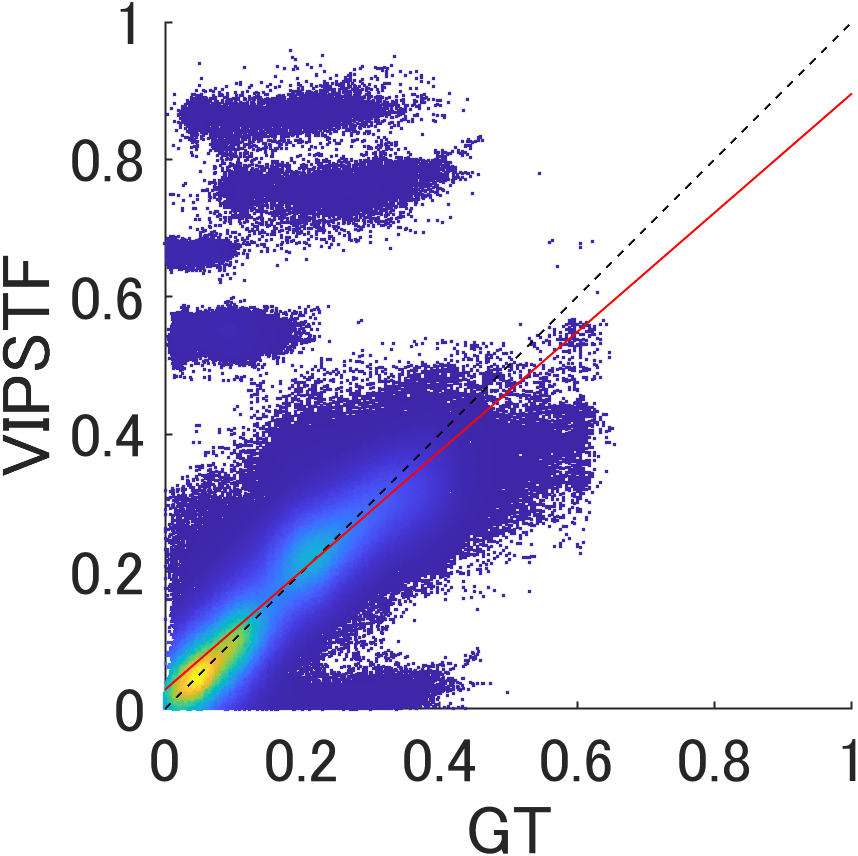}} 
        \end{minipage}
        \begin{minipage}{0.12\hsize}
            \centerline{\includegraphics[width=\hsize]{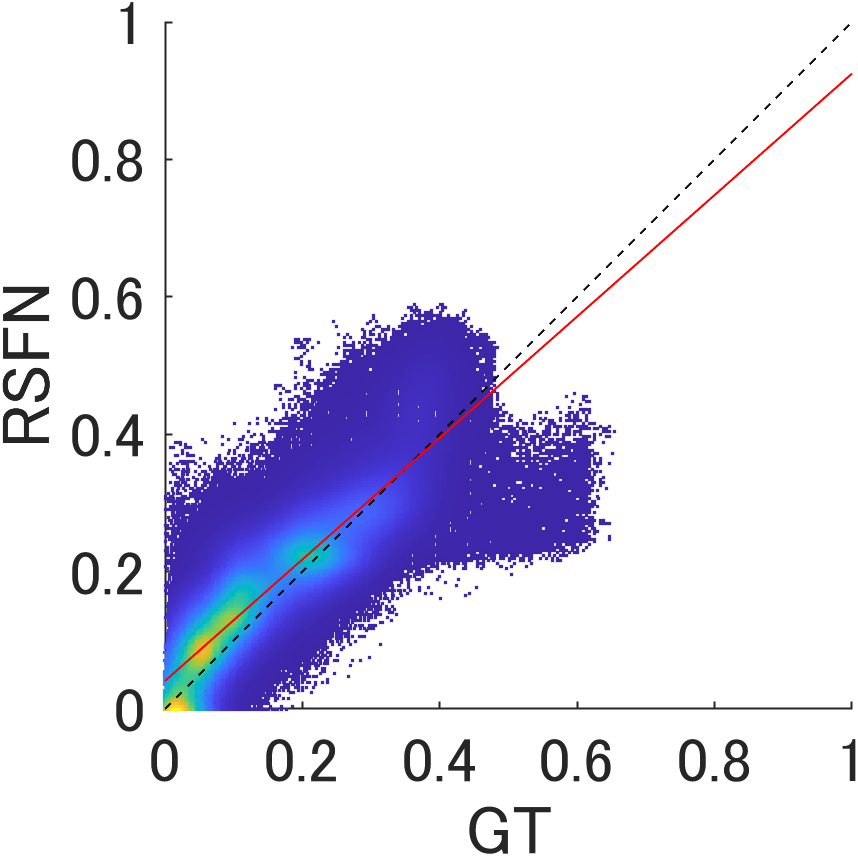}} 
        \end{minipage}
        \begin{minipage}{0.12\hsize}
            \centerline{\includegraphics[width=\hsize]{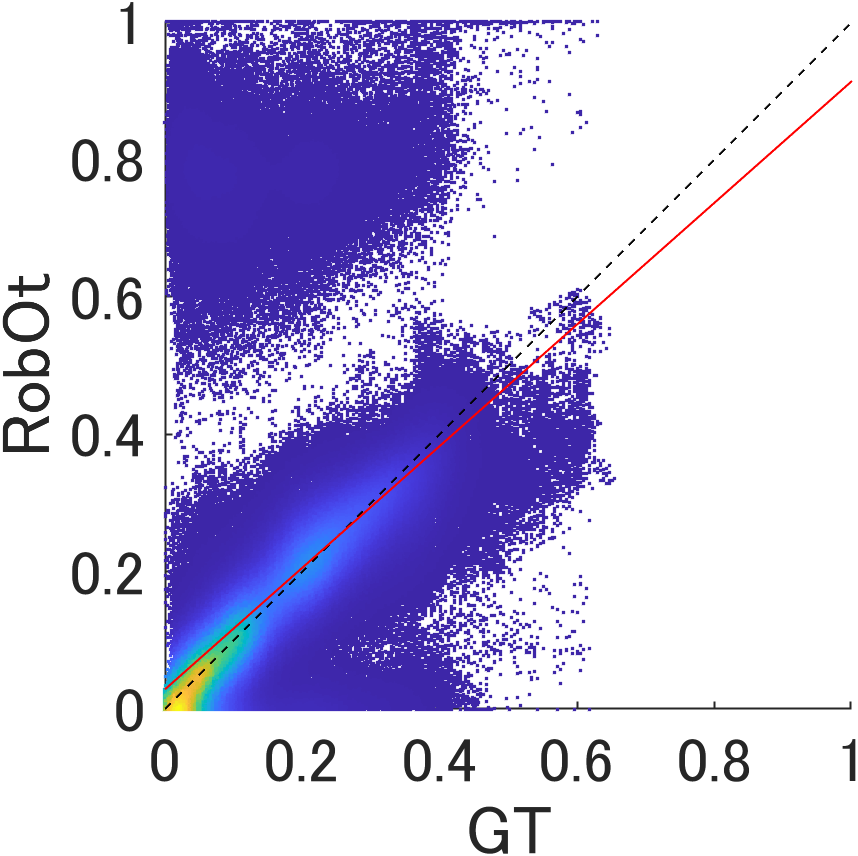}} 
        \end{minipage}
        \begin{minipage}{0.12\hsize}
            \centerline{\includegraphics[width=\hsize]{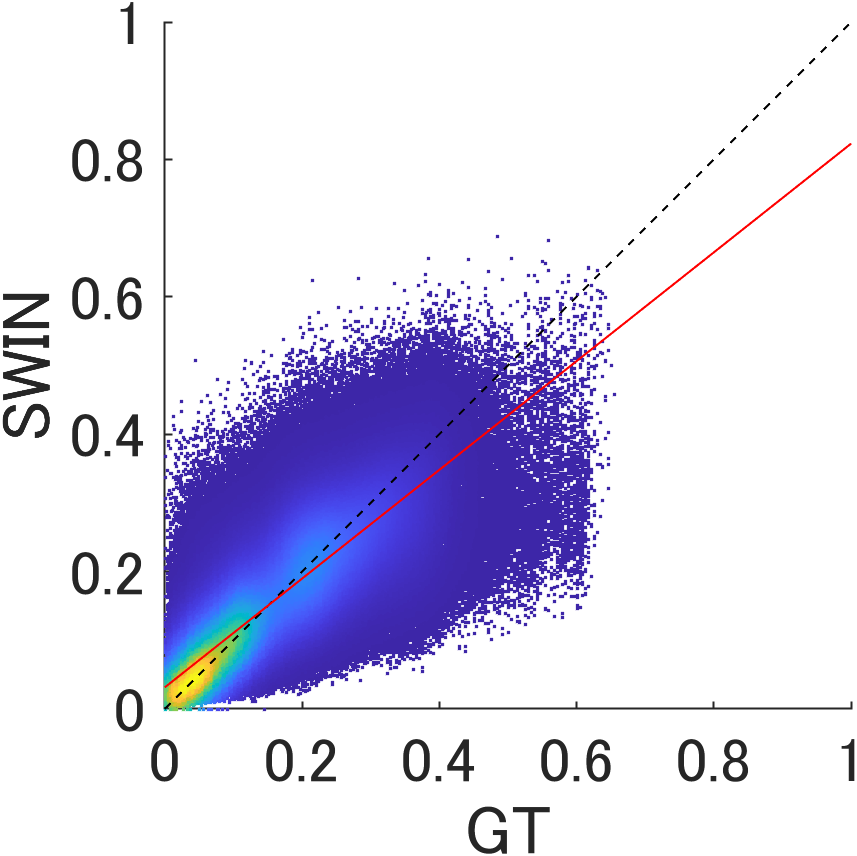}} 
        \end{minipage}
        \begin{minipage}{0.12\hsize}
            \centerline{\includegraphics[width=\hsize]{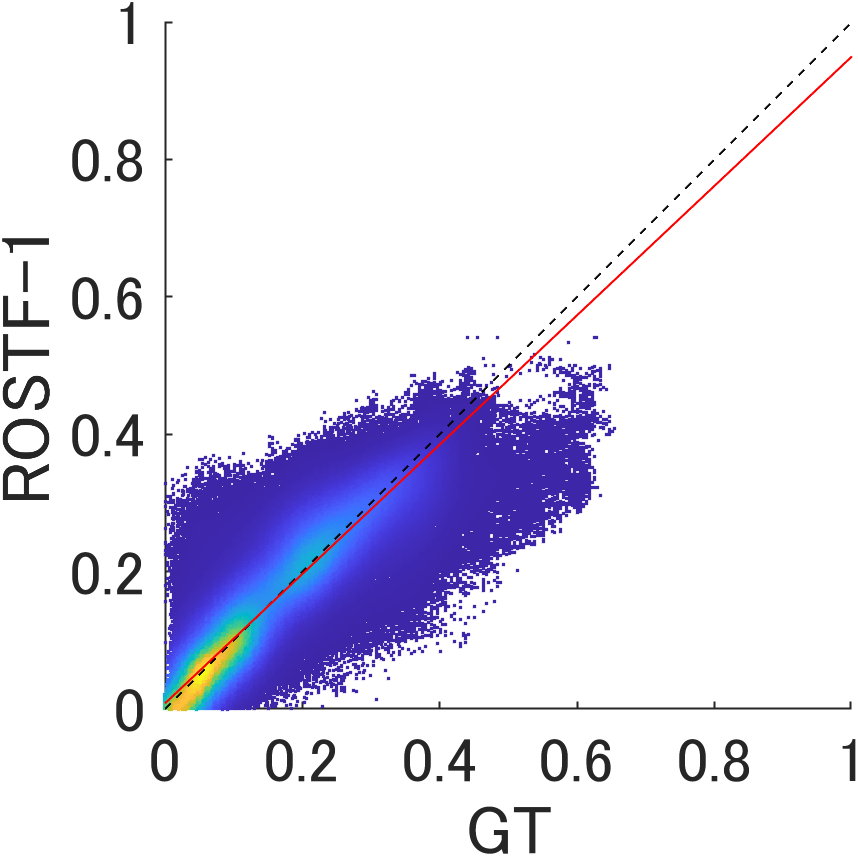}} 
        \end{minipage}
        \begin{minipage}{0.12\hsize}
            \centerline{\includegraphics[width=\hsize]{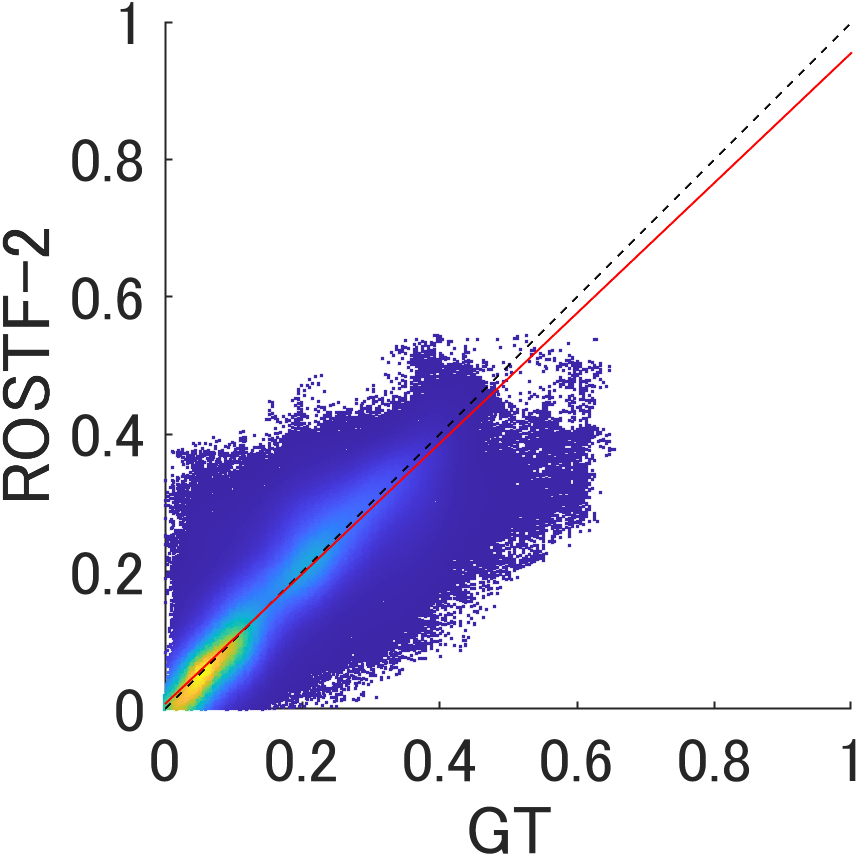}} 
        \end{minipage}  
        \begin{minipage}{0.05\hsize}
            \centerline{\includegraphics[height=70pt]{img/colorbar.png}} 
        \end{minipage} \\
  
        \begin{minipage}{0.01\hsize}
            \centerline{\rotatebox{90}{Case4}} 
        \end{minipage}
	\begin{minipage}{0.12\hsize}
            \centerline{\includegraphics[width=\hsize]{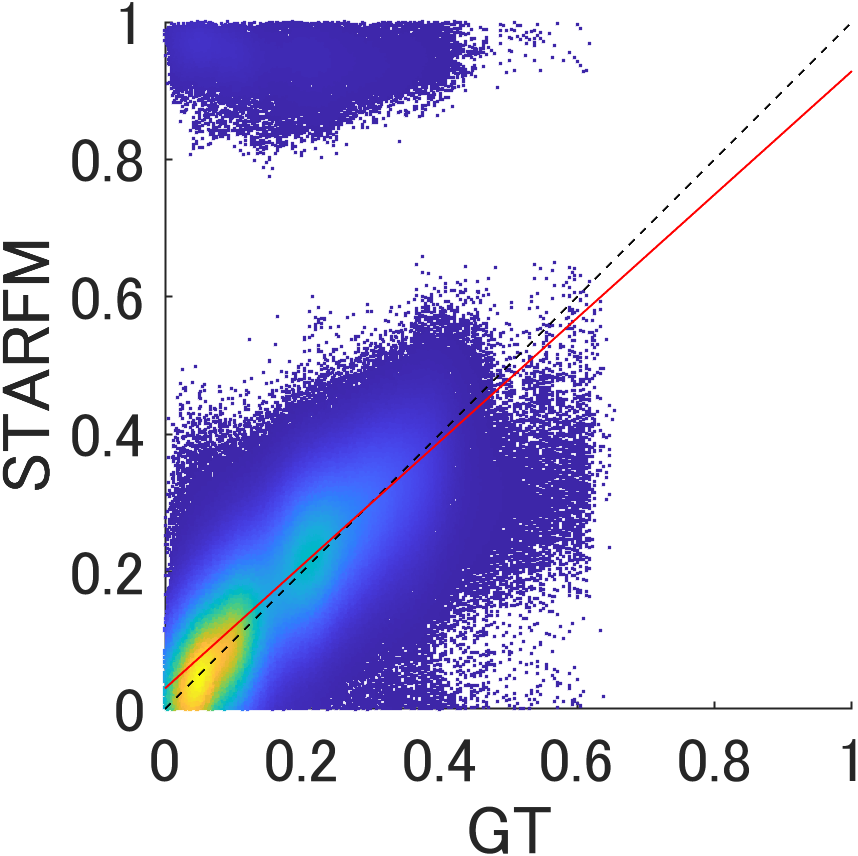}} 
        \end{minipage}
        \begin{minipage}{0.12\hsize}
            \centerline{\includegraphics[width=\hsize]{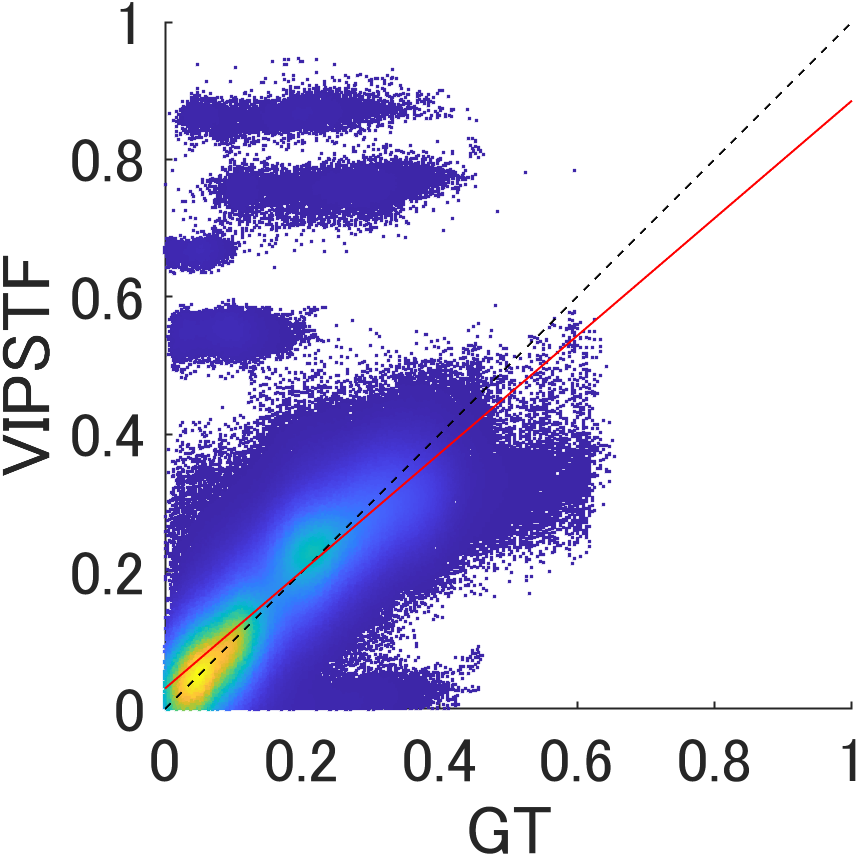}} 
        \end{minipage}
        \begin{minipage}{0.12\hsize}
            \centerline{\includegraphics[width=\hsize]{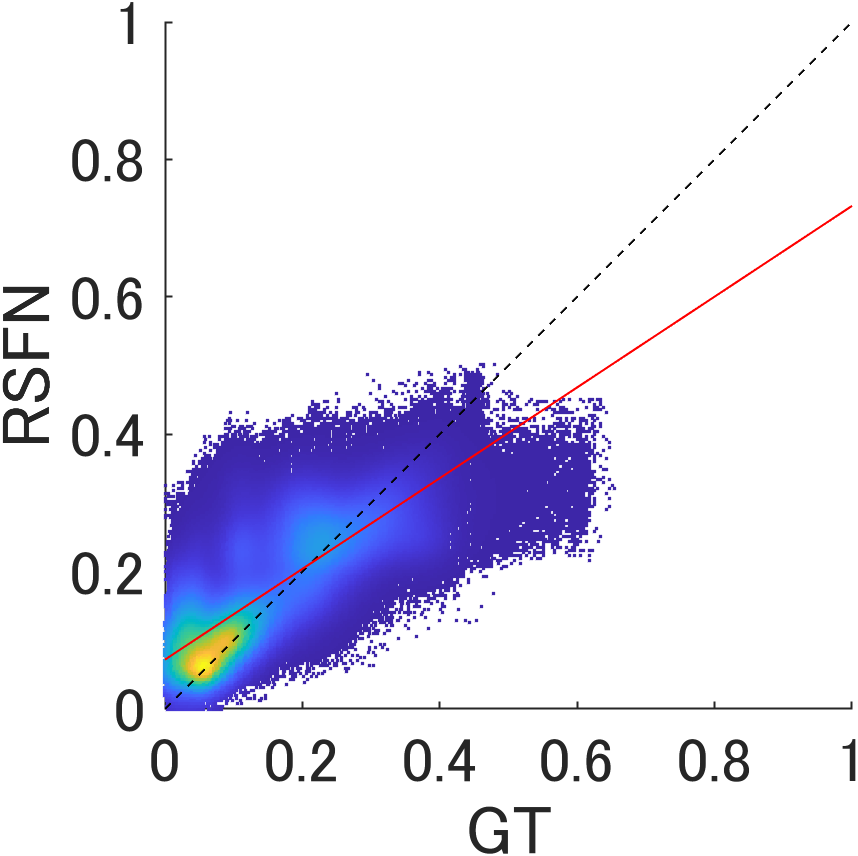}} 
        \end{minipage}
        \begin{minipage}{0.12\hsize}
            \centerline{\includegraphics[width=\hsize]{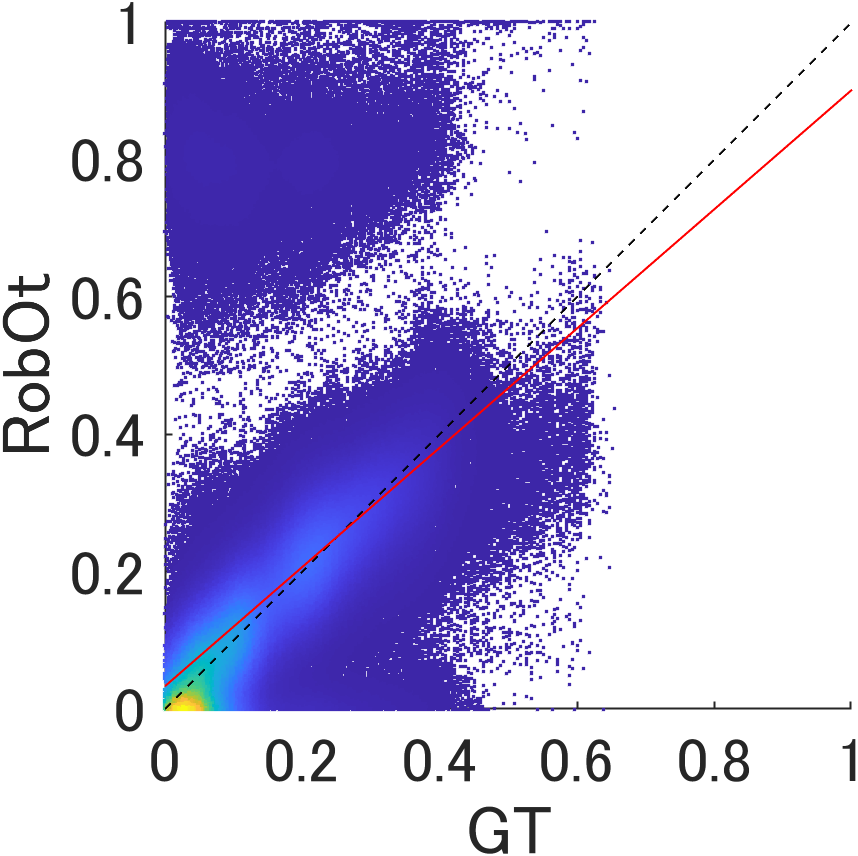}} 
        \end{minipage}
        \begin{minipage}{0.12\hsize}
            \centerline{\includegraphics[width=\hsize]{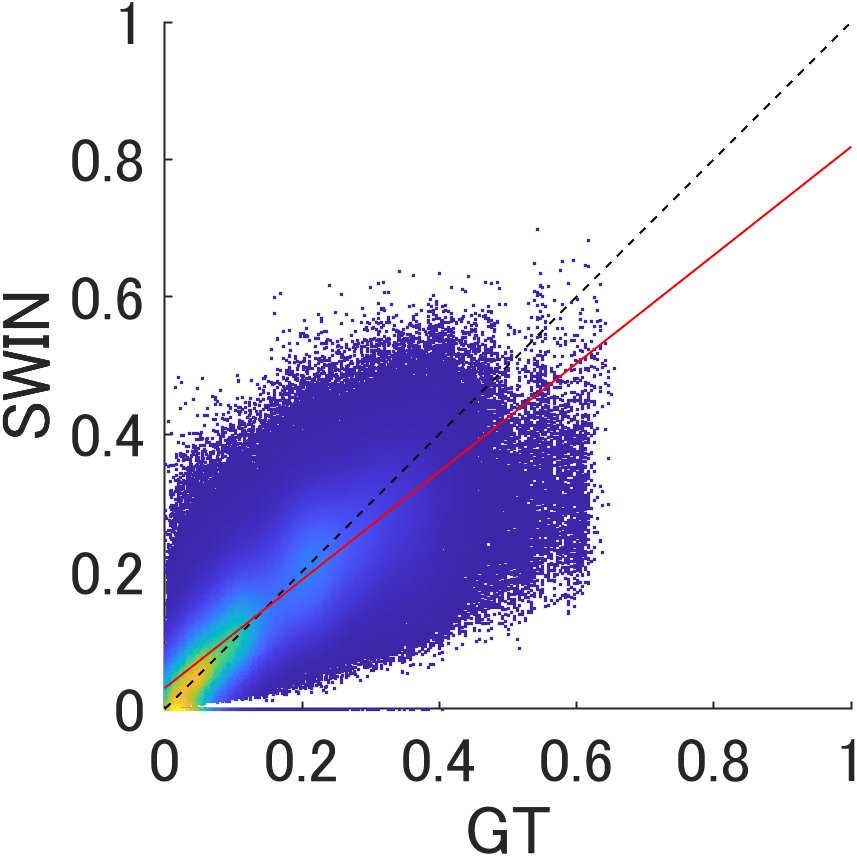}} 
        \end{minipage}
        \begin{minipage}{0.12\hsize}
            \centerline{\includegraphics[width=\hsize]{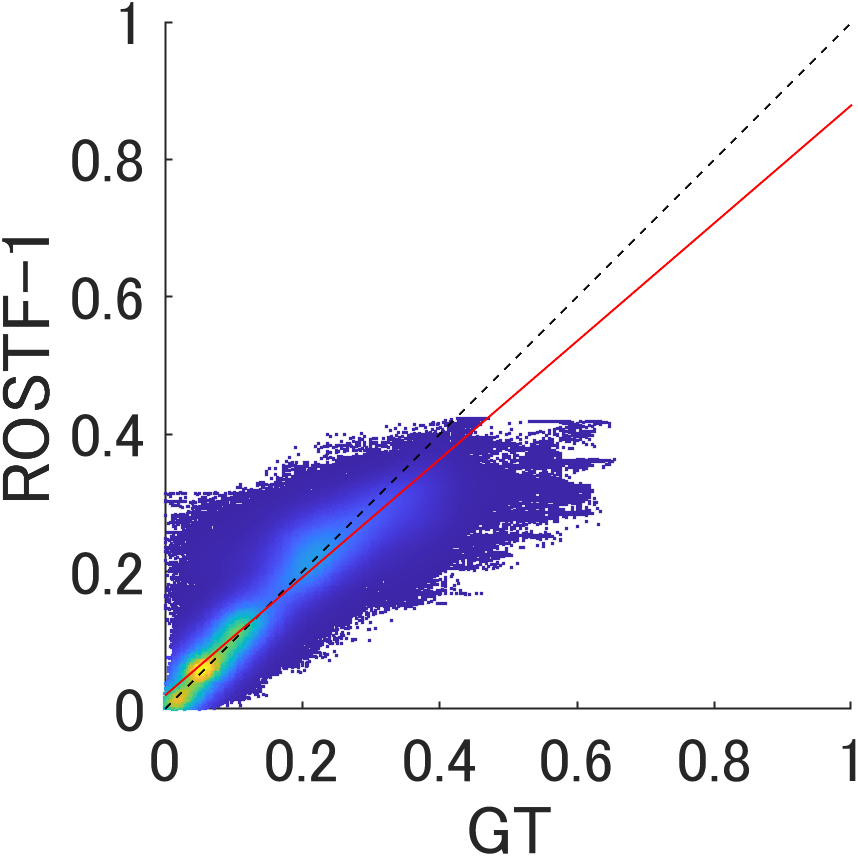}} 
        \end{minipage}
        \begin{minipage}{0.12\hsize}
            \centerline{\includegraphics[width=\hsize]{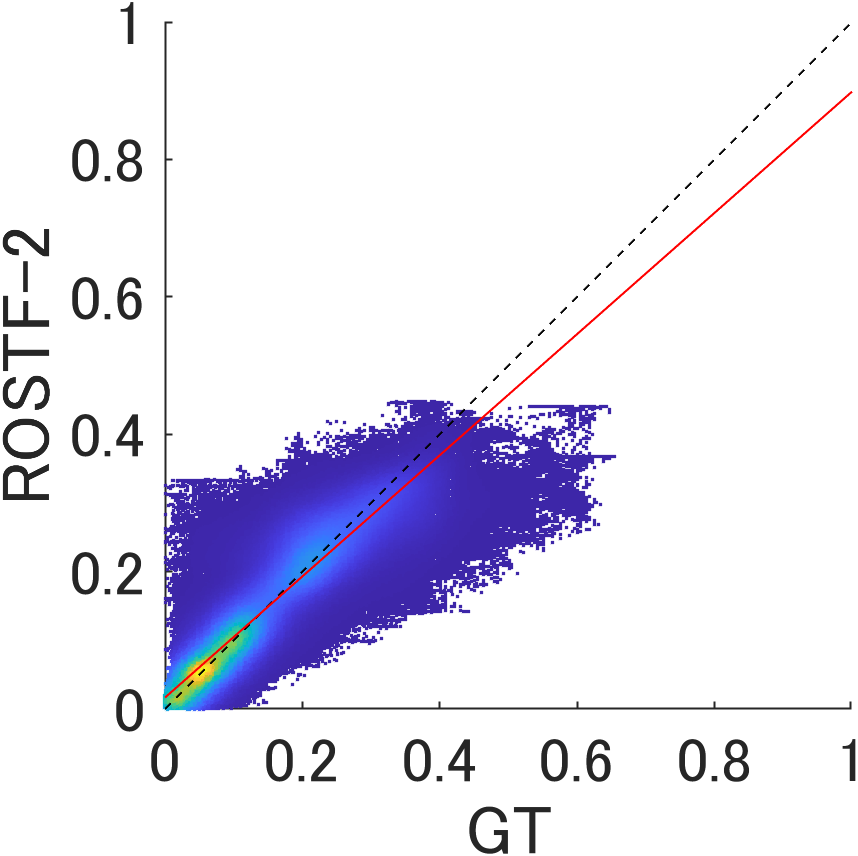}} 
        \end{minipage}  
        \begin{minipage}{0.05\hsize}
            \centerline{\includegraphics[height=70pt]{img/colorbar.png}} 
        \end{minipage} \\
        \begin{minipage}{0.01\hsize}
            \centerline{~} 
        \end{minipage}
        \begin{minipage}{0.12\hsize} 
            \centerline{\;\;\;\;STARFM}
        \end{minipage}
        \begin{minipage}{0.12\hsize} 
            \centerline{\;\;\;\;VIPSTF}
        \end{minipage}
        \begin{minipage}{0.12\hsize} 
            \centerline{\;\;\;\;RSFN}
        \end{minipage}
        \begin{minipage}{0.12\hsize} 
            \centerline{\;\;\;\;RobOt}
        \end{minipage}
        \begin{minipage}{0.12\hsize} 
            \centerline{\;\;\;\;SwinSTFM}
        \end{minipage}
        \begin{minipage}{0.12\hsize} 
            \centerline{\;\;\;\;ROSTF-1}
        \end{minipage}
        \begin{minipage}{0.12\hsize} 
            \centerline{\;\;\;\;ROSTF-2}
        \end{minipage} 
        \begin{minipage}{0.05\hsize}
            \centerline{~} 
        \end{minipage} \\
	\end{center}
        \vspace{-3mm}
	\caption{Scatter plots of the ground-truth and the estimated values for the Site2 simulated data.}
        \label{fig: Site2 SemiSim scatter}
\end{figure*}

%% file: Experiments/results/Site2_SemiSim_SpectralPlot.tex
\begin{figure}[ht]
		\begin{minipage}{0.47\hsize}
			\centerline{\includegraphics[width=\hsize]{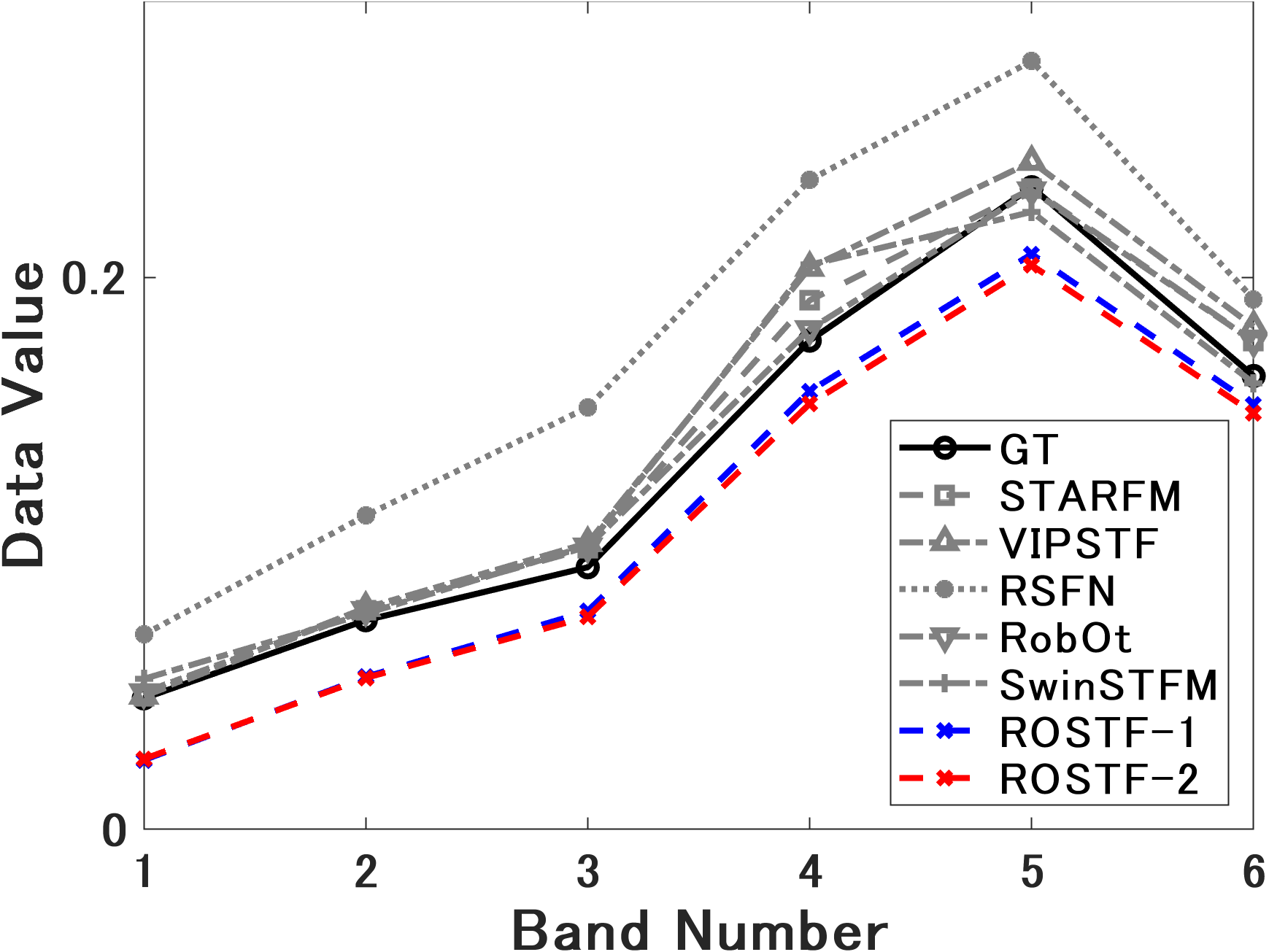}} 
		\end{minipage}
        \begin{minipage}{0.47\hsize}
			\centerline{\includegraphics[width=\hsize]{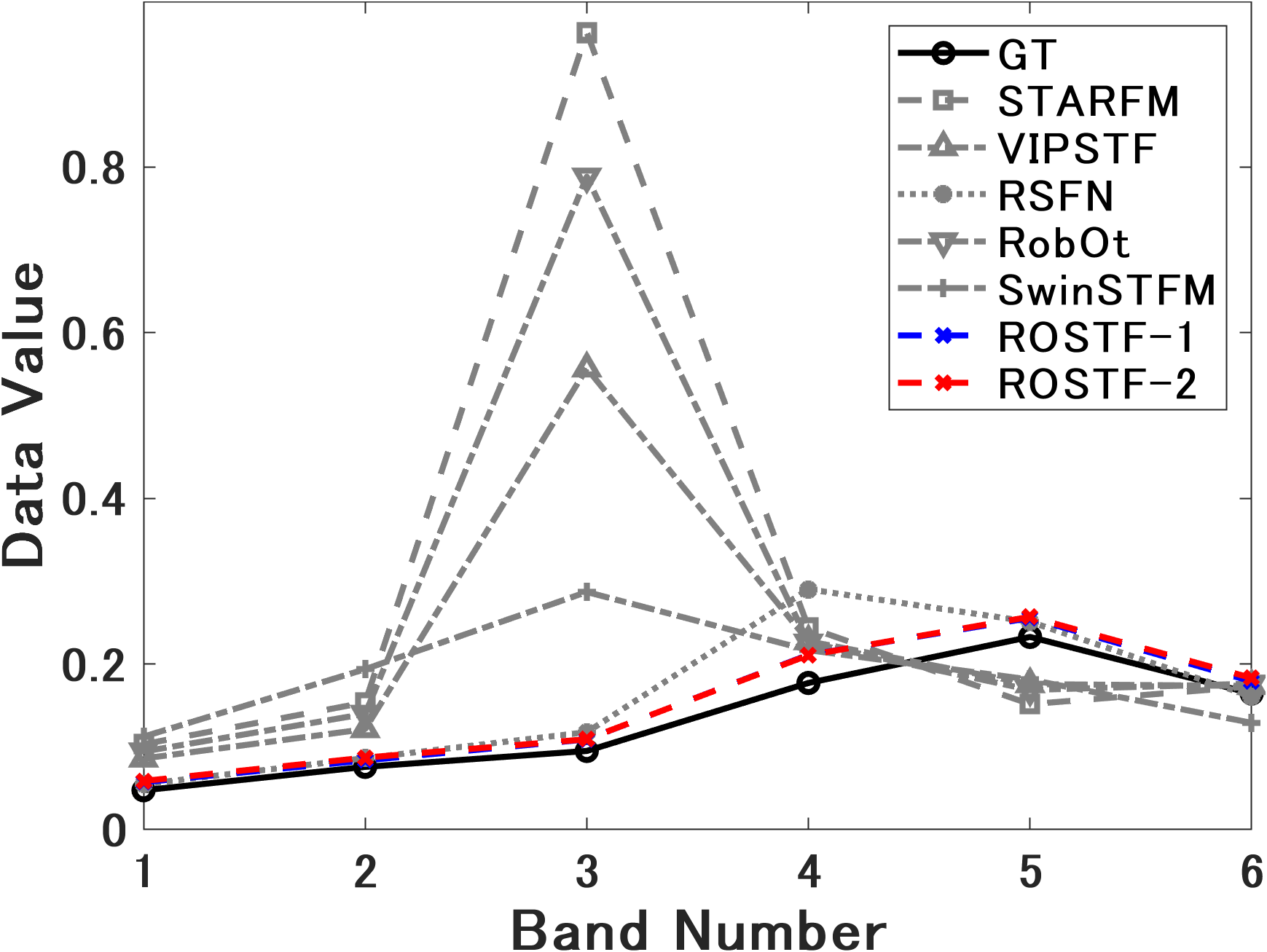}}
		\end{minipage} \\
        \begin{minipage}{0.47\hsize}
			\centerline{(a)}
		\end{minipage}
        \begin{minipage}{0.47\hsize}
			\centerline{(b)}
		\end{minipage} \\
        \vspace{-6mm}
	\caption{Spectral profiles of a specific pixel in the results of each method for the Site2 data in Case1~(a) and Case4~(b).}
        \label{fig: Spectral Plot Site1}
\end{figure}


%% file: Experiments/Real_experimental_result_metrics.tex
\begin{table*}[t]
	\begin{center}
		\caption{The RMSE, SAM, MSSIM, and CC Results in the Experiments With Real Data}
		
        \label{table: real metrics}
            \vspace{-2mm}
		\centering
            \scalebox{1}{
		\begin{tabular}{p{0.5cm} p{0.7cm} p{0.7cm} >{\centering\arraybackslash}p{1.3cm} >{\centering\arraybackslash}p{1.3cm} >{\centering\arraybackslash}p{1.3cm} >{\centering\arraybackslash}p{1.3cm} >{\centering\arraybackslash}p{1.3cm} >{\centering\arraybackslash}p{1.3cm} >{\centering\arraybackslash}p{1.3cm}}
			\toprule
                
                \multirow{2}{*}{Site} & \multirow{2}{*}{Noise} & \multirow{2}{*}{Metrics} & STARFM & VIPSTF & RSFN & RobOt & SWIN & \textbf{ROSTF-1} & \textbf{ROSTF-2} \\

                & & & \cite{STARFM} & \cite{VIPSTF} & \cite{RSFN} & \cite{RobOt} & \cite{SWIN} & \textbf{(Ours)} & \textbf{(Ours)} \\
                
                
                \midrule 
                \multirow{16}{*}{Site1} 
                & \multirow{4}{*}{Case1} 
                & RMSE 
                & \textbf{0.0263} & 0.0275 & 0.0523 & 0.0281 & 0.0297 & 0.0278 & 0.0266 \\ 
                
                & & SAM 
                & 0.0838 & 0.0897 & 0.1601 & 0.0906 & 0.0987 & 0.0937 & \textbf{0.0829} \\ 
                
                & & SSIM 
                & 0.9701 & 0.9666 & 0.8700 & 0.9667 & 0.9625 & 0.9652 & \textbf{0.9704} \\ 
                
                & & CC 
                & \textbf{0.9742} & 0.9719 & 0.9029 & 0.9710 & 0.9680 & 0.9711 & 0.9735  \\ 
                \cmidrule(lr){2-10} 
                & \multirow{4}{*}{Case2} 
                & RMSE 
                & 0.0565 & 0.0439 & 0.0692 & 0.0562 & 0.0398 & 0.0363 & \textbf{0.0333} \\ 
                
                & & SAM 
                & 0.2466 & 0.1826 & 0.2455 & 0.2436 & 0.1471 & 0.1277 & \textbf{0.1119} \\ 
                
                & & SSIM 
                & 0.8448 & 0.9024 & 0.7957 & 0.8465 & 0.9271 & 0.9347 & \textbf{0.9463} \\ 
                
                & & CC 
                & 0.8942 & 0.9282 & 0.8089 & 0.8927 & 0.9465 & 0.9501 & \textbf{0.9582}  \\ 
                \cmidrule(lr){2-10} 
                & \multirow{4}{*}{Case3} 
                & RMSE 
                & 0.1409 & 0.0991 & 0.0539 & 0.1402 & 0.0512 & 0.0362 & \textbf{0.0300} \\ 
                
                & & SAM 
                & 0.2234 & 0.1955 & 0.1696 & 0.2301 & 0.1699 & 0.1434 & \textbf{0.0998} \\ 
                
                & & SSIM 
                & 0.5165 & 0.6604 & 0.8679 & 0.5181 & 0.8742 & 0.9289 & \textbf{0.9608} \\ 
                
                & & CC 
                & 0.6095 & 0.7278 & 0.8998 & 0.6120 & 0.9039 & 0.9515 & \textbf{0.9664}  \\ 
                \cmidrule(lr){2-10} 
                & \multirow{4}{*}{Case4} 
                & RMSE 
                & 0.1490 & 0.1046 & 0.0700 & 0.1481 & 0.0551 & 0.0447 & \textbf{0.0382} \\ 
                
                & & SAM 
                & 0.3533 & 0.2706 & 0.2493 & 0.3521 & 0.1938 & 0.1608 & \textbf{0.1252} \\ 
                
                & & SSIM 
                & 0.4804 & 0.6321 & 0.7849 & 0.4800 & 0.8563 & 0.8988 & \textbf{0.9318} \\ 
                
                & & CC 
                & 0.5875 & 0.7085 & 0.8012 & 0.5848 & 0.8919 & 0.9238 & \textbf{0.9448}  \\ 
                \midrule 
                \multirow{16}{*}{Site2} 
                & \multirow{4}{*}{Case1} 
                & RMSE 
                & 0.0458 & \textbf{0.0371} & 0.0503 & 0.0428 & 0.0388 & 0.0483 & 0.0487 \\ 
                
                & & SAM 
                & 0.1406 & \textbf{0.1283} & 0.1585 & 0.1310 & 0.1285 & 0.1478 & 0.1345 \\ 
                
                & & SSIM 
                & 0.8762 & \textbf{0.9136} & 0.8713 & 0.8911 & 0.9041 & 0.8707 & 0.8725 \\ 
                
                & & CC 
                & 0.9051 & \textbf{0.9326} & 0.9064 & 0.9127 & 0.9267 & 0.8950 & 0.8957  \\ 
                \cmidrule(lr){2-10} 
                & \multirow{4}{*}{Case2} 
                & RMSE 
                & 0.0677 & 0.0485 & 0.0742 & 0.0588 & 0.0466 & 0.0470 & \textbf{0.0463} \\ 
                
                & & SAM 
                & 0.3067 & 0.2146 & 0.3121 & 0.2514 & 0.1668 & 0.1649 & \textbf{0.1355} \\ 
                
                & & SSIM 
                & 0.7047 & 0.8229 & 0.7184 & 0.7641 & 0.8527 & 0.8684 & \textbf{0.8821} \\ 
                
                & & CC 
                & 0.8191 & 0.8856 & 0.7425 & 0.8460 & 0.8944 & 0.8939 & \textbf{0.8996}  \\ 
                \cmidrule(lr){2-10} 
                & \multirow{4}{*}{Case3} 
                & RMSE 
                & 0.1419 & 0.0914 & 0.0539 & 0.1140 & 0.0568 & \textbf{0.0490} & 0.0494 \\ 
                
                & & SAM 
                & 0.2906 & 0.2248 & 0.1842 & 0.2533 & 0.2097 & 0.1558 & \textbf{0.1380} \\ 
                
                & & SSIM 
                & 0.3891 & 0.5789 & 0.8513 & 0.4821 & 0.7730 & 0.8618 & \textbf{0.8690} \\ 
                
                & & CC 
                & 0.5448 & 0.6911 & 0.8922 & 0.6180 & 0.8411 & 0.8920 & \textbf{0.8929}  \\ 
                \cmidrule(lr){2-10} 
                & \multirow{4}{*}{Case4} 
                & RMSE 
                & 0.1500 & 0.0963 & 0.0755 & 0.1222 & 0.0595 & 0.0509 & \textbf{0.0484} \\ 
                
                & & SAM 
                & 0.4229 & 0.2972 & 0.3177 & 0.3461 & 0.2266 & 0.2072 & \textbf{0.1546} \\ 
                
                & & SSIM 
                & 0.3526 & 0.5439 & 0.7096 & 0.4374 & 0.7524 & 0.8291 & \textbf{0.8724} \\ 
                
                & & CC 
                & 0.5214 & 0.6686 & 0.7338 & 0.5844 & 0.8252 & 0.8725 & \textbf{0.8863}  \\ 
                
                \bottomrule 
		\end{tabular}
            }
	\end{center}
\end{table*}

%% file: Experiments/results/Site1_Real_Case1.tex
\begin{figure}[t]
	\begin{center}
        \begin{minipage}{0.24\hsize} 
            \centerline{\includegraphics[width=\hsize]{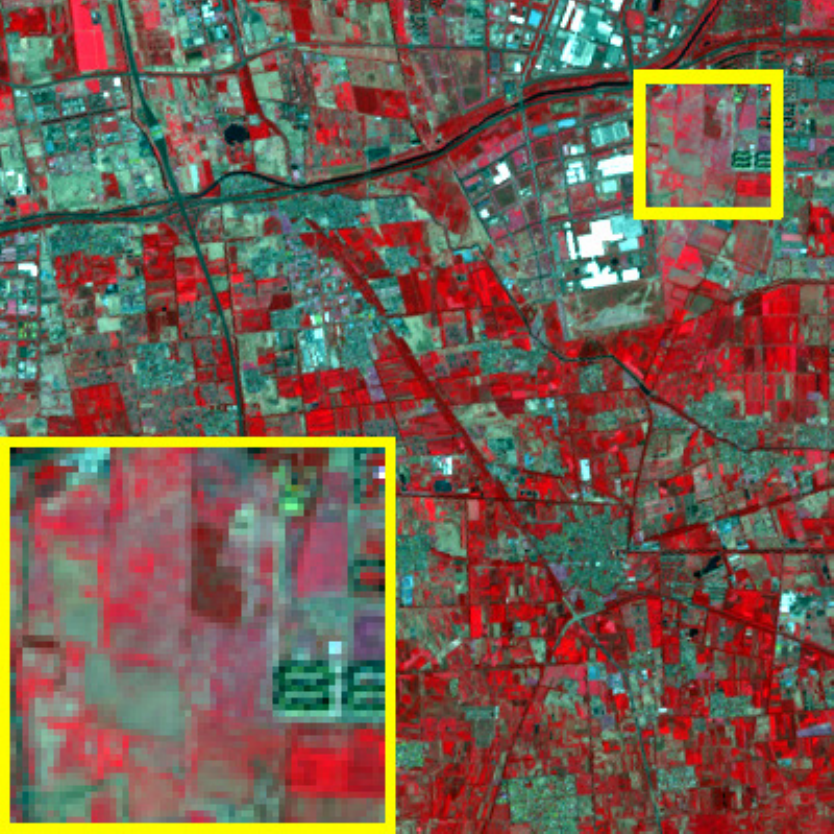}} 
        \end{minipage}
        \begin{minipage}{0.24\hsize}
            \centerline{\includegraphics[width=\hsize]{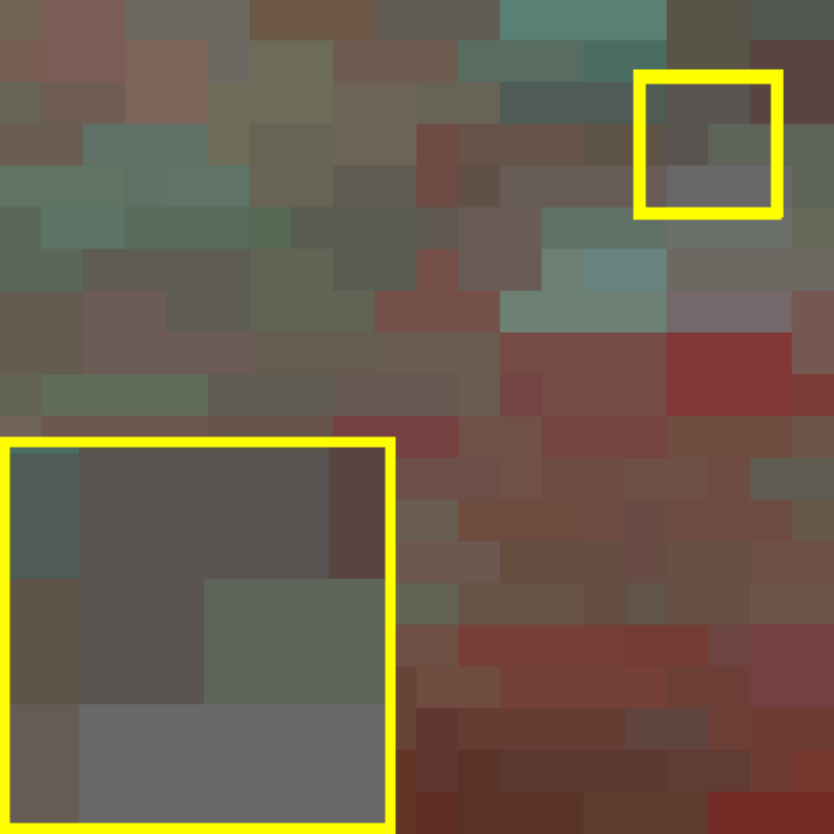}} 
        \end{minipage}
        \begin{minipage}{0.24\hsize}
            \centerline{\includegraphics[width=\hsize]{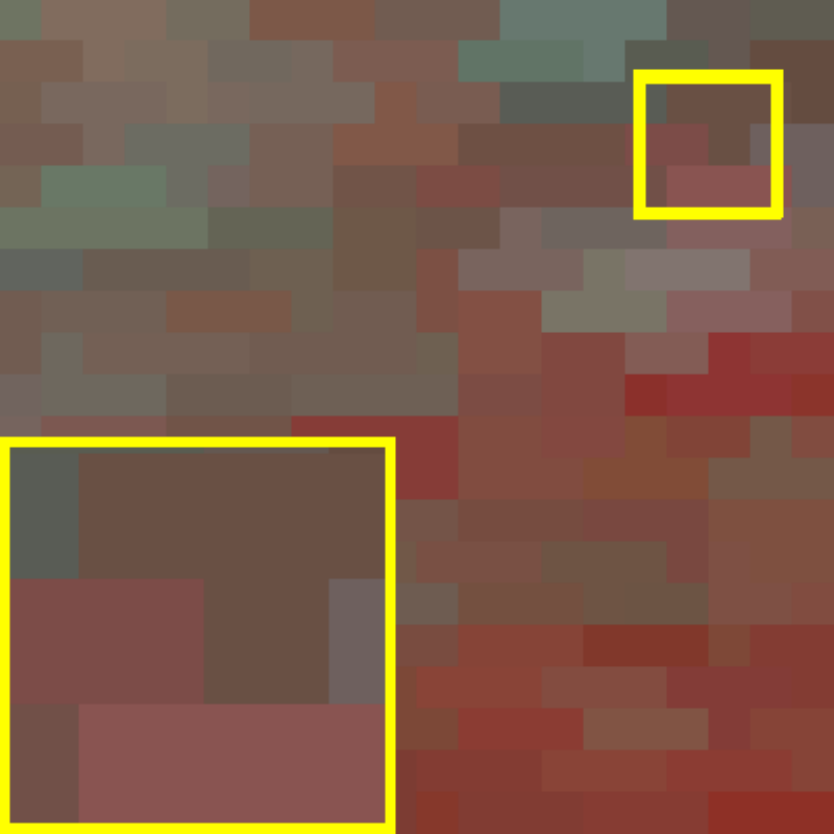}} 
        \end{minipage} \\
        \vspace{1mm}
        \begin{minipage}{0.24\hsize} 
            \centerline{$\Hr$}
        \end{minipage}
        \begin{minipage}{0.24\hsize} 
            \centerline{$\Lr$}
        \end{minipage}
        \begin{minipage}{0.24\hsize} 
            \centerline{$\Lt$}
        \end{minipage} \\
        \vspace{2mm}
        \begin{minipage}{0.24\hsize}
            \centerline{\includegraphics[width=\hsize]{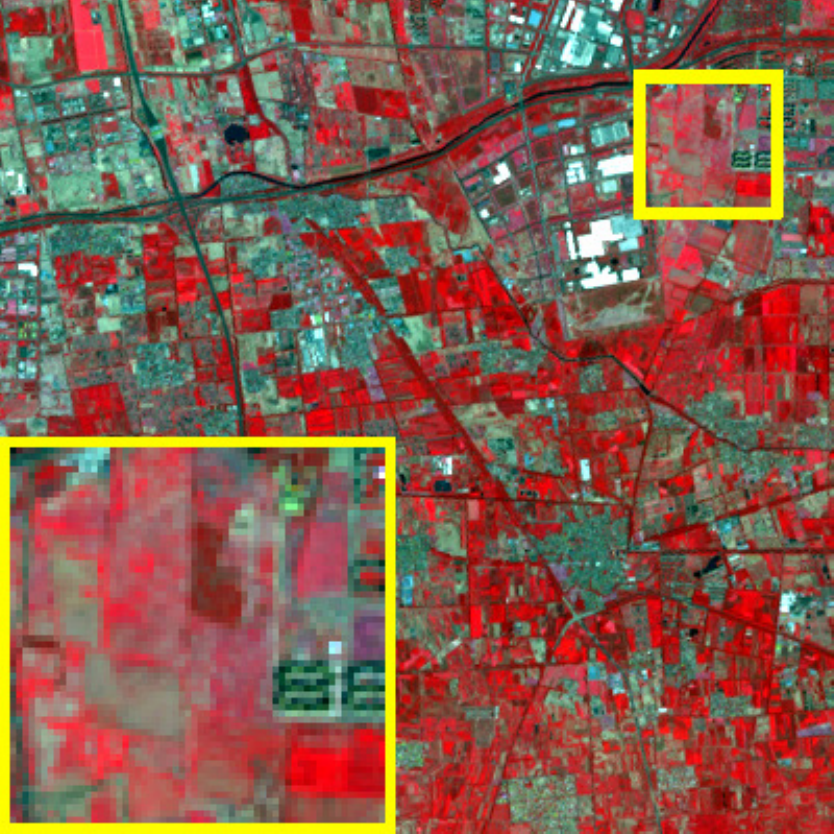}} 
        \end{minipage} 
        \begin{minipage}{0.24\hsize}
            \centerline{\includegraphics[width=\hsize]{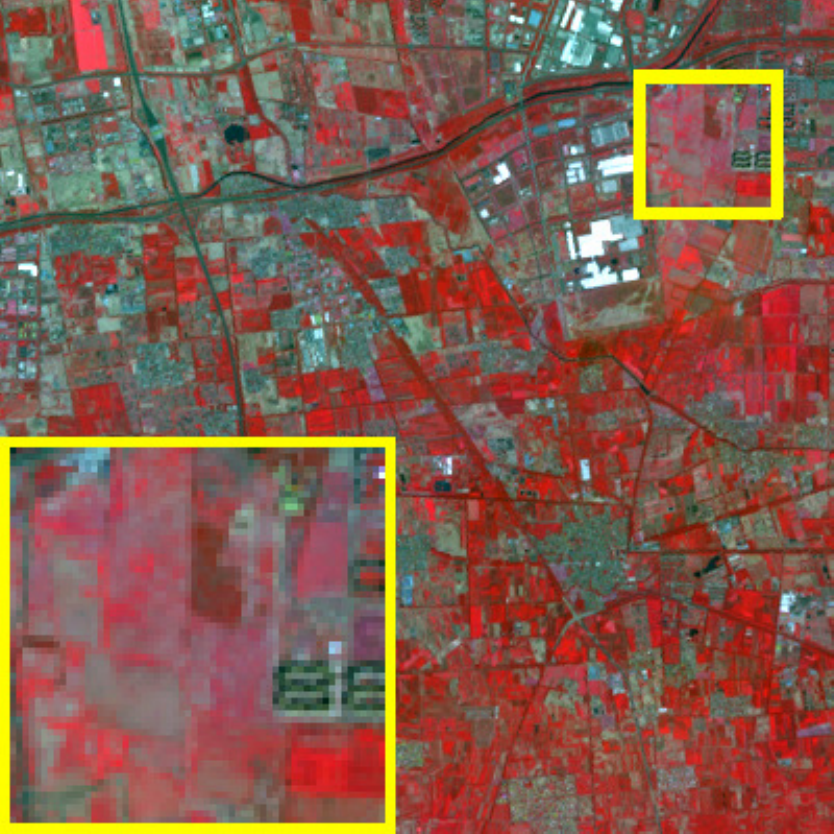}} 
        \end{minipage}
        \begin{minipage}{0.24\hsize}
            \centerline{\includegraphics[width=\hsize]{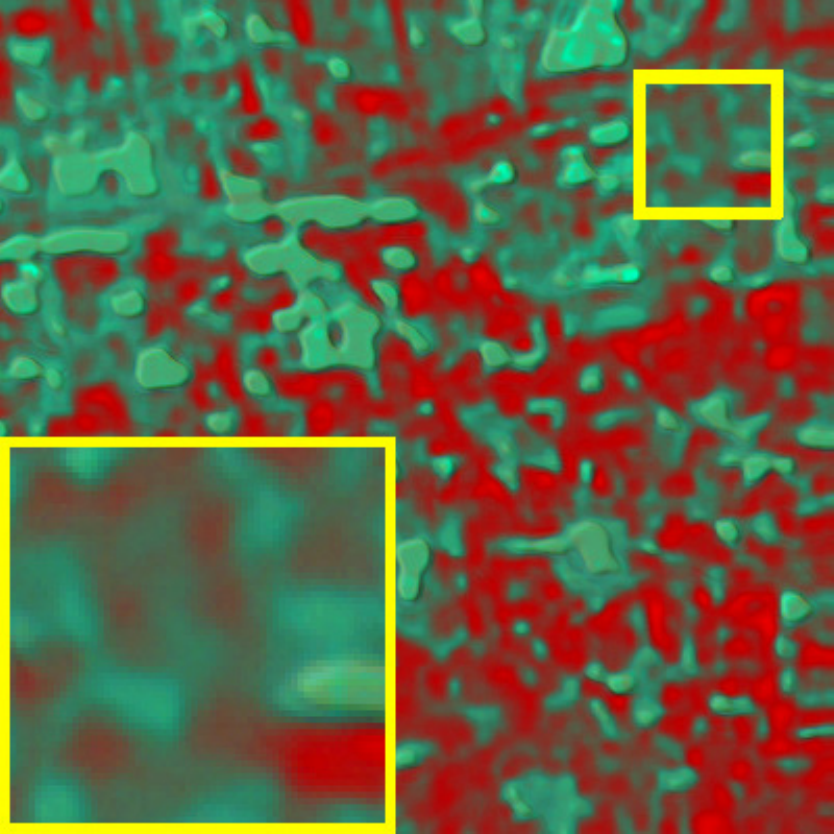}} 
        \end{minipage}
        \begin{minipage}{0.24\hsize}
            \centerline{\includegraphics[width=\hsize]{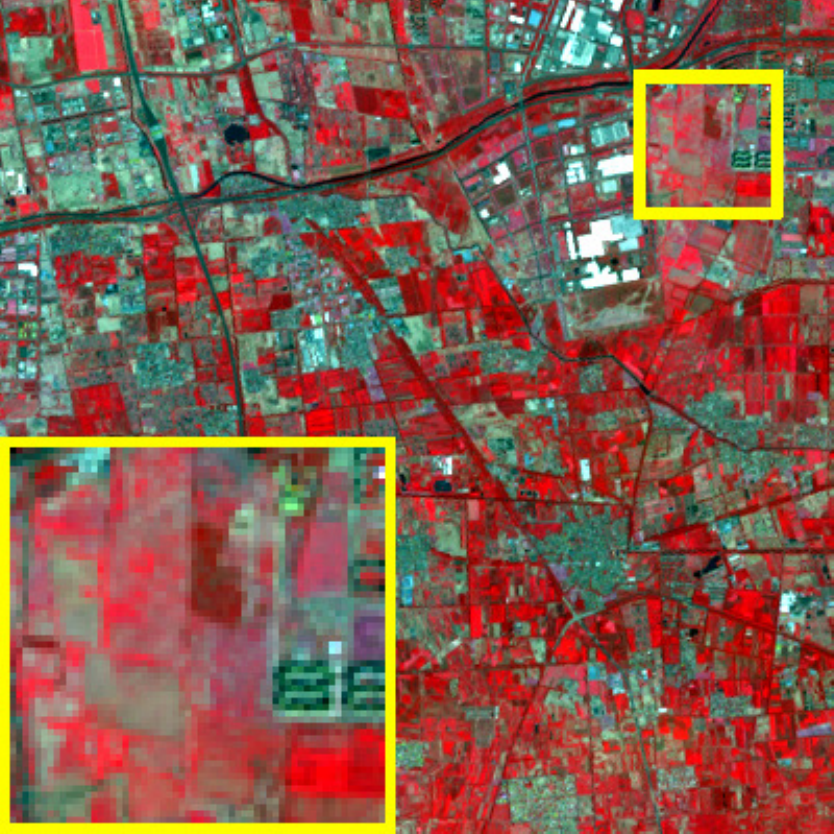}} 
        \end{minipage} \\
        \vspace{1mm}
        \begin{minipage}{0.24\hsize} 
            \centerline{STARFM}
        \end{minipage} 
        \begin{minipage}{0.24\hsize} 
            \centerline{VIPSTF}
        \end{minipage}
        \begin{minipage}{0.24\hsize} 
            \centerline{RSFN}
        \end{minipage}
        \begin{minipage}{0.24\hsize} 
            \centerline{RobOt}
        \end{minipage} \\
        \vspace{2mm}
        \begin{minipage}{0.24\hsize}
            \centerline{\includegraphics[width=\hsize]{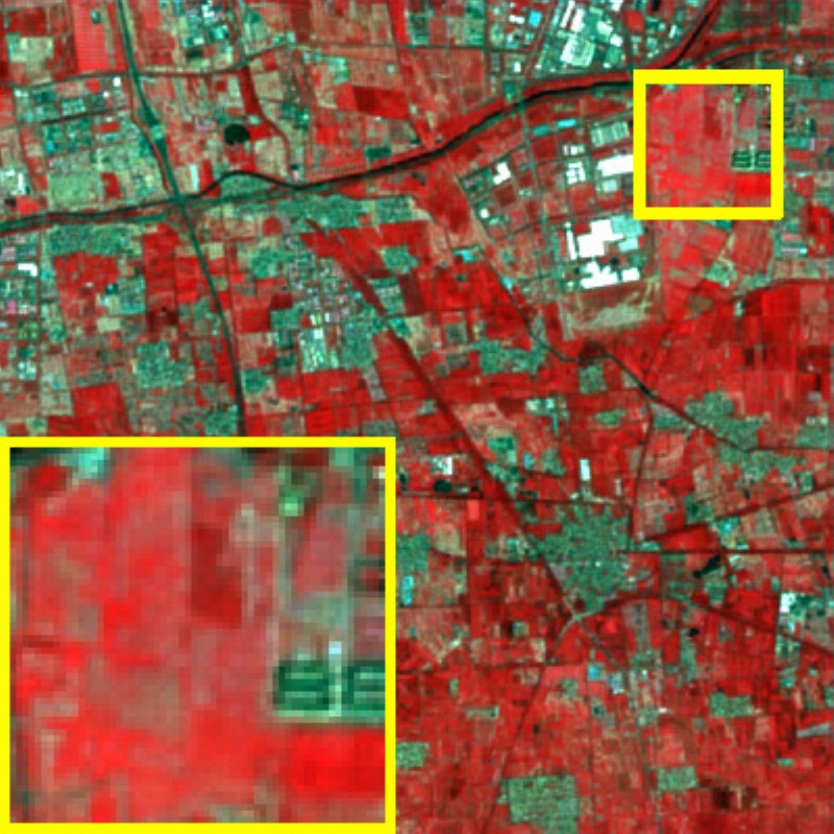}} 
        \end{minipage}
        \begin{minipage}{0.24\hsize}
            \centerline{\includegraphics[width=\hsize]{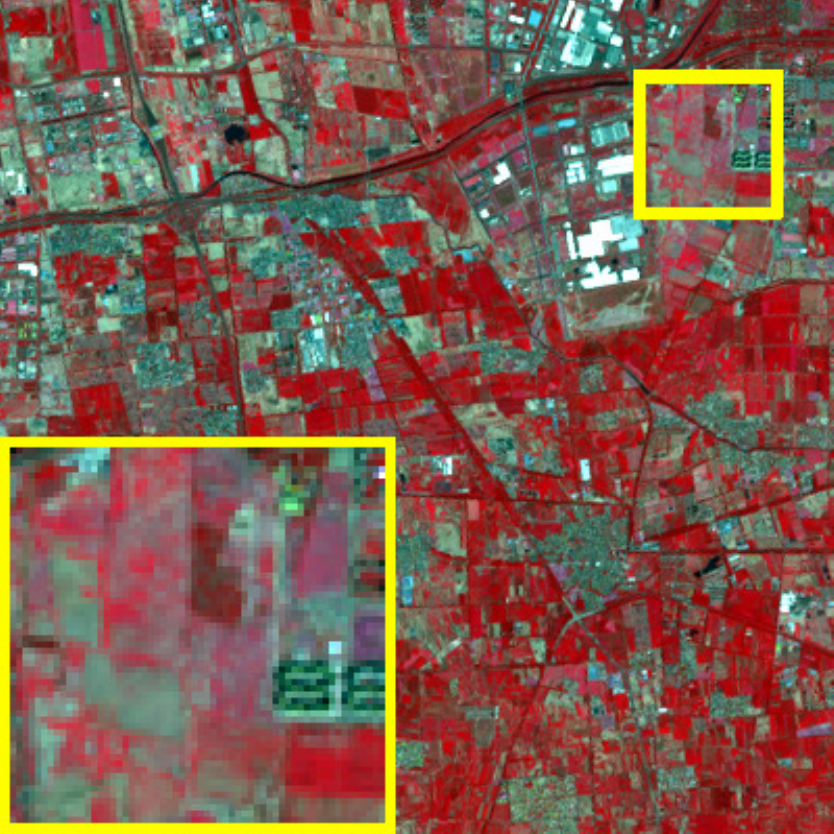}} 
        \end{minipage}
        \begin{minipage}{0.24\hsize}
            \centerline{\includegraphics[width=\hsize]{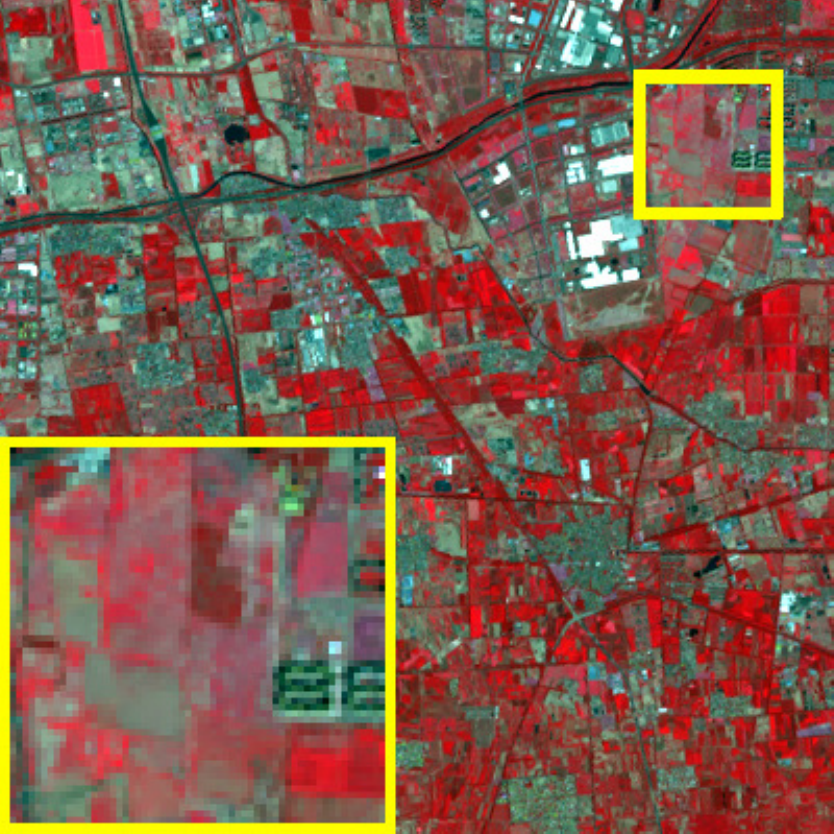}} 
        \end{minipage} 
        \begin{minipage}{0.24\hsize}
            \centerline{\includegraphics[width=\hsize]{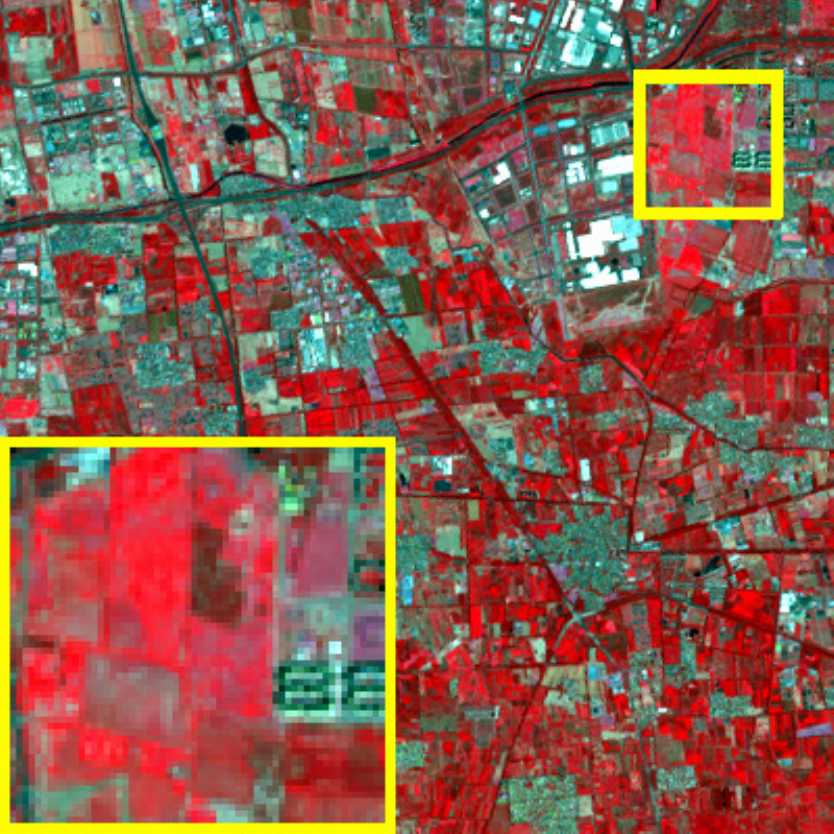}} 
        \end{minipage} \\
        \vspace{1mm}
        \begin{minipage}{0.24\hsize} 
            \centerline{SwinSTFM}
        \end{minipage}
        \begin{minipage}{0.24\hsize} 
            \centerline{\textbf{ROSTF-1}}
        \end{minipage}
        \begin{minipage}{0.24\hsize} 
            \centerline{\textbf{ROSTF-2}}
        \end{minipage}
        \begin{minipage}{0.24\hsize} 
            \centerline{Ground-truth}
        \end{minipage} \\
	\end{center}
        \vspace{-3mm}
	\caption{ST fusion results for the Site1 real data in Case1.}
 \label{fig: Site1 Real Case1 results}
\end{figure}

%% file: Experiments/results/Site2_Real_Case1.tex
\begin{figure}[t]
	\begin{center}
        \begin{minipage}{0.24\hsize} 
            \centerline{\includegraphics[width=\hsize]{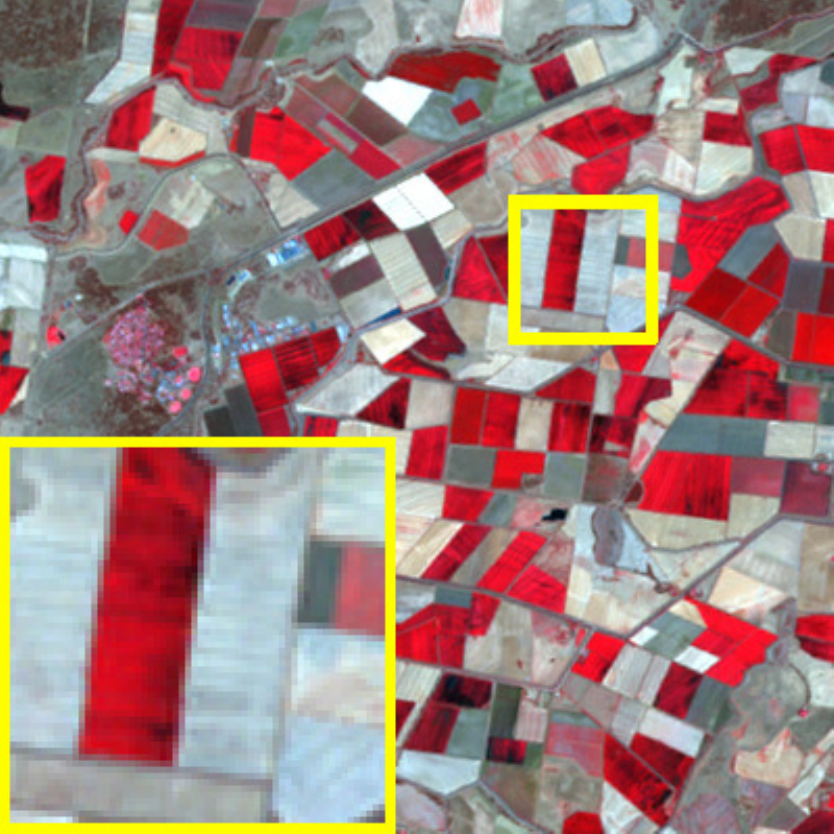}} 
        \end{minipage}
        \begin{minipage}{0.24\hsize}
            \centerline{\includegraphics[width=\hsize]{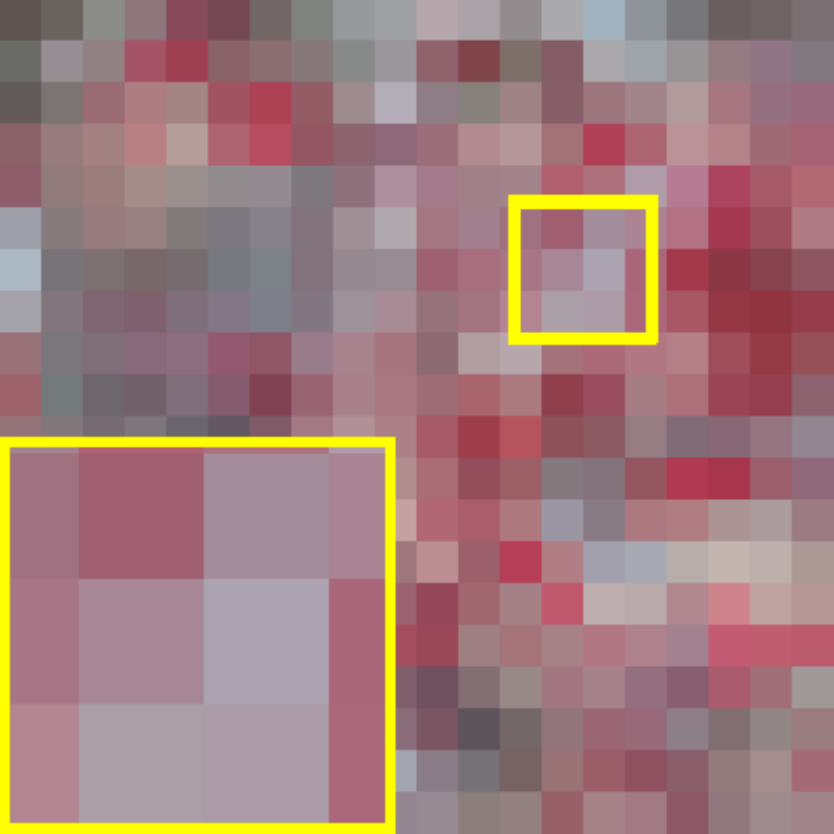}} 
        \end{minipage}
        \begin{minipage}{0.24\hsize}
            \centerline{\includegraphics[width=\hsize]{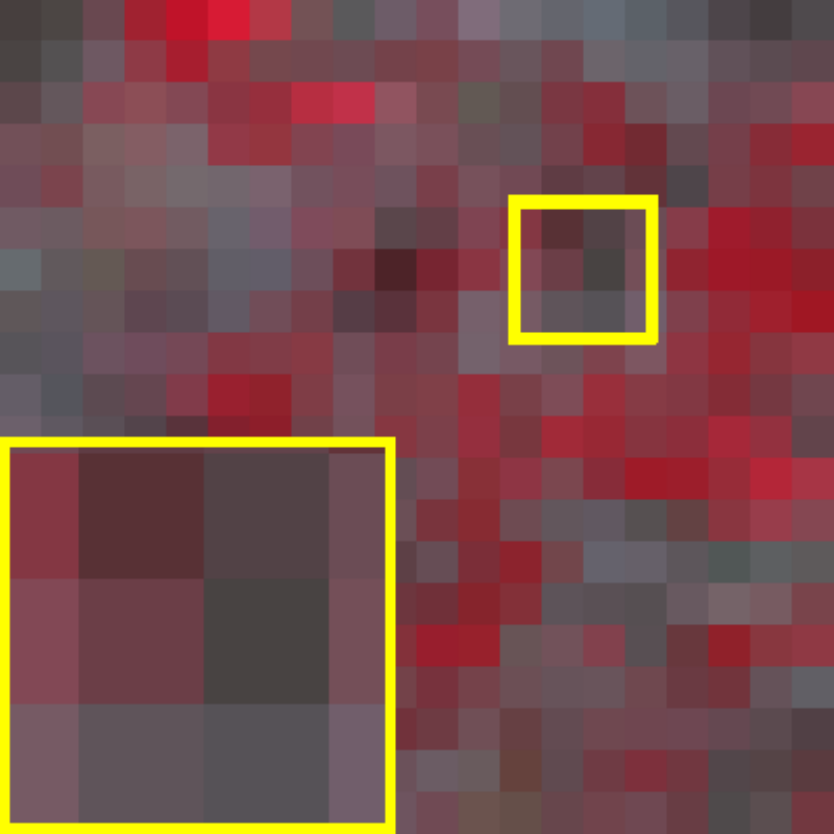}} 
        \end{minipage} \\
        \vspace{1mm}
        \begin{minipage}{0.24\hsize} 
            \centerline{$\Hr$}
        \end{minipage}
        \begin{minipage}{0.24\hsize} 
            \centerline{$\Lr$}
        \end{minipage}
        \begin{minipage}{0.24\hsize} 
            \centerline{$\Lt$}
        \end{minipage} \\
        \vspace{2mm}
        \begin{minipage}{0.24\hsize}
            \centerline{\includegraphics[width=\hsize]{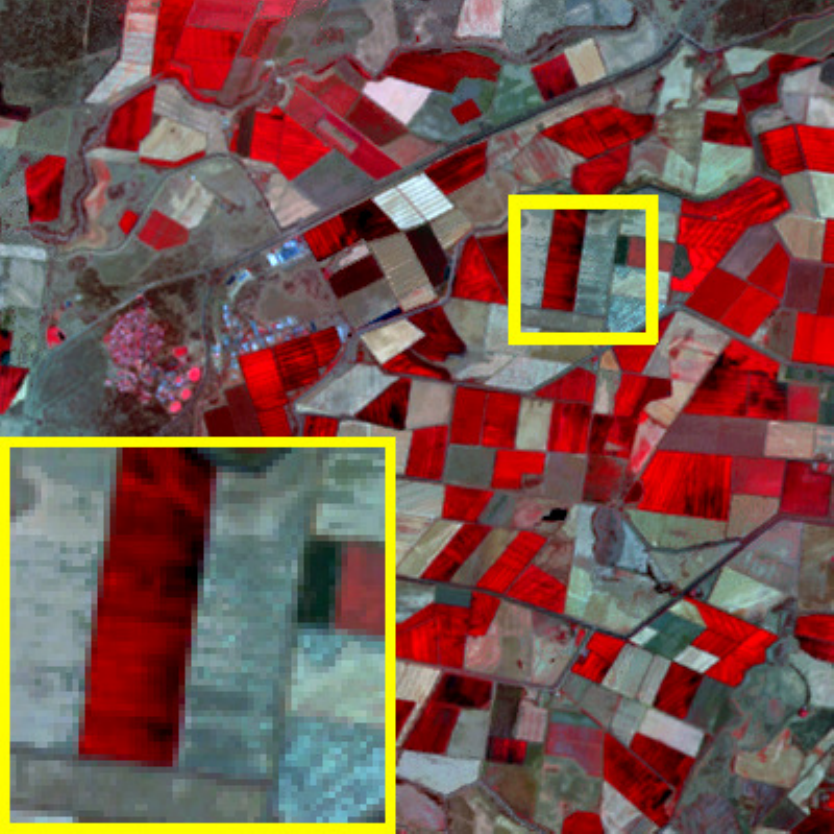}} 
        \end{minipage} 
        \begin{minipage}{0.24\hsize}
            \centerline{\includegraphics[width=\hsize]{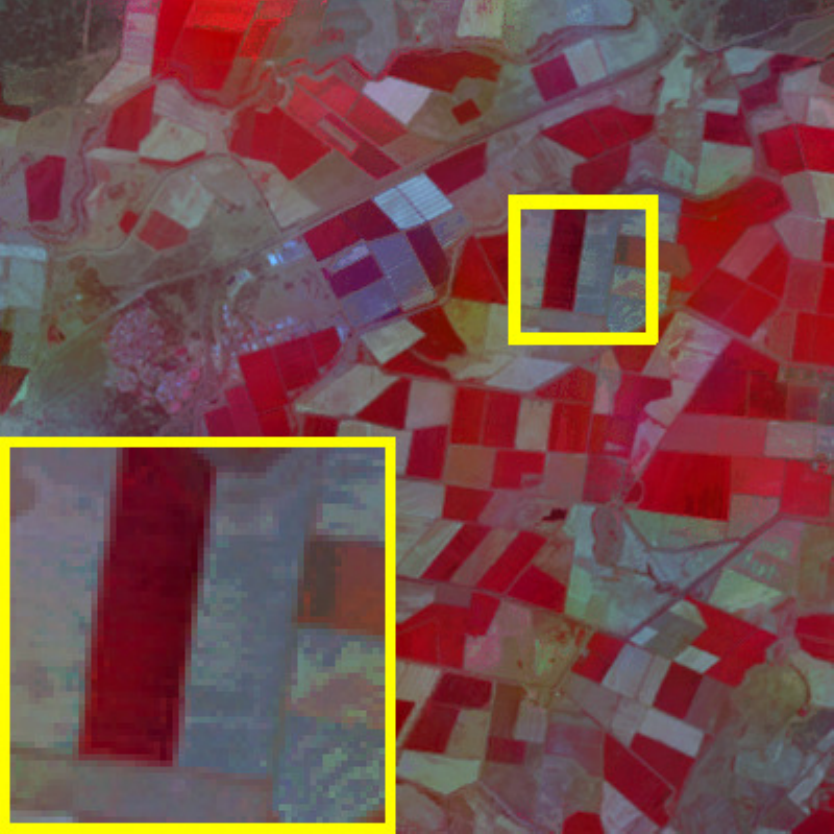}} 
        \end{minipage}
        \begin{minipage}{0.24\hsize}
            \centerline{\includegraphics[width=\hsize]{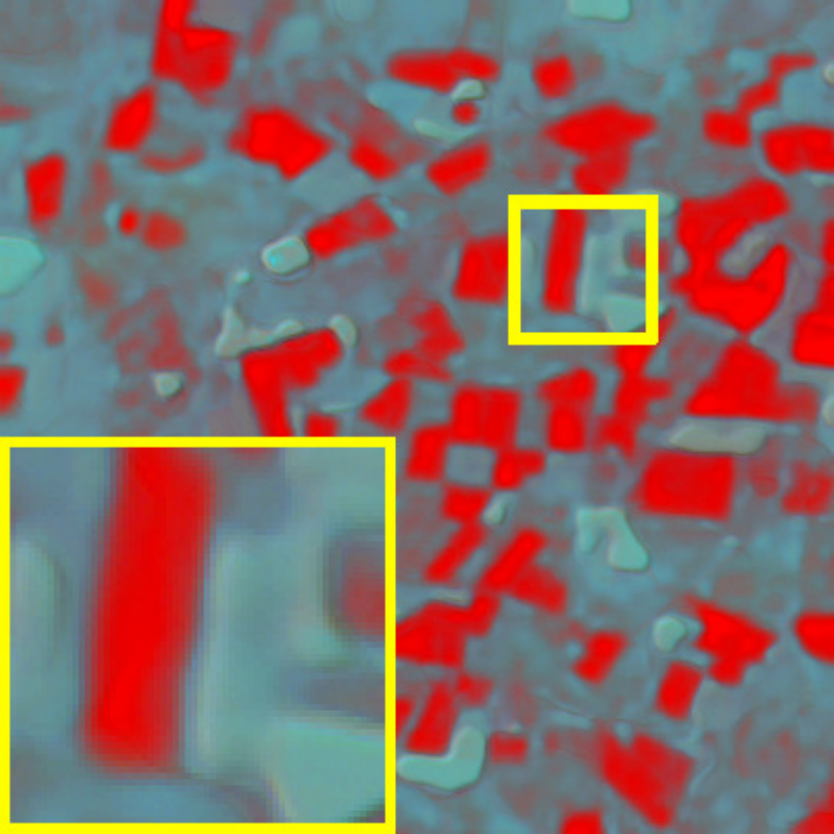}} 
        \end{minipage}
        \begin{minipage}{0.24\hsize}
            \centerline{\includegraphics[width=\hsize]{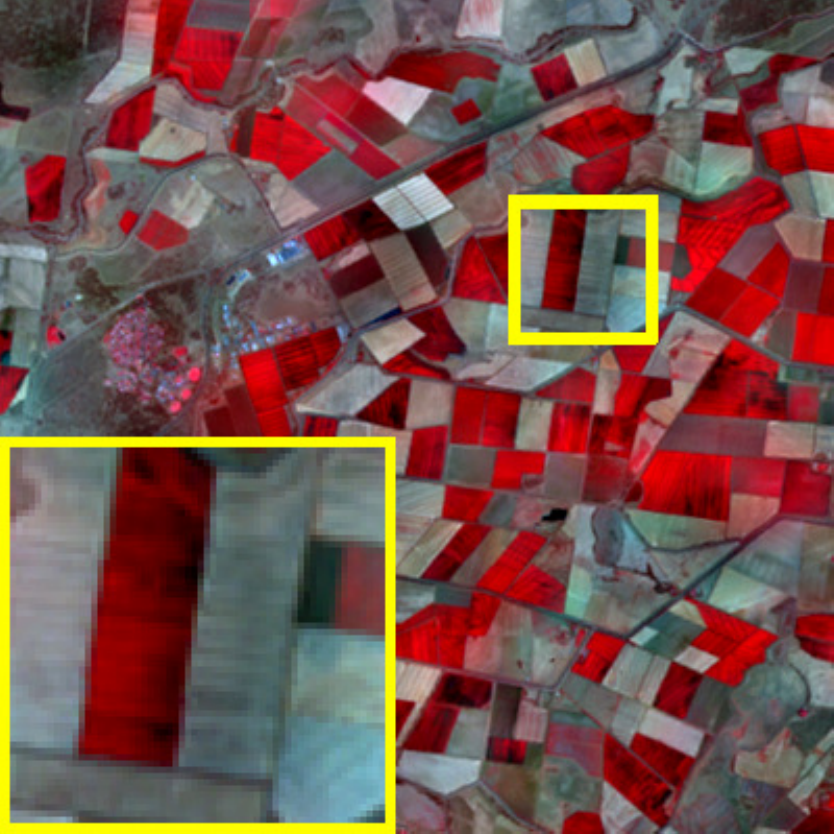}} 
        \end{minipage} \\
        \vspace{1mm}
        \begin{minipage}{0.24\hsize} 
            \centerline{STARFM}
        \end{minipage} 
        \begin{minipage}{0.24\hsize} 
            \centerline{VIPSTF}
        \end{minipage}
        \begin{minipage}{0.24\hsize} 
            \centerline{RSFN}
        \end{minipage}
        \begin{minipage}{0.24\hsize} 
            \centerline{RobOt}
        \end{minipage} \\
        \vspace{2mm}
        \begin{minipage}{0.24\hsize}
            \centerline{\includegraphics[width=\hsize]{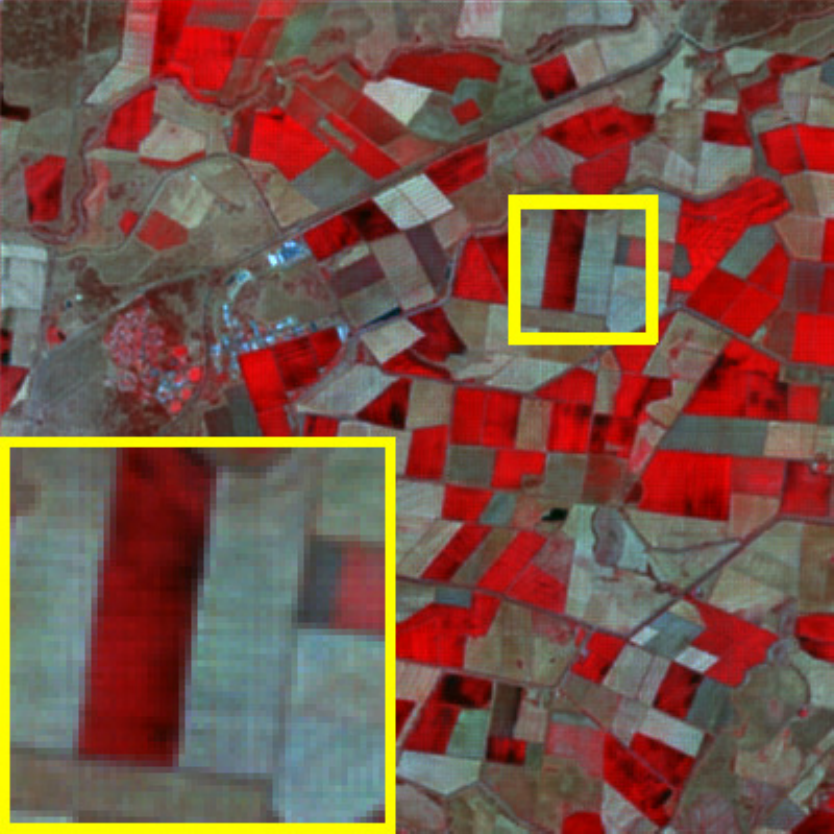}} 
        \end{minipage}
        \begin{minipage}{0.24\hsize}
            \centerline{\includegraphics[width=\hsize]{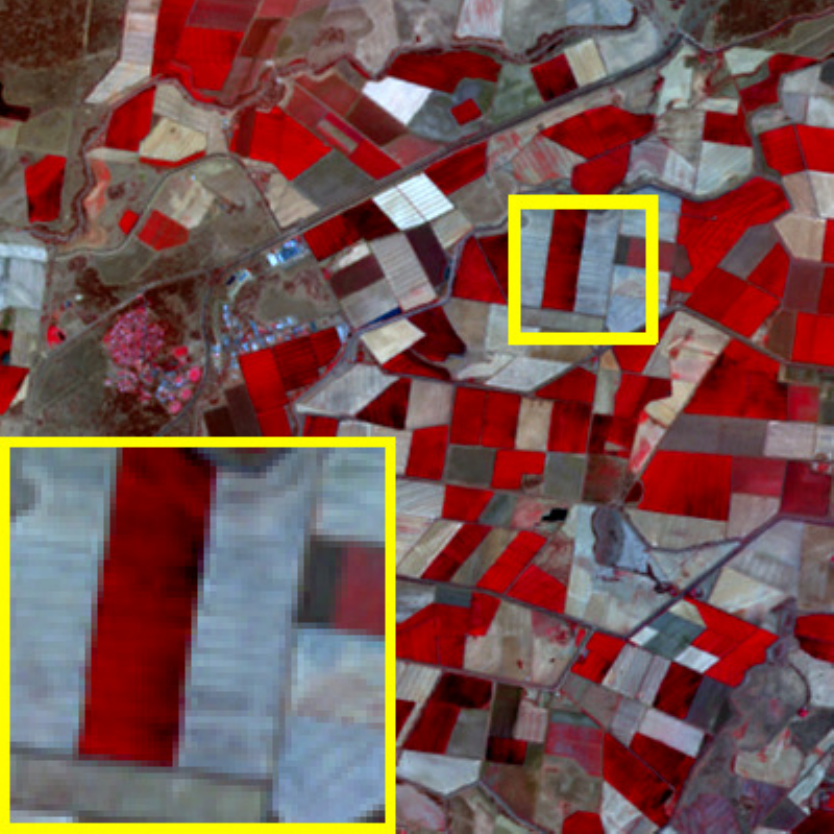}} 
        \end{minipage}
        \begin{minipage}{0.24\hsize}
            \centerline{\includegraphics[width=\hsize]{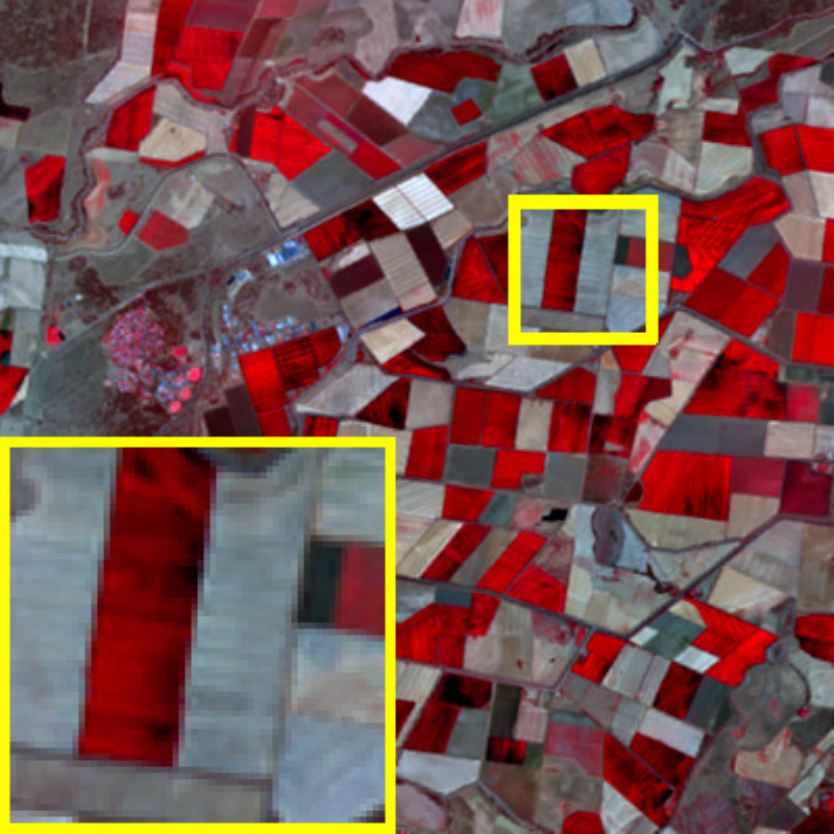}} 
        \end{minipage} 
        \begin{minipage}{0.24\hsize}
            \centerline{\includegraphics[width=\hsize]{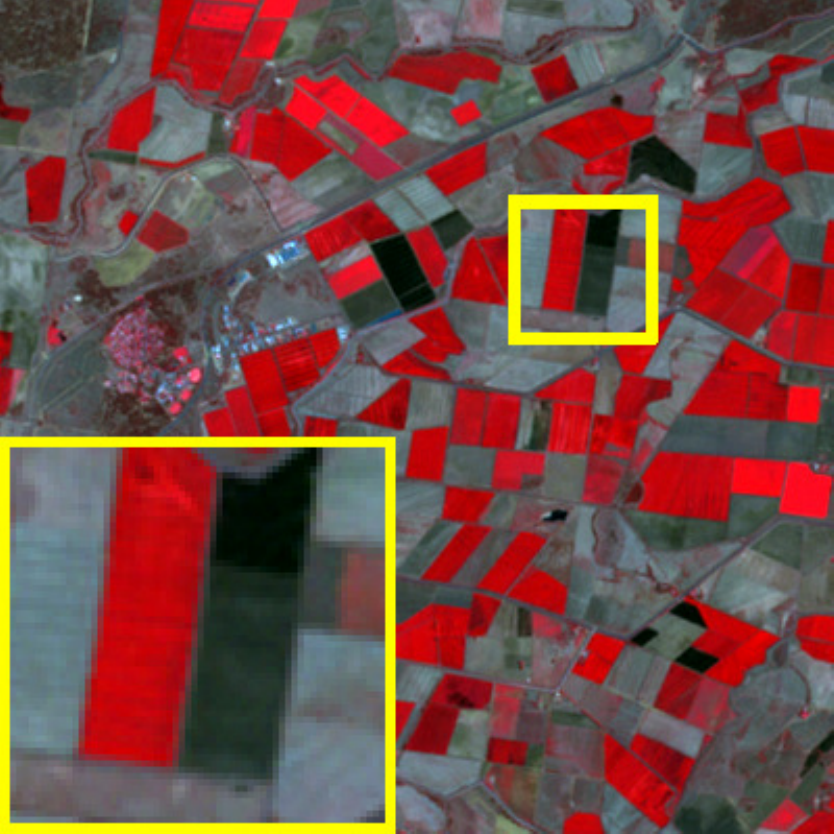}} 
        \end{minipage} \\
        \vspace{1mm}
        \begin{minipage}{0.24\hsize} 
            \centerline{SwinSTFM}
        \end{minipage}
        \begin{minipage}{0.24\hsize} 
            \centerline{\textbf{ROSTF-1}}
        \end{minipage}
        \begin{minipage}{0.24\hsize} 
            \centerline{\textbf{ROSTF-2}}
        \end{minipage}
        \begin{minipage}{0.24\hsize} 
            \centerline{Ground-truth}
        \end{minipage} \\
	\end{center}
        \vspace{-3mm}
	\caption{ST fusion results for the Site2 real data in Case1.}
 \label{fig: Site2 Real Case1 results}
\end{figure}

%% file: Experiments/results/Site1_Real_Case234.tex
\begin{figure*}[ht]
	\begin{center}
        \begin{minipage}{0.01\hsize}
            \centerline{\rotatebox{90}{Case2}}
        \end{minipage}
        \begin{minipage}{0.1\hsize} 
            \centerline{\includegraphics[width=\hsize]{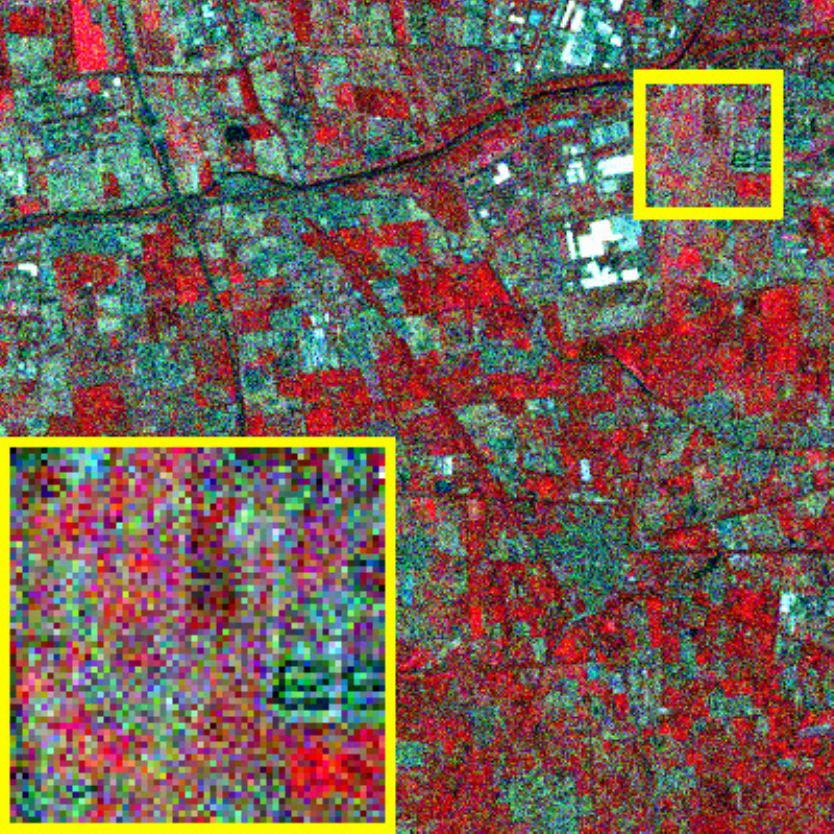}} 
        \end{minipage}
        \begin{minipage}{0.1\hsize}
            \centerline{\includegraphics[width=\hsize]{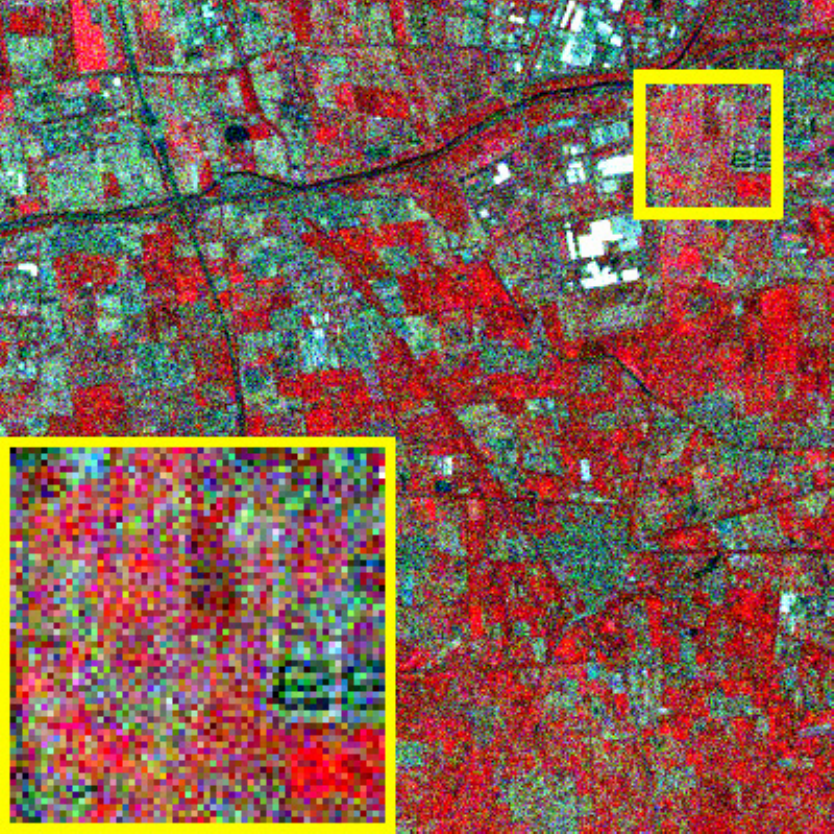}} 
        \end{minipage}
        \begin{minipage}{0.1\hsize}
            \centerline{\includegraphics[width=\hsize]{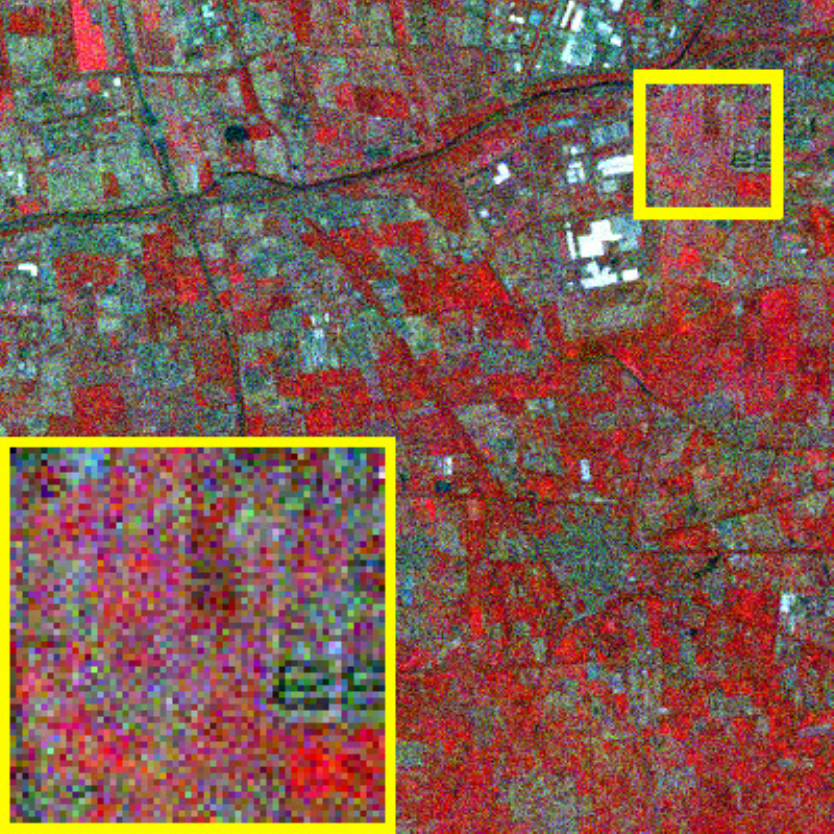}} 
        \end{minipage}
        \begin{minipage}{0.1\hsize}
            \centerline{\includegraphics[width=\hsize]{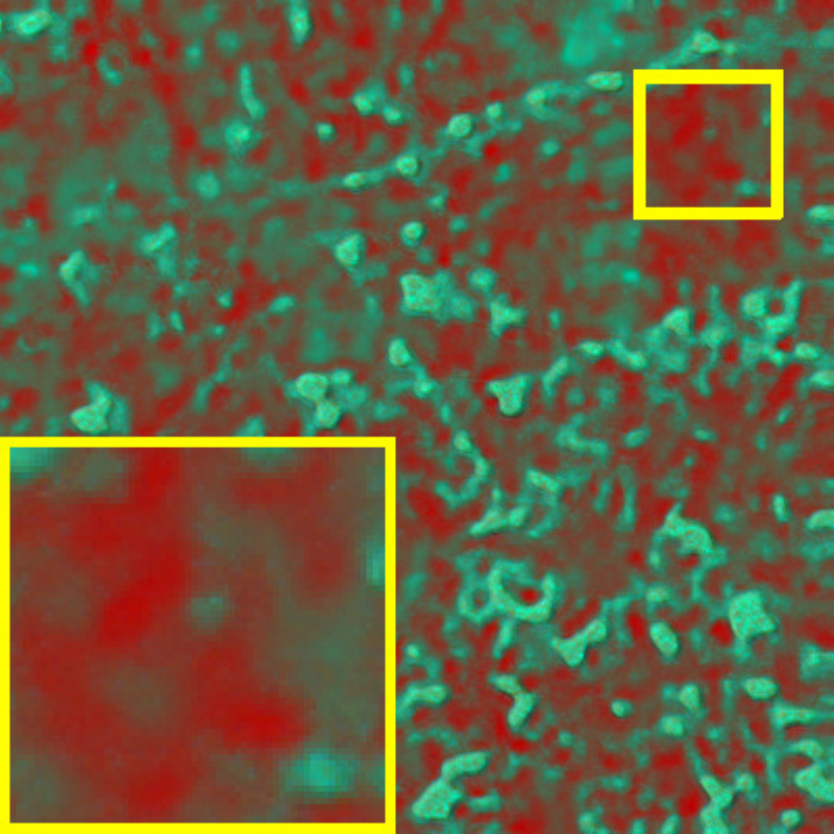}} 
        \end{minipage}
        \begin{minipage}{0.1\hsize}
            \centerline{\includegraphics[width=\hsize]{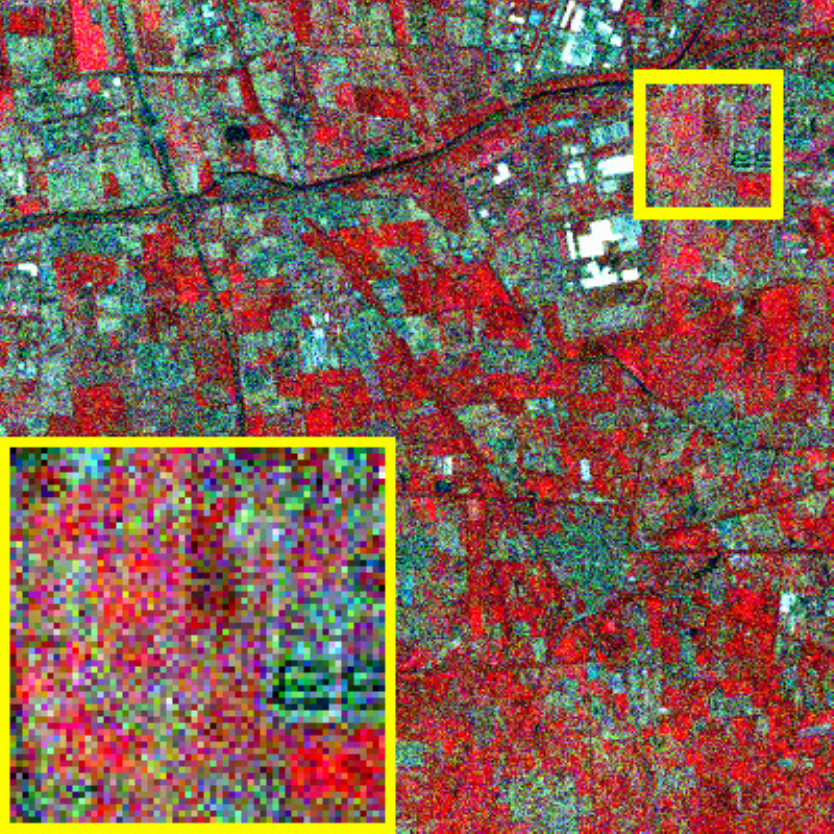}} 
        \end{minipage}
        \begin{minipage}{0.1\hsize}
            \centerline{\includegraphics[width=\hsize]{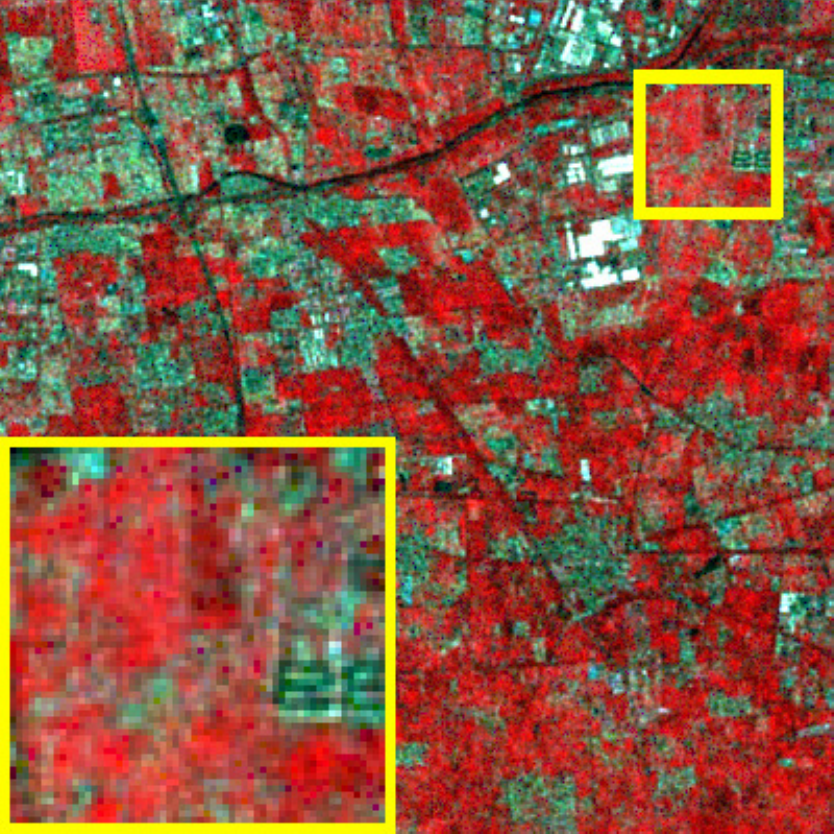}} 
        \end{minipage}
        \begin{minipage}{0.1\hsize}
            \centerline{\includegraphics[width=\hsize]{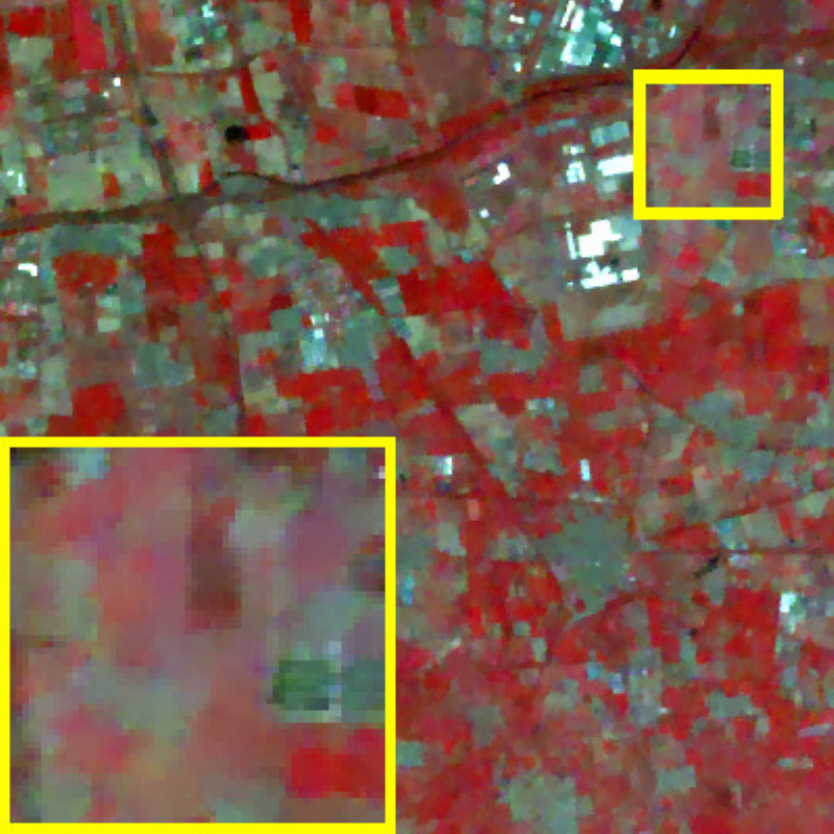}} 
        \end{minipage}
        \begin{minipage}{0.1\hsize}
            \centerline{\includegraphics[width=\hsize]{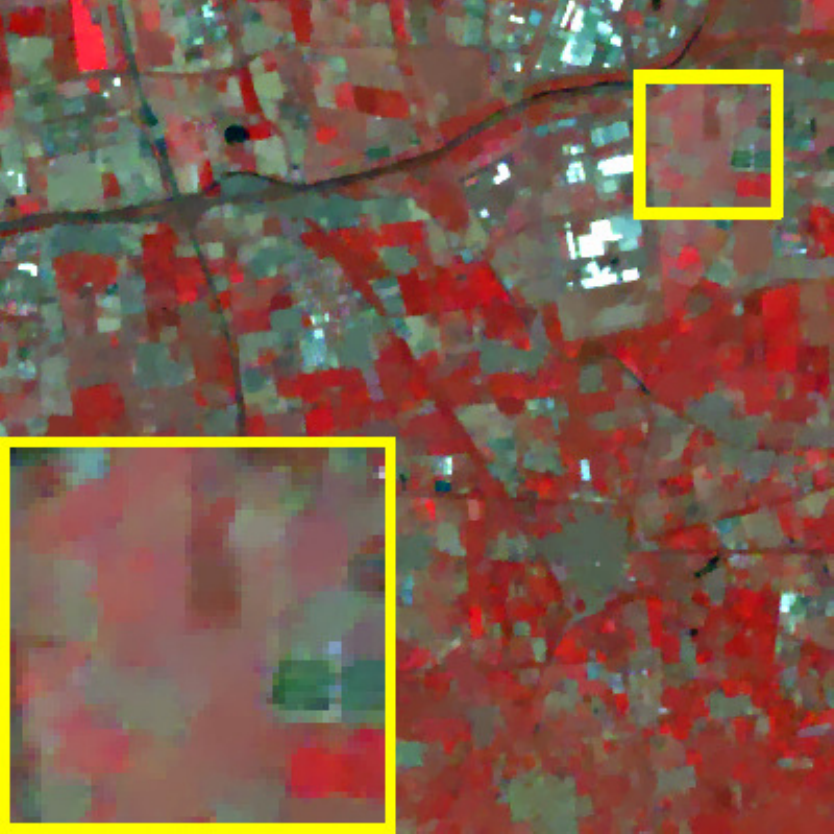}} 
        \end{minipage}
        \begin{minipage}{0.1\hsize}
            \centerline{\includegraphics[width=\hsize]{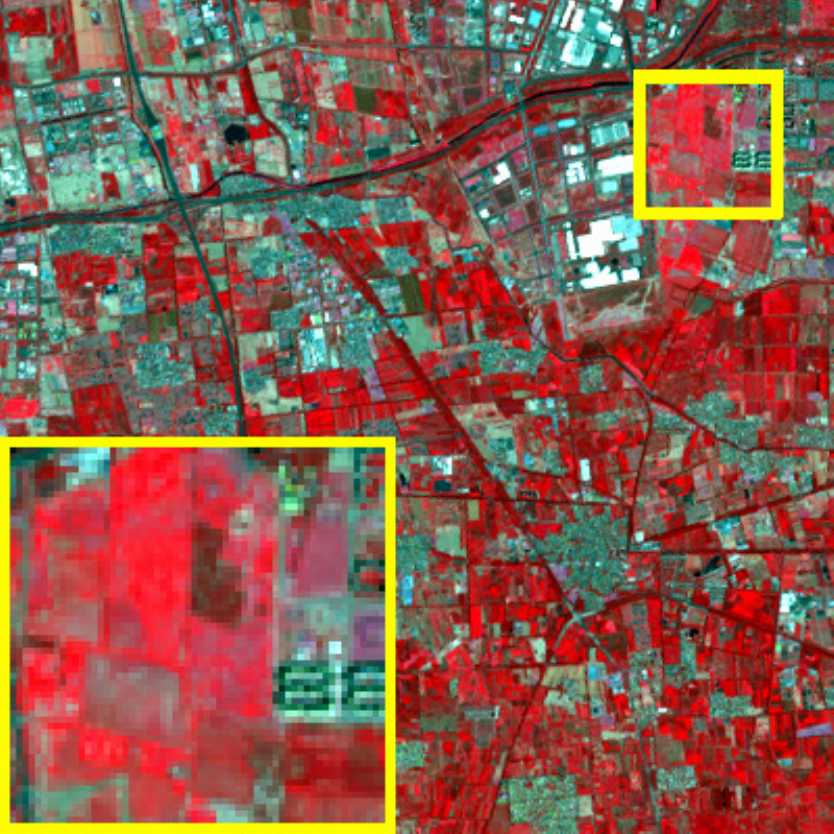}} 
        \end{minipage}  \\

        \vspace{2mm}
        
        \begin{minipage}{0.01\hsize}
			\centerline{\rotatebox{90}{Case3}}
	\end{minipage}
        \begin{minipage}{0.1\hsize} 
            \centerline{\includegraphics[width=\hsize]{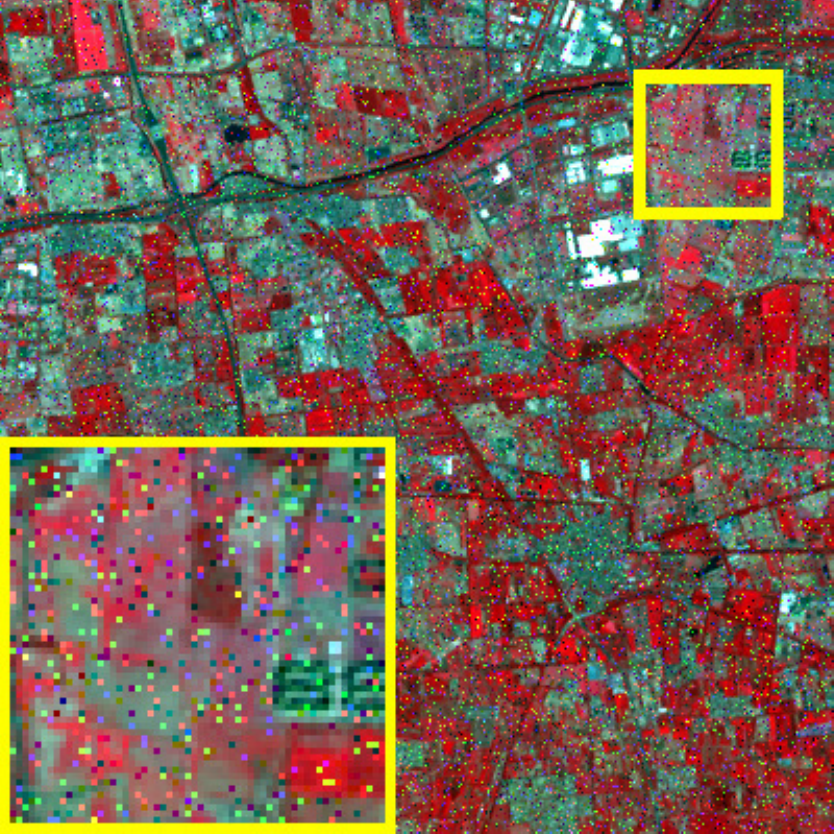}} 
        \end{minipage}
        \begin{minipage}{0.1\hsize}
            \centerline{\includegraphics[width=\hsize]{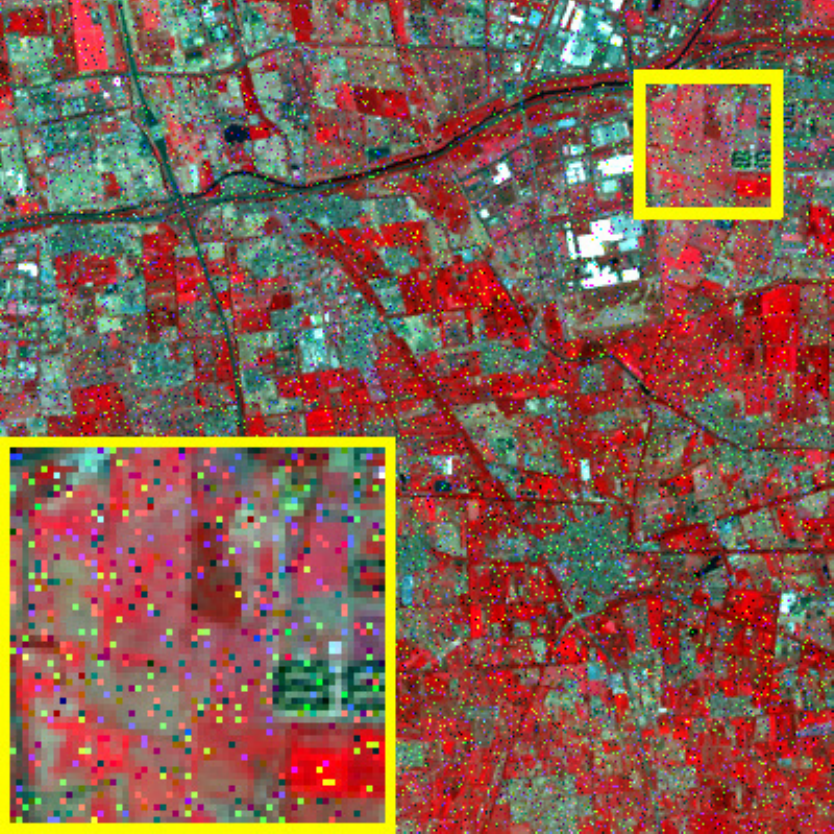}} 
        \end{minipage}
        \begin{minipage}{0.1\hsize}
            \centerline{\includegraphics[width=\hsize]{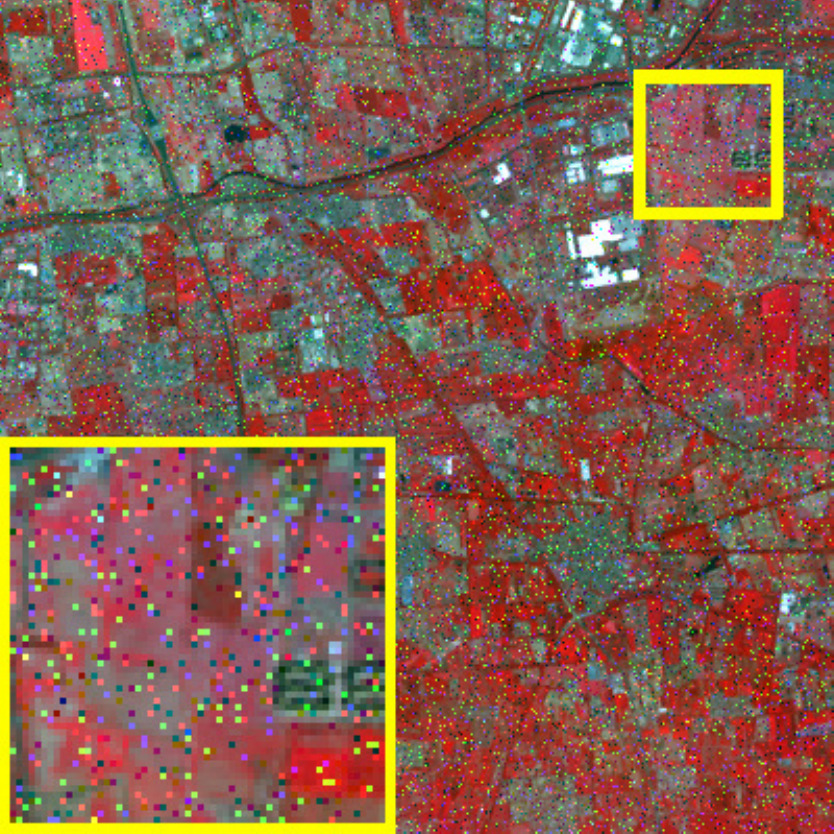}} 
        \end{minipage}
        \begin{minipage}{0.1\hsize}
            \centerline{\includegraphics[width=\hsize]{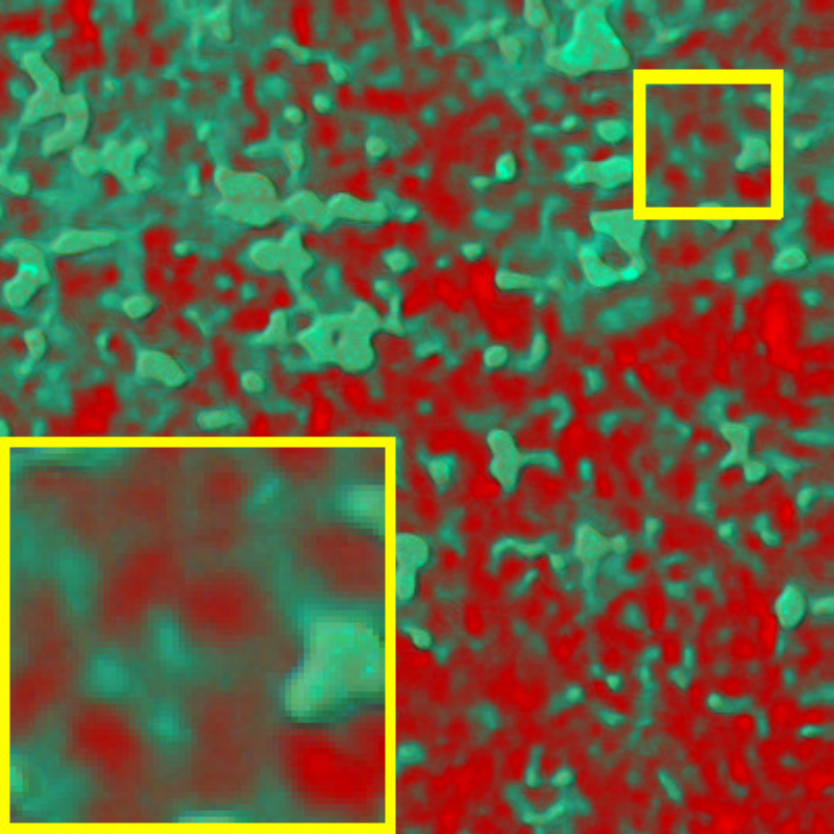}} 
        \end{minipage}
        \begin{minipage}{0.1\hsize}
            \centerline{\includegraphics[width=\hsize]{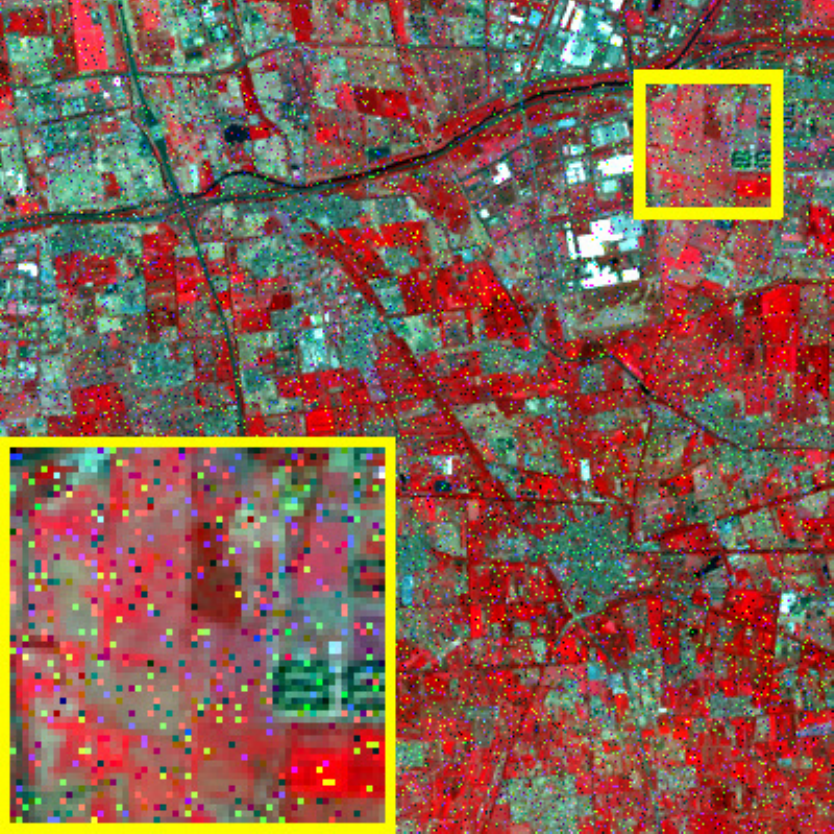}} 
        \end{minipage}
        \begin{minipage}{0.1\hsize}
            \centerline{\includegraphics[width=\hsize]{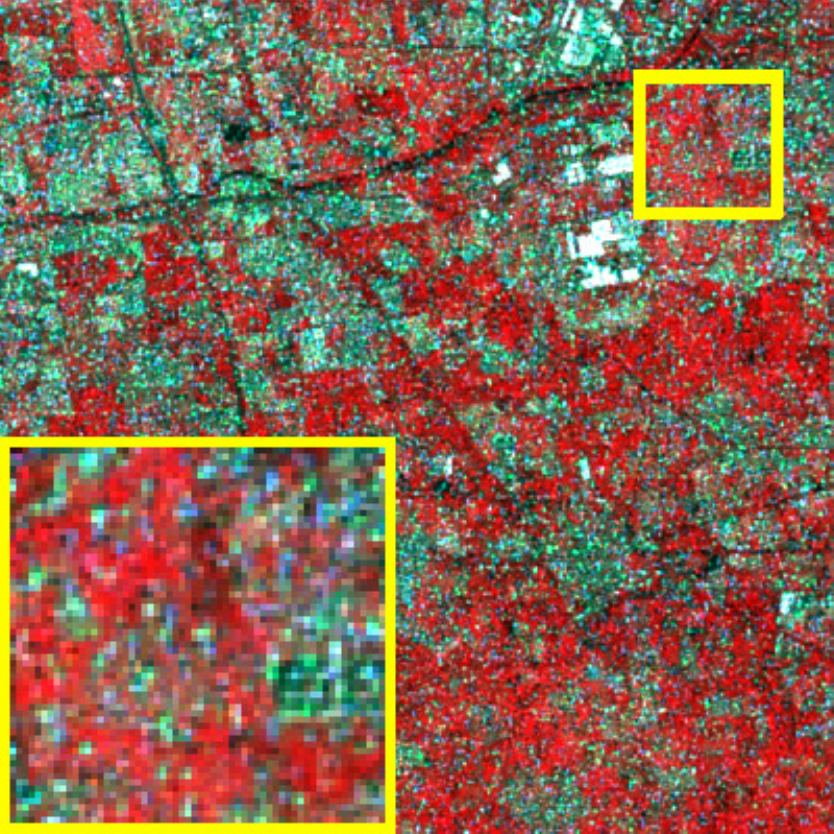}} 
        \end{minipage}
        \begin{minipage}{0.1\hsize}
            \centerline{\includegraphics[width=\hsize]{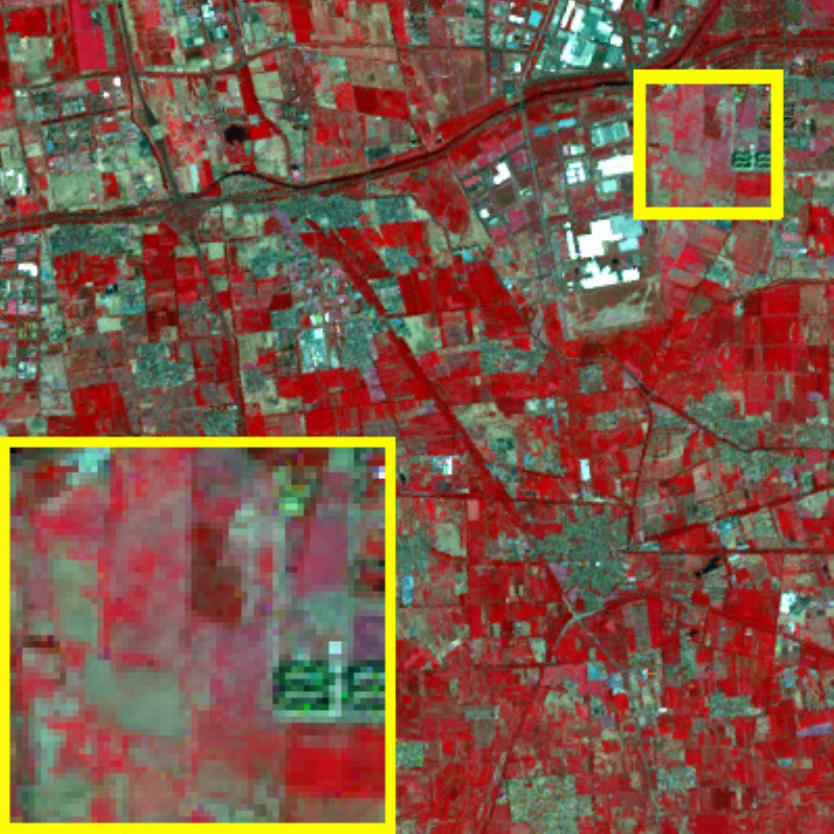}} 
        \end{minipage}
        \begin{minipage}{0.1\hsize}
            \centerline{\includegraphics[width=\hsize]{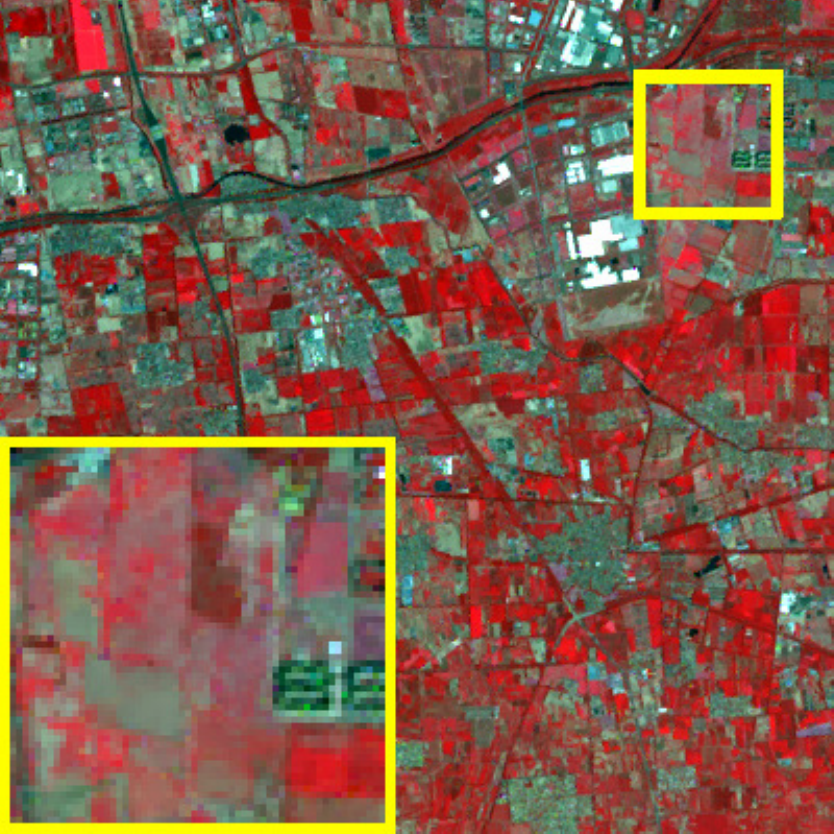}} 
        \end{minipage}
        \begin{minipage}{0.1\hsize}
            \centerline{\includegraphics[width=\hsize]{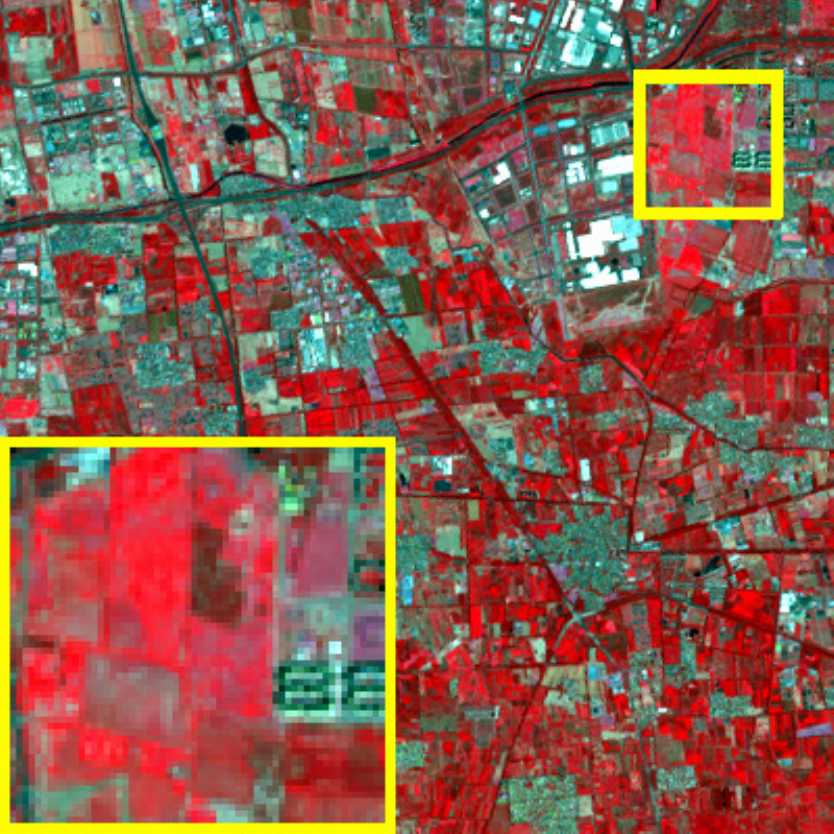}} 
        \end{minipage}  \\

        \vspace{2mm}
  
        \begin{minipage}{0.01\hsize}
			\centerline{\rotatebox{90}{Case4}}
	\end{minipage}
        \begin{minipage}{0.1\hsize} 
            \centerline{\includegraphics[width=\hsize]{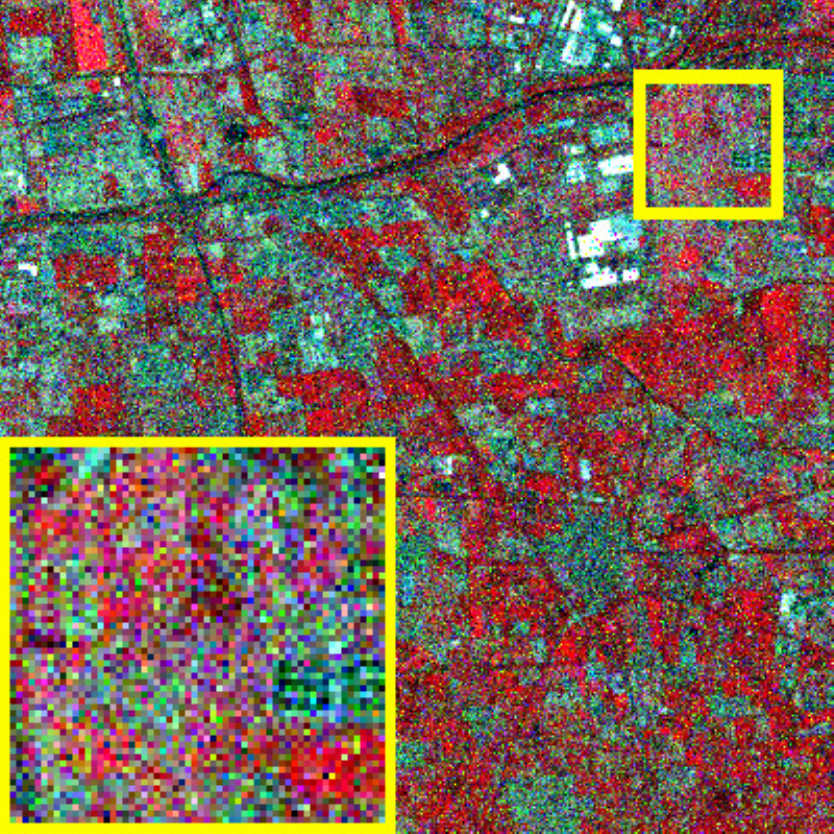}} 
        \end{minipage}
        \begin{minipage}{0.1\hsize}
            \centerline{\includegraphics[width=\hsize]{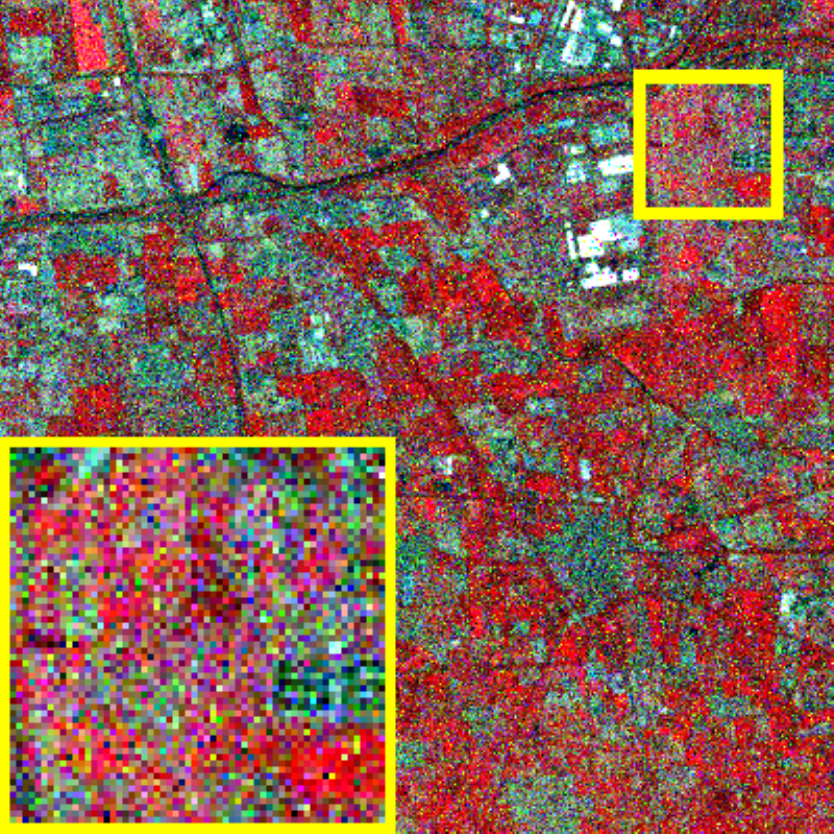}} 
        \end{minipage}
        \begin{minipage}{0.1\hsize}
            \centerline{\includegraphics[width=\hsize]{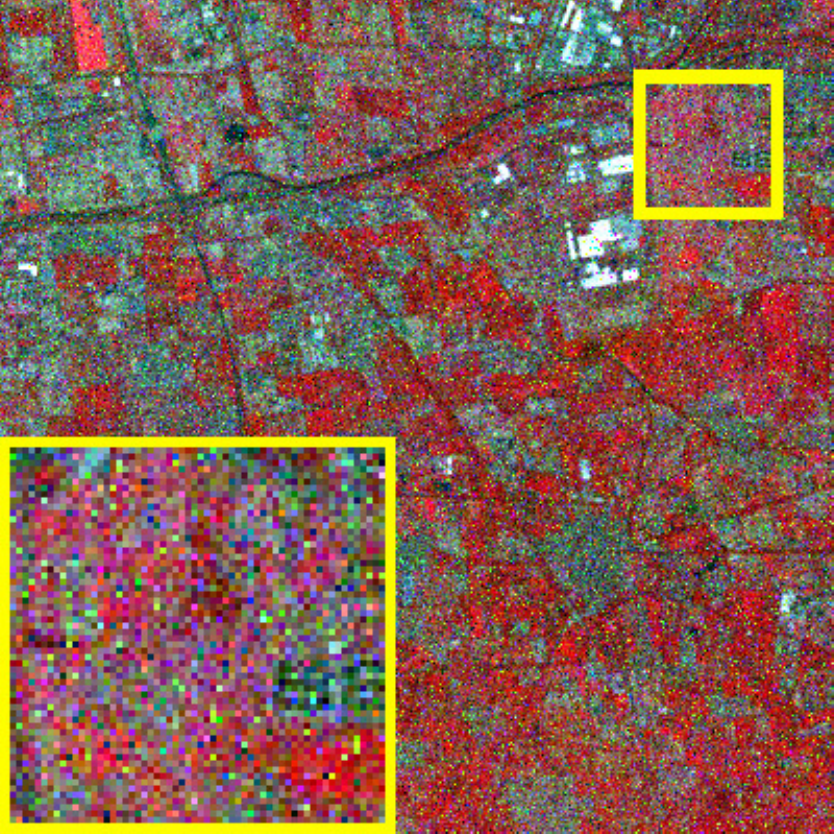}} 
        \end{minipage}
        \begin{minipage}{0.1\hsize}
            \centerline{\includegraphics[width=\hsize]{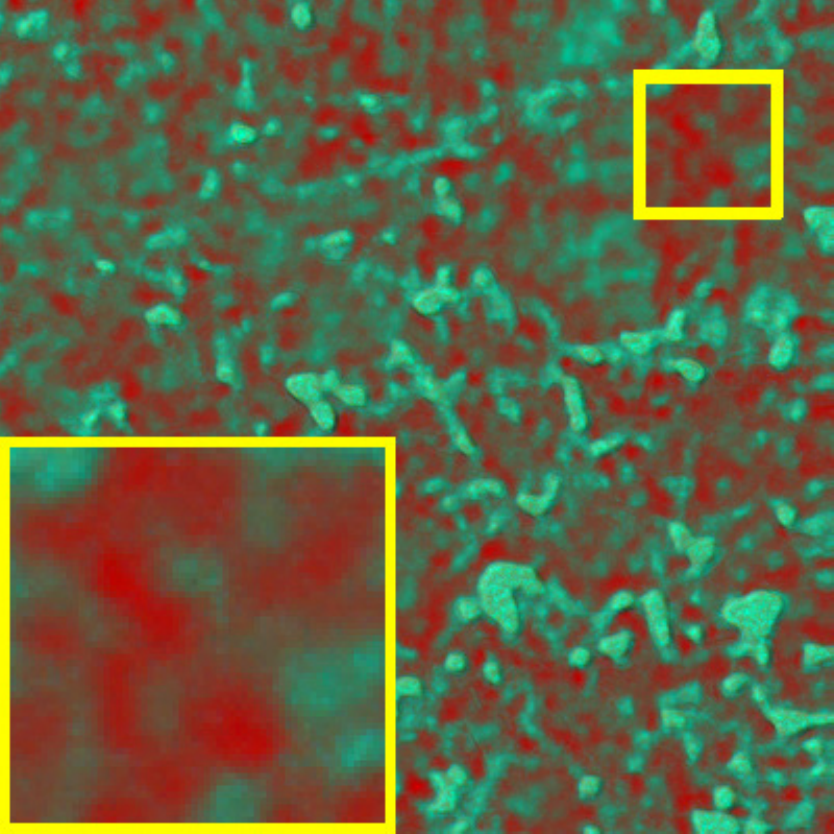}} 
        \end{minipage}
        \begin{minipage}{0.1\hsize}
            \centerline{\includegraphics[width=\hsize]{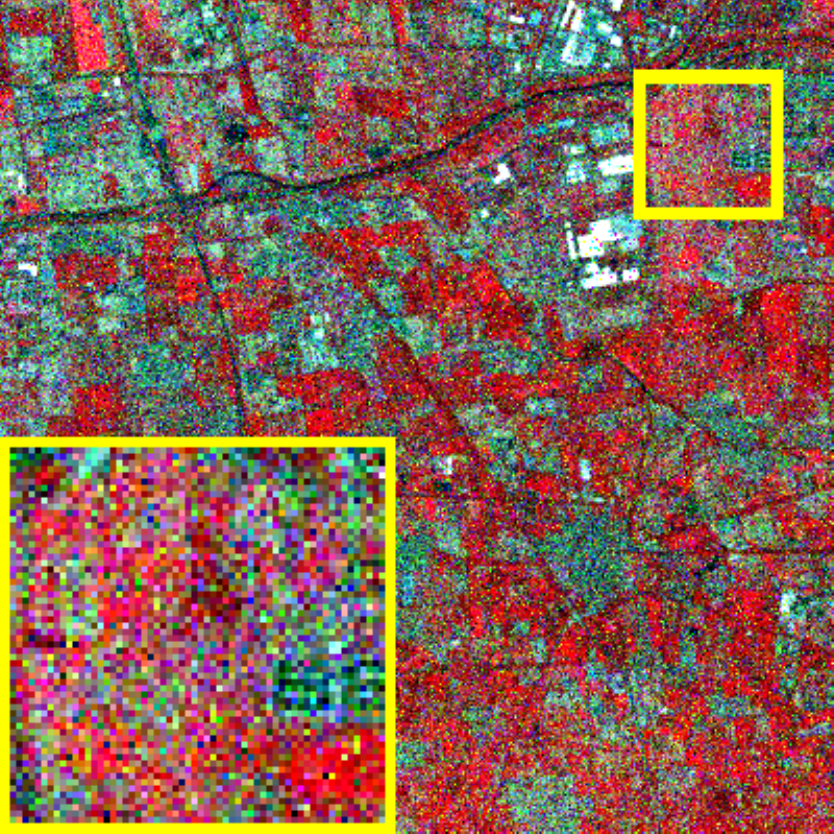}} 
        \end{minipage}
        \begin{minipage}{0.1\hsize}
            \centerline{\includegraphics[width=\hsize]{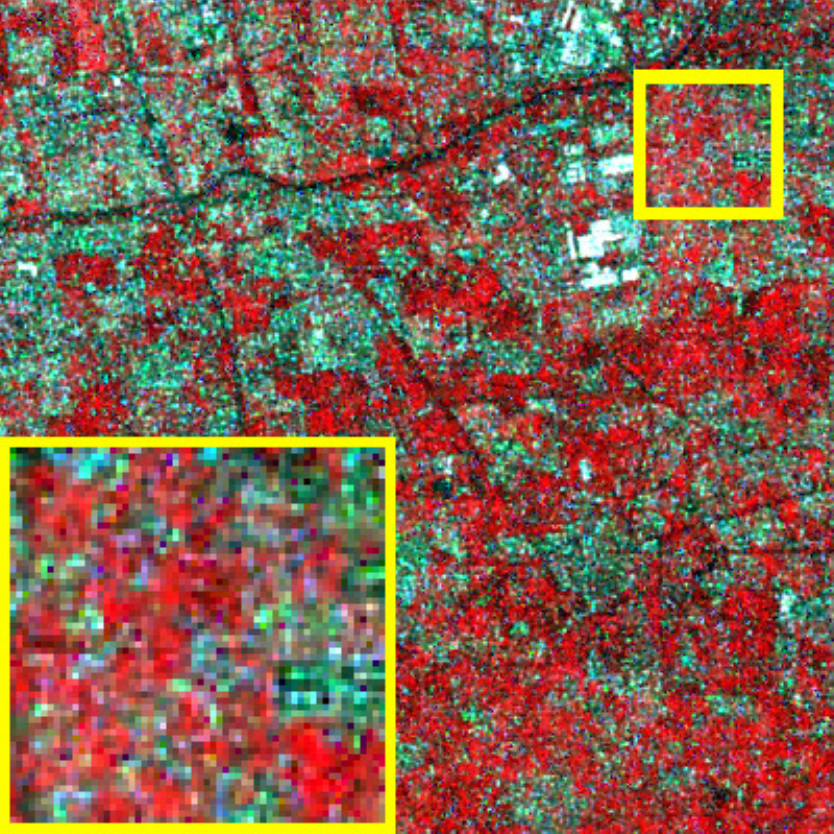}} 
        \end{minipage}
        \begin{minipage}{0.1\hsize}
            \centerline{\includegraphics[width=\hsize]{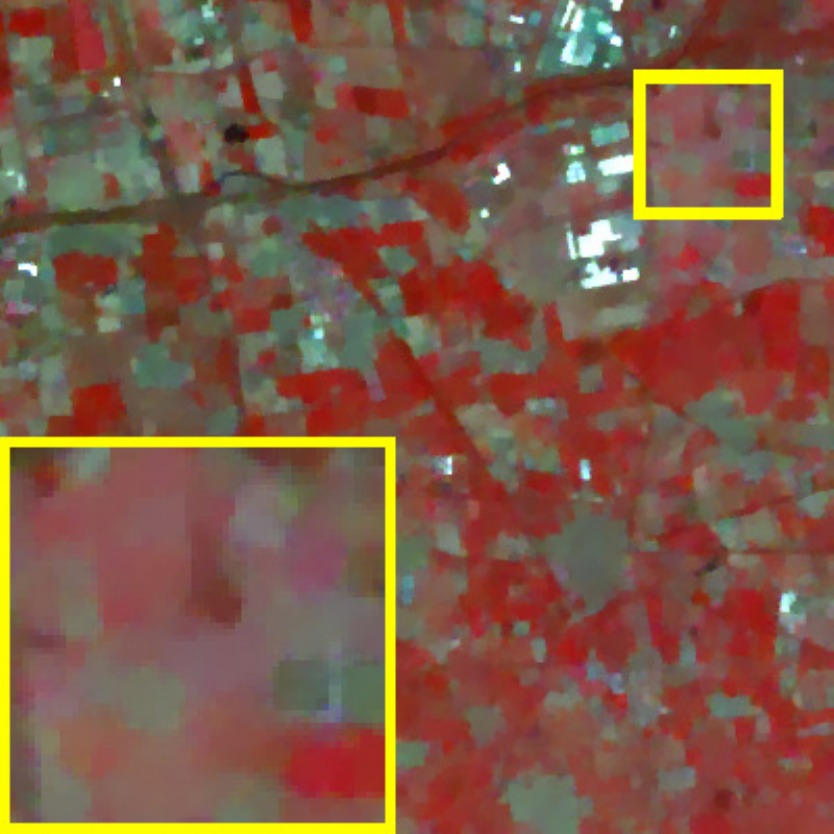}} 
        \end{minipage}
        \begin{minipage}{0.1\hsize}
            \centerline{\includegraphics[width=\hsize]{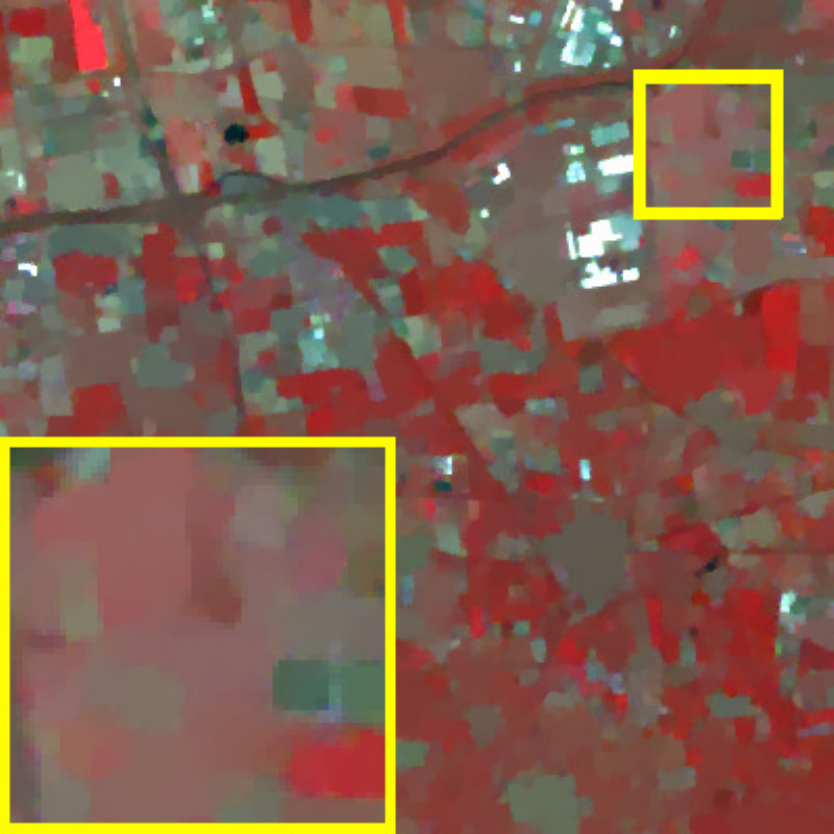}} 
        \end{minipage}
        \begin{minipage}{0.1\hsize}
            \centerline{\includegraphics[width=\hsize]{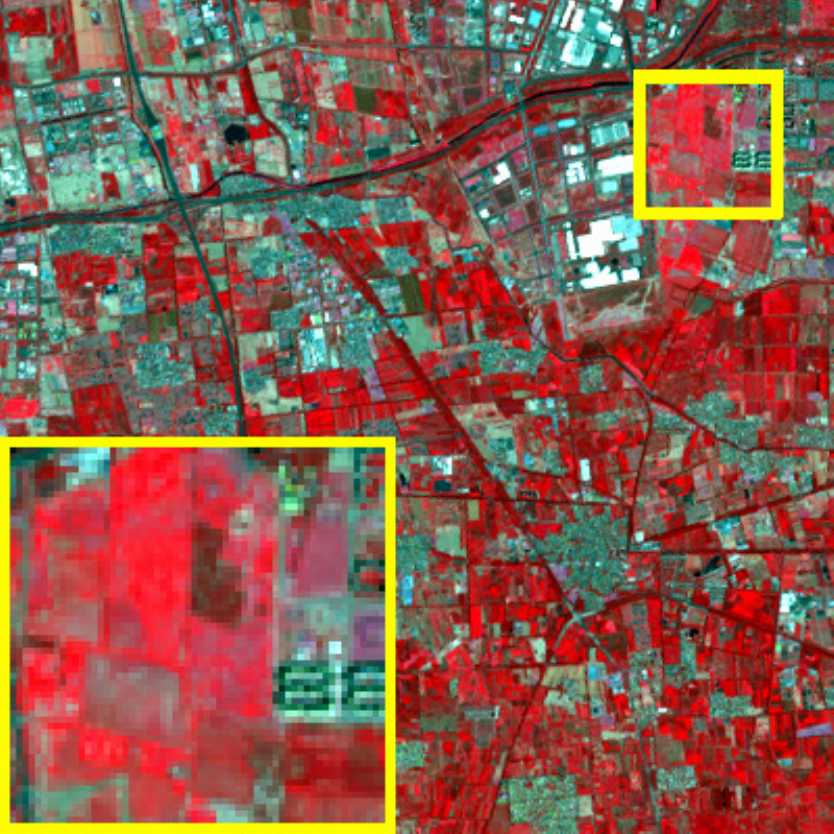}} 
        \end{minipage}  \\
        \vspace{1mm}
        \begin{minipage}{0.01\hsize}
            \centerline{}
        \end{minipage}
        \begin{minipage}{0.1\hsize} 
            \centerline{$\Hr$}
        \end{minipage}
        \begin{minipage}{0.1\hsize} 
            \centerline{STARFM}
        \end{minipage}
        \begin{minipage}{0.1\hsize} 
            \centerline{VIPSTF}
        \end{minipage}
        \begin{minipage}{0.1\hsize} 
            \centerline{RSFN}
        \end{minipage}
        \begin{minipage}{0.1\hsize} 
            \centerline{RobOt}
        \end{minipage}
        \begin{minipage}{0.1\hsize} 
            \centerline{SwinSTFM}
        \end{minipage}
        \begin{minipage}{0.1\hsize} 
            \centerline{\textbf{ROSTF-1}}
        \end{minipage}
        \begin{minipage}{0.1\hsize} 
            \centerline{\textbf{ROSTF-2}}
        \end{minipage}
        \begin{minipage}{0.1\hsize} 
            \centerline{Ground-truth}
        \end{minipage}\\
	\end{center}
        \vspace{-3mm}
	\caption{ST fusion results for the noisy Site1 simulated data. The top, middle, and bottom rows represent the results in Case2, Case3, and Case4, respectively.}
        \label{fig: Site1 Real Case234 results}
\end{figure*}

%% file: Experiments/results/Site1_Real_Case234_dif_crop.tex
\begin{figure*}[ht]
	\begin{center}
        \begin{minipage}{0.01\hsize}
			\centerline{\rotatebox{90}{Case2}}
		\end{minipage}
        \begin{minipage}{0.11\hsize} 
		  \centerline{\includegraphics[width=\hsize]{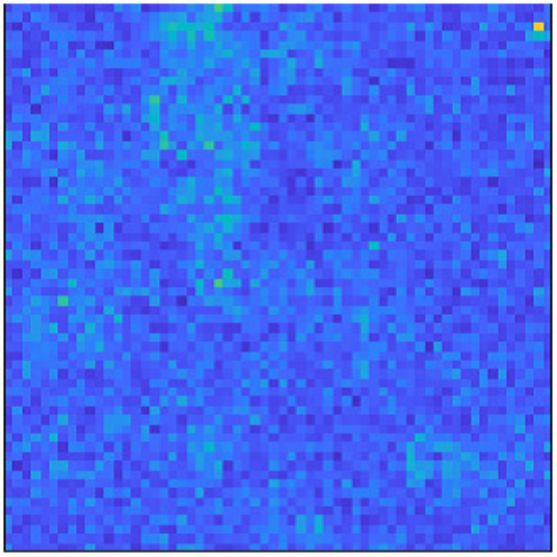}} 
	\end{minipage}
        \begin{minipage}{0.11\hsize} 
		  \centerline{\includegraphics[width=\hsize]{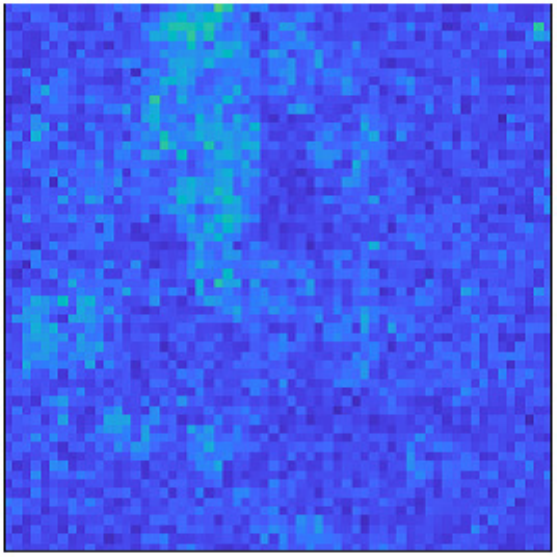}} 
	\end{minipage}
        \begin{minipage}{0.11\hsize} 
		  \centerline{\includegraphics[width=\hsize]{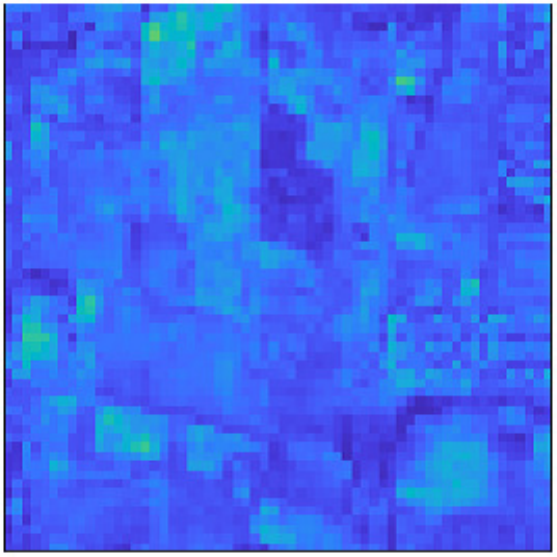}} 
	\end{minipage}
        \begin{minipage}{0.11\hsize}
            \centerline{\includegraphics[width=\hsize]{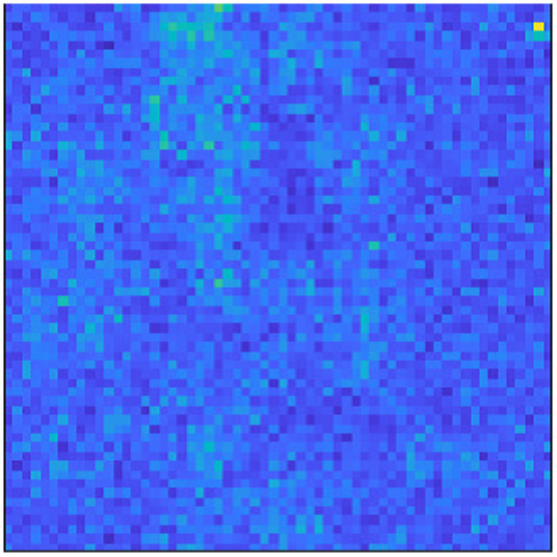}} 
        \end{minipage}
        \begin{minipage}{0.11\hsize}
            \centerline{\includegraphics[width=\hsize]{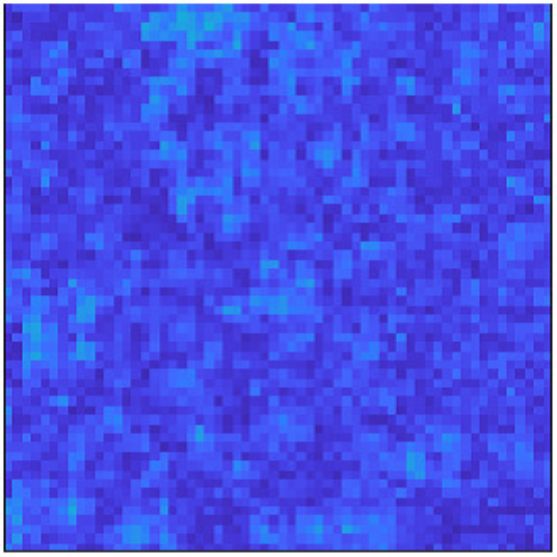}} 
        \end{minipage}
        \begin{minipage}{0.11\hsize}
			\centerline{\includegraphics[width=\hsize]{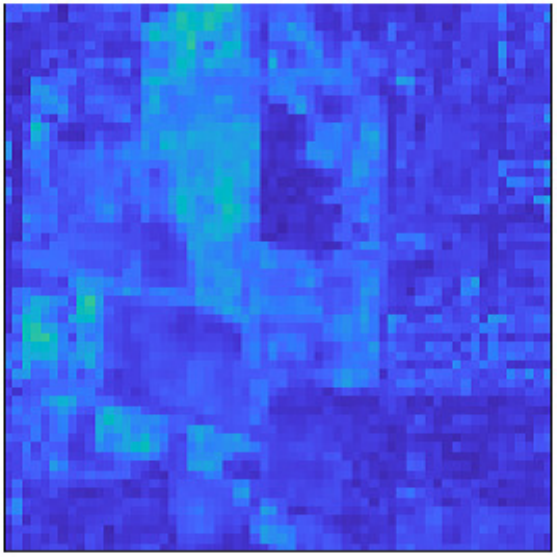}} 
		\end{minipage}
        \begin{minipage}{0.11\hsize}
			\centerline{\includegraphics[width=\hsize]{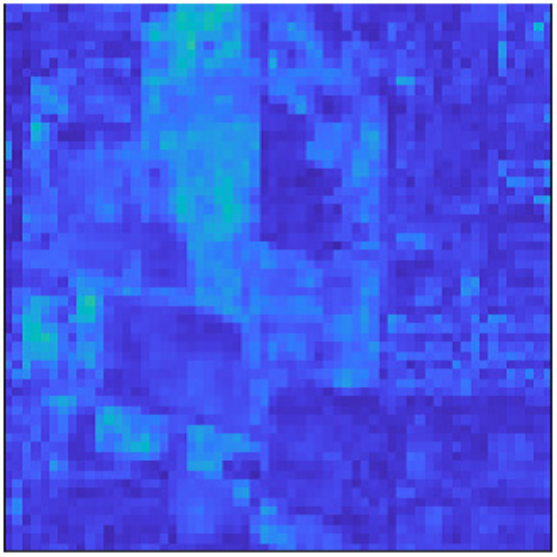}} 
	\end{minipage}
        \begin{minipage}{0.05\hsize}
			\centerline{\includegraphics[height=55pt]{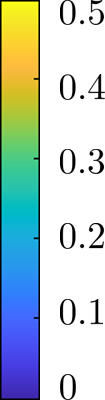}} 
	\end{minipage}\\

        \vspace{2mm}
        
        \begin{minipage}{0.01\hsize}
			\centerline{\rotatebox{90}{Case3}}
		\end{minipage}
        \begin{minipage}{0.11\hsize} 
		  \centerline{\includegraphics[width=\hsize]{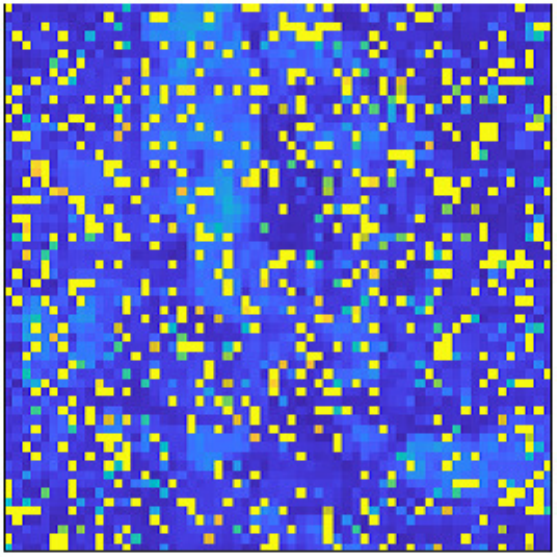}} 
	\end{minipage}
        \begin{minipage}{0.11\hsize} 
		  \centerline{\includegraphics[width=\hsize]{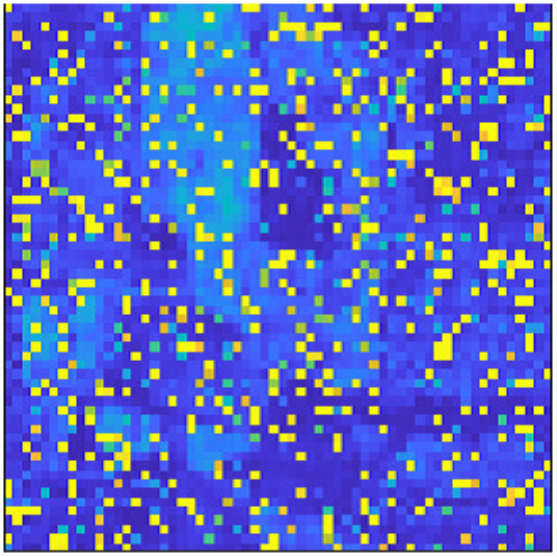}} 
	\end{minipage}
        \begin{minipage}{0.11\hsize} 
		  \centerline{\includegraphics[width=\hsize]{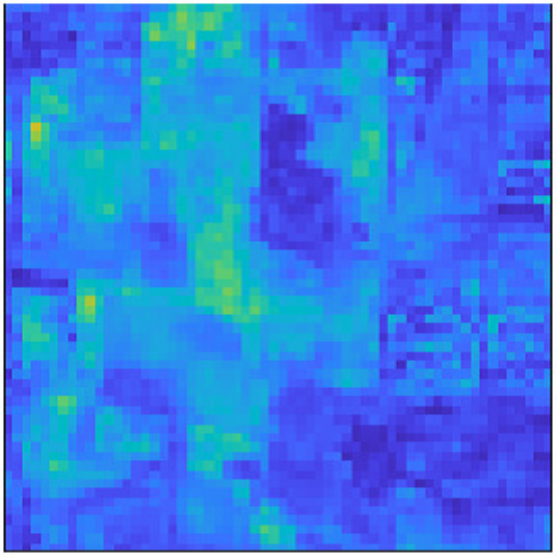}} 
	\end{minipage}
        \begin{minipage}{0.11\hsize}
            \centerline{\includegraphics[width=\hsize]{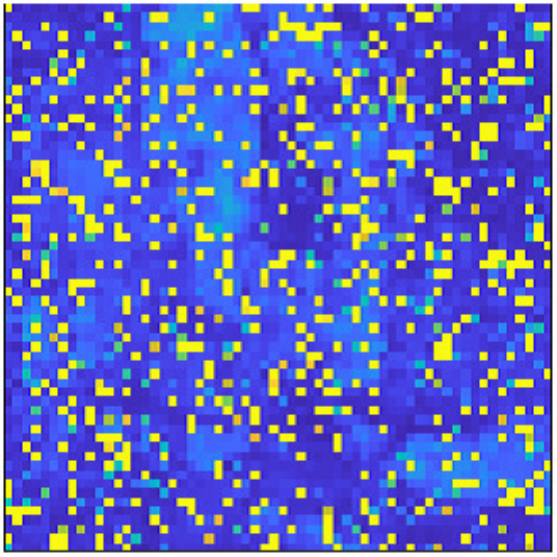}} 
        \end{minipage}
        \begin{minipage}{0.11\hsize}
            \centerline{\includegraphics[width=\hsize]{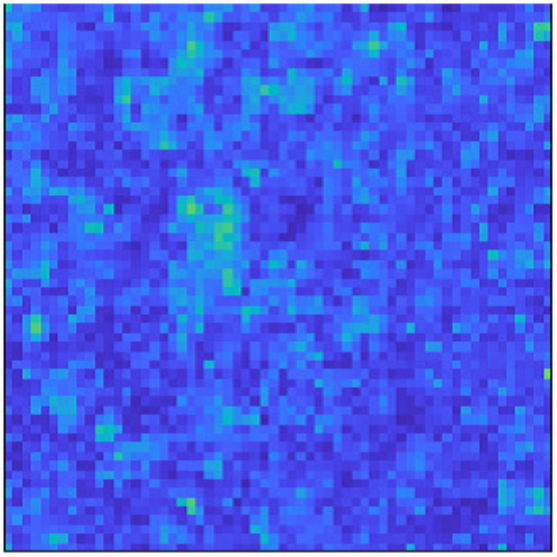}} 
        \end{minipage}
        \begin{minipage}{0.11\hsize}
			\centerline{\includegraphics[width=\hsize]{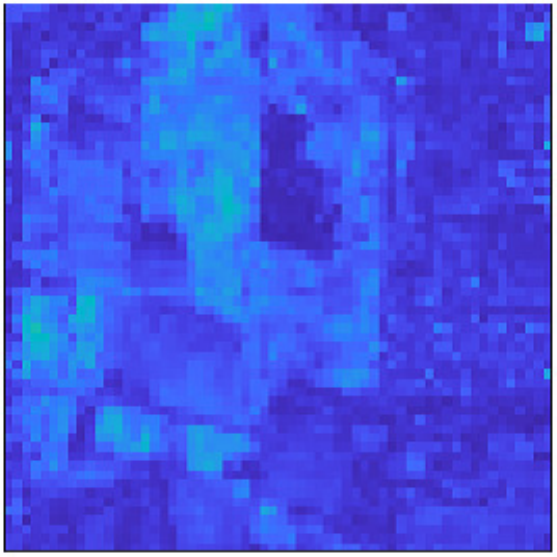}} 
		\end{minipage}
        \begin{minipage}{0.11\hsize}
			\centerline{\includegraphics[width=\hsize]{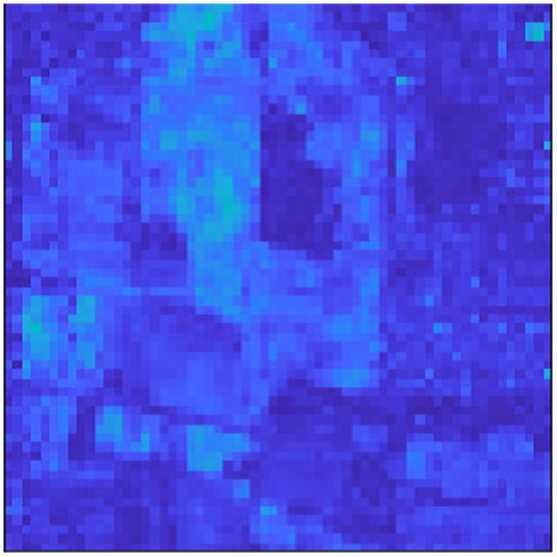}} 
	\end{minipage} 
        \begin{minipage}{0.05\hsize}
			\centerline{\includegraphics[height=55pt]{img/colorbar2.png}} 
	\end{minipage}\\

        \vspace{2mm}
  
        \begin{minipage}{0.01\hsize}
			\centerline{\rotatebox{90}{Case4}}
		\end{minipage}
        \begin{minipage}{0.11\hsize} 
		  \centerline{\includegraphics[width=\hsize]{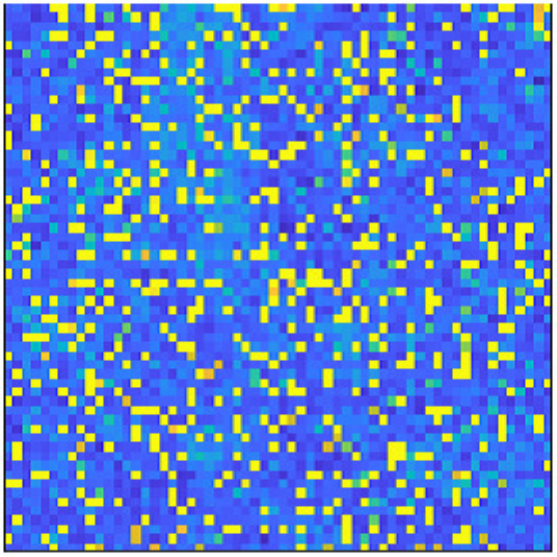}} 
	\end{minipage}
        \begin{minipage}{0.11\hsize} 
		  \centerline{\includegraphics[width=\hsize]{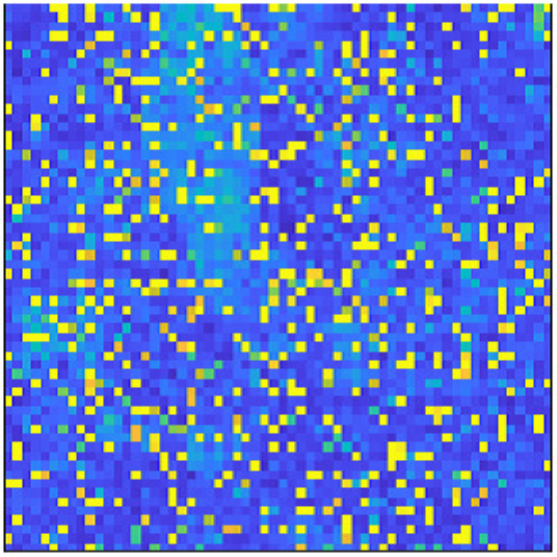}} 
	\end{minipage}
        \begin{minipage}{0.11\hsize} 
		  \centerline{\includegraphics[width=\hsize]{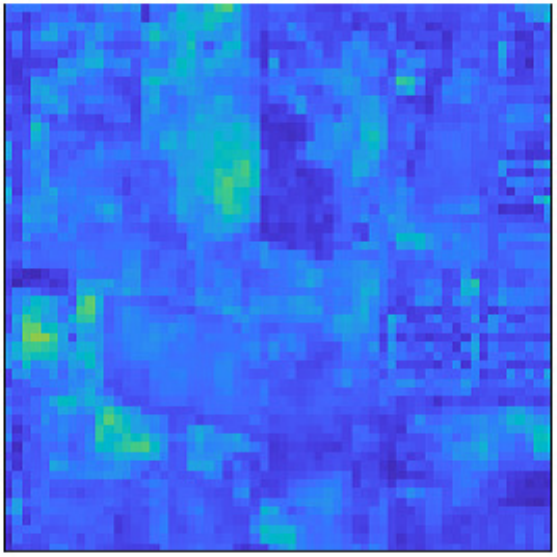}} 
	\end{minipage}
        \begin{minipage}{0.11\hsize}
            \centerline{\includegraphics[width=\hsize]{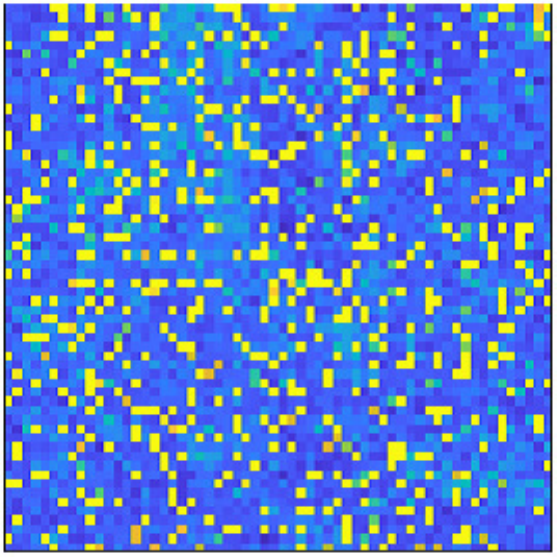}} 
        \end{minipage}
        \begin{minipage}{0.11\hsize}
            \centerline{\includegraphics[width=\hsize]{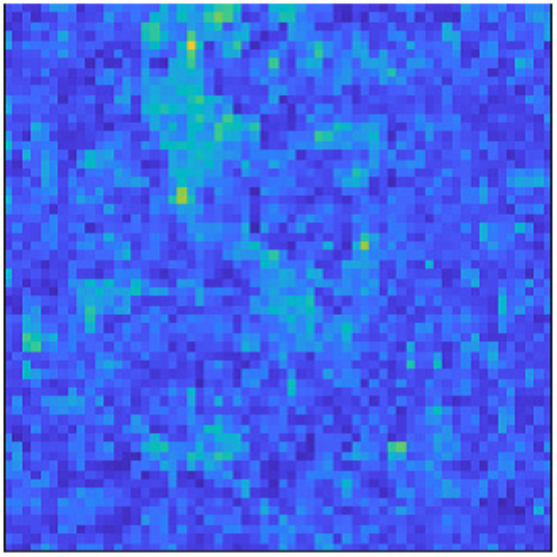}} 
        \end{minipage}
        \begin{minipage}{0.11\hsize}
			\centerline{\includegraphics[width=\hsize]{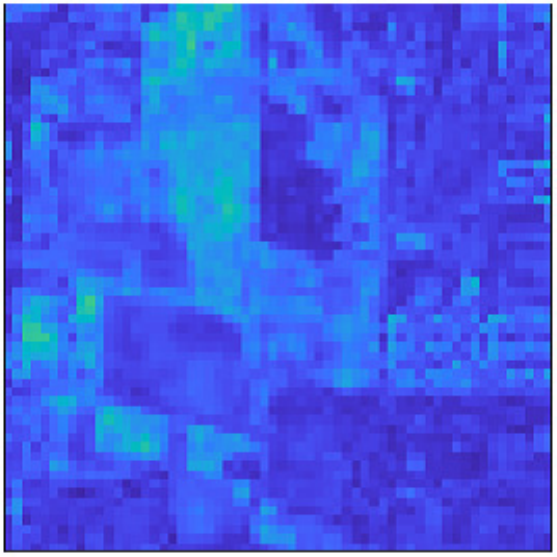}} 
		\end{minipage}
        \begin{minipage}{0.11\hsize}
			\centerline{\includegraphics[width=\hsize]{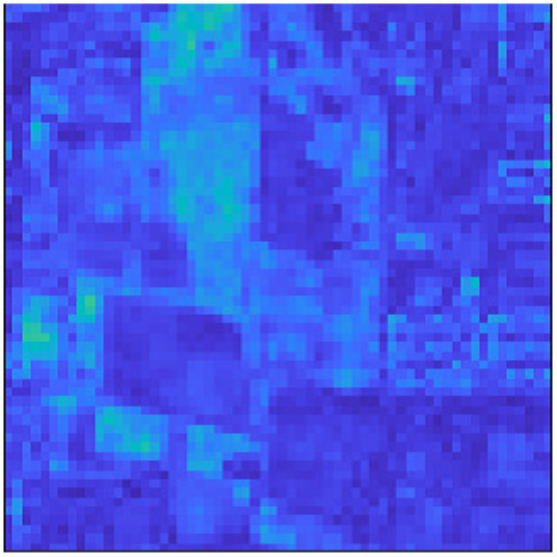}} 
	\end{minipage}
        \begin{minipage}{0.05\hsize}
			\centerline{\includegraphics[height=55pt]{img/colorbar2.png}} 
	\end{minipage}\\
        \vspace{1mm}
        \begin{minipage}{0.01\hsize}
			\centerline{}
		\end{minipage}
        \begin{minipage}{0.11\hsize} 
			\centerline{STARFM}
		\end{minipage}
        \begin{minipage}{0.11\hsize} 
			\centerline{VIPSTF}
		\end{minipage}
        \begin{minipage}{0.11\hsize} 
			\centerline{RSFN}
		\end{minipage}
        \begin{minipage}{0.11\hsize} 
			\centerline{RobOt}
        \end{minipage}
        \begin{minipage}{0.11\hsize} 
			\centerline{SwinSTFM}
		\end{minipage}
        \begin{minipage}{0.11\hsize} 
			\centerline{\textbf{ROSTF-1}}
		\end{minipage}
        \begin{minipage}{0.11\hsize} 
			\centerline{\textbf{ROSTF-2}}
	\end{minipage}
        \begin{minipage}{0.05\hsize} 
			\centerline{~}
	\end{minipage}\\
	\end{center}
        \vspace{-3mm}
	\caption{The difference map (absolute errors) of the fusion results in the zoomed-in area in the noisy Site1 real data. The top, middle, and bottom rows represent the results in Case2, Case3, and Case4, respectively.}
        \label{fig: difference map}
\end{figure*}

%% file: Experiments/AblationStudy/average_convergence_iterations_and_performance_metrics.tex
\begin{table}[t]
	\begin{center}
		\caption{The Average Number of Iterations Spent Before Each Algorithm Stopped and the Average Performance Results for All the Situations}
        \label{table: convergence iterations}
            \vspace{-2mm}
		\centering
		\begin{tabular}{p{0.9cm} >{\centering\arraybackslash}p{2.0cm} >{\centering\arraybackslash}p{2.0cm} >{\centering\arraybackslash}p{2.0cm}}
			\toprule
                & \multicolumn{2}{c}{ROSTF Without} & \\
                \cmidrule(lr){2-3}
                & The Edge Constraint & The Brightness Constraint & The Original\par ROSTF \\
                \midrule 
                iter 
                & 47788 & 28217 & \textbf{27208} \\
                \cmidrule(lr){1-4}
                RMSE 
                & 0.0578 & 0.0581 & \textbf{0.0381}  \\ 
                SAM 
                & 0.2071 & 0.1511 & \textbf{0.1195}  \\ 
                SSIM 
                & 0.8393 & 0.8780 & \textbf{0.9151}  \\ 
                CC 
                & 0.8484 & 0.9314 & \textbf{0.9341}  \\ 
                \bottomrule 
		\end{tabular}
	\end{center}
\end{table}

%% file: Experiments/AblationStudy/compare_convergence_figure.tex
\begin{figure}[t]
	\begin{center}
        \begin{minipage}{0.45\hsize} 	  
            \centerline{\includegraphics[width=\hsize]{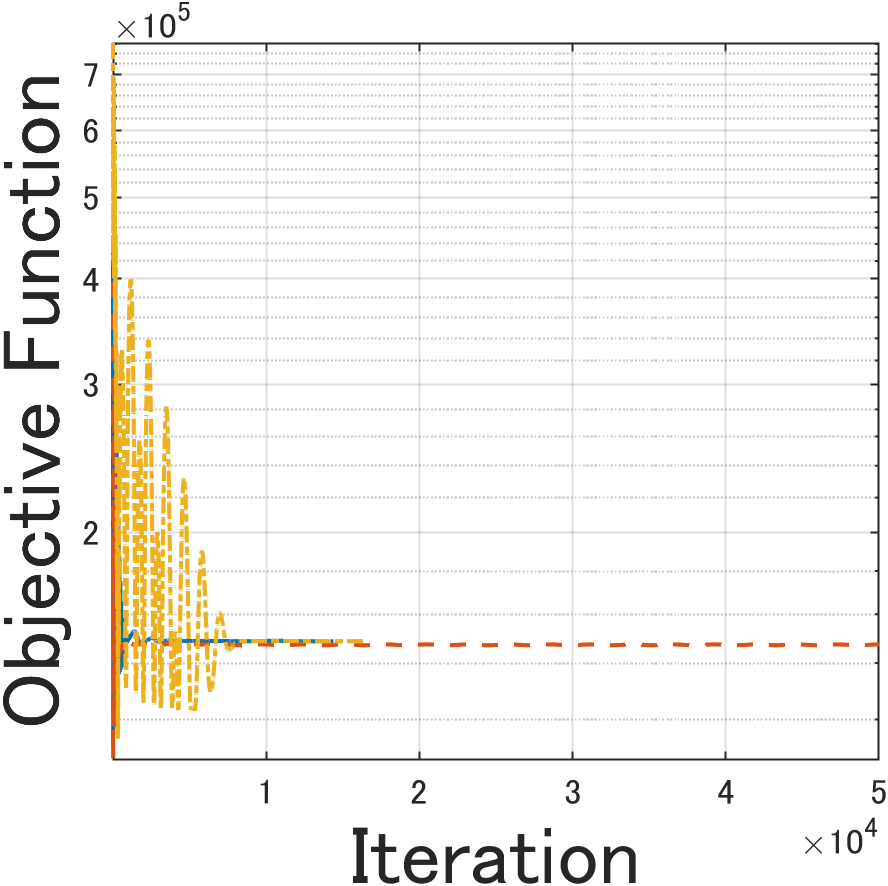}} 
	\end{minipage}
        \begin{minipage}{0.45\hsize}
            \centerline{\includegraphics[width=\hsize]{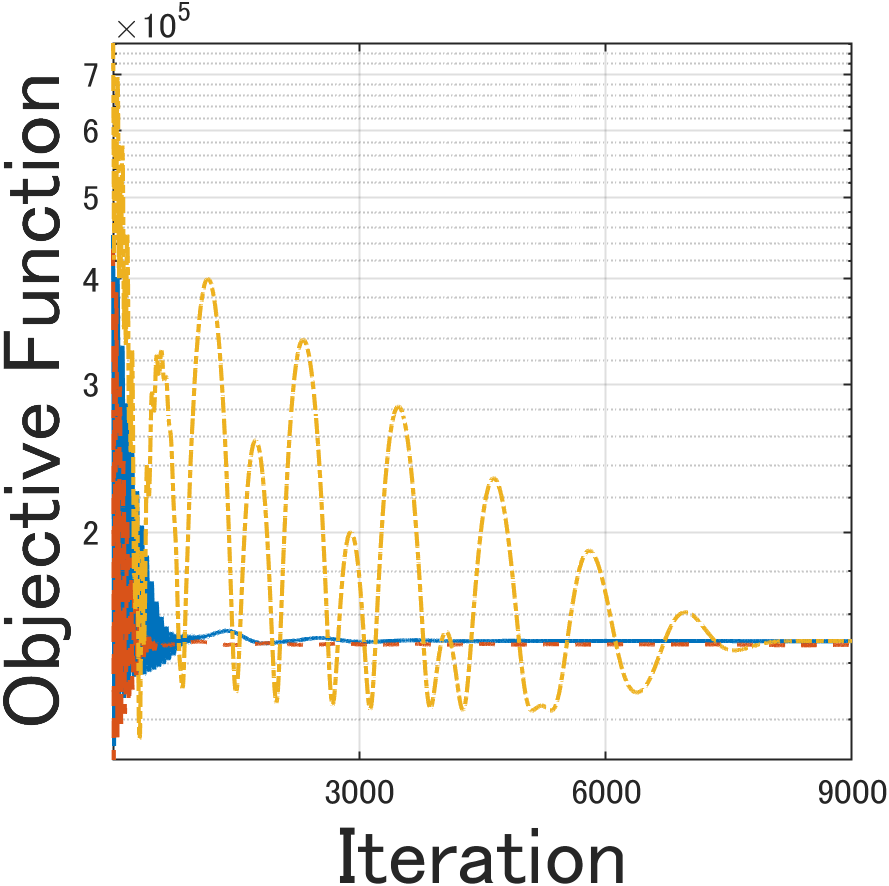}} 
        \end{minipage} \\ \vspace{1mm}
        \begin{minipage}{0.45\hsize} 	  
            \centerline{(a)} 
	\end{minipage} 
        \begin{minipage}{0.45\hsize} 	  
            \centerline{(b)} 
	\end{minipage} \\ \vspace{3mm}
        \begin{minipage}{0.45\hsize}
            \centerline{\includegraphics[width=\hsize]{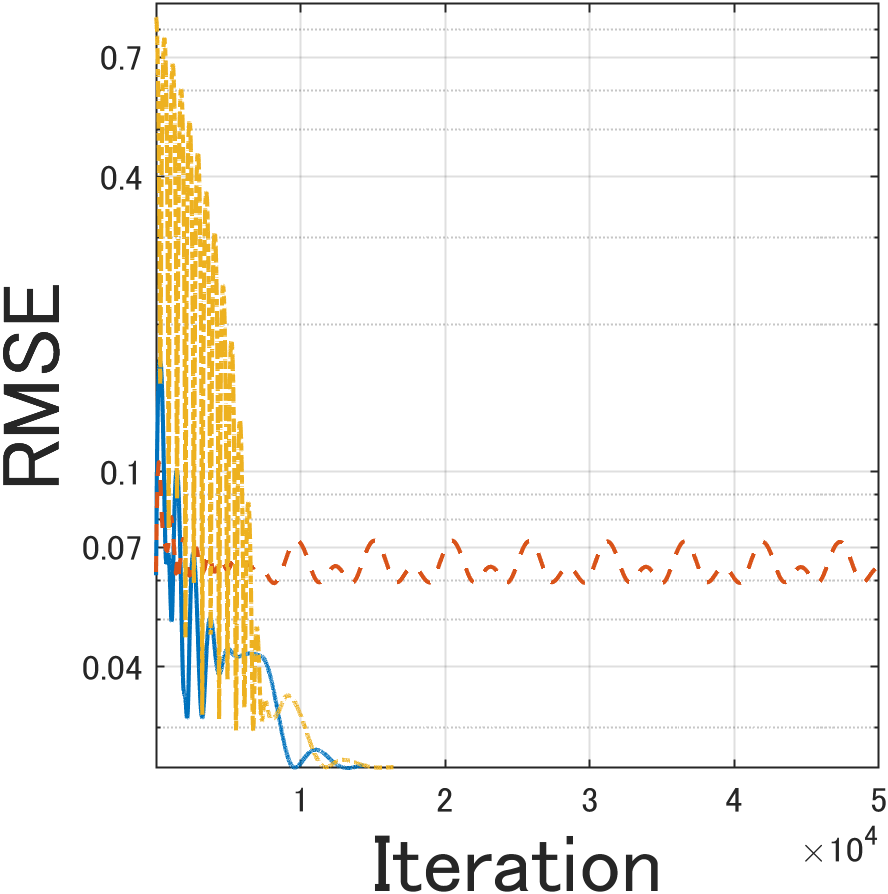}} 
        \end{minipage}
        \begin{minipage}{0.45\hsize}
		\centerline{\includegraphics[width=\hsize]{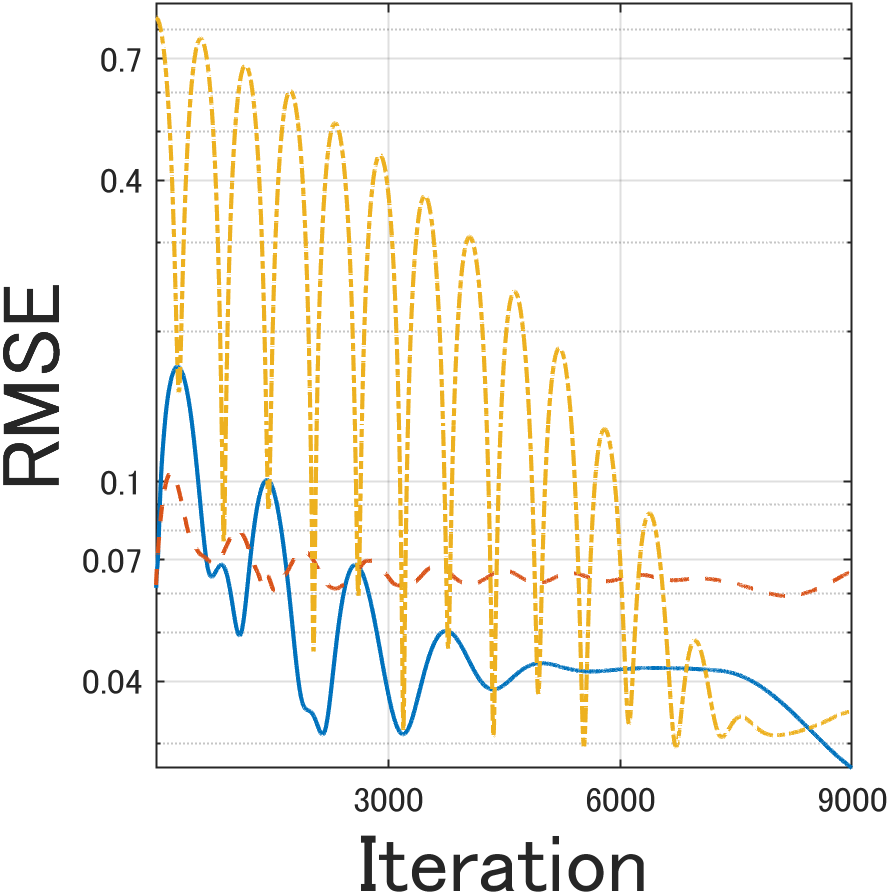}} 
	\end{minipage} 
        \\ \vspace{1mm}
        \begin{minipage}{0.45\hsize} 	  
            \centerline{(c)} 
	\end{minipage}
        \begin{minipage}{0.45\hsize} 	  
            \centerline{(d)} 
	\end{minipage} 
        \\
        \vspace{2mm}
        \begin{minipage}{0.93\hsize} 	  
            \centerline{\includegraphics[width=0.77\hsize]{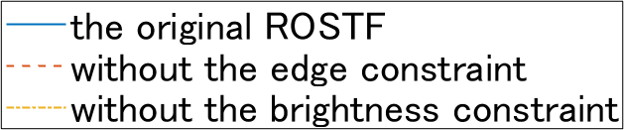}}
	\end{minipage}\\
	\end{center}
	\caption{The behavior of the original ROSTF-2, ROSTF-2 without the edge constraint, and ROSTF-2 without the brightness constraint in the experiment on the Site1 simulated data in Case1. (a), (b) The transition of the objective function values until the algorithms stopped and in early iterations, respectively. (c), (d) The transition of the RMSE values until the algorithms stopped and in early iterations, respectively.}
        \label{fig: convergence results}
\end{figure}

%% file: Experiments/AblationStudy/without_12constraint_figure.tex
\begin{figure}[ht]
	\begin{center}
        \begin{minipage}{0.24\hsize} 	  
            \centerline{\includegraphics[width=\hsize]{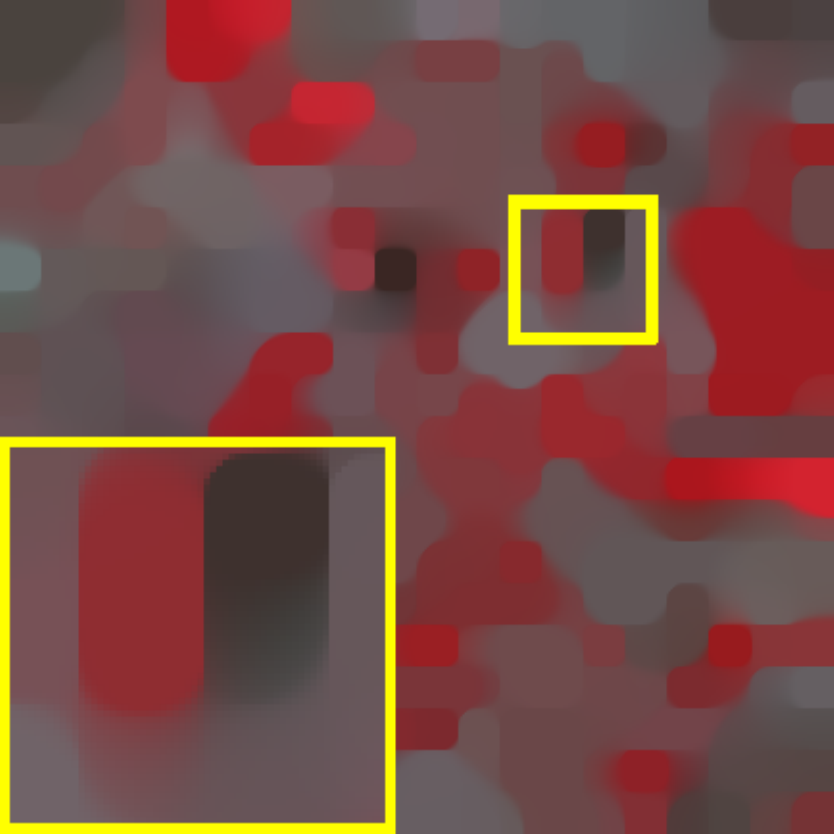}} 
	\end{minipage}
        \begin{minipage}{0.24\hsize}
            \centerline{\includegraphics[width=\hsize]{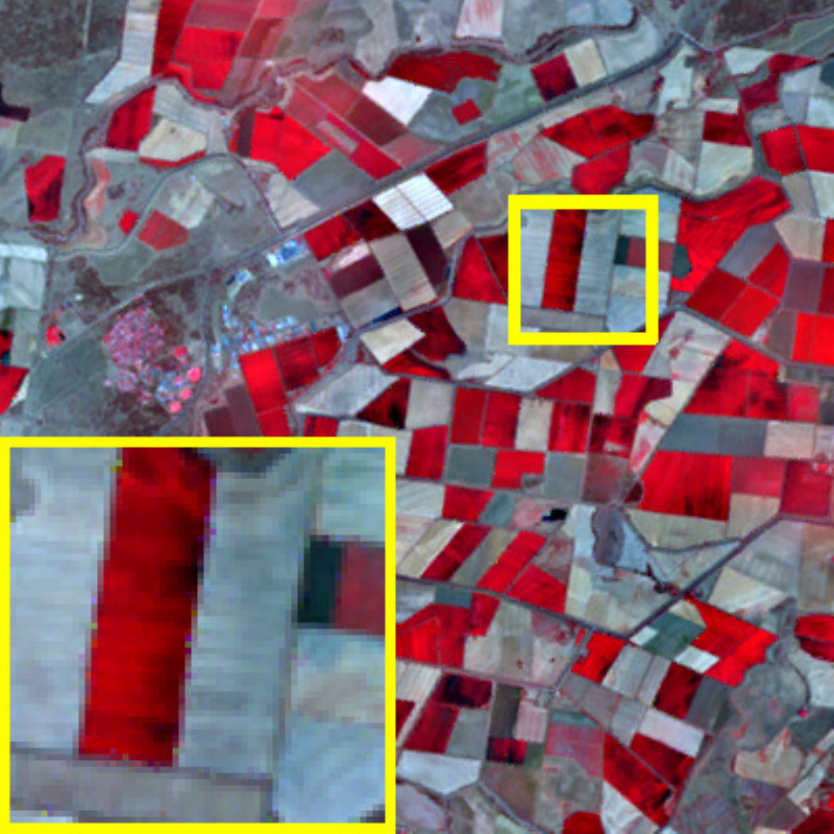}} 
        \end{minipage} 
        \begin{minipage}{0.24\hsize}
            \centerline{\includegraphics[width=\hsize]{img/results/AustraliaDataset/SemiSim/Case3/proposed2-eps-converted-to.pdf}} 
        \end{minipage}
        \begin{minipage}{0.24\hsize}
		  \centerline{\includegraphics[width=\hsize]{img/results/AustraliaDataset/SemiSim/Case3/GT-eps-converted-to.pdf}} 
	\end{minipage} \\
        \vspace{1mm}
        \begin{minipage}{0.24\hsize} 	  
            \centerline{(a)} 
	\end{minipage}
        \begin{minipage}{0.24\hsize} 	  
            \centerline{(b)} 
	\end{minipage} 
        \begin{minipage}{0.24\hsize} 	  
            \centerline{(c)} 
	\end{minipage}
        \begin{minipage}{0.24\hsize} 	  
            \centerline{(d)} 
	\end{minipage} \\
	\end{center}
        \vspace{-3mm}
	\caption{ST fusion results of (a)~ROSTF-2 without Edge Constraint, (b)~ROSTF-2 without Brightness Constraint, and (c)~the original ROSTF-2 for the Site2 simulated data in Case3. (d)~Ground-truth.}
        \label{fig: without 12const results}
\end{figure}

%% file: Experiments/AblationStudy/without_denoising_metrics.tex
\begin{table}[t]
	\begin{center}
		\caption{The Average Performance Results for Each Noise Case in the Ablation Study of the Denoising Mechanism}
            \label{table: without denoising}
            \vspace{-2mm}
		\centering
		\begin{tabular}{p{0.9cm} p{0.9cm} >{\centering\arraybackslash}p{1.7cm} >{\centering\arraybackslash}p{1.7cm}}
			\toprule
                \multirow{2}{*}{Noise} & \multirow{2}{*}{Metrics} & Without denoising & \multirow{2}{*}{ROSTF} \\
                \midrule 
                \multirow{4}{*}{Case1} 
                & RMSE 
                & \textbf{0.0355} & 0.0363  \\ 
                & SAM 
                & \textbf{0.1049} & 0.1101  \\ 
                & SSIM 
                & \textbf{0.9244} & 0.9212  \\ 
                & CC 
                & \textbf{0.9404} & 0.9384  \\ 
                \cmidrule{1-4}
                \multirow{4}{*}{Case2} 
                & RMSE 
                & 0.0614 & \textbf{0.0382}  \\ 
                & SAM 
                & 0.2702 & \textbf{0.1218}  \\ 
                & SSIM 
                & 0.7772 & \textbf{0.9154}  \\ 
                & CC 
                & 0.8596 & \textbf{0.9356}  \\ 
                \cmidrule{1-4}
                \multirow{4}{*}{Case3} 
                & RMSE 
                & 0.1391 & \textbf{0.0370}  \\ 
                & SAM 
                & 0.2891 & \textbf{0.1137}  \\ 
                & SSIM 
                & 0.4597 & \textbf{0.9192}  \\ 
                & CC 
                & 0.5976 & \textbf{0.9364}  \\ 
                \cmidrule{1-4}
                \multirow{4}{*}{Case4} 
                & RMSE 
                & 0.1471 & \textbf{0.0408}  \\ 
                & SAM 
                & 0.4140 & \textbf{0.1324}  \\ 
                & SSIM 
                & 0.4238 & \textbf{0.9046}  \\ 
                & CC 
                & 0.5767 & \textbf{0.9259}  \\ 
                \bottomrule 
		\end{tabular}
	\end{center}
\end{table}

%% file: Experiments/AblationStudy/without_denoising_figure.tex
\begin{figure}[t]
	\begin{center}
        \begin{minipage}{0.24\hsize} 	  
            \centerline{\includegraphics[width=\hsize]{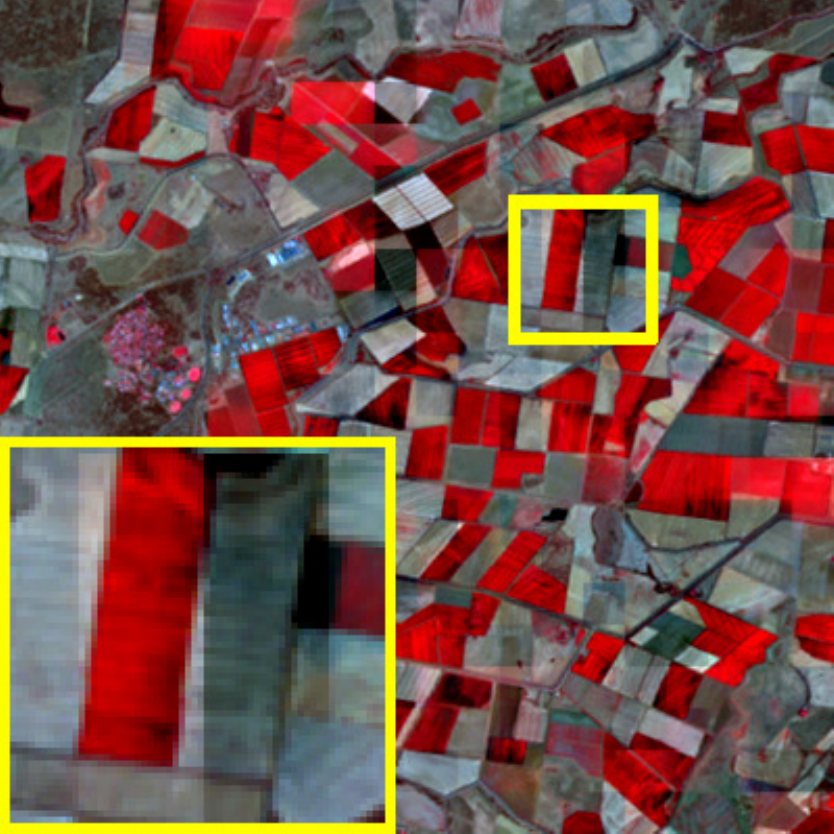}} 
	\end{minipage}
        \begin{minipage}{0.24\hsize}
            \centerline{\includegraphics[width=\hsize]{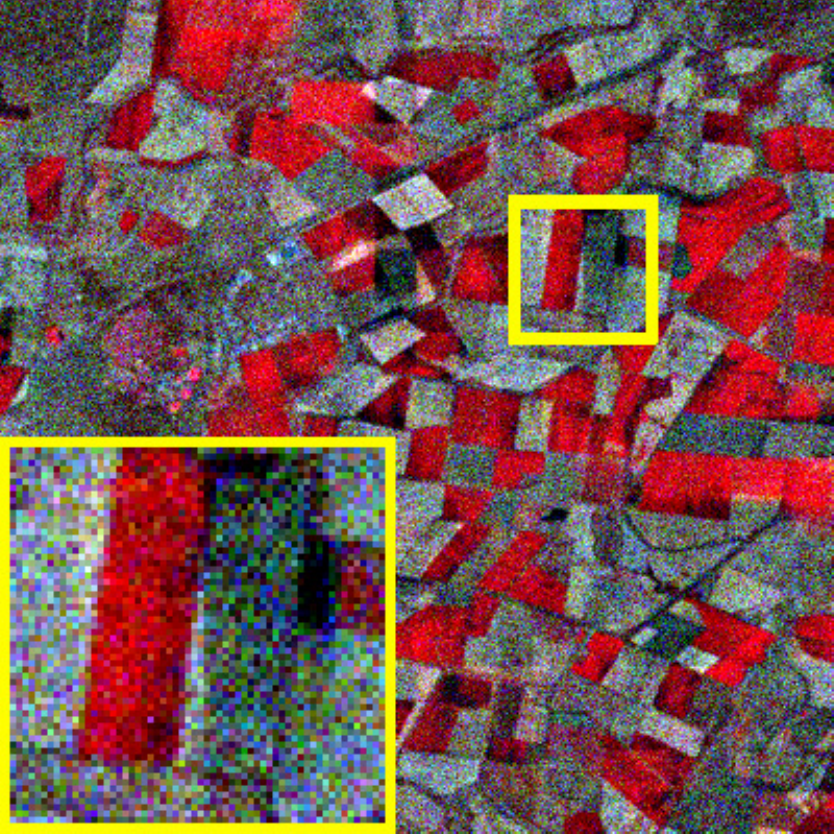}} 
        \end{minipage} 
        \begin{minipage}{0.24\hsize}
            \centerline{\includegraphics[width=\hsize]{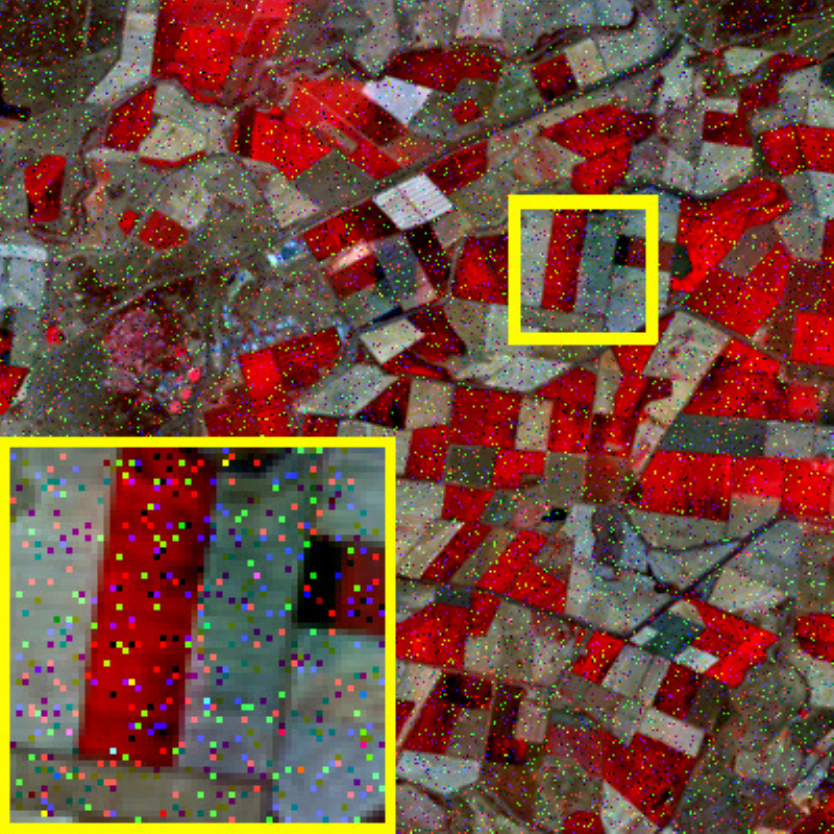}} 
        \end{minipage}
        \begin{minipage}{0.24\hsize}
		\centerline{\includegraphics[width=\hsize]{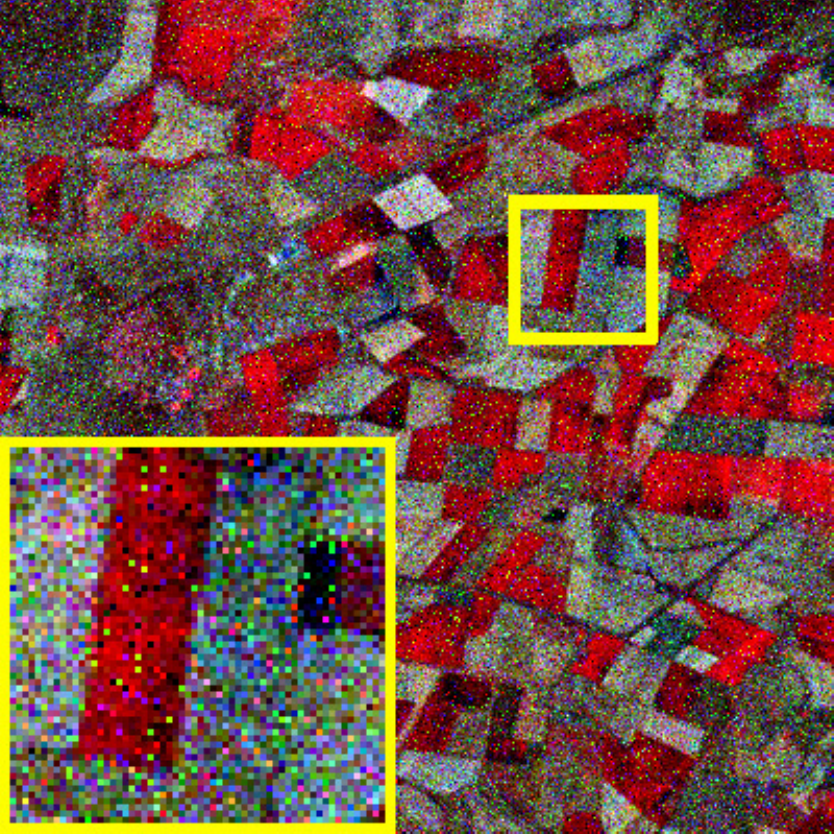}} 
	\end{minipage} \\
        \vspace{1mm}
        \begin{minipage}{0.24\hsize} 	  
            \centerline{Case1} 
	\end{minipage}
        \begin{minipage}{0.24\hsize} 	  
            \centerline{Case2} 
	\end{minipage} 
        \begin{minipage}{0.24\hsize} 	  
            \centerline{Case3} 
	\end{minipage}
        \begin{minipage}{0.24\hsize} 	  
            \centerline{Case4} 
	\end{minipage} \\
	\end{center}
        \vspace{-3mm}
	\caption{ST fusion results of ROSTF-1 without the denoising mechanism for the Site2 simulated data.}
        \label{fig: without denoising}
\end{figure}

%% file: Experiments/results/running_time.tex
\begin{table}[t]
	\begin{center}
		\caption{The Average Running Time~[s] Spent Before Each Algorithm Stopped}
        \label{table: running time}
            \vspace{-2mm}
		\centering
		\begin{tabular}{p{1.5cm} >{\centering\arraybackslash}p{2.0cm} >{\centering\arraybackslash}p{2.0cm}}
			\toprule
                & Site1 & Site2 \\
                \midrule 
                STARFM & $1.841\times10^{3}$ & $6.237\times10^{2}$ \\
                \cmidrule(lr){1-3}
                VIPSTF & $1.279\times10^{4}$ & $1.301\times10^{4}$ \\
                \cmidrule(lr){1-3}
                RobOt & $2.750\times10^{-1}$ & $2.740\times10^{-1}$ \\
                \cmidrule(lr){1-3}
                \multirow{2}{*}{ROSTF-1} & $4.536\times10^{2}$ & $5.743\times10^{2}$ \\
                & (26048~ite) &  (33796~ite) \\
                \cmidrule(lr){1-3}
                \multirow{2}{*}{ROSTF-2} & $2.358\times10^{2}$ & $6.003\times10^{2}$ \\
                & (13893~ite) &  (35093~ite) \\
                \bottomrule 
		\end{tabular}
	\end{center}
\end{table}

%% file: conclusion.tex
We have proposed an optimization-based ST fusion method, named ROSTF, which is robust to mixed Gaussian and sparse noise in observed satellite images. We have formulated the fusion problem as a constrained optimization problem and have developed the optimization algorithm based on P-PDS with OVDP. ROSTF was tested through experiments using both simulated and real data. The experimental results demonstrate that ROSTF achieves performance comparable to state-of-the-art ST fusion methods in noiseless cases and significantly better in noisy cases. ROSTF will have a strong impact on the field of remote sensing, including the estimation of satellite image series with high spatial and temporal resolution from observed image series taken in measurement environments with severe degradation.